\documentclass[PhD]{uclathes}
\usepackage{epsf,epsfig}
\usepackage{amsmath}
\usepackage{amssymb}
\usepackage{shigedef}

\makeatletter

\@addtoreset{equation}{section}

\newcommand{\BeginAppendix}{%
  \setcounter{section}{0}
  \renewcommand{\thesection}{\arabic{chapter}.\Alph{section}}
  \renewcommand{\theequation}{\arabic{chapter}.\Alph{section}.\arabic{equation}}
  \section*{\LARGE\bf Appendix}
}
\newcommand{\EndAppendix}{%
  \setcounter{section}{0}
  \renewcommand{\thesection}{\arabic{chapter}.\arabic{section}}
  \renewcommand{\theequation}{\arabic{chapter}.\arabic{section}.\arabic{equation}}
}

\def\unit{{1\kern-.65ex {\mathrm l}}}
\def\CF{\mathcal{F}}
\def\CM{\mathcal{M}}
\def\CN{\mathcal{N}}
\def\CO{\mathcal{O}}
\def\CR{\mathcal{R}}
\def\CW{\mathcal{W}}
\def\Dt{\widetilde{D}}
\def\Pt{\widetilde{P}}
\def\St{\widetilde{S}}
\def\Wt{\widetilde{W}}
\def\Sh{\widehat{S}}
\def\zh{\widehat{z}}
\def\chit{\widetilde{\chi}}
\def\RPtwo{{\mathbb{RP}^2}}
\def\Stwo{{S^2}}
\def\Pone{\mathbb{P}^1}
\def\Ztwo{{\mathbb{Z}_2}}
\def\Dtwo{{D_2}}
\def\so{{\therefore}}

\newcommand{\MM}[1]{{\mathbf{#1}}}
\newcommand{\mg}{\MM{g}}
\newcommand{\mf}{\MM{f}}
\newcommand{\mF}{\MM{F}}
\newcommand{\mA}{\MM{A}}
\newcommand{\mM}{\MM{M}}
\newcommand{\mN}{\MM{N}}
\newcommand{\mQ}{\MM{Q}}
\newcommand{\mR}{\MM{R}}
\newcommand{\mS}{\MM{S}}
\newcommand{\mZ}{\MM{Z}}
\newcommand{\mRt}{\tilde{\MM R}}
\newcommand{\mQt}{\tilde{\MM Q}}
\newcommand{\mPhi}{{\boldsymbol{\Phi}}}
\newcommand{\mPsi}{{\boldsymbol{\Psi}}}

\newcommand{\bf}{\bfseries}
\newcommand{\sl}{\slshape}
\newcommand{\rm}{\rmfamily}
\newcommand{\it}{\itshape}

\makeatother

\title          {The Geometry/Gauge Theory Duality\\[1ex]
                 and the Dijkgraaf--Vafa Conjecture}
\author         {Masaki Shigemori}
\department     {Physics}
\degreeyear     {2004}

\dedication{\textsl{To my parents}}

\acknowledgments{%

First of all, I would like to express my gratitude to my thesis advisor
Per Kraus, for sharing with me his insights and intuition, and for
teaching me attitude toward physics --- how to understand and think
about things physically.
I also wish to thank my other advisor Eric D'Hoker, who always explained
to me any physical concepts in his usual crystal-clear manner.
It is also my pleasure to thank Changhyun Ahn, Bo Feng, Yutaka Ookouchi
and Anton Ryzhov for exciting collaborations, and Iosif Bena, Brian
Forbes, Michael Gutperle and Sebastian de Haro for enlightening
discussions.

I am grateful to my parents, my sister, and my wife for their unfailing
support and for giving me the opportunity to carry out my Ph.D. studies.

\bigskip\bigskip

Chapter \ref{KS} is a version of P.~Kraus and M.~Shigemori, ``On the
matter of the Dijkgraaf--Vafa conjecture,'' JHEP \textbf{0304}, 052
(2003) [arXiv:hep-th/0303104] \cite{Kraus:2003jf}.  We would like to
thank Iosif Bena, Eric D'Hoker, and Anton Ryzhov for helpful
discussions.

Chapter \ref{KRS} is a version of P.~Kraus, A.~V.~Ryzhov
and M.~Shigemori, ``Loop equations, matrix models, and ${\mathcal N}=1$
supersymmetric gauge theories,'' JHEP \textbf{0305}, 059 (2003)
[arXiv:hep-th/0304138] \cite{Kraus:2003jv}.

Chapter \ref{IKRSV} is a version of K.~Intriligator, P.~Kraus,
A.~V.~Ryzhov, M.~Shigemori and C.~Vafa, ``On low rank classical groups
in string theory, gauge theory and matrix models,'' Nucl.\ Phys.\ B {\bf
682}, 45 (2004) [arXiv:hep-th/0311181] \cite{Intriligator:2003xs}.  I
would like to thank F. Cachazo for valuable discussions.

Chapter \ref{AFOS} is a version of C.~Ahn, B.~Feng, Y.~Ookouchi, and
M.~Shigemori, ``Supersymmetric Gauge Theories with Flavors and Matrix
Models,'' Nucl.\ Phys.\ B {\bf 698}, 3 (2004) [arXiv:hep-th/0405101]
\cite{Ahn:2004ym}. I would like to thank Freddy Cachazo, Eric D'Hoker,
Ken Intriligator, Romuald Janik, Per Kraus, and Hirosi Ooguri for
enlightening discussions.

This research  was partly supported  by
Heiwa Nakajima Foundation Scholarship and
NSF grant PHY-0099590.
}

\abstract       {%

In this dissertation we discuss various issues concerning application of
the Dijkgraaf--Vafa conjecture to the study of supersymmetric gauge
theories.  The conjecture states that for a large class of $\CN=1$
supersymmetric gauge theories, the exact effective superpotential for
the glueball superfield can be computed by matrix models.  This approach
is very powerful in that it provides a systematic way of computing the
nonperturbative, sometimes even exact, superpotential of the system,
which was possible only on a case-by-case basis in the more traditional
approach based on holomorphy and symmetry.

This conjecture has been checked for many nontrivial examples, but the
range of applicability of the conjecture remained unclear.  In Chapter
\ref{KS}, we give an explicit example, $Sp(N)$ theory with antisymmetric
tensor, which challenges the applicability of the conjecture.  We will
show that, the superpotential obtained by the Dijkgraaf--Vafa approach
starts to disagree with the standard gauge theory result at $N/2+1$
loops.  Thus we present a relatively simple example for which a
straightforward application of the Dijkgraaf--Vafa conjecture leads to a
different result from the known result.  Furthermore, in Chapter
\ref{KRS}, we will reproduce the same discrepancy in the generalized
Konishi anomaly method, an alternative approach to computing the
glueball superpotential.

In order to look for the physical origin of the discrepancy, in Chapter
\ref{IKRSV}, we consider the string theory realization of the gauge
theories by certain  Calabi--Yau compactifications, on which the
Dijkgraaf--Vafa conjecture is based.  By closely analyzing the geometric
transition of the Calabi--Yau space, we will uncover the following
prescription regarding when to include a glueball field: one should not
include a glueball field for $U(0)$, $SO(0)$, $SO(2)$ while for all
other gauge groups, including $U(1)$ and $Sp(0)$, one should consider an
associated glueball field.  We will explicitly show that the discrepancy
found in Chapters \ref{KS} and \ref{KRS} is resolved if we follow this
prescription and introduce a glueball field for the ``$Sp(0)$'' group.

In Chapter \ref{AFOS}, we consider generalizing the result in Chapter
\ref{IKRSV} to include flavors; we give a prescription regarding when to
include glueball fields based on string theory realization, and
demonstrate that the matrix model computations along with the
generalized prescription correctly reproduce the gauge theory results.

}

\begin{document}
\makeintropages



\EndAppendix

\chapter{Introduction}

\section{Supersymmetric gauge theory}

The importance of the study of supersymmetric gauge theory cannot be
emphasized enough.  Supersymmetric gauge theory is widely regarded as
the most promising and natural candidate for a framework which  extends the
Standard Model and which may be able to explain the outstanding issues
of the Standard Model such as the hierarchy problem, grand unification,
proton decay, and possibly the cosmological constant problem
\cite{Weinberg:cr}.
Supersymmetric gauge theory is attractive not only from such
phenomenological viewpoints but also from theoretical viewpoints.  In
particular, supersymmetry makes the theory more tractable than the
theory without supersymmetry.  For example, in the ordinary,
non-supersymmetric QCD, we do not have an analytic way to study even the
most fundamental properties of the theory such as confinement or chiral
symmetry breaking, because of its notorious strongly coupled dynamics.
However, in its supersymmetric versions, those properties can be proven
exactly.  The reason why we have analytic control over supersymmetric
theory is closely related to the holomorphy of the superpotential which
supersymmetric theories possess.\footnote{For reviews on supersymmetric
gauge theories, see e.g.\ \cite{Peskin:1997qi, Argyres:lectureNotes,
Intriligator:1995au}.}  Therefore, supersymmetric gauge theory is
important in that it serves as a toy model which helps us to understand
the properties of non-supersymmetric theory in a simpler setting.

Many powerful techniques have been developed for analytically studying
the properties of supersymmetric gauge theories.  Among the physical
quantities of the theory, superpotential is of fundamental importance
because it determines the structure of the vacua of the theory.  The
superpotential can sometimes be computed exactly using holomorphy and
symmetry considerations, however there was no systematic way of
computing it --- it has been worked out only on a case-by-case basis.

\section{The Dijkgraaf--Vafa conjecture}

Recently Dijkgraaf and Vafa \cite{Dijkgraaf:2002fc, Dijkgraaf:2002vw,
Dijkgraaf:2002dh} proposed a powerful, systematic way of computing
superpotential --- the so-called Dijkgraaf--Vafa conjecture.  For a
large class of $\CN=1$ supersymmetric gauge theories, this conjecture
states: i) at low energies, the holomorphic physics, namely the
superpotential term in the action, is captured by the glueball
superfield $S$, ii) the exact effective superpotential of the glueball
$S$ can be computed by $0+0$ dimensional field theory, i.e., matrix
model.
As we will review later (section
\ref{IKRSV:geo_trans_U(N)_SO/Sp(N)}), this conjecture was proposed
based on the geometric transition (conifold transition) duality
\cite{Gopakumar:2002wx,Vafa:2001wi,Cachazo:2001jy} in the string theory
realization of the models by Calabi--Yau compactification, and on the
fact that the four-dimensional superpotential of such a compactified
theory can be computed by topological strings \cite{Bershadsky:1993cx}.

The first part of the conjecture has a clear physical meaning, although
it is hard to prove: at low energies, the strongly coupled gauge theory
confines and the vacuum acquires nonzero expectation value of the
glueball field, $S=-{(1/32\pi^2)}\Tr[\CW_\alpha \CW^\alpha]$.  Here
$\CW_\alpha$ is the field strength superfield, and the lowest component
of $S$ is the fermion (gluino) bilinear whose nonvanishing vacuum
expectation value (vev) signals chiral symmetry breaking.  The low energy
excitation around this vacuum should be the glueball field $S$ changing
slowly over spacetime.

The second part of the conjecture claims that the momentum integration
in the computation of Feynman diagrams in supersymmetric gauge theory is
in fact trivial, and reduces simply to computation of combinatorial
factors.  Furthermore, almost all such diagrams actually vanish --- for
example, for $U(N)$ theory with a chiral superfield in the adjoint
representation, only diagrams whose topology in the 't Hooft double line
notation \cite{'tHooft:1974hx} is $S^2$ are nonvanishing.  Later, this
second part of the conjecture was proven \cite{Dijkgraaf:2002xd,
Ita:2002kx, Aganagic:2003xq} totally in the standard framework of the
super-space Feynman diagram expansion \cite{Wess:cp}.  It is surprising
that this tremendous simplification had been overlooked until the work
of Dijkgraaf and Vafa.

For example, for $U(N)$ theory with an adjoint chiral superfield $\Phi$,
this conjecture states that the exact effective glueball superpotential
is given by
\begin{align}
 W_{\text{eff}}(S) &= N{\partial \CF_{S^2} \over \partial S},\label{ndsz5May04}
\end{align}
where $\CF_{S^2}$ is the contribution to the matrix model free energy
from diagrams of $S^2$ topology.

This conjecture was checked for many nontrivial examples (see
\cite{Argurio:2003ym} for a list of references) including the
cases with fundamentals, baryonic interaction, multi-trace interaction;
in some theories one can obtain the exact superpotential by more
traditional techniques based on holomorphy and symmetry considerations,
and compare the results with the ones predicted by the Dijkgraaf--Vafa
conjecture.  The agreement was perfect.

\section{Generalized Konishi anomaly approach}

Inspired by the work of Dijkgraaf and Vafa, an alternative approach for
obtaining the glueball superpotential was developed in
\cite{Cachazo:2002ry, Seiberg:2003jq}, based on the generalized Konishi
anomaly\footnote{For a recent review on the generalized Konishi anomaly
approach and the diagrammatic approach to the Dijkgraaf--Vafa
conjecture, see \cite{Argurio:2003ym}.}.

In the matrix model approach, one can compute the glueball superpotential
from the expectation values of matrices of the form $\ev{\Tr[\mPhi^n]}$,
where $\mPhi$ is an $\mN\times \mN$ matrix which corresponds to the
adjoint chiral superfield $\Phi$ in gauge theory.  It is well known that
these matrix model expectation values satisfy certain relations called
the loop equations, which are analogues of the Schwinger--Dyson
equations in field theory.  These loop equations are so powerful 
that they determine those expectation values $\ev{\Tr[\mPhi^n]}$
completely, up to a finite number of undetermined parameters.

Because matrix model and gauge theory are closely related according to
the Dijkgraaf--Vafa conjecture, there should be some relations also in
the gauge theory framework, which are powerful enough to determine the
glueball superpotential.  In gauge theory, the glueball superpotential
can be determined by the vev of the chiral operators, which are defined
to be operators annihilated by supersymmetry generators.  For example,
$\Tr[\Phi^n]$, $\Tr[\CW_\alpha \CW^\alpha \Phi^n]$ are chiral operators.
Chiral operators form a ring called the chiral ring, and their vev are
independent of spacetime coordinates.  Furthermore, the vev of the
product of chiral operators is equal to the product of the vev of the
chiral operators.  Therefore, chiral operators are just as ordinary
numbers, as far as their expectation values are concerned.
Obviously, the vev of these chiral operators are natural candidate of
the quantities that play the role of $\ev{\Tr[\mPhi^n]}$ in matrix
model.  Indeed, in \cite{Cachazo:2002ry, Seiberg:2003jq}, the
Schwinger--Dyson equations for these operators were derived and it was
shown that these equations determine all vev's of the chiral operators
up to a finite number of undetermined parameters.  These
Schwinger--Dyson equations are called the generalized Konishi anomaly
equations, because these generalize the Konishi anomaly equation
\cite{Konishi:1983hf} which is the superfield version of the ordinary
anomaly equations.

The generalized Konishi anomaly approach established a firm connection
between the standard gauge theory formalism and the Dijkgraaf--Vafa
conjecture which appeared to be rather out of a hat from the traditional
gauge theory point of view, although it of course is a very natural
conjecture from the viewpoint of string theory.

\section{$Sp(N)$ theory with antisymmetric tensor}

As mentioned above, the Dijkgraaf--Vafa conjecture passed many
nontrivial\break checks.  However, it is not clear how large a class of
supersymmetric theories the conjecture is applicable to; it is very
important to investigate the applicability of the conjecture.

$Sp(N)$ theory\footnote{Our convention for $Sp(N)$ is such that $N$ is
an even integer, so that $Sp(N)\subset U(N)$ and $Sp(2) \cong SU(2)$.}
with a chiral superfield in the antisymmetric tensor is among those
theories for which the exact superpotential can be obtained by
traditional techniques \cite{Cho:1996bi, Csaki:1996eu}.
As equation \eqref{ndsz5May04} exemplifies, the Dijkgraaf--Vafa
conjecture generally predicts a superpotential with a simple pattern in
$N$, the rank of the gauge group.  However, in this $Sp(N)$ theory, the
dynamical superpotential obtained by the traditional techniques does not
appear to have any obvious pattern in $N$.  Therefore, this theory is an
ideal touchstone for checking the applicability of the conjecture.

In Chapter \ref{KS}, we apply the Dijkgraaf--Vafa conjecture to this
$Sp(N)$ theory with antisymmetric tensor based on the diagrammatic
approach \cite{Dijkgraaf:2002xd, Ita:2002kx, Aganagic:2003xq}, and the
effective glueball superpotential is computed using $Sp(N)$ matrix
model, for the cubic tree level superpotential and trivial breaking
pattern $Sp(N)\to Sp(N)$.  Comparing the resulting superpotential with the one
computed using the traditional techniques for $Sp(4)$, $Sp(6)$ and
$Sp(8)$, we find that the two superpotentials agree up to $N/2$-th order
in the matrix model perturbation theory, with discrepancy setting in at
the next order.  Thus, this $Sp(N)$ theory gave the first explicit
example for which a straightforward application of the Dijkgraaf--Vafa
conjecture leads to a different result from the known result.  We will
discuss possible origin of the discrepancy, e.g., possible
ambiguity in defining high powers of the glueball superfield $S^n$ with
$n\ge h= N/2+1$.

In Chapter \ref{KRS}, we apply the method of the generalized Konishi
anomaly to theories based on the classical gauge groups with various
two-index tensors and fundamentals, including $Sp(N)$ theory with
antisymmetric tensor.  We will see that the generalized Konishi anomaly
approach gives the same discrepancy as was observed in the matrix model
approach.  Because the generalized Konishi anomaly approach is not based
on diagram expansion, we can discuss the aforementioned $S^h$ problem
from a different point of view.  Moreover, we present a general formula
which expresses the solutions to the generalized Konishi anomaly
equation in terms of the solutions to the loop equations of the
corresponding matrix model.

\section{String theory prescription}

The discovery of the above ``counterexample'' to the Dijkgraaf--Vafa
conjecture by \cite{Kraus:2003jf} stimulated active research
\cite{Aganagic:2003xq, Cachazo:2003kx, Ahn:2003ui, Matone:2003bx,
Naculich:2003ka, Gomez-Reino:2004rd, Landsteiner:2003ph} to look for the
physical origin of the discrepancy.
In particular, Cachazo \cite{Cachazo:2003kx} showed that the generalized
Konishi anomaly equations governing the $Sp(N)$ theory with
antisymmetric tensor matter \cite{Alday:2003dk, Kraus:2003jv} can be
mapped to those of $U(N+2K)$ gauge theory with adjoint matter, with the
breaking pattern $Sp(N) \rightarrow \prod_{i=1}^K Sp(N_i)$ mapped to
$U(N+2K)\rightarrow \prod_{i=1}^K U(N_i+2)$.  Here, $K$ is the number of
the critical points in the tree level superpotential.  
Especially, the trivial breaking pattern studied \cite{Kraus:2003jf} for
the cubic superpotential, which can be written as $Sp(N)\to Sp(N)\times
Sp(0)$, is mapped to $U(N+4)\to U(N+2)\times U(2)$.  It was shown
\cite{Cachazo:2003kx} that this $U(N+2)\times U(2)$ theory correctly
reproduces the superpotential of the $Sp(N)$ theory and resolves the
discrepancy found in \cite{Kraus:2003jf}.
Because $U(2)$ has a glueball dynamics in it, this map implies that we
should consider a glueball field for the ``$Sp(0)$''.  In other words, in
the matrix model context, one has to consider glueball dynamics 
where one does not expect any in the standard gauge theory context.

Then, what about other low rank groups, such as $U(0)$, $SO(0)$ and
$SO(2)$, where one does not consider a strongly coupled glueball
dynamics in the standard gauge theory context?  Should we include
glueball fields also for them in the matrix model context, or not?  If
so (or if not), why?  Clearly, these questions cannot be answered within
the standard gauge theory framework; we should go back and investigate
the string theory realization of these gauge theories, on which the
Dijkgraaf--Vafa matrix model conjecture is based.

To answer these questions, in Chapter \ref{IKRSV}, we consider the
string theory realization (geometric engineering) of supersymmetric
$U(N)$, $SO(N)$, and $Sp(N)$ gauge theories with various two-index
tensor matter fields and added tree-level superpotential, for general
breaking patterns of the gauge group.  These theories are realized in
type IIB superstring theory compactified on certain noncompact
Calabi--Yau 3-folds, as the world-volume theory on the D5-branes which
wrap compact $2$-cycles in the Calabi--Yau space and fill four
dimensional Minkowski spacetime.
At low energies these gauge theories confine, developing a nonzero
expectation value of the glueball field.  This confinement transition is
described in string theory by the geometric transition
\cite{Gopakumar:2002wx,Vafa:2001wi,Cachazo:2001jy} where the 2-cycles in
the Calabi--Yau space are blown down and 3-cycles are blown up instead.
In string theory, the glueball field is interpreted as the size of these
3-cycles.  This is an example of the geometry/gauge theory duality,
where gravity in some background is dual to gauge theory in one less
dimensions.  Another related example of the duality is the celebrated
AdS/CFT duality \cite{Maldacena:1997re, Witten:1998qj, Gubser:1998bc}.

By closely studying the string theory physics near the blown up 3-cycles
after the geometric transition, we clarify when glueball fields should
be included and treated as dynamical fields, or rather set to zero.
The resulting string theory prescription is the following: one should
not include a glueball field for $U(0)$, $SO(0)$, $SO(2)$ while for all
other gauge groups, including $U(1)$ and $Sp(0)$, one should consider an
associated glueball field.

In particular, these string theory considerations give a clear physical
explanation of the origin of the apparent discrepancy observed in
\cite{Kraus:2003jf} for $Sp(N)$ theory with antisymmetric tensor.
Furthermore, we will give more examples in which this string theory
prescription is crucial for obtaining the correct superpotential.

\section{Adding flavors}

So far, we discussed the string theory prescription when to
include a glueball field, in theories with a matter field in the adjoint
representation.  It is a natural generalization to include matter fields
in the fundamental and  anti-fundamental representations, i.e.,
flavors.

For example, consider $\CN=1$ $U(N)$ theory with an adjoint and $N_f$
flavors.  Classically, this theory has two kinds of vacua: the
pseudo-confining vacua in which the gauge group breaks as $U(N)\to
\prod_{i=1}^K U(N_i)$, $\sum_{i=1}^K N_i=N$, and the Higgs vacua in
which $U(N)\to \prod_{i=1}^K U(N_i)$, $\sum_{i=1}^K N_i<N$.  In the
Higgs vacua, the total rank of the unbroken groups is reduced (Higgsed
down), motivating the name of the vacua.

In the formalism of the generalized Konishi anomaly
\cite{Cachazo:2003yc}, this theory is described on a Riemann surface
which is a double cover of the complex $z$-plane. On the $z$-plane,
there are $K$ cuts $A_i$, $i=1,\dots,K$, corresponding to the $K$
unbroken gauge group factors $U(N_i)$, $i=1,\dots,K$.  The glueball
field $S_i$ associated with the $U(N_i)$ group is related to a contour
integral around the $i$-th cut, $A_i$.  Moreover, there are $N_f$ poles
on the Riemann surface, one for each flavor.  The position of a pole is
determined by the mass of the corresponding flavor.

In this setup, the pseudo-confining vacua correspond to having all the
$N_f$ poles on the second sheet of the Riemann surface.  On the other
hand, the Higgs vacua is obtained as follows.  One begins with the
pseudo-confining vacua with all the poles on the second sheet, and starts
varying the mass of a flavor smoothly.  Then the position of the pole
changes smoothly, and in particular, one can pass the pole through the
cut $A_i$ to the first sheet.  In this process, the gauge group factor
$U(N_i)$ becomes $U(N_i-1)$, i.e., it is Higgsed down.  In this way, by
passing poles on the second sheet through cuts to the first sheet, one
can obtain the Higgs vacua.  Different choices of the cuts through which
the poles pass correspond to different Higgsing pattern.

Clearly, there should be a limit to this process of passing poles
through a given cut, since as one passes poles one by one through the
cut $A_i$ associated with the unbroken $U(N_i)$ group, the group gets
Higgsed down as $U(N_i)\to U(N_i-1)\to U(N_i-2)\to \cdots$, and one
eventually ends up with a $U(0)$, which cannot be Higgsed down any
further.  Therefore, one expects that the cut shrinks as one passes more
and more poles and it eventually closes up, when one has Higgsed down
the $U(N_i)$ group completely.
As mentioned above, the glueball $S_i$ is related to the contour
integral around the cut $A_i$.  Therefore, when the $U(N_i)$ group has
completely broken down and the cut $A_i$ has closed up, the glueball
vanishes: $S_i=0$.  This is consistent with the fact that there is no
strongly coupled dynamics any more which gives rise to nonzero
$S_i$.    

One may think that the above picture is very reasonable and the $S_i\to
0$ limit should be describable in the matrix model framework in terms of
the glueball $S_i$.  However, that is not quite correct.  Note that the
Higgsing $U(N_i\neq 0)\to U(0)$ is not a smooth process, because in this
process the number of massless $U(1)$ photons changes from one to zero,
discontinuously.  This is possible only by condensation of some charged
massless particle \cite{Seiberg:1994rs, Seiberg:1994aj} which makes the
photon massive by the Meissner effect and which is clearly missing in
the glueball description.

In Chapter \ref{AFOS}, we will closely study the above process of
passing $N_f$ poles through a cut associated with $U(N_c)$ group, and
demonstrate that the cut indeed becomes arbitrarily small if one tries
to pass too many poles through the cut.  Furthermore, we will see that
the situation where the cut has completely closed up, namely $S=0$, is
not describable in matrix model; we need some extra charged massless
degree of freedom, as we discussed above.
To find out what this massless field is, we will consider the string
theory realization of the theory, and argue that this massless field
should be a compact D3-brane which emanates fundamental strings.  In
addition, we will argue that the effect of this charged massless field
can be taken care of by simply setting the glueball $S_i$ to zero by
hand in matrix model, and demonstrate by explicit computation that the
prescription indeed gives the correct superpotential obtained by the
factorization method in gauge theory.

\EndAppendix

\chapter{$\boldsymbol{Sp(N)}$ theory with antisymmetric tensor
}
\label{KS}

With the aim of extending the gauge theory -- matrix model connection to
more general matter representations, we prove that for various two-index
tensors of the classical gauge groups, the perturbative contributions to
the glueball superpotential reduce to matrix integrals.  Contributing
diagrams consist of certain combinations of spheres, disks, and
projective planes, which we evaluate to four and five loop order.  In
the case of $Sp(N)$ with antisymmetric matter, independent results are
obtained by computing the nonperturbative superpotential for $N=4,6$ and
$8$. Comparison with the Dijkgraaf--Vafa approach reveals agreement up
to $N/2$ loops in matrix model perturbation theory, with disagreement
setting in at $h=N/2+1$ loops, $h$ being the dual Coxeter number. At
this order, the glueball superfield $S$ begins to obey nontrivial
relations due to its underlying structure as a product of fermionic
superfields.  We therefore find a relatively simple example of an
${\mathcal N}=1$ gauge theory admitting a large $N$ expansion, whose
dynamically generated superpotential fails to be reproduced by the
matrix model approach.

\section{Introduction}

The methods of Dijkgraaf and Vafa \cite{Dijkgraaf:2002fc,
Dijkgraaf:2002vw, Dijkgraaf:2002dh} represent a potentially powerful
approach to obtaining nonperturbative results in a wide class of
supersymmetric gauge theories.  Their original conjecture consists of
two parts. First, that holomorphic physics is captured by an effective
superpotential for a glueball superfield, with nonperturbative effects
included via the Veneziano--Yankielowicz superpotential
\cite{Veneziano:1982ah}. Second, that the Feynman diagrams contributing to
the perturbative part of the glueball superpotential reduce to matrix
model diagrams.

The second part of the conjecture has been proven for a few choices of
matter fields and gauge groups, namely $U(N)$ with adjoint
\cite{Dijkgraaf:2002xd, Cachazo:2002ry} and fundamental
\cite{Seiberg:2003jq} matter, and $SO/Sp(N)$ with adjoint matter
\cite{Ita:2002kx, Ashok:2002bi, Janik:2002nz}.  Combining this with the
first part of the conjecture has then been shown to reproduce known
gauge theory results.  Some examples of ``exotic'' tree-level
superpotentials have also been considered successfully, such as multiple
trace \cite{Balasubramanian:2002tm}\ and baryonic
\cite{Argurio:2003hk,Bena:2002ua,Bena:2003vk} 
interactions.

One naturally wonders how far this can be pushed.  Generic
$\mathcal{N}=1$ theories possess intricate dynamically generated
superpotentials which are difficult or (nearly) impossible to obtain by
traditional means, and so a systematic method for computing them would
be most welcome.  The promise of the DV approach is that these perhaps
can be obtained to any desired order by evaluating matrix integrals.
With this in mind, we will demonstrate the reduction to matrix integrals
for some new matter representations.  Comparing with known gauge theory
results will turn out to illustrate some apparent limitations of the DV
approach.

In particular, it is straightforward to generalize the results of
\cite{Dijkgraaf:2002xd, Ita:2002kx} to more general two-index tensors of
$U(N)$ and $SO/Sp(N)$, with or without tracelessness conditions imposed.
The relevant $0+0$ dimensional Feynman diagrams which one needs to compute
consist of various spheres, disks and projective planes, and
disconnected sums of these.  We evaluate these to five-loop order.

For comparison with gauge theory we focus on the particular case of
$Sp(N)$\footnote{Our convention for $Sp(N)$ is such that $N$ is an even
integer, and $Sp(2) \cong SU(2)$.}  with an antisymmetric tensor chiral
superfield.  The dynamically generated superpotentials for such theories
are highly nontrivial, and cannot be obtained via the ``integrating in''
approach of \cite{Intriligator:1994uk}.  Furthermore, the results display no
simple pattern in $N$.  Nevertheless, a method is known for computing
these superpotentials on a case-by-case basis
\cite{Cho:1996bi,Csaki:1996eu}.  Results for $Sp(4)$ and $Sp(6)$ were
obtained in \cite{Cho:1996bi,Csaki:1996eu}, and here we extend this to
$Sp(8)$ as well (partial results for $Sp(8)$ appear in
\cite{Csaki:1996eu}).  We believe that these examples illustrate the
main features of generic $\mathcal{N}=1$ superpotentials, and so are a
good testing ground for the DV approach.

For our $Sp(N)$ examples, we will demonstrate agreement between
gauge theory and the DV approach up to $N/2$ loops in perturbation
theory, with a disagreement setting in at $N/2+1$ loops.  In terms
of the glueball superpotential, we thus find a disagreement at
order $S^h$, where $h=N/2+1$ is the dual Coxeter number of
$Sp(N)$.

The fact that discrepancies set in at order $S^h$ is not a surprise, for
it is at this order that $S$ begins to obey relations due to its being a
product of two fermionic superfields
\cite{Cachazo:2002ry, Witten:2003ye}.  Furthermore, at this order
contributions to the effective action for $W_\alpha$ of the schematic
form $\Tr (W_\alpha)^{2h}$ can be reexpressed in terms of lower traces,
including $S^h$.  Unfortunately, it is not clear how to ascertain these
relations \textit{a priori}, since they receive corrections from
nonperturbative effects (see \cite{Witten:2003ye} for a recent
discussion).  These complications do not arise for theories with purely
adjoint matter, since the results are known to have a simple pattern in
$N$, and so $N$ can be formally taken to infinity to avoid having to
deal with any relations involving the $S$'s.  But in the more generic
case, it seems that additional input is required to make progress at $h$
loops and beyond.

The remainder of this paper is organized as follows.  In section
\ref{KS:red2MM} we isolate the field theory diagrams that contribute to
the glueball superpotential, derive the reduction of these diagrams to
those of a matrix model, and discuss their computation.  These results
are used in section \ref{KS:results_mm} to derive effective
superpotentials for $Sp(N)$ with matter in the antisymmetric tensor
representation.  In section \ref{KS:gt} we state the corresponding
results derived from a nonperturbative superpotential for these
theories.  Comparison reveals a discrepancy, which we discuss in section
\ref{KS:comparison}.  Appendix \ref{KS:diagram} gives more details on
diagram calculations; appendix \ref{KS:summary_pert} collects results
from matrix model perturbation theory; and appendix \ref{KS:GT_results}
concerns the computation of dynamically generated superpotentials for
the $Sp(N)$ theories.

\section{Reduction to matrix model}
\label{KS:red2MM}

In this section we will extend the results of
\cite{Dijkgraaf:2002xd, Ita:2002kx} to include the following matter
representations: 
\begin{itemize}
 \item $U(N)$ adjoint.
 \item $SU(N)$ adjoint.
 \item $SO(N)$ antisymmetric tensor
 \item $SO(N)$ symmetric tensor, traceless or traceful.
 \item $Sp(N)$ symmetric tensor.
 \item $Sp(N)$ antisymmetric tensor, traceless or traceful.
\end{itemize}
 We will use $\Phi_{ij}$ to denote the matter
superfield.    In the case of $Sp(N)$, $\Phi_{ij}$ is defined as
\begin{align}
 \Phi=
 \begin{cases} S
   J & \text{$S_{ij}$: symmetric tensor,}\\
  AJ & \text{$A_{ij}$: antisymmetric tensor.}
 \end{cases}
\end{align}
Here $J$ is the invariant antisymmetric tensor of $Sp(N)$, namely
\begin{align}
 J_{ij}= \begin{pmatrix} 0 & \unit_{N/2}\\ -\unit_{N/2} & 0\end{pmatrix}~.
\end{align}
The tracelessness of the $Sp$ antisymmetric tensor is defined
with respect to this $J$, i.e., by $\Tr[AJ]=0$.

The fact that allows us to treat the above cases in parallel to
those considered in \cite{Dijkgraaf:2002xd, Ita:2002kx} is that gauge
transformations act by commutation,  $\delta_\Lambda \Phi \sim
[\Lambda, \Phi]$.  A separate analysis is needed for, say, $U(N)$
with a symmetric tensor.

\subsection{Basic setup}

Following \cite{Dijkgraaf:2002xd}, we consider a supersymmetric gauge theory
with chiral superfield $\Phi$ and field strength $\CW^\alpha$.
Treating $\CW^\alpha$ as a fixed background, we integrate out $\Phi$
to all orders in perturbation theory.  We are interested in the
part of the effective action which takes the form of a
superpotential for the glueball superfield $S={1\over 32\pi^2} \Tr
[\CW^\alpha \CW_\alpha]$. In \cite{Dijkgraaf:2002xd}, using the superspace
formalism, it was shown that this  can be obtained from a simple
action involving only chiral superfields:
\begin{align}
  S(\Phi)= \int \!d^4p\, d^2\pi\,\left[
 \frac{1}{2}\Phi(p^2+\CW^\alpha \pi_\alpha)\Phi +W_{\mathrm{tree}}(\Phi)
 \right]~. 
\end{align}
We choose the tree level superpotential to be
\begin{align}
 W_{\textrm{tree}}=\frac{m}{2}\Tr[\Phi^2]+{\textrm{interactions}}~,
\end{align}
where the interactions are single trace terms,  and include the
mass in the propagator:
\begin{align}
 \frac{1}{p^2+m+\CW^\alpha \pi_\alpha}~.
\end{align}

Actually, we have to be a little more precise here.  Displaying
all indices, we can write the quadratic action as
\begin{align}
 \frac{1}{2}\int d^4p \, d^2\pi\, \Phi_{ji} G^{-1}_{ijkl} \Phi_{kl}
\end{align}
with
\begin{align}
 G^{-1}_{ijkl}
 =
 \left[ (p^2+m)\delta_{im}\delta_{jn}+(\CW^\alpha)_{ijmn}\pi_\alpha \right]P_{mnkl}~.
\end{align}
Here the $P$'s are projection operators appropriate for the gauge
group and  matter representation under consideration:
\begin{align}
  P_{ijkl}= 
 \begin{cases}
  \delta_{ik}\delta_{jl} & \text{$U(N)$ adjoint,}\\
  \delta_{ik}\delta_{jl}- {1 \over N}\delta_{ij}\delta_{kl} & \text{$SU(N)$ adjoint,}\\
  {1\over 2}(\delta_{ik}\delta_{jl} - \delta_{il}\delta_{jk}) & \text{$SO(N)$ antisymmetric,}\\
  {1\over 2}(\delta_{ik}\delta_{jl} + \delta_{il}\delta_{jk}) & \text{$SO(N)$ traceful symmetric,}\\
  {1\over 2} (\delta_{ik}\delta_{jl} + \delta_{il}\delta_{jk} - {2 \over N}\delta_{ij}\delta_{kl}) 
    & \text{$SO(N)$  traceless symmetric,} \\
  {1\over 2}(\delta_{ik}\delta_{jl} - J_{il}J_{jk}) & \text{$Sp(N)$ symmetric,}\\
  {1\over 2}(\delta_{ik}\delta_{jl} + J_{il}J_{jk}) & \text{$Sp(N)$ traceful antisymmetric,}\\
  {1\over 2}\left(\delta_{ik}\delta_{jl} + J_{il}J_{jk} - \frac{2}{N}\delta_{ij}\delta_{kl}\right)
    & \text{$Sp(N)$ traceless antisymmetric.}
 \end{cases}
 \label{msye21Mar04}
\end{align}
The propagator is then given by the inverse of $G^{-1}$ in the
subspace spanned by $P$:
\begin{align}
 \langle \Phi_{ji} \Phi_{kl}\rangle = \left[ { P \over p^2
 +m + \CW^\alpha \pi_\alpha}\right]_{ijkl}= \left[ \int_0^\infty \! ds ~
 e^{-s(p^2+m + \CW^\alpha \pi_\alpha)} P\right]_{ijkl}~. 
\end{align}
Our rule for multiplying four-index objects is
$(AB)_{ijkl}=\sum_{mn} A_{ijmn} B_{mnkl}$.  The fact that gauge
transformations act by commutation means that we can write
\begin{align}
 (\CW^\alpha)_{ijkl} = (\CW^\alpha)_{ik}\delta_{jl} - (\CW^\alpha)_{lj}\delta_{ik}~,
 \label{mtnp21Mar04}
\end{align}
where on the right hand side $(\CW^\alpha)_{ij}$ are field strengths
in the defining representation of the gauge group.

\subsection{Diagrammatics}

\begin{figure}
  \begin{center}
   \vspace{-.5cm}
  \epsfxsize=5cm \epsfbox{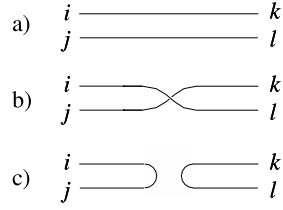}
   \vspace{-1.75cm}
  \caption{Propagators.  a) untwisted; b) twisted; c) disconnected \label{fig:propagators} }
 \end{center}
\end{figure}

\begin{figure}
  \begin{center}
   \vspace{-1cm}
  \epsfxsize=3.5cm \epsfbox{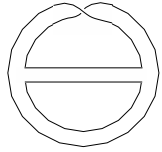}
   \vspace{-1.35cm}
  \caption{Typical diagram \label{fig:typical diagram} }
 \end{center}
\end{figure}

\begin{figure}
  \begin{center}
   \vspace{-.5cm}
  \epsfxsize=5cm \epsfbox{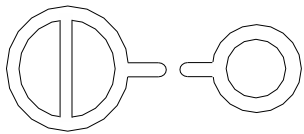}
   \vspace{-1.5cm}
  \caption{Contributing diagram \label{fig:contributing diagram} }
 \end{center}
\end{figure}

\begin{figure}
  \begin{center}
   \vspace{-1cm}
  \epsfxsize=3.5cm \epsfbox{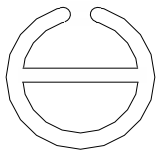}
   \vspace{-1.5cm}
  \caption{Non-contributing diagram \label{fig:noncontributing diagram} }
 \end{center}
\end{figure}

The presence of the three sorts of terms in the projection
operators \eqref{msye21Mar04} means that in double line notation we have three
types of propagators, displayed in Fig.\ \ref{fig:propagators}.
%
%
Note in particular the disconnected propagator, which allows us to draw
Feynman diagrams which have disconnected components in index space (All
diagrams are connected in momentum space since we are computing the free
energy).  A typical diagram involving cubic interactions is shown in
Fig.\ \ref{fig:typical diagram}.
%
%
Since we are computing the superpotential for $S$, we include either
zero or two insertions of $\CW^\alpha$ on each index loop.\footnote{Note
that we are explicitly {\em not} including the contributions coming from
more than two $\CW^\alpha$'s on an index loop, even if for a particular
$N$ these can be expressed in terms of $S$'s.  We will come back to this
point in section 5.} We will now prove that the diagrams which
contribute are those consisting of some number of sphere, disk, and
projective plane components.  Furthermore, the total number of
disconnected components must be one greater than the number of
disconnected propagators.  Fig.\ \ref{fig:contributing diagram} is an
example of a contributing diagram,
%
%
while Fig.\ \ref{fig:noncontributing diagram} is a
diagram which does not contribute.
%
%

The proof is similar to that given in \cite{Dijkgraaf:2002xd, Ita:2002kx},
so we mainly focus on the effect of the new disconnected propagator. In
double line notation we associate each Feynman diagram to a
two-dimensional surface.  Let $F$ by the number of faces (index loops);
$P$ be the number of edges; and $V$ be the number of vertices.  The
Feynman diagram also has some number $L$ of momentum loops.  Euler's
theorem tells us that
\begin{align}
 F= P- V + \chi~,
 \label{mwgs21Mar04}
\end{align}
where $\chi_{\Stwo} = 2$, $\chi_{\Dtwo}=\chi_{\RPtwo} = 1$.  We also
have the relation
\begin{align}
 F =L -1 +\chi~.
 \label{ncib21Mar04}
\end{align}
In a diagram with $L$ loops we need to bring down $L$ powers of $S$ to
saturate the fermion integrals, and we allow at most one $S$ per index
loop.  Therefore, for a graph to be nonvanishing we need $F \geq L$.
Graphs on $\Stwo$ with no disconnected propagators have $F= L+1$, and
those on $\RPtwo$ have $F=L$.

To proceed we will make use of the following operation.  Considering
some diagram $D$ that includes some number of disconnected propagators.
To each $D$ we associate a diagram $\tilde{D}$, obtained by replacing
each disconnected propagator of $D$ by an untwisted propagator.  Each
$\Dt$ diagram thus consists of a single connected component.  $D$ and
$\Dt$ have the same values of $L$ and $V$, but can have different values
of $F$, $P$, and $\chi$.  We use $\Ft$, $\Pt$, and $\chit$ to denote the
number of faces, edges, and the Euler number of $\Dt$.  To see which
diagrams can contribute we consider various cases. 

\medskip\noindent
\textbf{Case 1:} $D$ has no disconnected propagators, so $\Dt=D$.  This
case reduces to that of \cite{Dijkgraaf:2002xd, Ita:2002kx}, and so we know
that only $\Stwo$ and $\RPtwo$ graphs contribute (since no $\Dtwo$ graphs can
arise without disconnected propagators, these are the only graphs for
which $F\geq L$.)

\medskip
The remaining cases to consider are those for which we have at
least one disconnected propagator.

\medskip\noindent
\textbf{Case 2:} $\chi =\chit \leq 1 $ \\ In this case $\Ft \leq L$,
from \eqref{ncib21Mar04}. Each time we take a disconnected
propagator and replace it by an untwisted propagator we are
increasing $P$ by $1$ but keeping $L$ unchanged. Therefore,
from \eqref{mwgs21Mar04}, this operation increases $F$ by
$1$. So we see that in this case $F < L$. This means that
the diagram $D$ does not contribute.

\medskip\noindent
\textbf{Case 3:}  $\chi= \chit =2$

$\Dt$ has $\Ft=L+1$.  In this case, if $D$ has a single disconnected
propagator, we will have $F=L$, and so the diagram might seem to
contribute. But we will now show that the fermion determinant vanishes
for such diagrams.

We follow the conventions of \cite{Ita:2002kx}, where the reader is referred
for more details.  The fermion contribution is proportional to $[\det
N(s)]^2$ where
\begin{align}
 N(s)_{ma} = \sum_i s_i K^T_{mi}L_{ia}~.
\end{align}
Here, $i$ labels propagators; $m$ labels ``active'' index loops on which
we insert an $S$; and $a$ labels momentum loops. In the present case,
since $F=L$, all index loops are active and so $N$ is a square matrix.
To show that the determinant vanishes, we will show that the rectangular
matrix
\begin{align}
 s_i K_{im}\label{ndlb21Mar04}
\end{align}
has a nontrivial kernel.

Recall the definition of $K_{im}$. For each oriented propagator
labeled by $i$, the $m$th index loop can do one of three things: 1)
coincide and be parallel, giving $K_{im}=1$; 2) coincide and be
anti-parallel, giving $K_{im}=-1$; 3) not coincide, giving $K_{im}=0$.
Consider $K_{im}$ acting on the vector $b_m$ whose components are all
equal to $1$. It should be clear that
\begin{align}
 \sum_m K_{im} b_m = 1-1=0~.
\end{align}

The intuitive way to think about this is that $b_m$ are the index loop
momenta and $\sum_m K_{im}b_m$ are the propagator momenta. By setting
all index loop momenta equal, one makes all propagator momenta vanish,
and this corresponds to an element of the kernel of \eqref{ndlb21Mar04}.
This finally implies $\det N(s)=0$, which is what we wanted to show.

\medskip\noindent
\textbf{Case 4:}  $\chi \neq \chit$

This can only happen when $D$ has two or more disconnected components.
In this case, $\chi > \chit$ and so \eqref{mwgs21Mar04} still allows $F
\geq L$ even when $P<\Pt$.

In order to have a nonvanishing fermion integral, each component
of $D$ must have $F \geq L$, so each component must be an $\Stwo$, a
$\Dtwo$, or an $\RPtwo$.   Suppose $D$ has $N_{\Stwo}$ $\Stwo$ components,
$N_{\Dtwo}$ $\Dtwo$ components, and $N_{\RPtwo}$ $\RPtwo$ components, so
that
\begin{align}
 \chi = 2N_{\Stwo}+ N_{\RPtwo} +N_{\Dtwo}~.
\end{align}

Next consider the relation between $P$ and $\Pt$. The number of
disconnected propagators must be at least the number of disconnected
components of $D$ minus one, so
\begin{align}
 P = \Pt -(N_{\Stwo}+ N_{\RPtwo} + N_{\Dtwo} -1) - a = \Pt +1 +N_{\Stwo} -\chi -a ~,
\end{align}
where $a$ is a nonnegative integer.  Now use
\begin{align}
 F  = P-V+\chi =\Pt +1 +N_{\Stwo}- V -a  ~.
\end{align}
$\Dt$ satisfies
\begin{align}
 \Pt+1 -V = L ~,
\end{align}
so we get
\begin{align}
  F = L+N_{\Stwo} -a~. 
\end{align}

Now, in order to have a nonzero fermion determinant we need to
have at least one inactive index loop (no $\CW^a$ insertions) per
$\Stwo$ component after choosing $L$ active index loops.  In other
words, a nonvanishing fermion determinant requires
\begin{align}
 F \geq L+N_{\Stwo}~.
\end{align}
Putting these two conditions together, we clearly need $a=0$. This says
that the number of disconnected propagators in $D$ must be precisely
equal to the number of disconnected components of $D$ minus one.

\medskip\noindent
\textbf{Summary:}  Diagrams which contribute to the
glueball superpotential have any number of disconnected $\Stwo$,
$\Dtwo$, and $\RPtwo$ components.  The number of disconnected
propagators must be one less than the number of disconnected
components.\footnote{It is easy to convince oneself that disconnected
diagrams will therefore never contribute in theories with only
even powers in the tree level superpotential, thus giving the same 
glueball superpotential in the traceful and traceless cases.}

\subsection{Computation of diagrams}

Now that we have isolated the class of diagrams which contribute
to the glueball superpotential, we turn to their computation. This
turns out to be a simple extension of what is already known.  In
particular, the contribution from a general disconnected diagram
is simply equal to an overall combinatorial factor times the
product of the contributions of the individual components.   This
follows from the fact that, for the diagrams we are considering,
the disconnected propagators carry vanishing momentum, so the
diagrams are actually disconnected in both momentum space and
index space.

Next, we observe that the stubs from the disconnected propagators
can be neglected in the computation; it is easily checked that the
sum over  $\CW^\alpha$ insertions on the stubs  gives zero due to the
minus sign in \eqref{mtnp21Mar04}.

So we just need rules for treating each component individually,
and then we multiply the contributions together to get the total
diagram. The rules consist of relating the gauge theory
contribution to a corresponding matrix contribution.  The cases of
interest are:

\medskip\noindent
\textbf{$\Stwo$ components:}~ From the work of \cite{Dijkgraaf:2002xd},
we know that if $N^{L+1} F_{\Stwo}^{(L)}(g_k)$ is the contribution
in the matrix model diagram from an $L$ loop $\Stwo$ graph, then the
contribution in the gauge theory diagram is
\begin{align}
  W_{\Stwo}^{(L)}(S,g_k) = (L+1)N S^L F_{\Stwo}^{(L)}(g_k)~.
\end{align}
The prefactor $(L+1)N$ comes from the choice of, and trace
over, a single  inactive index loop.

\medskip\noindent
\textbf{$\RPtwo$ components:}~  From the work of \cite{Ita:2002kx}, we
know that if $N F_{\RPtwo}^{(L)}(g_k)$ is the contribution in the
matrix model diagram from an $L$ loop $\RPtwo$ graph, then the
contribution in the gauge theory diagram is
\begin{align}
  W_{\RPtwo}^{(L)}(S,g_k) = \pm 4 S^L F_{\RPtwo}^{(L)}(g_k)~.
\end{align}
The prefactor of $\pm
4$ comes from the fermion determinant, and is equal to $+4(-4)$ for
symmetric(antisymmetric) tensors.

\medskip\noindent
\textbf{$\Dtwo$ components:}~ These have $L=0$ and hence no
$\CW^a$ insertions.  So if the contribution to the matrix model is
$N F_{\Dtwo}^{(L)}(g_k)$ then
\begin{align}
 W_{\Dtwo}^{(L)}(g_k)=N F_{\Dtwo}^{(L)}(g_k) ~.
\end{align}

With the above rules in hand, it is a simple matter to convert a given
matrix model Feynman diagram into a contribution to the glueball
superpotential.  The example given in Appendix \ref{KS:diagram} should help to
clarify this.  We should emphasize that the above procedure must by done
diagram by diagram --- there is no obvious way to directly relate the
entire glueball superpotential to the matrix model free energy; the
situation is similar to \cite{Balasubramanian:2002tm} in this respect.

\section{Results from matrix integrals}
\label{KS:results_mm}

The considerations thus far apply to any single trace, polynomial,
tree level superpotential.
We now restrict attention to cubic interactions,
\begin{align}
  W_{\textrm{tree}}=\frac{m}{2}\Tr\,\Phi^2 +\frac{g}{3} \Tr\,\Phi^3~,
\end{align}
(which are of course trivial in the case of $SO/Sp$ with adjoint
matter.) In Appendix \ref{KS:summary_pert} we collect our matrix model results for
the various matter representations.  In this section we focus on
two particular cases, which  will be compared to gauge theory
results in the next section.

\subsection{$Sp(N)$ with traceful antisymmetric matter}

The perturbative part of the glueball superpotential for $Sp(N)$
with traceful antisymmetric matter is
\begin{multline}
 W_{\textrm{traceful}}^{\textrm{pert}}(S,\alpha)
 =
 \left(-N+3\right)\alpha \Stwo
 +\left(-{16\over 3}N+{59\over 3}\right)\alpha^2 S^3\\
 +\left(-{140\over 3}N+197\right)\alpha^3 S^4
 +\left(-{512}N+{4775\over 2}\right)\alpha^4 S^5
 +\cdots
 \label{ngpw21Mar04}
\end{multline}
with 
\begin{align}
 \alpha\equiv {g^2 \over 2 m^3}~.
 \label{ngtf21Mar04}
\end{align}
In terms of diagrams,
\eqref{ngpw21Mar04} represents the contribution from $2,3,4$ and $5$ loops.
According to the DV conjecture, the full glueball superpotential
is then $W^{\textrm{eff}} = W^{VY}+W^{\textrm{pert}}$, where $W^{VY}$ is the
Veneziano--Yankielowicz superpotential:
\begin{align}
 W^{VY}=(N/2+1) S[1-\log(S/\Lambda^3)]~.
\end{align}
We are now instructed to extremize $W^{\textrm{eff}}$ with respect to
$S$ and substitute back in.  We call the result
$W^{\textrm{DV}}$. Working in a power series in $g$, we obtain
\begin{multline}
  W_{\textrm{traceful}}^{\textrm{DV}}(\Lambda,m,g)
 =
 (N/2+1)\Lambda^3
 \Bigg[
 1
 -\frac{2(N-3)}{N+2}\Lambda^3\alpha
 -\frac{2(4N^2+45N-226)}{3(N+2)^2}\Lambda^6\alpha^2\\
 -\frac{2(12N^3+293N^2+368N-8340)}{3(N+2)^3}\Lambda^9\alpha^3\\
 -\frac{96N^4+3803N^3+25868N^2-85092N-744768}{3(N+2)^4}\Lambda^{12}\alpha^4
 -\cdots
 \Bigg]~.
\end{multline}
For $N=4,6,8$, this yields
\begin{equation}
\begin{split}
 W^{{\textrm{DV}},Sp(4)}_{\textrm{traceful}}(\Lambda,\alpha) &=
 3\Lambda^3
 -{}\Lambda^6\alpha
 -{}\Lambda^9\alpha^2
 -{353 \over 27}\Lambda^{12}\alpha^3
 -{25205 \over 81}\Lambda^{15}\alpha^4
 -\cdots~,
 \\
 W^{{\textrm{DV}},Sp(6)}_{\textrm{traceful}}(\Lambda,\alpha) &=
 4\Lambda^3
 -{3}\Lambda^6\alpha
 -{47\over 6}\Lambda^9\alpha^2
 -{73 \over 2}\Lambda^{12}\alpha^3
 -{ 6477 \over 32}\Lambda^{15}\alpha^4
 -\cdots~,
 \\
 W^{{\textrm{DV}},Sp(8)}_{\textrm{traceful}}(\Lambda,\alpha) &=
 5\Lambda^3
 -{5}\Lambda^6\alpha
 -{13}\Lambda^9\alpha^2
 -{65}\Lambda^{12}\alpha^3
 -{2142\over 5}\Lambda^{15}\alpha^4
 -\cdots~.
\end{split} 
\label{ngzz21Mar04}
\end{equation}

\subsection{$Sp(N)$ with traceless antisymmetric matter}

Including the contribution from the disconnected propagator, the
perturbative part of the glueball superpotential for $Sp(N)$ with
traceless antisymmetric matter is
\begin{multline}
 W_{\textrm{traceless}}^{\textrm{pert}}(S,\alpha)
 =
 \left(-1+{4\over N}\right)\alpha \Stwo
 +\left(-{1\over 3}-{8\over N}+{160\over 3N^2}\right)\alpha^2 S^3\\
 +\left(-{1\over 3}-{12\over N} -{256 \over 3N^2}+{3584 \over 3N^3}\right)\alpha^3 S^4+\cdots~.
 \label{nhtb21Mar04}
\end{multline}
The presence of many disconnected diagrams makes this case more
complicated than the traceful case, and we have correspondingly
worked to one lower order than in \eqref{ngpw21Mar04}.

Adding the Veneziano--Yankielowicz superpotential and integrating out
the glueball superfield, we obtain
\begin{multline}
  W_{\textrm{traceless}}^{\textrm{DV}}(\Lambda,m,g)
 =
 (N/2+1)\Lambda^3
 \Bigg[
 1
 -\frac{2(N-4)}{N(N+2)}\Lambda^3\alpha\\
 -\frac{2(N^3+14N^2-16N-512)}{3N^2(N+2)^2}\Lambda^6\alpha^2\\
 -\frac{2(N+8)^2(N^3+12N^2-52N-528)}{3N^3(N+2)^3}\Lambda^9\alpha^3
 -\cdots
 \Bigg]~.
\end{multline}
This yields
\begin{equation}
\begin{split}
  W^{{\textrm{DV}},Sp(4)}_{\textrm{traceless}}(\Lambda,\alpha)
 &=
 3\Lambda^3
 +{}\Lambda^9\alpha^2
 +{10}\Lambda^{12}\alpha^3
 +\cdots~,
 \\
 W^{{\textrm{DV}},Sp(6)}_{\textrm{traceless}}(\Lambda,\alpha)
 &=
 4\Lambda^3
 -{1\over 3}\Lambda^6\alpha
 -{7\over 54}\Lambda^9\alpha^2
 +{49 \over 54}\Lambda^{12}\alpha^3
 +\cdots~,
 \\
 W^{{\textrm{DV}},Sp(8)}_{\textrm{traceless}}(\Lambda,\alpha)
 &=
 5\Lambda^3
 -{1\over 2}\Lambda^6\alpha
 -{2\over 5}\Lambda^9\alpha^2
 -{14\over 25}\Lambda^{12}\alpha^3
 -\cdots~.
\end{split} 
\label{nhwr21Mar04}
\end{equation}

\section{Gauge theory example: $Sp(N)$ with antisymmetric matter}
\label{KS:gt}

Dynamically generated superpotentials can be determined for ${\mathcal
N}=1$ theories with gauge group $Sp(N)$ and a chiral superfield $A_{ij}$
in the antisymmetric tensor representation. The general procedure was
given in \cite{Cho:1996bi,Csaki:1996eu}, and is reviewed in Appendix C.
Since these superpotentials cannot be obtained by the integrating in
procedure of \cite{Intriligator:1994uk}, they are more difficult to
establish, and the results are correspondingly more involved, than more
familiar examples.
A separate computation is required for each $N$, and the results display
no obvious pattern in $N$.  $N=4$ is a simple special case (since $Sp(4)
\cong SO(5)$ and $A_{ij} \cong $ vector); $N=6$ was worked out in
\cite{Cho:1996bi,Csaki:1996eu}, and in Appendix C we extend this to
$Sp(8)$ (\cite{Csaki:1996eu} gives the result for $Sp(8)$ with some
additional fundamentals, which need to be integrated out for our
purposes).  In this section we state the results, and integrate out
$A_{ij}$ to obtain formulas that we can compare with the DV approach.

The moduli space of the classical theory is parameterized by the
gauge invariant operators
\begin{align}
  O_n=\Tr[(A J)^n],\qquad n = 1, 2, \ldots , N/2~,
\end{align}
the upper bound coming from the characteristic equation of the
matrix $AJ$.

From the gauge theory point of view, it is natural to demand
tracelessness, and this will be denoted by a tilde:  $\Tr [\At J]=0$,
\begin{align}
  \Ot_n=\Tr[(\At J)^n],\qquad n =  2, \ldots , N/2~.
\end{align}
In comparing with the DV approach, we will consider both the traceless
and traceful cases.

\subsection{Traceless case}

The $Sp(4)$ and $Sp(6)$ dynamical superpotentials for these
fields are \cite{Cho:1996bi,Csaki:1996eu}:
\begin{align}
  W_{\textrm{dyn}}^{Sp(4)}={2\Lambda_0{}^4 \over \Ot_2^{1/2}}~,
 \label{nivn21Mar04}
\end{align}
\begin{align}
  W_{\textrm{dyn}}^{Sp(6)}
 ={4\Lambda_0{}^5 \over \Ot_2 [(\sqrt{R}+\sqrt{R+1})^{2/3}+
 (\sqrt{R}+\sqrt{R+1})^{-2/3}-1]}~,
 \label{niyl21Mar04}
\end{align}
with $R= -12 \Ot_3^{~2}/\Ot_2^{~3}$.

Also, as derived in Appendix \ref{KS:GT_results}, the $Sp(8)$ superpotential is
\begin{multline}
  W_{\textrm{dyn}}^{Sp(8)}
 = \frac{6\sqrt{2}\,\Lambda_0^6}{\Ot_2^{3/2}}
 \Bigl[
 -36 R_4
 +144 b^2 R_4
 +288 c R_4
 +8 R_3^2\\
 +192 bc R_3
 +1152b^2c^2
 -36 b^2
 -72 c
 +9
 \Bigr]^{-1}~,
 \label{njbi21Mar04}
\end{multline}
where $R_3\equiv \Ot_3/\Ot_2{}^{3/2}$, $R_4\equiv \Ot_4/\Ot_2{}^2$, and
$b$ and $c$ are determined by
\begin{equation}
\begin{split}
  12R_4  +16bR_3 -192b^2c +24b^2 +96c^2 -3  &=0~,\\
 12bR_4 +8b^2R_3 +8R_3c -96bc^2  +24bc -3b  &=0~.
\end{split}
 \label{njda21Mar04}
\end{equation}
We choose the root which gives $R_3=0$ as the solution of the
$F$-flatness condition.

Now let us integrate out the antisymmetric matter.
We add the tree level superpotential
\begin{align}
  W_{\textrm{tree}}=\frac{m}{2}\Ot_2+\frac{g}{3}\Ot_3
\end{align}
to the dynamical part, solve the F-flatness equations, and substitute
back in.  We do this perturbatively in $g$, and obtain
\begin{equation}
\begin{split}
   W^{{\textrm{gt}}, Sp(4)}_{\textrm{traceless}}
 &=
 3\Lambda^3,
 \\
 W^{{\textrm{gt}}, Sp(6)}_{\textrm{traceless}}
 &=
 4\Lambda^3
 -{1 \over 3}\Lambda^6\alpha
 -{7 \over 54}\Lambda^9\alpha^2
 -{5 \over 54}\Lambda^{12}\alpha^3
 -{221 \over 2592}\Lambda^{15}\alpha^4
 -\cdots~,
 \\
 W^{{\textrm{gt}}, Sp(8)}_{\textrm{traceless}}
 &=
 5\Lambda^3
 -{1 \over 2}\Lambda^6\alpha
 -{2 \over 5}\Lambda^9\alpha^2
 -{14 \over 25}\Lambda^{12}\alpha^3
 -{}\Lambda^{15}\alpha^4
 -\cdots~,
\end{split}
\label{njoe21Mar04}
\end{equation}
where $\alpha$ is defined in \eqref{ngtf21Mar04}, and the low-energy scales
are defined from the usual matching conditions as
\begin{equation}
 \begin{split}
  Sp(4): & \quad \Lambda^9=(\frac{m}{2})\Lambda_0^8~,\\
  Sp(6): & \quad \Lambda^6=(\frac{m}{2})\Lambda_0^5 ~,\\
  Sp(8): & \quad \Lambda^{15}=(\frac{m}{2})^3\Lambda_0^{12}~.
 \end{split}
\end{equation}

\subsection{Traceful case}

For $Sp(N)$ theory with a traceful antisymmetric tensor $A_{ij}$,
we separate out the trace part as
\begin{align}
 A_{ij}=\At_{ij}-\frac{1}{N}J_{ij}\phi,\qquad
 \Tr[\At J]=0,\qquad
 \Tr[AJ]=\phi~.
\end{align}
$\Ot_n$ are related to their traceful counterparts $O_n\equiv\Tr[(AJ)^n]$ by
\begin{align}
 O_2=\Ot_2+{\phi^2 \over N},\qquad
 O_3=\Ot_3
 +{3 \over N}\Ot_2 \phi
 +{1 \over N^2} \phi^3.
\end{align}
The dynamical superpotential of this traceful theory is the same as the
traceless theory, since $\phi$ has its own $U(1)_{\phi}$ charge and
hence cannot enter in $W_{\textrm{dyn}}$.

Integrating out $\At_{ij}$ and $\phi$ in the presence of the tree level
superpotential
\begin{align}
  W_{\textrm{tree}}=\frac{m}{2}O_2+\frac{g}{3}O_3,
\end{align}
we obtain
\begin{equation}
\begin{split}
   W^{{\textrm{gt}}, Sp(4)}_{\textrm{traceful}}
 &=
 3\Lambda^3
 -{}\Lambda^6\alpha
 -{2}\Lambda^9\alpha^2
 -{187 \over 27}\Lambda^{12}\alpha^3
 -{2470 \over 27}\Lambda^{15}\alpha^4
 -\cdots~,
 \\
 W^{{\textrm{gt}}, Sp(6)}_{\textrm{traceful}}
 &=
 4\Lambda^3
 -{3}\Lambda^6\alpha
 -{47 \over 6}\Lambda^9\alpha^2
 -{75 \over 2}\Lambda^{12}\alpha^3
 -{7437 \over 32}\Lambda^{15}\alpha^4
 -\cdots~,
 \\
 W^{{\textrm{gt}}, Sp(8)}_{\textrm{traceful}}
 &=
 5\Lambda^3
 -{5}\Lambda^6\alpha
 -{13}\Lambda^9\alpha^2
 -{65}\Lambda^{12}\alpha^3
 -{2147 \over 5}\Lambda^{15}\alpha^4
 -\cdots~.
\end{split}
\label{nkju21Mar04}
\end{equation}

\section{Comparison and discussion}
\label{KS:comparison}

According to the general conjecture, we are supposed to compare
\eqref{nhwr21Mar04} with \eqref{njoe21Mar04}, and \eqref{ngzz21Mar04}
with \eqref{nkju21Mar04}.  We write $\bigtriangleup W\equiv
W^{\textrm{DV}}-W^{\textrm{gt}}$, and find
\begin{equation}
\begin{split}
  \bigtriangleup W_{\textrm{traceless}}^{Sp(4)}
 &= 0 \cdot \Lambda^6 \alpha+ \Lambda^9 \alpha^2 +\cdots~, \\
 \bigtriangleup W_{\textrm{traceless}}^{Sp(6)} &=0 \cdot \Lambda^6 \alpha
 +  0 \cdot \Lambda^9 \alpha^2+ \Lambda^{12}\alpha^3 +\cdots~, \\
 \bigtriangleup W_{\textrm{traceless}}^{Sp(8)} &=0 \cdot \Lambda^6 \alpha
 +  0 \cdot \Lambda^9 \alpha^2+ 0 \cdot \Lambda^{12} \alpha^3 + \CO(\Lambda^{15 }\alpha^4)~.
\end{split}
\end{equation}
and
\begin{equation}
 \begin{split}
 \bigtriangleup W_{\textrm{traceful}}^{Sp(4)}
 &= 0 \cdot \Lambda^6 \alpha+ \Lambda^9 \alpha^2 +\cdots~,\\
 \bigtriangleup W_{\textrm{traceful}}^{Sp(6)} &=0 \cdot \Lambda^6 \alpha
 +  0 \cdot \Lambda^9 \alpha^2+ \Lambda^{12}\alpha^3 +\cdots~,\\
 \bigtriangleup W_{\textrm{traceful}}^{Sp(8)} &=0 \cdot \Lambda^6 \alpha
 +  0 \cdot \Lambda^9 \alpha^2+ 0 \cdot \Lambda^{12} \alpha^3 + \Lambda^{15 }\alpha^4+\cdots~.
\end{split}
\end{equation}
We have indicated the terms that canceled nontrivially by including
them with a coefficient of zero. From these examples, we see that a
disagreement sets in at order $(\Lambda^3)^h\alpha^{h-1}$, where $h=N/2+1$
is the dual Coxeter number.  We also observe that the coefficient of the
disagreement at this order is unity.  We now discuss the implications of
this result.

First, it is very unlikely that the discrepancy is due to a
computational error, such as forgetting to include a diagram.  This is
apparent from the fact that the mismatch arises at a different order in
perturbation theory for different rank gauge groups.  So adding a new
contribution to the $Sp(4)$ result at order $\Lambda^9 \alpha^2$, say,
would generically destroy the agreement for $Sp(6)$ and $Sp(8)$ at this
order.  Instead, it is much more likely that our results indicate a
breakdown of the underlying approach.

Let us return to the two basic elements of the DV conjecture.  The first
part asserts that the perturbative part of the glueball superpotential
can be computed from matrix integrals, and the second part assumes that
nonperturbative effects are captured by adding the
Veneziano--Yankielowicz superpotential.  We have proven the perturbative
part of the conjecture for the relevant matter fields, but there is one
subtlety which we have so far avoided but now must discuss.

In our perturbative computations we inserted no more than two $\CW_\alpha$'s
on any index loop, since we were interested in a superpotential for
$S \sim \Tr \CW^2$, and not in operators such as $\Tr ~\CW^{2n}$, $n>1$.  
However, for a given gauge group, it may be possible to use Lie algebra 
identities to express such ``unwanted'' operators in terms of other operators,
including $S$.  Should we then include these new $S$ terms along with our
previous results?

This issue seems especially pertinent given
that our discrepancy sets in at  order $S^h$, which is when we begin
to find nontrivial relations involving $S$ due to its underlying
structure as a product of the fermionic field $\CW_\alpha$.  For example,
for $Sp(4)$ there are relations such as
\begin{align}
 \Tr[(\CW^2)^3]=
 \frac{3}{4}\Tr[\CW^2]\Tr[(\CW^2)^2]
 -\frac{1}{8}(\Tr[\CW^2])^3.
 \label{fh22Mar04}
\end{align}
So a naive guess is that the discrepancy can be accounted for if we keep
all contributions coming from more than two $\CW_\alpha$'s on an index
loop, and re-express the traces of the form $\Tr(\CW_\alpha)^{2n}$
($n\ge h$) in favor of $S$ using relations like \eqref{fh22Mar04},
setting all traces to zero that are not re-expressible in terms of $S$.
Such considerations are indeed necessary in order to avoid getting
nonsensical results in certain cases, e.g.\ antisymmetric matter for
$Sp(2)$.  Such a matter field is uncharged, and so should certainly
contribute a vanishing result for the glueball superpotential, but this
is seen only if we compute all the trace structures.  We should
emphasize that if we keep all of these contributions then perturbation
theory will not reduce to matrix integrals since the Schwinger parameter
dependence will not cancel; nevertheless we can try this procedure and
see what we get.

In order to check if the above guess is correct, we took a $\Phi^{2p}$
interaction and evaluated the perturbative superpotential explicitly
keeping all the traces.  For this interaction, a discrepancy arises at
the first order if we take $p>N$.  Specifically, we considered a
$\Phi^{6}$ interaction in $Sp(4)$ with antisymmetric matter.  After a
tedious calculation, we have found that this does {\em not} account for
the discrepancy.

It therefore seems more likely that the problem lies in the
nonperturbative part of the conjecture, involving truncating to just $S$
(and dropping other operators like $\Tr[(\CW^2)^2]$), and just adding
the Venziano--Yankielowicz superpotential.  There is no solid motivation
for this procedure beyond the fact that it seems to give sensible
results in various cases.  Our results indicate that for generic
theories this recipe is valid only up to $(h-1)$ loops.  On the other
hand, the fact that our discrepancies arise in a very simple fashion ---
always with a coefficient of unity --- suggests that perhaps there
exists a way of generalizing the DV recipe to enable us to go to $h$
loops and beyond.

Clearly, it is important to resolve these issues in order to determine
the range of validity of the DV approach.  One might have hoped that the
approach would be useful for any $\mathcal{N}=1$ theory admitting a
large $N$ expansion.  Our $Sp(N)$ theories are certainly in this class,
and so seem to provide an explicit counterexample.


\BeginAppendix

\section{diagrammatics for traceless matter field}
\label{KS:diagram}

In this appendix, we sketch the diagrammatics for evaluating the
perturbative glueball superpotential, focusing on the case with a
traceless tensor.  To be specific, we consider the cubic interaction
below:
\begin{align}
 e^{-W^{\textrm{pert}}(S)}=\int \mathcal{D}\Phi\, e^{
 -\int d^4x\, d^2\theta\,\Tr\left[
 -\frac{1}{2}\Phi(\partial^2-i\CW^\alpha D_\alpha)\Phi
 +\frac{m}{2}\Phi^2
 +\frac{g}{3}\Phi^3
 \right]
 }.
\end{align}
Namely, we consider $SO$ with traceless symmetric matter, or
$Sp$ with traceless antisymmetric matter. 

%
%
\begin{figure}
  \begin{center}
   \vspace{-.25cm}
  \epsfxsize=14cm \epsfbox{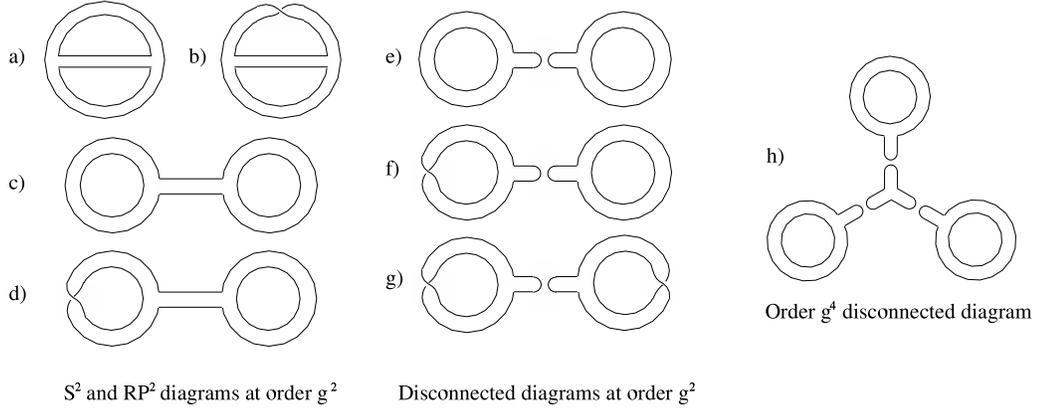}
   \vspace{-1.25cm}
  \caption{Diagrams for traceless tensor matter field \label{fig:diagrams for traceless tensor} }
 \end{center}
\end{figure}

At order $g^2$, there are four $\Stwo$ and $\RPtwo$ diagrams without
disconnected propagators that contribute, as shown in Fig.\
\ref{fig:diagrams for traceless tensor} a)--d).  These can be evaluated
by combinatorics.  For a) and b) there are $6$ ways to contract legs.
For c) and d), on the other hand, there are $3^2$ ways to contract legs
and 2 choices for the middle propagator (untwisted or twisted).  Since
there are two loop momenta, the glueball $S\sim \CW^\alpha \CW_\alpha$
should be inserted in two index loops, and we may insert only up to one
glueball on each index loop. For $\Stwo$ graphs a) and c), there are 3
ways to do so, and a trace over the remaining index loop contributes
$N$.  For $\RPtwo$ graphs b) and d), there is only one way to insert the
glueball, but the fermionic determinant gives an extra factor $(\pm 4)$.
The sign depends on the matter field under consideration {see section
2.)  Finally, for b) and d) there are respectively 3 and 2 ways to
choose which propagator to twist. Therefore, the contributions are
\begin{equation}
\begin{split}
  f_{\textrm{a}}&=6\cdot 3\cdot N \Stwo,\quad
 f_{\textrm{b}}=6 \cdot 3\cdot (\pm 4) \Stwo,\\
 f_{\textrm{c}}&=3^2\cdot 2\cdot 3\cdot N \Stwo,\quad
 f_{\textrm{d}}=3^2\cdot 2\cdot 2 \cdot (\pm 4) \Stwo.
\end{split}
\end{equation}
Including factors coming from propagators and coefficients from Taylor
expansion, we obtain
\begin{align}
 W^{(2)}_{\textrm{conn}}=-{g^2 \over (2m)^3}{1 \over 2! \,3^2}(f_{\textrm{a}}+f_{\textrm{b}}+f_{\textrm{c}}+f_{\textrm{d}})
 = (-N\mp 3)\alpha \Stwo.
\end{align}
where we defined  $\alpha =\frac{g^2}{2m^3}$ as before. This
reproduces the first term of the traceful result \eqref{ngpw21Mar04}.

For a traceless  tensor, there are three additional
diagrams e), f) and g),  with disconnected propagators that give
nonvanishing contributions.

These can be evaluated similarly to the connected ones.  First, there
are factors common to all three graphs; $(-2/N)$ from the disconnected
propagator, and $3^2=9$ from the ways to contract legs.  In addition,
the particular graphs have the additional factors;\ \ e): $(2N)^2$ from 
the ways of inserting a glueball in one of two index loops in each
$\Stwo$ component, and the trace on the remaining index loop.  f): $(\pm 4)$ from
the fermionic determinant of the $\RPtwo$ component, and $2N$ from the
glueball insertion into the $\Stwo$ component.  Also, there is the same
contribution from the $\Stwo\times \RPtwo$ graph.  g): $(\pm 4)^2$ from two
$\RPtwo$ components.  Altogether we obtain
\begin{align}
 W^{(2)}_{\textrm{disconn}}
 &=
 -{g^2 \over (2m)^3}{1 \over 2! \,3^2}
 \left(-\frac{2}{N}\right)\cdot 9\cdot
 [(2N)^2+2\cdot 2N\cdot(\pm 4)+(\pm 4)^2]\Stwo
 \notag\\
 &={(2N\pm 4)^2\over 4N}\alpha \Stwo.
\end{align}
Summing the connected and disconnected contributions, we obtain
\begin{align}
 W^{(2)}_{\textrm{conn}}+
 W^{(2)}_{\textrm{disconn}}
 =
 \left(\pm 1+\frac{4}{N}\right)\alpha \Stwo
\end{align}
which is the first term of the traceless result \eqref{nhtb21Mar04}.

Higher order diagrams can be worked out in much the same way, although
the number of diagrams increases rapidly.  For the disconnected
diagrams, we only have to consider the diagrams which are
one-particle-reducible with respect to the disconnected propagator.
Therefore, we basically just splice lower order diagrams with the
disconnected propagator.  The contribution is just the product of the
contributions from the lower order pieces, multiplied by the ways to
insert the disconnected propagator into them, and by $(-2/N)^n$ from the
disconnected propagator itself.  However, note that one should also
consider diagrams such as h) of Fig.\ \ref{fig:diagrams for traceless
tensor}.  In this case, the central $D_2$ piece contributes $N$ from its
index loop.

\section{Summary of results from perturbation theory}
\label{KS:summary_pert}

In this appendix we state our results for $W^{\textrm{pert}}(S,\alpha)$,
the perturbative contributions to the glueball superpotential.  These
correspond to evaluating certain diagrams in the matrix model.  We
consider cubic interactions only,
\begin{align}
  W_{\textrm{tree}}=\frac{m}{2}\Tr\,\Phi^2 +\frac{g}{3} \Tr\,\Phi^3~,
\end{align}
which means that we will not consider the case of $SO/Sp$ with adjoint
matter. In any event, it is not necessary to compute the perturbative
superpotential for the latter cases, since closed form expressions for
even power interactions are already known
\cite{Ita:2002kx, Ashok:2002bi, Janik:2002nz}.  The case of $U(N)$ with
adjoint matter is also well known \cite{Brezin:1978sv}, but for convenience
we include it in the list below.

For traceful matter fields, instead of evaluating individual Feynman
diagrams, there exists a much simpler method for computing which we have
used to obtain the results below.  We can simply compute the matrix
model free energy by computer for certain low values of $N$.  Since
$\Stwo$ and $\RPtwo$ diagrams scale as $N^{L+1}$ and $N^L$ at $L$ loops
in perturbation theory, we can easily read off the $\Stwo$ and $\RPtwo$
contributions to any desired order.  For traceless fields things are not
so simple, since the $N$ dependence becomes more complicated, and
certain diagrams must be discarded (as discussed in Section 2.)

We define
\begin{align}
 \alpha = 
 \begin{cases}  
  g^2/m^3   & \qquad U(N)~, \\
  2g^2 /m^3 & \qquad SO/Sp(N)~.
 \end{cases}
\end{align}

\subsection{$U(N)$ with adjoint matter}
\begin{align}
 W^{\textrm{pert}}(S,\alpha) 
 & = 
 N {\partial F_{\chi=2} \over \partial S}~,
 \quad\quad  F_{\chi=2}= -{\Stwo \over 2} \sum_{k=1}^\infty
 { (8 \alpha S)^k \over (k+2)!}{\Gamma\left({3k \over 2}\right) \over
 \Gamma\left({k \over 2}+1 \right)}~, \\
 W^{\textrm{pert}}(S,\alpha)& = -2N\alpha \Stwo
 -{32 \over 3}N \alpha^2 S^3 - {280 \over 3} N \alpha^3 S^4 - 1024 N \alpha^4 S^5 - \cdots~.
\end{align}

\subsection{$SU(N)$ with adjoint matter}
\begin{align}
 W^{\textrm{pert}}(S,\alpha) =0\cdot \alpha \Stwo+0\cdot \alpha^2 S^3+
 0\cdot \alpha^3 S^4 + \cdots  ~.
\end{align}

\subsection{$SO(N)$ with traceful symmetric matter}
\begin{align}
 W^{\textrm{pert}}(S,\alpha)
 &=
 -\left(N+3\right)\alpha \Stwo
 -\left({16\over 3}N+{59\over 3}\right)\alpha^2 S^3\cr
 &-\left({140\over 3}N+197\right)\alpha^3 S^4
 -\left({512}N+{4775\over 2}\right)\alpha^4 S^5
 -\cdots
\end{align}

\subsection{$SO(N)$ with traceless symmetric matter}
\begin{multline}
 W^{\textrm{pert}}(S,\alpha)
 =
 \left(1+{4\over N}\right)\alpha \Stwo
 +\left({1\over 3}-{8\over N}-{160\over 3N^2}\right)\alpha^2 S^3\\
 +\left({1\over 3}-{12\over N} +{256 \over 3N^2}
 +{3584 \over 3N^3}\right)\alpha^3 S^4+\cdots~.
\end{multline}

\subsection{$Sp(N)$ with traceful antisymmetric matter}
\begin{multline}
 W^{\textrm{pert}}(S,\alpha)
 =
 \left(-N+3\right)\alpha \Stwo
 +\left(-{16\over 3}N+{59\over 3}\right)\alpha^2 S^3\\
 +\left(-{140\over 3}N+197\right)\alpha^3 S^4
 +\left(-{512}N+{4775\over 2}\right)\alpha^4 S^5
 +\cdots~.
\end{multline}

\subsection{$Sp(N)$ with traceless antisymmetric matter}
\begin{multline}
 W^{\textrm{pert}}(S,\alpha)
 =
 \left(-1+{4\over N}\right)\alpha \Stwo
 +\left(-{1\over 3}-{8\over N}+{160\over 3N^2}\right)\alpha^2 S^3\\
 +\left(-{1\over 3}-{12\over N} -{256 \over 3N^2}
 +{3584 \over 3N^3}\right)\alpha^3 S^4+\cdots~.
\end{multline}

These results exhibit some remarkable cancellations.   We find a
vanishing result for $SU(N)$ with adjoint matter, and a
cancellation of the terms linear in $N$ for $Sp(N)$ with traceless
antisymmetric matter.  In both cases the cancellation seems to
involve all the diagrams at a given order.   We do not have a
proof of cancellation beyond the order indicated; it would be nice
to provide one and to better understand the significance of this
fact.

\section{Gauge theory results}
\label{KS:GT_results}

In \cite{Cho:1996bi,Csaki:1996eu}, a systematic method for determining
the dynamical superpotential of the $Sp(N)$ gauge theory with a
traceless antisymmetric matter $\At_{ab}$ was proposed.  In this
appendix, we briefly review the strategy, focusing on the $Sp(8)$ case.

First, we add to the theory $2N_F$ fundamentals $Q_i$.  The
moduli space of this enlarged $(N_{\At},N_F)$ theory is parameterized by
\begin{align}
  \Ot_n=\Tr[(\At J)^n],\qquad n=2,3,\cdots, N/2
\end{align}
as well as the antisymmetric matrices
\begin{equation}
\begin{split}
 M_{ij}&=Q_i^T J Q_j~,\\
 N_{ij}&=Q_i^T J\At J Q_j~,\\
 P_{ij}&=Q_i^T J(\At J)^2 Q_j~,\\
 &\cdots\\
 R_{ij}&=Q_i^T J(\At J)^{k-1} Q_j~.
\end{split}
\end{equation}
The basic observation is that for $N_F=3$,  symmetry and
holomorphy considerations restrict the dynamical superpotential to
be of the form
\begin{align}
  W_{(1,3)}^{\textrm{dyn}}
 = \frac{{\textrm{Some\ polynomial\ in\ }}\Ot_n, M_{ij}, \dots, R_{ij}}{\Lambda_{(1,3)}^{b_0}}~,
\end{align}
where $b_0=N-N_F+4=N+1$ and the subscript $(1,3)$ denotes the matter
content $(N_{\At},N_F)$.  The polynomial must of course respect the
various symmetries of the theory.  More significantly, the F-flatness
equations following from $W_{(1,3)}^{\textrm{dyn}}$ can be written in a
$\Lambda$ independent form.  By setting $\Lambda=0$, one sees that the equations
must reduce to the classical constraints which follow upon expressing
the gauge invariant field $\Ot_n$, $M_{ij}$, $\dots$, $R_{ij}$ in terms
of their constituents $Q_i$ and $\At$.  Quantum corrections to these
classical constraints are forbidden by symmetry and holomorphy. These
requirements fix $W_{(1,3)}^{\textrm{dyn}}$ up to an overall
normalization.  Once we have obtained $W_{(1,3)}^{\textrm{dyn}}$, we can
derive the desired $W_{(1,0)}^{\textrm{dyn}}$ by giving mass to $Q_i$ and
integrating them out.

For $Sp(8)$, the above procedure uniquely determines the superpotential
to be (this result appears in \cite{Csaki:1996eu})
\begin{multline}
  W_{(1,3)}^{\textrm{dyn}}
 =\frac{1}{\Lambda^9_{(1,3)}}
 \Big[
 1152(PPP)+6912(RPN)+3456(RRM) -864 \Ot_2 (PNN)
 \\
 -1728 \Ot_2(RNM)+108 \Ot_2^2(NNM) -108 \Ot_2^2(PMM)+9 \Ot_2^3 (MMM)\\
 +192 \Ot_3(NNN)-576 \Ot_3 (RMM) + 144 \Ot_2 \Ot_3 (NMM) +32
 \Ot_3^2(MMM)\\
 +432 \Ot_4 (NNM)+432 \Ot_4(PMM)-36 \Ot_2 \Ot_4 (MMM) \Big],
\end{multline}
up to normalization, where $(ABC)\equiv
\epsilon^{ijklmn}A_{ij}B_{kl}C_{mn}$.  Now that we have obtained
$W_{(1,3)}^{\textrm{dyn}}$, we can integrate out $Q_i$ by adding a mass term
\begin{align}
  W_{(1,3)}^{\textrm{dyn}}\to W_{(1,3)}^{\textrm{dyn}}+\frac{\mu^{ij}}{2} M_{ij}.
\end{align}
When solving the $F$-flatness condition, we can assume that $M_{ij},
N_{ij}, P_{ij}, R_{ij}\propto (\mu^{-1})_{ij}$ since $\mu^{ij}$ is the
only quantity they can depend on.  Plugging back in, we obtain the
$Sp(8)$ superpotential \eqref{njbi21Mar04} and \eqref{njda21Mar04}.  The
same procedure leads to the superpotential \eqref{nivn21Mar04} and
\eqref{niyl21Mar04} for $Sp(4)$ and $Sp(6)$, respectively
\cite{Cho:1996bi,Csaki:1996eu}.


\EndAppendix

\chapter{The generalized Konishi anomaly approach}
\label{KRS}

We derive the Konishi anomaly equations for $\CN=1$ supersymmetric gauge
theories based on the classical gauge groups with matter in 
two-index tensor and fundamental representations, thus 
extending the existing results for $U(N)$.
A general formula is obtained which expresses  solutions to the Konishi
anomaly equation in terms of solutions to the loop equations of the
corresponding matrix model.
This provides an alternative to the diagrammatic proof that the
perturbative part of the glueball superpotential $W_{\textrm{eff}}$ for these
matter representations can be computed from matrix model integrals, and
further shows that the two approaches always give the same result.  The
anomaly approach is found to be computationally more efficient in the
cases we studied.
%
Also, we show in the anomaly approach how theories with a traceless
two-index tensor can be solved using an associated theory with a
traceful tensor and appropriately chosen coupling constants.

\section{Introduction}

The recently established connection
\cite{Dijkgraaf:2002fc,Dijkgraaf:2002vw,Dijkgraaf:2002dh} between matrix
models and the effective superpotentials of certain ${\mathcal N}=1$ gauge
theories provides us with a new tool for studying supersymmetric field
theories.  The connection, originally formulated in the context of
$U(N)$ gauge theories with adjoint matter, has been established
following two distinct approaches, one based on superspace diagrammatics
\cite{Dijkgraaf:2002xd}, and the other on generalized Konishi anomalies
\cite{Cachazo:2002ry}.  These derivations were subsequently generalized
to a few more gauge groups and matter representations, but the list of
examples is actually quite short at present.  In particular, the
diagrammatic approach has been applied to the classical gauge groups
with matter in arbitrary two-index representations
\cite{Ita:2002kx,Ashok:2002bi,Janik:2002nz,Kraus:2003jf,Naculich:2003cz},
while the anomaly approach has so far been used for $U(N)$ with matter
in the adjoint and fundamental representations
\cite{Cachazo:2002ry,Seiberg:2002jq}, in (anti)symmetric tensor
representations \cite{Naculich:2003cz}, and to quiver theories
\cite{Naculich:2003cz,Casero:2003gf}.  \footnote{The Konishi anomalies
have also been applied without direct reference to a matrix model in
\cite{Brandhuber:2003va}.}  So basic questions remain regarding the
general applicability of these ideas, and also whether matrix models can
in fact successfully reproduce the known physics of supersymmetric gauge
theories.

In \cite{Kraus:2003jf}, theories based on the classical gauge groups
with two-index tensor matter were considered using the diagrammatic
approach. In the case\footnote{Our convention is such that $Sp(2) \cong
SU(2)$.}  of $Sp(N)$ with anti-symmetric matter, a comparison was made
against an independently derived dynamical superpotential
\cite{Cho:1996bi,Csaki:1996eu} governing these theories.  The comparison
revealed agreement up to $h-1$ loops in perturbation theory ($h$ is the
dual Coxeter number), and a disagreement at $h$ loops and beyond.
Although it seemed most likely that the disagreement was due to
nonperturbative effects, even at the perturbative level there were a
number of subtleties deserving of further scrutiny. These subtleties
mainly concern the class of diagrams which should be kept in the
evaluation of the superpotential, and whether one is allowed to use Lie
algebra identities to express objects of the form $\Tr ({\mathcal
W}_\alpha )^{2h}$ in terms of lower traces including the glueball
superfield $S \sim \Tr ({\mathcal W}_\alpha)^2$.  Since these subtleties
arise at the same order in perturbation theory as the observed
discrepancies, it seems important to gain a better understanding of
them.  One motivation for the present work was to rederive the results
of \cite{Kraus:2003jf} in the anomaly approach to see if this gives the
same result, and if so, to see which diagrams are effectively being
computed.  We will see that the anomaly approach corresponds to keeping
at most two ${\mathcal W}_\alpha$'s per index loop and not using Lie
algebra identities. So using these rules, whether one computes using
diagrams or anomalies, one finds the same agreements/discrepancies
between the gauge theory and the matrix model.

Another motivation for this work was to apply the anomaly approach to a
wider class of theories.  For the classical gauge groups with certain
two-index tensors plus fundamentals, we will show how solutions to the
Konishi anomaly equations can be obtained from solutions to the loop
equations of the corresponding matrix model.  This leads to the
following general formula for the perturbative contribution to the
effective glueball superpotential
\begin{align}
 W_{\textrm{eff}}
 =
 N\frac{\partial}{\partial S}\CF_{S^2}
 +\frac{w_\alpha w^\alpha}{2}\frac{\partial^2}{\partial S^2}\CF_{S^2}
 +4\CF_{\RPtwo} +\CF_{D^2}
\end{align}
where the ${\mathcal F}$'s are matrix model contributions of a given
topology to the free energy.  This formula generalizes the $U(N)$
results of \cite{Cachazo:2002ry,Seiberg:2002jq,Naculich:2003cz}, as well
as results
\cite{Ita:2002kx,Ashok:2002bi,Janik:2002nz,Kraus:2003jf,Naculich:2003cz}
found using the diagrammatic approach.

In fact, the above formula is only directly applicable to cases in which
no tracelessness condition is imposed on the two-index tensors.  In
\cite{Kraus:2003jf} it was shown that imposing a tracelessness condition
requires one to include additional disconnected matrix model diagrams,
and there was no simple formula relating the superpotential to the free
energy of the traceless matrix model. On the other hand, one expects
that the traceful theory should contain all the information about the
traceless case provided one includes a Lagrange multiplier field to set
the trace to zero. We will show how this works in detail, and find that
indeed, the superpotential of the traceless theory can be extracted from
the free energy of the traceful matrix model. We use this to rederive
and extend some results from \cite{Kraus:2003jf} in a much more
convenient fashion.

The remainder of this paper is organized as follows.  In sections
\ref{KRS:loop_eq_GT} and \ref{section: mm loop equations} we derive the
gauge theory Konishi anomaly equations and the matrix model loop
equations for the theories of interest.  The theories can all be treated
in a uniform way by using appropriate projection operators.  In section
\ref{KRS:conn_bt_GT_and_MM_rslvnt} we discuss some of the subtleties
alluded to above, and then go on to show that solutions to the gauge
theory anomaly equations follow from those of the matrix model loop
equations.  Section \ref{sec:traceless_cases} concerns the effects of
tracelessness.  Details of some of our calculations are given in
appendices \ref{app:loop_eq_gt} and \ref{app:loop_eq_mm}.

Note: As we were preparing the manuscript, \cite{Alday:2003dk} 
appeared which overlaps with some of our discussion. 

\section{Loop equations on the gauge theory side}
\label{KRS:loop_eq_GT}

In this section we derive the gauge theory loop equations for various
gauge groups and matter representations, extending the $U(N)$ result of
\cite{Cachazo:2002ry,Seiberg:2002jq}.

\subsection{Setup}

We consider an $\CN=1$ supersymmetric gauge theory with  tree level
superpotential
\begin{align}
 W_{\textrm{tree}}=
  \Tr[W(\Phi)]+\Qt_\ft m_{\ft f}(\Phi) Q_f,
 \label{eq:TL_superpot_def}
\end{align}
where the two-index tensor $\Phi_{ij}$ is in one of the following
representations:
\begin{itemize}
 \item $U(N)$ adjoint.
 \item $SU(N)$ adjoint.
 \item $SO(N)$ antisymmetric tensor.
 \item $SO(N)$ symmetric tensor, traceful or traceless.
 \item $Sp(N)$ symmetric tensor.
 \item $Sp(N)$ antisymmetric tensor, traceful or traceless.
\end{itemize}
For other $U(N)$ representations, see \cite{Klemm:2003cy,Naculich:2003cz}.
In the $Sp$ cases, the object with the denoted symmetry is related to $\Phi$ by
\begin{align}
 \Phi=
 \begin{cases}
  SJ & \text{$S_{ij}$: symmetric tensor},\\
  AJ & \text{$A_{ij}$: antisymmetric tensor}.
 \end{cases}
\end{align}
Here $J$ is the invariant antisymmetric tensor of $Sp(N)$, namely
\begin{align}
 J_{ij}=\begin{pmatrix}0&\unit_{N/2}\\-\unit_{N/2}&0\end{pmatrix}.
\end{align}
The tracelessness of the $Sp$ antisymmetric tensor is defined with
respect to this $J$, i.e., by $\Tr[AJ]=0$.

Also, $Q_f$ and $\Qt_f$ are fundamental matter fields, with $f$ and
$\ft$ being flavor indices.  In the $U(N)$ case we have $N_f$
fundamentals $Q_f$ and $N_f$ anti-fundamentals $\Qt_{\ft}$, while in the
$SO/Sp$ case we have 
$N_f$ fundamentals $Q_f$.  In the $SO/Sp$ case, $\Qt_\ft$ is
not an independent field but related to $Q_f$ by
\begin{align}
 (\Qt_{\ft})_i
 =
 \begin{cases}
  (Q_{\ft})_i & SO(N),\\
  (Q_{\ft})_j J_{ji} & Sp(N).
 \end{cases}
 \label{eq:def_Qtilde}
\end{align}
In the $Sp$ case, $N_f$ should be taken to be even to avoid the Witten
anomaly \cite{Witten:fp}.

$W$ and $m$ are taken to be polynomials
\begin{align}
 W(z)=\sum_{p=1}^n\frac{g_p}{p}z^p, \qquad
 m_{\ft f}(z)=\sum_{p=1}^{n'} \frac{(m_p)_{\ft f}}{p}z^p,
\end{align}
where in the traceless cases the $p=1$ term is absent from $W(z)$.  
Further, due to
the symmetry properties of the matrix $\Phi$, some  $g_p$ vanish for
certain representations:
\begin{align}
 g_{2p+1}=0 \quad (p=0,1,2,\cdots)
 \qquad\text{for $SO$ antisymmetric / $Sp$ symmetric}.
\end{align}
The symmetry properties of $\Phi$ also imply that the matrices
$(m_p)_{\ft f}$ have the following symmetry properties:
\begin{align}
 (m_p)_{f'f}
 =
\begin{cases}
  (-1)^p(m_p)_{ff'}     & \text{$SO$ antisymmetric,} \\
  (m_p)_{ff'}           & \text{$SO$ symmetric,} \\
  (-1)^{p+1}(m_p)_{ff'} & \text{$Sp$ symmetric,} \\
  -(m_p)_{ff'}          & \text{$Sp$ antisymmetric.}
\end{cases}
 \label{eq:sym_prop_m}
\end{align}

In this and the next few sections, we discuss traceful cases only,
postponing the traceless cases to section \ref{sec:traceless_cases} 
(we regard the $SU(N)$ case as the traceless $U(N)$ case).

\subsection{The loop equations}
\label{section: loop equations}

%

We will be interested in expectation values of chiral operators.  
As in \cite{Cachazo:2002ry,Seiberg:2002jq}, 
\begin{align}
 \{\CW_\alpha,\CW_\beta\}
=
 [\Phi,\CW_\alpha]
=
 \CW_\alpha Q
=
 \Qt \CW_\alpha
=
 0
 \label{eq:chiral_ring_relations}
\end{align}
in the chiral ring. 
Therefore, the complete list of independent single-trace chiral
operators are $\Tr[\Phi^p]$, $\Tr[\CW_\alpha\Phi^p]$,
$\Tr[\CW^2\Phi^p]$, and $\Qt_\ft \Phi^p Q_f$. 
As is standard, we define
\begin{align}
 S= -\frac{1}{32\pi}\Tr[\CW_\alpha \CW^\alpha],
\quad
w_\alpha=\frac{1}{4\pi}\Tr[\CW_\alpha].
\end{align}
The chiral operators can be packaged concisely in terms of the
resolvents
\begin{equation}
\begin{split}
  R(z)
 &\equiv
 -\frac{1}{32\pi^2} \Bracket{\Tr\left[\frac{\CW^2}{z-\Phi}\right]},\qquad
 w_\alpha(z) \equiv  \frac{1}{4\pi}
 \Bracket{\Tr\left[\frac{\CW_\alpha}{z-\Phi}\right]},
 \\
 T(z)
 &\equiv
 \Bracket{\Tr\left[\frac{1}{z-\Phi}\right]},\qquad
 M_{f\ft}(z)
 \equiv
 \Bracket{\Qt_\ft \frac{1}{z-\Phi} Q_f}~.
\end{split} 
\label{eq:def_resolvent_gt}
\end{equation}
Note that the indices of $M_{f\ft}$ are reversed relative to
$\Qt_\ft$, $Q_f$.  The resolvent $w_\alpha(z)$ is nonvanishing only
for $U(N)$; in all other cases $w_\alpha(z)\equiv 0$. This can be
understood as follows.  In these semi-simple cases the Lie algebra
generators are traceless, so we cannot have a nonzero background field
$w_\alpha$.  There being no preferred
spinor direction specified by the background $w_\alpha$, the spinor
$w_\alpha(z)$ can be nothing but zero.  Alternatively, if we integrate
out $\Phi$, then $w_\alpha(z)$ should be of the form 
$\Bracket{\Tr[\CW_\alpha](\Tr[\CW^2])^n}$ by  the chiral ring relations
(\ref{eq:chiral_ring_relations}).  If we use the factorization property
of chiral operator expectation values, this is proportional to $w_\alpha
S^n$, which vanishes.

The resolvents defined in equation (\ref{eq:def_resolvent_gt}) 
provide sufficient data to determine the effective
superpotential up to a coupling independent part, because of the
relation
\begin{align}
 \Bracket{\Tr[\Phi^p]}
 = p\frac{\partial}{\partial g_p}W_{\textrm{eff}},
 \qquad
 \Bracket{\Qt_{\ft} \Phi^p Q_f}
 =
 p\frac{\partial}{\partial (m_p)_{\ft f}}W_{\textrm{eff}}.
 \label{eq:rel_Weff_and_trPhi_etc}
\end{align}

The generalized Konishi \cite{Konishi:1983hf} anomaly equation
\cite{Cachazo:2002ry,Seiberg:2002jq,Brandhuber:2003va} is obtained by
considering the divergence of the current associated with the variation
of a particular field $\Psi_a$:
\begin{align}
 \delta \Psi_a=f_a,
 \label{eq:gK_anomaly_variation}
\end{align}
where $a$ is a gauge index.  Then the anomaly equation reads
\begin{align}
 \Bracket{\frac{\partial W_{\textrm{tree}}}{\partial \Psi_a} f_a}
 +
 \frac{1}{32\pi^2}
 \Bracket{[\CW_\alpha \CW^\alpha]_a^b \frac{\partial f_b}{\partial \Psi_a}}
 =0,
 \label{eq:gK_anomaly}
\end{align}
where $\CW_\alpha$ is in the representation furnished by  $\Psi$.
The first term in (\ref{eq:gK_anomaly}) 
represents the classical change of the action under the variation
(\ref{eq:gK_anomaly_variation}), 
while the second term in (\ref{eq:gK_anomaly}) corresponds to
the quantum variation due to the change in the functional measure.

In the $U(N)$ case considered in \cite{Cachazo:2002ry,Seiberg:2002jq}, 
there is no additional symmetry imposed on the field $\Phi$, 
so $\delta \Phi_{ij}=f_{ij}$ can be any function of
$\CW_\alpha$ and $\Phi$. 
In general, the tensor $\Phi$ will have some symmetry properties 
(symmetric or antisymmetric tensor in the present $SO/Sp$ study),
and $f_{ij}$ should
be chosen to reflect those. 
Similarly, the derivative 
$\partial/\partial \Psi_a=\partial/\partial \Phi_{ij}$ 
should be defined in accord with the symmetry property of $\Phi_{ij}$.
To this end, 
we define a projector $P$ 
appropriate to
each case: 
\begin{align}
 P_{ij,kl}=
  \frac{1}{2}(\delta_{ik}\delta_{jl}+\sigma t_{il} t_{jk}) ,
 \label{eq:def_projector}
\end{align}
where
\begin{align}
 \begin{cases}
  t_{ij}=\delta_{ij},~ \sigma=-1 & \text{$SO$ antisymmetric,}\\
  t_{ij}=\delta_{ij},~ \sigma=+1 & \text{$SO$ symmetric,    }\\
  t_{ij}=J_{ij}     ,~ \sigma=-1 & \text{$Sp$ symmetric,    }\\
  t_{ij}=J_{ij}     ,~ \sigma=+1 & \text{$Sp$ antisymmetric }.
 \end{cases}
 \label{eq:def_sigma_and_t}
\end{align}
The tensor $\Phi_{ij}$ satisfies $P_{ij,kl}\Phi_{kl}=\Phi_{ij}$.  Then,
the symmetry property of $\delta \Phi$ discussed above is implemented by
the replacements
\begin{align}
 f_{a} = f_{ij}\to  P_{ij,kl}f_{kl},
 \qquad
 \frac{\partial}{\partial \Psi_{a}}=
 \frac{\partial}{\partial \Phi_{ij}}\to
 P_{ij,kl}\frac{\partial}{\partial \Phi_{kl}}.
 \label{eq:act_with_projectors_gt}
\end{align}
With this replacement, $f_{ij}$ can be any function of $\CW_\alpha$ and
$\Phi$ as in the $U(N)$ case.  The derivative can be treated as in
the $U(N)$ case also.

There is no such  issue for the $Q$ and $\Qt$ fields, although we have to
remember that they are not independent for $SO/Sp$.

With the projectors in hand, there is no difficulty in
deriving the loop equations for $SO/Sp$.  Here we just present the
resulting loop equations, leaving the details to Appendix
\ref{app:loop_eq_gt}:
\begin{equation}
\begin{split}
 [W'R]_-&= \tfrac{1}{2} R^2, \\
 [W'T+\tr(m'M)]_-&=
 \begin{cases}
  \left(T-\frac{2}{z}\right)R    & \text{$SO$ antisymmetric}, \\
  \left(T-2\frac{d}{dz}\right)R  & \text{$SO$ symmetric    }, \\
  \left(T+\frac{2}{z}\right)R    & \text{$Sp$ symmetric    }, \\
  \left(T+2\frac{d}{dz}\right)R  & \text{$Sp$ antisymmetric},
 \end{cases}
 \\
 2[(Mm)_{ff'}]_-     &=  R\delta_{ff'    }, \\
 2[(mM)_{\ft\ft'}]_- &=  R\delta_{\ft\ft'},
\end{split}
\label{eq:loop_eq_gt}
\end{equation}
where $[F(z)]_-$ means to drop non-negative powers in a Laurent
expansion in $z$.  The last two equations are really the same equation
due to the symmetry properties of 
$m$ (see equation (\ref{eq:sym_prop_m})), and $\Phi$.  
Note that there is no $w_\alpha(z)$ in these
cases as explained below Eq.\ (\ref{eq:def_resolvent_gt}).
For the sake of comparison, the $U(N)$ loop equations  are
\cite{Cachazo:2002ry,Seiberg:2002jq}
\begin{equation}
\begin{split}
 [W'R]_-&=   R^2,\\
 [W'w_\alpha]_-&=2w_\alpha R,\\
 [W'T+\tr(m'M)]_-&=  2TR+w_\alpha w^\alpha,\\
 [(Mm)_{ff'}]_-    &= R\delta_{ff'},\\
 [(mM)_{\ft\ft'}]_- &= R\delta_{\ft\ft'}.
\end{split}
\label{eq:loop_eq_gt_U(N)}
\end{equation}
One observes some extra numerical factors in the $SO/Sp$ case as
compared to the $U(N)$ case.  The $\frac{1}{2}$ in the first equation is
from the $\frac{1}{2}$ in the definition of $P_{ij,kl}$, while the
factor $2$ in the last two equations is because in the $SO/Sp$ case $Q$
and $\Qt$ are really the same field, so the variation of $\Qt mQ$ under
$\delta Q$ for $SO/Sp$ is twice as large as that for $U(N)$.  Finally,
the $\frac{1}{z}R(z)$ and $\frac{d}{dz} R(z)$ 
terms in the second equation of
(\ref{eq:loop_eq_gt}) come from the second term of $P_{ij,kl}$.

The solution to the loop equations (\ref{eq:loop_eq_gt}) or
(\ref{eq:loop_eq_gt_U(N)}) is determined uniquely \cite{Cachazo:2002ry}
given the condition
\begin{align}
 S
 =\oint_{C} \frac{dz}{2\pi i} R(z),\qquad
 w_\alpha
 =\oint_{C} \frac{dz}{2\pi i} w_\alpha(z),\qquad
 N
 =\oint_{C} \frac{dz}{2\pi i} T(z),
 \label{eq:requirement_for_gt_resolvents}
\end{align}
where the second equation is only for the $U(N)$ case.  The contour $C$
goes around the critical point of $W(z)$.  Therefore, if we
recall the relation (\ref{eq:rel_Weff_and_trPhi_etc}), we can say that
the loop equations are all we need to determine the superpotential
$W_{\textrm{eff}}$.

\section{Loop equations on the matrix model side}
\label{section: mm loop equations}

Let us consider the matrix model which corresponds to the gauge theory
in the previous section.  
Its partition function is 
\begin{align}
 \mZ
 =
 e^{-\frac{1}{\mg^2}\mF(\mS)}
 =
 \int d\mPhi d\mQ d\mQt \,
   e^{-\frac{1}{\mg}W_{\textrm{tree}}(\mPhi,\mQ,\mQt)}.
 \label{eq:mm_def}
\end{align}
We denote matrix model quantities by boldface letters.  Here, $\mPhi$ is an
$\mN\times\mN$ matrix with the same symmetry property as the
corresponding matter field in the gauge theory.  $\mQ_f$ and $\mQt_{\ft}$
are defined in a similar way to their gauge theory counterparts
(therefore $d\mQt$ in (\ref{eq:mm_def}) is not included for $SO/Sp$).
The function (or the ``action'') $W_{\textrm{tree}}$ is the one defined in
(\ref{eq:TL_superpot_def}).  We will take the $\mN\to\infty$, $\mg\to 0$
limit with the 't Hooft coupling $\mS=\mg \mN$ kept fixed.  The dependence
of the free energy $\mF(\mS)$ on $\mN$ is eliminated using the relation
$\mN=\mS/\mg$, and we expand $\mF(\mS)$ as
\begin{align}
 \mF(\mS)
 =\sum_{\CM} \mg^{2-\chi(\CM)} \CF_{\CM}(\mS)
 =\CF_{S^2}+\mg\CF_{\RPtwo}+\mg\CF_{D^2}+\cdots,
\end{align}
where the sum is over all compact topologies $\CM$ of the matrix model
diagrams written in the 't Hooft double-line notation, and $\chi(\CM)$ is
the Euler number of $\CM$.  The cases which will be of interest to us are
the sphere $S^2$, projective plane $\RPtwo$, and  disk $D^2$, with
$\chi=2,1$, and 1, respectively.  All other contributions have
$\chi\le 0$.

We define matrix model resolvents as follows:
\begin{align}
 \mR(z)
 \equiv
 \mg\Bracket{\Tr\left[\frac{1}{z-\mPhi}\right]}
 ,\qquad
 \mM_{f\ft}(z)
 \equiv
 \mg\Bracket{\mQt_\ft \frac{1}{z-\mPhi} \mQ_f}.
 \label{eq:def_mm_resolvents}
\end{align}
These resolvents provide sufficient data to determine the free energy $\mF$
up to a coupling independent part since
\begin{align}
 \mg\Bracket{\Tr[\mPhi^p]}
 = p \frac{\partial}{\partial g_p}\mF,\qquad
 \mg\Bracket{\mQt_{\ft}\mPhi^p\mQ_f}
 =  p \frac{\partial}{\partial (m_p)_{\ft f}}\mF.
\end{align}
We expand the resolvents in topologies just as we did for $\mF$:
\begin{align}
 \mR(z)=\sum_{\CM} \mg^{2-\chi(\CM)} \mR_\CM(z),
 \qquad
 %
 \mM(z)=\sum_{\CM} \mg^{2-\chi(\CM)} \mM_\CM(z).
 \label{eq:mm_resolvents_expansion}
\end{align}
Although $\mR_{\RPtwo}$ and $\mR_{D^2}$ are of the same order in $\mg$,
they can be distinguished unambiguously because all terms in $\mR_{D^2}$
contains coupling constants $m_{\ft f}$, while $\mR_{\RPtwo}$ does not
depend on them at all.  This is easily seen in the  diagrammatic expansion of
$\mF$.  Also, because $\mF_{S^2}$ and $\mF_{\RPtwo}$ do not contain $m$,
the expansion of $\mM$ starts from the disk contribution, $\mM_{D^2}$.

Now we can derive the matrix model loop equations.  Consider changing the
integration variables as
\begin{align}
 \delta \mPsi_a=\mf_a.
 \label{eq:mm_variation}
\end{align}
Since the partition function is invariant under this variation, we
obtain
\begin{align}
 0=
 -\frac{1}{\mg}
 \frac{\partial W_{\textrm{tree}}}{\partial \mPsi_a} \mf_a
 +\frac{\partial \mf_a}{\partial \mPsi_a}.
 \label{eq:mm_anomaly_eq}
\end{align}
The first term came from the change in the ``action'' and corresponds to
the first term (the classical variation) of the generalized Konishi
anomaly equation (\ref{eq:gK_anomaly}).  On the other hand, the second
term came from the Jacobian and corresponds to the second term (the
anomalous variation) of Eq.\ (\ref{eq:gK_anomaly}).

The derivation of the loop equations now can be done exactly in parallel
to the derivation of the gauge theory loop equations.  In the $SO/Sp$
case, we again have to consider the projector $P_{ij,kl}$.  Here we
leave details of the derivation to Appendix \ref{app:loop_eq_mm} and present
the results.  For $SO/Sp$, they are
\begin{align}
 \mg \Bracket{ \Tr {W'(\mPhi) \over z - \mPhi} }
 +& \mg \Bracket{\mQt {m'(\mPhi) \over z - \mPhi} \mQ }\notag\\
 &=
 {1\over 2}
 \Bracket{ \left(\mg\Tr {1 \over z - \mPhi}\right)^{\!\!2} }
 \pm
 {\sigma\over 2} \mg^2
 \Bracket{ \Tr {1 \over (z - \mPhi)(z - \sigma \mPhi)}},
 \nonumber\\
 &\hspace*{-5ex}
 2\Bracket{ \mQt_{\ft} {m_{\ft f}(\mPhi)  \over z - \mPhi} \mQ_{f'} }
 =
 \mg
 \Bracket{  \Tr {1 \over z - \mPhi} }
 \delta_{ff'},\notag\\
 &\hspace*{-5ex}
 2\Bracket{ \mQt_{\ft} {m_{\ft' f}(\mPhi)  \over z - \mPhi} \mQ_{f} }
 =
 \mg
 \Bracket{  \Tr {1 \over z - \mPhi} }
 \delta_{\ft\ft'},
 \label{eq:exact_mm_loop_eqs}
\end{align}
in the $SO$ and $Sp$ cases, respectively.
The last two
equations are really the same because of the symmetry properties of
$\mPhi$ and $m_{\ft f}$.

Equations (\ref{eq:exact_mm_loop_eqs}) 
include terms of all orders in $\mg$.  
Expanding the matrix model expectation values in powers 
of $\mg$, 
plugging in the expansion
(\ref{eq:mm_resolvents_expansion}) and comparing the $\CO(1)$ and
$\CO(\mg^1)$ terms, we obtain the $SO/Sp$ loop equations%
\footnote{In
the $SO$ antisymmetric and $Sp$ symmetric cases, $\mR_{\RPtwo}$ can be
expressed \cite{Janik:2002nz,Ashok:2002bi} in terms of $\mR_{S^2}$,
which leads to the expression
\begin{align}
 \CF_{S^2}(\mS)=\mp \frac{1}{2} \frac{\partial}{\partial \mS}\CF_{\RPtwo}
\end{align}
in the $SO$ and $Sp$ cases, respectively.
}.
This is done in Appendix \ref{app:loop_eq_mm}, 
and the results are: 
\begin{equation}
 \begin{split}
 [W'\mR_{S^2}]_- &= \tfrac{1}{2}(\mR_{S^2}{})^2
 \\
 [W'\mR_{\RPtwo}]_-
 &=
 \begin{cases}
  \left(\mR_{\RPtwo}-\frac{1}{2z}\right)\mR_{S^2}           & \text{$SO$ antisymmetric} \\
  \left(\mR_{\RPtwo}-\frac{1}{2}\frac{d}{dz}\right)\mR_{S^2}& \text{$SO$ symmetric} \\
  \left(\mR_{\RPtwo}+\frac{1}{2z}\right)\mR_{S^2}           & \text{$Sp$ symmetric} \\
  \left(\mR_{\RPtwo}+\frac{1}{2}\frac{d}{dz}\right)\mR_{S^2}& \text{$Sp$ antisymmetric}
 \end{cases}
  \\
  [W'\mR_{D^2}+\tr(m'\mM_{D^2})]_-  &= \mR_{D^2} \mR_{S^2}
  \\
  2[(\mM_{D^2} m)_{ff'}]_-     &= \mR_{S^2} \delta_{ff'}
  \\
  2[(m \mM_{D^2})_{\ft\ft'}]_- &= \mR_{S^2} \delta_{\ft \ft'} ,
\end{split}
\label{eq:loop_eq_mm}
\end{equation}
We separated the $\mR_{\RPtwo}$ and $\mR_{D^2}$ contributions using the
difference in their dependence on $m_{\ft f}$ (see the argument below
Eq.\ (\ref{eq:mm_resolvents_expansion})).  
Again, the last two equations are really the same equation.  
For comparison, the $U(N)$ loop
equations are
\begin{equation}
 \begin{split}
 [W'\mR_{S^2}]_- &= (\mR_{S^2}{})^2
 \\
 [W'\mR_{D^2}+\tr(m'\mM_{D^2})]_-  &= 2\mR_{D^2} \mR_{S^2}
 \\
 [(\mM_{D^2} m)_{ff'}]_-     &= \mR_{S^2} \delta_{ff'}
  \\
 [(m \mM_{D^2})_{\ft\ft'}]_- &= \mR_{S^2} \delta_{\ft \ft'} ,
\end{split}
\label{eq:loop_eq_mm_U(N)}
\end{equation}
Note that there is no $\RPtwo$ contribution for $U(N)$.

The solutions to  equations (\ref{eq:loop_eq_mm}) or
(\ref{eq:loop_eq_mm_U(N)}) are determined uniquely given the condition
\begin{align}
 \mS=\oint_{C} \frac{dz}{2\pi i} \mR_{S^2}(z),\qquad
 0=\oint_{C} \frac{dz}{2\pi i} \mR_{\RPtwo}(z),\qquad
 0=\oint_{C} \frac{dz}{2\pi i} \mR_{D^2}(z).
 \label{eq:requirement_for_mm_resolvents}
\end{align}
In this sense, the loop equations are all we need to determine the free
energy $\mF$.

\section{Connection between gauge theory and matrix model resolvents}
\label{KRS:conn_bt_GT_and_MM_rslvnt}

On the gauge theory side we have arrived at the loop equations
(\ref{eq:loop_eq_gt}).  If we can solve these equations for the
resolvents, in particular for $T(z)$, we will have sufficient data to
determine the glueball superpotential $W_{\textrm{eff}}(S)$ up to a coupling
independent part. In \cite{Cachazo:2002ry}, it was shown for $U(N)$ with
adjoint matter that the solution can be obtained with the help of an
auxiliary matrix model. On the other hand, in
\cite{Dijkgraaf:2002xd,Ita:2002kx,Kraus:2003jf,Naculich:2003cz} it was
proved by perturbative diagram expansion that, for $U(N)$ and $SO/Sp$
with two-index tensor matter, if one only inserts up to two field
strength superfield $\CW_\alpha$'s per index loop then the calculation
of $W_{\textrm{eff}}(S)$ reduces to matrix integrals.


However, there are a number of reasons to study further the relation
between the gauge theory and matrix model loop equations.
First, as pointed out in \cite{Kraus:2003jf} (see also p.11 of
\cite{Cachazo:2002ry}, and \cite{Witten:2003ye}), there are subtleties
in using chiral ring relations at order $S^{h}$ and higher, where $h$ is
the dual Coxeter number of the gauge group, and these could be related
to the discrepancies observed in \cite{Kraus:2003jf}.  Since traces of
schematic form $\Tr[(\CW_\alpha^2)^n]$ $(n\ge h)$ can be rewritten in
terms of lower power traces at these orders, imposing chiral ring
relations {\em before\/} using the equation of motion of $S$ is not
necessarily justified.  So, it is important to clarify how this subtlety
is treated in the Konishi anomaly approach.
%
Second, as a practical matter, the anomaly approach is more efficient
than the diagrammatic approach in the cases we studied.


So, let us adopt the following point of view
(some related  ideas were explored in \cite{Brandhuber:2003va}).
Let us not assume the reduction to a matrix model \textit{a priori}.  Then 
the gauge
theory resolvents $R$, $T$, and $M$ are just unknown functions that
enable us to determine the coupling dependent part of the glueball
effective action.  We do know that we can evaluate the perturbative
contribution to them by Feynman diagrams, but we do not know
whether they are affected by nonperturbative effects or whether they can
be calculated using a matrix model.  These resolvents satisfy the loop
equations (\ref{eq:loop_eq_gt}), and given the conditions
(\ref{eq:requirement_for_gt_resolvents}), they are determined uniquely.
Similarly, the matrix model resolvents $\mR_{S^2}$,
$\mR_{\RPtwo}$, $\mR_{D^2}$, and $\mM_{D^2}$ are now just functions
satisfying matrix model loop equations (\ref{eq:loop_eq_mm}).  If we
impose the condition (\ref{eq:requirement_for_mm_resolvents}), these
resolvents are also determined uniquely, and by definition can be 
evaluated in matrix model perturbation theory.

Now, let us ask what the relation between the two sets of resolvents is.
Actually it is simple: if we know the matrix model resolvents, we can
construct the gauge theory resolvents as follows.  In the $SO/Sp$ case,
\begin{equation}
 \begin{split}
 R(z)
 &=
 \mR_{S^2}(z),
 \\
 T(z)
 &=
 N\frac{\partial}{\partial \mS}\mR_{S^2}(z)
 +4\mR_{\RPtwo}(z) +\mR_{D^2}(z),
 \\
 M(z)&=\mM_{D^2}(z)
 \end{split}
 \label{eq:relation_b/t_resolvents}
\end{equation}
with $S$ and $\mS$ identified; 
in the $U(N)$ case, we get 
\begin{equation}
 \begin{split}
 R(z) &= \mR_{S^2}(z),
 \qquad
 w_\alpha(z) = w_\alpha\frac{\partial}{\partial \mS}\mR_{S^2}(z),
 \\
 T(z)
 &=
 N\frac{\partial}{\partial \mS}\mR_{S^2}(z)
 +\frac{w_\alpha w^\alpha}{2}\frac{\partial^2}{\partial \mS^2}\mR_{S^2}(z)
 +\mR_{D^2}(z),
 \\
 M(z)&=\mM_{D^2}(z)
 \end{split}
 \label{eq:relation_b/t_resolvents_U(N)}
\end{equation}
with the same $S = \mS$ identification.\footnote{Some of these relations
have been written down in
\cite{Gopakumar:2002wx,Naculich:2002hi,Naculich:2002hr}.}  One can
easily check that if the matrix model resolvents satisfy the matrix
model loop equations (\ref{eq:loop_eq_mm}) or
(\ref{eq:loop_eq_mm_U(N)}), then the gauge theory resolvents satisfy the
gauge theory loop equations (\ref{eq:loop_eq_gt}) or
(\ref{eq:loop_eq_gt_U(N)}).  The requirement
(\ref{eq:requirement_for_gt_resolvents}) is also satisfied provided that
the matrix model resolvents satisfy the requirement
(\ref{eq:requirement_for_mm_resolvents}).  Further, these relations lead
to
\begin{align}
 \bracket{\Tr[\Phi^p]}_\text{gauge theory}
 &=p\frac{\partial}{\partial g_p}W_{\textrm{eff}}\notag\\
 &=p\frac{\partial}{\partial g_p}
 \left[
 N\frac{\partial}{\partial S}\CF_{S^2}
 +\frac{w_\alpha w^\alpha}{2}\frac{\partial^2}{\partial S^2}\CF_{S^2}
 +4\CF_{\RPtwo} +\CF_{D^2}
 \right],
\end{align}
which implies a relation between the effective superpotential and the
matrix model quantities:
\begin{align}
 W_{\textrm{eff}}
 =
 N\frac{\partial}{\partial S}\CF_{S^2}
 +\frac{w_\alpha w^\alpha}{2}\frac{\partial^2}{\partial S^2}\CF_{S^2}
 +4\CF_{\RPtwo} +\CF_{D^2}
 \label{eq:Weff_ito_mm_F}
\end{align}
up to a coupling independent additive part.  This proves that the gauge
theory diagrams considered in the Konishi anomaly approach reduce to
matrix model integrals for all matter representations considered.
Further, we do not have to take into account nonperturbative effects,
since we can assume a perturbative expansion in the matrix model
(although, strictly speaking, one should also verify that the Konishi
anomalies receive no nonperturbative corrections
\cite{Brandhuber:2003va}).

The relations (\ref{eq:relation_b/t_resolvents}) and 
(\ref{eq:relation_b/t_resolvents_U(N)})  are consistent with inserting
at most two $\CW_\alpha$'s per index loop, but not with 
inserting more than two and then using Lie algebra relations.  For instance,
this can be seen from the diagrammatic expansion of $\mR_{S^2}(z)$.  So
this shows us explicitly which diagrams are being computed in the
Konishi anomaly approach.  

In the $U(N)$ case \cite{Cachazo:2002ry}, it was convenient to collect all
the gauge theory resolvents into a ``superfield'' $\CR$, because of the
``supersymmetry'' under a shift of $\CW_\alpha$ by a Grassmann number, and
one could relate $\CR$ to the matrix model resolvent $\mR_{S^2}$.  This
fact enabled one to extract all the gauge theory resolvents solely from
$\mR_{S^2}$. However, in more general cases this trick does not work,
and we have to relate the two sets of resolvents directly as in
(\ref{eq:relation_b/t_resolvents}).

\section{Traceless cases}
\label{sec:traceless_cases}

So far, we considered two-index traceful matter $\Phi_{ij}$, and
discussed the relation between the gauge theory and the corresponding
matrix model.  In this section, we consider traceless\footnote{In this
section, we denote traceless quantities by tildes to distinguish them
from their traceful counterparts.}  tensors $\Phit_{ij}$.
These traceless tensors were studied in \cite{Kraus:2003jf}, and a
method of evaluating the glueball effective superpotential
$\Wt_{\textrm{eff}}(S)$ from the combinatorics of the matrix model
diagrams was given. However, the precise connection between the gauge
theory and the matrix model quantities was not transparent, since one
had to keep some of the matrix model diagrams and drop others in a way
that seemed rather arbitrary from the matrix model point of view.
Instead, here we show that the calculation of $\Wt_{\textrm{eff}}(S)$ in
gauge theory with {\em traceless\/} matter reduces to a {\em traceful\/}
matrix model\footnote{A connection between traceful and traceless gauge
theories in the Leigh--Strassler deformed $\CN=4$ $SU(N)$ theory was
discussed in \cite{Dorey:2002pq}.}.

\subsection{Traceless gauge theory vs.\ traceful matrix model}

To derive the generalized Konishi anomaly equation for a traceless
tensor we have to use the appropriate projector
\begin{align}
 \Pt_{ij,kl}
 \equiv
 P_{ij,kl}-\frac{1}{N}\delta_{ij}P_{mm,kl}
 =P_{ij,kl}-\frac{1}{N}\delta_{ij}\delta_{kl},
\end{align}
where $P$ is the projector of the corresponding traceful theory; 
the second equality holds for any projector defined in 
(\ref{eq:def_projector}).
The anomaly term (the second term of Eq.\ (\ref{eq:gK_anomaly}))
is the same as in the traceful case, since the trace
part is a singlet and does not couple to the gauge field.  Therefore, the
only difference in the anomaly equation between traceful and traceless
cases is in the
classical variation (the first term of Eq.\ (\ref{eq:gK_anomaly})), namely
\begin{align}
 \Tr[(Pf) \Wt'(\Phi)]
 \to
 \Tr[(\Pt f) \Wt'(\Phi)]
 =
 \Tr[(P f) \Wt'(\Phi)] - \frac{1}{N}\Tr[f]\Tr[\Wt'(\Phi)].
 \label{eq:diff_b/t_traceful_and_traceless}
\end{align}

For definiteness, let us focus on $SU(N)$ adjoint matter, which can be
thought of as traceless $U(N)$ adjoint matter, without fundamentals
added;  
we will generalize the discussion to other groups and 
matter representations afterward. 
In this case, the last term of Eq.\
(\ref{eq:diff_b/t_traceful_and_traceless}) changes the $U(N)$ loop
equation (the first and the third lines of (\ref{eq:loop_eq_gt_U(N)}))
to
\begin{align}
 [\Wt'(z)\Rt(z)]_-+g_1 \Rt(z)= \Rt(z)^2,
 \qquad
 [\Wt'(z)\Tt(z)]_-+g_1 \Tt(z)= 2\Rt(z)\Tt(z).
\end{align}
Note that $w_\alpha(z)=0$ for $SU(N)$.  The constant $g_1$ is
\begin{align}
 g_1\equiv -\frac{1}{N}\Bigl\langle\Tr[\Wt'(\Phit)]\Bigr\rangle.
 \label{eq:def_g1t}
\end{align}
If we define
\begin{align}
 W(z)\equiv \Wt(z)+g_1 z,
 \label{eq:shifted_supoerpot_traceless}
\end{align}
the above equations are
\begin{align}
 [W'(z)\Rt(z)]_- = \Rt(z)^2,\qquad
 [W'(z)\Tt(z)]_- = 2\Rt(z)\Tt(z).
 \label{eq:traceless_loop_eq_gt}
\end{align}
These are of the same form as the loop equations with
{\em traceful\/} matter and the tree level superpotential $W$.  
Therefore, in order to obtain the
effective glueball superpotential $\Wt_{\textrm{eff}}(S)$ for traceless
matter, we can instead solve the traceful theory with the shifted tree
level superpotential $W$, choosing the value of $g_1$ appropriately.
The solution to these loop equations is determined uniquely given the
condition
\begin{align}
 S=\oint_{C}\frac{dz}{2\pi i}\Rt(z),\qquad
 N=\oint_{C}\frac{dz}{2\pi i}\Tt(z).
 \label{eq:contour_cond}
\end{align}
In the case of traceful matter, the contour is around a critical
point of the tree level superpotential.  However, for traceless
matter, the loop equations above tell us that the contour should
be taken around the critical point of the shifted superpotential
(\ref{eq:shifted_supoerpot_traceless}), rather than the original
$\Wt$. This is because we cannot change all the eigenvalues of
$\Phit$ independently due to the tracelessness condition
$\Tr[\Phit]=0$.

Let the resolvents of the traceful theory with tree level
superpotential $W(\Phi)$ be $R$ and $T$, with $g_1$ treated as an
independent variable.  $R$ and $T$ are functions of $z$, $g_{p\ge
1}$ as well as $S$, $N$: $R=R(z;g_{p\ge 1},S)$, $T=T(z;g_{p\ge
1},S,N)$.  We will often omit $S$ and $N$ in the arguments
henceforth to avoid clutter.  Since $R$ and $T$ satisfy the same
loop equations as $\Rt$ and $\Tt$  provided $g_1$ is chosen
appropriately, i.e.\ $g_1=g_1(g_{p\ge 2},S,N)\equiv \gt_1$, it
should be that
\begin{align}
 \Rt(z;g_{p\ge 2})=R(z;g_{p\ge 1}) \big|_{g_1=\gt_1},\qquad
 \Tt(z;g_{p\ge 2})=T(z;g_{p\ge 1}) \big|_{g_1=\gt_1}.
 \label{eq:R=Rt,T=Tt}
\end{align}
These satisfy the conditions (\ref{eq:contour_cond}) given that
$R$ and $T$ satisfy the conditions (\ref{eq:contour_cond}) without
tildes. Expanding these  in $z$, we find
\begin{equation}
\begin{split}
  \bracket{\Tr[\CW^2\Phit^p]}{}_{g_{p\ge 2}}^{\textrm{traceless}}
 &=
 \bracket{\Tr[\CW^2\Phi^p]}{}_{g_{p\ge 1}}^{\textrm{traceful}} \big|_{g_1=\gt_1},
 \\
 \bracket{\Tr[\Phit^p]}{}_{g_{p\ge 2}}^{\textrm{traceless}}
 &=
 \bracket{\Tr[\Phi^p]}{}_{g_{p\ge 1}}^{\textrm{traceful}} \big|_{g_1=\gt_1}.
\end{split} 
\label{eq:equivalence_of_trPhi}
\end{equation}
In particular, setting $p=1$ in the second equation,
\begin{align}
 \bracket{\Tr[\Phi]}{}_{g_{p\ge 1}}^{\textrm{traceful}} \big|_{g_1=\gt_1}
 =
 \left.\left[\frac{\partial}{\partial g_1}T(z;g_{p\ge 1})\right]\right|_{g_1=\gt_1}
 =
 0,
 \label{eq:tracelessness_Rt}
\end{align}
which can be used for determining $g_1$ in terms of all other parameters.%
\footnote{
One might have expected that $g_1$ can be determined by Eq.\
(\ref{eq:def_g1t}).
However, it is easy to show using the relation
(\ref{eq:equivalence_of_trPhi}) that the equation is just the equation
of motion of the traceful theory, which is identically satisfied for
any $g_1$:
$
 0
 \equiv \Bracket{\Tr[W'(\Phi)]}
 =\bracket{\Tr[\Wt'(\Phit)]}+N g_1.
$
}
We infer from Eq.\ (\ref{eq:R=Rt,T=Tt}) equality between the traceless
and traceful effective superpotentials:
\begin{align}
 \Wt_{\textrm{eff}}(g_{p\ge 2},S,N)
 =
 W_{\textrm{eff}}(g_{p\ge 1},S,N)|_{g_1=\gt_1(g_{p\ge 2},S,N)}.
\end{align}
As long as we impose the tracelessness condition
(\ref{eq:tracelessness_Rt}), this correctly reproduces the
relation (\ref{eq:equivalence_of_trPhi}).  Note that $\gt_1$
depends on $N$; this is the origin of the complicated $N$
dependence of $\Wt_{\textrm{eff}}$ found in \cite{Kraus:2003jf}.

Because we know that the traceful theory can be solved by the associated
traceful matrix model, we can calculate the effective superpotential
using that matrix model.  Specifically, in the present case, it is given
in terms of the free energy of the traceful matrix model by
\begin{align}
 \Wt_{\textrm{eff}}(g_{p\ge 2},S,N)
 =
 \left.\left[
 N\frac{\partial}{\partial S}\CF_{S^2}
 \right]\right|_{g_1=\gt_1}.
\end{align}
The function $\gt_1(g_2,g_3,\cdots,S,N)$ is determined by
\begin{align}
 \bracket{\Tr[\Phit]}
 =
 \left.\left[
 N\frac{\partial}{\partial S}\frac{\partial}{\partial g_1}
 \CF_{S^2}
 \right]\right|_{g_1=\gt_1}
 =
 0.
\end{align}

If we add fundamental fields, the shift constant $g_1$ is changed to
\begin{align}
 g_1 \equiv -\frac{1}{N}\Bracket{\Tr[\Wt'(\Phit)]}
 -\frac{1}{N}\Bracket{\Qt_\ft m_{\ft f}Q_f},
\end{align}
but everything else remains the same; we just have to work with
the traceful theory and the shifted tree level superpotential.
$g_1$ is determined by the tracelessness condition.

We only discussed the $SU(N)$ case in the above, but the
generalization to other tensors, i.e., $SO$ traceless symmetric
tensor and $Sp$ traceless antisymmetric tensor, is
straightforward.  We just shift the tree level superpotential as
(\ref{eq:shifted_supoerpot_traceless}), and work with the traceful
theory instead.

\subsection{Examples}

Here we explicitly demonstrate how the method outlined above works 
in the case of a cubic tree level superpotential, 
\begin{align}
 \Wt(\Phit)=\frac{m}{2}\Phit^2+\frac{g}{3}\Phit^3.
\end{align}
The associated traceful tree level superpotential is
\begin{align}
 W(\Phi)=\lambda \Phi+\frac{m}{2}\Phi^2+\frac{g}{3}\Phi^3
\end{align}
($g_1=\lambda,g_2=m,g_3=g$).

\subsubsection{$SU(N)$ adjoint}

We first  consider $SU(N)$ with adjoint matter and no
fundamentals. 
In \cite{Kraus:2003jf} it was found by perturbative computation
to order $g^6$ that the corresponding $W_{{\textrm{eff}}}$ vanishes due
to a cancellation among diagrams.  We will now prove that 
$W_{{\textrm{eff}}} = 0$ to all orders in $g$.

The planar contributions to the free energy of the 
traceful matrix model can be
computed exactly by the standard method \cite{Brezin:1978sv}:
\begin{align}
\CF_{S^2}
 =& S W_0 +\frac{1}{2} S^2  \ln\left( \frac{\tilde{m}}{ \sqrt{1+y}m}\right)
-\frac{2}{ 3} \frac{S^2 }{ y}
\left[1+\frac{3}{ 2} y
+\frac{1}{ 8} y^2-(1+y)^{3/2}\right]
\end{align}
with
\begin{align}
& \tilde{m}  = \sqrt{m^2-4 \lambda g} \nonumber \\
& W_0  = \frac{1}{2g} (\tilde{m}-m)\left(\lambda +\frac{1}{12g}
(\tilde{m}-m)(\tilde{m}+2m)\right) \nonumber \\
&\frac{y}{(1+y)^{3/2}}  = \frac{8 g^2 S}{m^3}.
\end{align}
We discarded some $g$ independent contributions.  The $W_0$ term 
arises from shifting $\Phi$ to eliminate the linear term in $W(\Phi)$. 
The superpotential is therefore
\begin{align}
W_{{\textrm{eff}}}= N \frac{\partial \CF_{S^2}}{ \partial S}
=  N W_0  +  \frac{NS}{ 6 y}
 \left[-4-6y+6y \ln\left(\frac{\tilde{m}}{ m
\sqrt{1+y}}\right)+4(1+y)^{3/2}\right].
\label{Weff}
\end{align}
Imposing $\partial W_{{\textrm{eff}}} / \partial \lambda =0$ leads to,
after some algebra,
\begin{align}
\lambda = - \frac{2 g S}{m}, \quad y = \frac{8 g^2 S}{m^3}.
\end{align}
Substituting back into (\ref{Weff}) and doing some more algebra, we find
\begin{align}
W_{{\textrm{eff}}} = 0.
\end{align}
This vanishing of the perturbative contribution to the effective
superpotential is consistent with the gauge theory analysis of
\cite{Ferrari:2002jp}.   In fact, it is shown there that $W_{{\textrm{eff}}} = 0$
for any tree level superpotential with only odd power interactions. 


\subsubsection{$Sp(N)$ antisymmetric tensor}

Now consider $Sp(N)$ with an antisymmetric tensor and no fundamentals.
By diagram calculations or by computer, the planar and $\RPtwo$
contributions to the free energy of the traceful matrix model are
\begin{align}
 \CF_{S^2}
 =&
 -\frac{\lambda^2S}{2 m}
 -
 \left(\frac{\lambda S^2}{2m^2}+\frac{\lambda^3 S}{3m^3}\right)g
 -
 \left(\frac{S^3}{6m^3}+\frac{\lambda^2 S^2}{m^4}+\frac{\lambda^4S}{2m^3}\right)
 g^2
 \nonumber\\
 &-
 \left(\frac{\lambda S^3}{m^5}+\frac{8\lambda^3 S^2}{3m^6}
   +\frac{\lambda^5 S}{m^7}\right)
 g^3
 -
  \left(
   \frac{S^4}{3m^6}
  +\frac{5 \lambda^2S^3}{m^7}
  +\frac{8 \lambda^4S^2}{m^8}
  +\frac{7\lambda^6S}{3m^9}
 \right)g^4
 \nonumber\\
 &
 -
  \left(
   \frac{4 \lambda S^4 }{m^8}
  +\frac{70 \lambda^3 S^3}{3 m^9}
  +\frac{128 \lambda^5 S^2 }{5 m^{10}}
  +\frac{6\lambda^7 S}{m^{11}}
 \right)
 g^5
 \nonumber\\
 &
 -
  \left(
   \frac{7S^5}{6m^9}
  + \frac{32\lambda^2 S^4}{m^{10}}
  + \frac{105 \lambda^4 S^3}{m^{11}}
  + \frac{256 \lambda^6 S^2}{3m^{12}}
  + \frac{33 \lambda^8 S}{2m^{13}}
 \right)
 g^6
 \nonumber\\
 &
 -
  \left(
   \frac{21 \lambda S^5}{m^{11}}
  + \frac{640 \lambda^3 S^4}{3m^{12}}
  + \frac{462 \lambda^5 S^3}{m^{13}}
  + \frac{2048 \lambda^7 S^2}{7m^{14}}
  + \frac{143 \lambda^9 S}{3m^{15}}
 \right)
 g^7
 \nonumber\\
 &
 -
  \biggl(
    \frac{16              S^6}{3 m^{12}}
   + \frac{231 \lambda^2   S^5}{ m^{13}}
   + \frac{1280 \lambda^4  S^4}{ m^{14}}\notag\\
 &\qquad\qquad\qquad
   + \frac{2002 \lambda^6  S^3}{ m^{15}}
   + \frac{1024 \lambda^8  S^2}{ m^{16}}
   + \frac{143  \lambda^{10}S^1}{ m^{17}}
 \biggr)
 g^8
 +\CO(g^9),
 \label{eq:corr_traceful_F_S2_Sp(N)_mm}
\end{align}
\begin{align}
 \CF_{\RPtwo}
 =&
 \frac{\lambda S}{2 m^2}g
 +
 \left(\frac{3\lambda S^2}{8m^3}+\frac{\lambda^2 S}{m^4}\right)g^2
 +
 \left(\frac{9\lambda S^2}{4m^5}+\frac{8\lambda^3 S}{3m^6}\right)
 g^3
 \nonumber\\
 &+
  \left(
   \frac{59 S^3}{48 m^6}
  +\frac{45\lambda^2 S^2}{4m^7}
  +\frac{8 \lambda^4 S}{m^8}
 \right)g^4
 +
  \left(
   \frac{59 \lambda S^3}{4m^8}
   +\frac{105 \lambda^3 S^2}{2m^9}
   +\frac{128 \lambda^5 S}{5m^{10}}
 \right)g^5
 \nonumber\\
 &+
  \left(
   \frac{197 S^4}{32m^9}
   +\frac{118\lambda^2 S^3}{m^{10}}
   +\frac{945 \lambda^4 S^2}{4m^{11}}
   +\frac{256 \lambda^6 S}{3m^{12}}
 \right)g^6
 \nonumber\\
 &+
  \left(
   \frac{1773 \lambda S^4}{16m^{11}}
   +\frac{2360 \lambda^3 S^3}{3 m^{12}}
   +\frac{2079 \lambda^5 S^2}{2m^{13}}
   +\frac{2048\lambda^7 S}{7m^{14}}
 \right)g^7
 \nonumber\\
 &+
  \left(
    \frac{4775           S^5}{128 m^{12}}
   +\frac{19503\lambda^2 S^4}{16  m^{13}}
   +\frac{4720 \lambda^4 S^3}{    m^{14}}
   +\frac{9009 \lambda^6 S^2}{2   m^{15}}
   +\frac{1024 \lambda^8 S  }{    m^{16}}
 \right)g^8\notag\\
 &+\CO(g^9),
 \label{eq:corr_traceful_F_RP2_Sp(N)_mm}
\end{align} 
up to a $\lambda$ and $g$ independent part.
From the tracelessness (\ref{eq:tracelessness_Rt}), we find
\begin{align}
 \lambda
 =&
 \left(-1+\frac{2}{N}\right)\frac{S}{m}g
 +\left(-{3\over N}+{12\over N^2}\right)\frac{S^2}{m^4}g^3
 +\left(-{1\over N}-{24\over N^2}+{160\over N^3}\right)\frac{S^3}{m^7}g^5
 \nonumber\\
 &+
 \left(-{3\over 4 N}-{27\over N^2}-{192\over N^3}+{2688\over N^4}
  \right)\frac{S^4}{m^{10}}g^7
 +\CO(g^9)
 \equiv \lambdat.
\end{align}
Therefore, the effective superpotential is, up to an $\alpha$
independent additive part,
\begin{align}
 \Wt_{\textrm{eff}}
 =&
 W_{\textrm{eff}}|_{\lambda=\lambdat}
 =
 \left.\left[
   N\frac{\partial}{\partial S}\CF_{S^2} + 4\CF_{\RPtwo}
 \right]\right|_{\lambda=\lambdat}
 \nonumber\\
 =&
 \left(-1+\frac{4}{N}\right)S^2\alpha
 +\left(-\frac{1}{3}-\frac{8}{N}+\frac{160}{3N^2}\right)S^3\alpha^2
 \nonumber\\
 &\qquad\qquad+
 \left(-\frac{1}{3}-\frac{12}{N}-\frac{256}{3N^2}+\frac{3584}{3N^3}\right)S^4\alpha^3
 \nonumber\\
 &\qquad\qquad
 +\left(-\frac{1}{2}-\frac{24}{N}-\frac{352}{N^2}+\frac{33792}{N^4}\right)S^5\alpha^4
  +\cdots,
\end{align}
where $\alpha\equiv\frac{g^2}{2m^3}$.  This reproduces the result of
\cite{Kraus:2003jf} up to $\CO(\alpha^3)$ and extends it further to
$\CO(\alpha^4)$.

From these examples, the advantage of the present approach over the
traceless diagram approach of \cite{Kraus:2003jf} should be clear.  In
that approach, one has to evaluate contributing diagrams order by order
and evaluating the combinatorics gets very cumbersome.  On the other
hand, in this traceful approach, there is no issue of keeping and
dropping diagrams, and calculations can be done more systematically.
Therefore, being able to reduce the traceless problem to a traceful
problem is a great advantage.

\subsection{Traceless matrix model}

We saw that the traceless gauge theory can be solved by the traceful
matrix model, not the traceless matrix model.  In the following, we
argue that the traceless matrix model is not useful in determining the
effective superpotential of the traceless gauge theory, $\Wt_{\textrm{eff}}$.
The relation among traceless and traceful theories, as far as the
effective superpotential is concerned, is shown in Fig.\
\ref{fig:rel_traceless/traceful_gt_mm}.

%
%
\begin{figure}[h]
 \begin{align}
 \begin{array}{ccc}
  \fbox{traceful gauge theory} & \leftrightarrow & \fbox{traceful matrix model}
   \\[1ex]
  \downarrow && \downarrow \\[1ex]
  \fbox{traceless gauge theory} &\not\leftrightarrow & \fbox{traceless
  matrix model}
 \end{array}
 \end{align}
 \caption{Relation among traceful and traceless theories}
 \label{fig:rel_traceless/traceful_gt_mm}
\end{figure}

The matrix model loop equation for traceless matter can be derived
almost in parallel to the traceless gauge theory loop equation derived
in the previous subsection.  Again, we replace the projector $P$ with
the appropriate traceless version $\Pt$.  For example, in the case of
$SU(N)$ adjoint without fundamentals, which was considered in the
previous section on the gauge theory side, the loop equation is
\begin{align}
 [W'\mRt_{S^2}]_- &= (\mRt_{S^2}{})^2.
 \label{eq:traceless_loop_eq_mm}
\end{align}
Here $W$ is the shifted superpotential defined in
(\ref{eq:shifted_supoerpot_traceless}), with $g_1$ defined in
(\ref{eq:def_g1t}) and the gauge theory expectation values replaced by
the matrix model expectation values.

Eq.\ (\ref{eq:traceless_loop_eq_mm}) is of the same form as the traceful
matrix model loop equation, and the first equation of the traceless gauge
theory loop equations (\ref{eq:traceless_loop_eq_gt}).  Finally, 
using the equivalence of the traceful gauge theory and matrix model, we
conclude that
\begin{align}
 \mRt(z;g_{p\ge 2})
 = \mR(z;g_{p\ge 1}) \big|_{g_1=\gt_1}
 = R(z;g_{p\ge 1}) \big|_{g_1=\gt_1}
 = \Rt(z;g_{p\ge 2}).
\end{align}
However, what we need to determine $\Wt_{\textrm{eff}}$ is $\Tt$, which we
saw in the last subsection to be obtainable from the traceful theory as
\begin{align}
 \Tt(z;g_{p\ge 2},S,N)
 &=
 T(z;g_{p\ge 2},S,N) \big|_{g_1=\gt_1(g_2,g_3,\cdots,S,N)}\notag\\
 &= \left.\left[
 N\frac{\partial}{\partial S}R(z;g_{p\ge 2})
 \right]\right|_{g_1=\gt_1(g_2,g_3,\cdots,S,N)}.
 \label{eq:Tt_ito_R}
\end{align} From the standpoint of the traceless matrix model, the only thing we
know is $\mRt=\Rt=R|_{g_1=\gt_1}$, and we have no information about the
$g_1$ dependence of $R$.  In the framework of the traceless matrix
model, there is no way of performing the derivative $\partial/\partial
S$ in (\ref{eq:Tt_ito_R}) before making the replacement $g_1=\gt_1$,
because $\gt_1$ depends on $S$ also.

Therefore, it is impossible to obtain the effective superpotential for
the traceless gauge theory directly, just by using the data from the
corresponding traceless matrix model.  We really need to invoke the
traceful matrix model.

%
%
%

\BeginAppendix

\section{Loop equations on the gauge theory side}
\label{app:loop_eq_gt}

In this appendix, we are going to calculate the gauge theory loop
equations using the approach of \cite{Cachazo:2002ry, Seiberg:2002jq}.  We start with generalized Konishi currents and
corresponding transformations of the fields \begin{eqnarray} \label{eq: most
generalized currents}
\begin{matrix}
J_f &\equiv& 
\Tr \Phi^\dagger e^{V_{\textrm{adj}}} f (\CW_\alpha, \Phi) 
\hfill
&
\quad
\so 
&
\delta \Phi &=& f (\CW_\alpha, \Phi) 
\hfill
\cr
J_g &\equiv& 
Q^\dagger_f e^{V_{\textrm{fund}}} g_{f f'} (\Phi) Q_{f'} 
&
\quad
\so 
&
\delta Q_{f} &=& g_{f f'} (\Phi) Q_{f'} 
\end{matrix}
\end{eqnarray} 
The explicitly written indices on the $Q_f$'s and $g_{f f'}$ are 
flavor indices, and gauge indices are suppressed.  
We find the generalized anomaly equations 
\begin{eqnarray}
\label{eq: generalized anomaly}
\bar D^2 J_f
&=& 
\Tr f(\CW_\alpha , \Phi) W'(\Phi)
+
\tilde Q f(\CW_\alpha , \Phi) m'(\Phi) Q
+
\sum_{jklm} 
A_{jk,lm} 
{\partial f_{kj}
\over
\partial \Phi_{lm} }
\nonumber\\
\bar D^2 J_g
&=& 
2 
\tilde Q m(\Phi) g(\Phi) Q
+
\Tr A^{\textrm{fund}} g(\Phi)
\end{eqnarray}
and $\bar D^2 J_f$ and $\bar D^2 J_g$ 
vanish in the chiral ring.

The field $\Phi$ being considered transforms by commutation 
under gauge transformations, 
so the elementary anomaly coefficient 
is the same as the one appearing in \cite{Cachazo:2002ry},  
\begin{eqnarray}
\label{eq: elementary anomaly coefficient}
A_{jk,lm} 
&=& 
{1 \over 32 \pi^2} 
\left[
(\CW_\alpha \CW^\alpha)_{jm} \delta_{lk} 
+
(\CW_\alpha \CW^\alpha)_{lk} \delta_{jm} 
-
2 (\CW_\alpha)_{jm} (\CW^\alpha)_{lk} 
\right]
\hspace{-1.5ex}
\nonumber\\
&\equiv&
{1 \over 32 \pi^2} 
\left\{
\CW_\alpha , \left[ 
\CW^\alpha, e_{ml} 
\right]
\right\}_{jk}
\end{eqnarray}
where $e_{ml}$ is the basis matrix with the single non-zero entry
$(e_{ml})_{jk} = \delta_{mj} \delta_{lk}$. 
For fields transforming 
in the fundamental representation we should use 
\begin{eqnarray}
\label{eq: elementary anomaly coefficient: fundamental}
A_{jk}^{\textrm{fund}} 
&=& 
{1 \over 32 \pi^2} 
(\CW_\alpha \CW^\alpha)_{jk} 
\end{eqnarray}

There is one modification in the treatment of fundamental fields,
as compared to the $U(N)$ case studied in \cite{Seiberg:2002jq}.
Since the fundamental representation is real for $SO$ and
pseudo-real for $Sp$, the fields $Q$ and $\tilde Q$ are not
independent; instead, they are related by (\ref{eq:def_Qtilde}).
This results in the factor of 2 in the second equation in
(\ref{eq: generalized anomaly}), but otherwise the discussion
proceeds as in \cite{Seiberg:2002jq}. In the rest of the Appendix
we omit reference to fundamentals.

Next we consider the symmetries of $\Phi$. 
In equation (\ref{eq: most generalized currents}), 
$f = \delta \Phi$ 
must have the same symmetry properties as $\Phi$ itself. 
The tensor field will be taken either symmetric or antisymmetric.
We can discuss all four cases in a uniform fashion 
by using the notation 
\begin{eqnarray}
\Phi^T = 
\left\{
\begin{matrix}
\sigma \Phi
\hfill
&
\quad
\mbox{for groups $SO(N)$,}
\cr
\sigma J \Phi J^{-1}
&\quad
\mbox{for groups $Sp(N)$,}
\end{matrix}
\right.
\end{eqnarray}
and $\sigma = \pm 1$. 
The gauge field satisfies 
$\CW_\alpha{}^T = - \CW_\alpha$ for $SO$ groups, and 
$\CW_\alpha{}^T = - J \CW_\alpha J^{-1}$ for $Sp$ groups. 
%
%
As discussed in Subsection \ref{section: loop equations}, $\Phi$
has the property \begin{eqnarray} \label{eq: projector: general} \Phi = P \Phi
,\quad\mbox{or explicitly}\quad \Phi_{ab} = P_{ab,ij} \Phi_{ij}
\end{eqnarray} with the projectors defined in (\ref{eq:def_projector}). To
ensure that $f$ has the same symmetry as $\Phi$, we should replace
$f \to P f$ in (\ref{eq: generalized anomaly}). Specifically, we
will take $\delta \Phi$ of the form
\begin{align} 
\label{eq: delta Phi:so}
f^{SO} &= P^{SO} {B \over z - \Phi} 
= \left( {B \over z - \Phi}
\right) + \sigma \left( {B \over z - \Phi} \right)^T = \left( {B
\over z - \Phi} \right) + \sigma \left( {B^T \over z - \sigma
\Phi} \right) ,
\\
\label{eq: delta Phi:sp}
f^{Sp} &= 
P^{Sp} {B \over z - \Phi} =
\left( {B \over z - \Phi} \right) 
+ \sigma J \left( {B \over z - \Phi} \right)^T J
= 
\left( {B \over z - \Phi} \right) 
+ \sigma \left( {J B^T J \over z -\sigma \Phi} \right),
\end{align}
with $B = 1$ or $B = \CW^2 \equiv \CW_\beta \CW^\beta$. 
Using the symmetry of the gauge field
and the chiral ring relations, 
both (\ref{eq: delta Phi:so}) and (\ref{eq: delta Phi:sp}) 
reduce to 
\begin{eqnarray}
\label{eq: delta Phi: in CR only}
f &=& 
{B \over z - \Phi}  
+ \sigma 
{B \over z - \sigma \Phi} 
.
\end{eqnarray}
%
%
Also, to take derivatives with respect to matrix elements%
\footnote{
    In the case of $U(N)$ of \cite{Cachazo:2002ry},
    one had $P_{lm,ab} = (e_{lm})_{ab}$
    which satisfies $(e_{lm})_{ab} B_{ab} = B_{lm}$, for any matrix $B$.
    }
correctly we should set 
\begin{eqnarray}
\label{eq: change of matrix elements}
\partial_{lm} \Phi_{ab} 
= 
P_{lm,ab}
\end{eqnarray}
Then the tensor field anomaly term becomes 
\begin{eqnarray}
\label{eq: anomaly: general}
A_{jk,lm} 
\; 
\partial_{lm} f_{kj}
&=& {1 \over 32 \pi^2} \left[ (\CW^2)_{jm} \delta_{lk} +
\delta_{jm} (\CW^2)_{lk} - 2 (\CW_\alpha)_{jm} (\CW^\alpha)_{lk}
\right] \nonumber\\&&\hspace{1.4em}\times \left[ \left( {B \over z
- \Phi} \right)_{ra} \left( {1 \over z - \Phi} \right)_{bs}
P_{kj,rs} P_{lm,ab} \right]. \end{eqnarray} After using the projectors
(\ref{eq:def_projector}), the identity $\Tr \CW_\alpha \Phi^k =
0$, the symmetry properties of $\Phi$ and $\CW_\alpha$, and the
chiral ring relations, we find
\begin{eqnarray} \label{eq: anomaly: simplified}
A_{jk,lm} \;
\partial_{lm} f_{kj}
&=& {1 \over 32 \pi^2} \left[ \left( \Tr {\CW^2 \over z - \Phi}
\right) \left( \Tr {B \over z - \Phi} \right) + \left( \Tr {1
\over z - \Phi} \right) \left( \Tr {\CW^2 B \over z - \Phi}
\right) \right.\nonumber\\&&\left.\hspace{3em} + 4 k \sigma \; \Tr
{\CW^2 B \over (z - \Phi) (z - \sigma \Phi)} \right] \end{eqnarray} The only
difference in (\ref{eq: anomaly: simplified}) between the two
types of gauge groups is that the sign in front of the single
trace term is $k = +1$ for $SO$, and $k = -1$ for $Sp$.
Taking $B = 1$ and $B = \CW^2$ in (\ref{eq: anomaly: simplified}) we find 
\begin{align}
\label{eq: anomaly: simplified: B=1}
0 &=
\Tr { W'(\Phi) \over z - \Phi}
+ \sigma \,
\Tr { W'(\Phi) \over z - \sigma \Phi}
\nonumber\\&
\quad+
{2 \over 32 \pi^2} 
\left[
\left( \Tr {\CW^2 \over z - \Phi} \right) 
\left( \Tr {1 \over z - \Phi} \right) 
+ 2 k \sigma 
\left( \Tr {\CW^2 \over (z - \Phi)(z - \sigma \Phi)} \right) 
\right]
\\
\label{eq: anomaly: simplified: B=W-squared}
0
&=
\Tr { \CW^2 W'(\Phi) \over z - \Phi}
+ \sigma \,
\Tr { \CW^2 W'(\Phi) \over z + \Phi}
+ {1 \over 32 \pi^2} \left[ \left( \Tr {\CW^2 \over z - \Phi}
\right) \left( \Tr {\CW^2 \over z - \Phi} \right) \right] \end{align} Now
recall that $W(\Phi)^T = W(\Phi)$ for $SO(N)$, and $W(\Phi)^T = J
W(\Phi) J^{-1}$ for $Sp(N)$ since it only appears inside a trace;
so \begin{eqnarray} \Tr { W'(\Phi) \over z -\sigma \Phi} &=& \sigma \Tr {
W'(\Phi) \over z - \Phi} ,\quad \Tr { \CW^2 W'(\Phi) \over z
-\sigma \Phi} = \sigma\Tr { \CW^2 W'(\Phi) \over z - \Phi} . \end{eqnarray}
The single trace terms have to be treated separately: when $\sigma
= -1$, \begin{eqnarray} \Tr { \CW^2 \over z^2 - \Phi^2} &=& {1\over 2 z} \; \Tr
\left[ \CW^2 \left( { 1 \over z - \Phi} + { 1 \over z + \Phi}
\right) \right] = {1\over z} \; \Tr { \CW^2 \over z - \Phi} \end{eqnarray}
while for $\sigma = +1$, we should use \begin{eqnarray} \Tr { \CW^2 \over (z -
\Phi)^2} &=& - {d\over dz} \; \Tr { \CW^2 \over z - \Phi} . \end{eqnarray}

Putting everything together, we find the loop equations written in
equation (\ref{eq:loop_eq_gt}).

\section{Loop equations on the matrix model side}
\label{app:loop_eq_mm}

Here we derive the matrix model loop equations for $SO/Sp$
following Seiberg \cite{Seiberg:2002jq}, who discussed the $U(N)$
case.  Start with the matrix model partition function
\begin{eqnarray}
\label{eq: mm partition function}
Z = \int d \mPhi d \mQ\,
\exp 
\left\{
- {1 \over \mg} \left[
\Tr[W(\mPhi)]
+ 
\mQt_{\tilde f} m_{\tilde f f}(\mPhi) \mQ_f\right]
\right\}.
\end{eqnarray}
Because the fundamental matter is real for $SO(N)$ and pseudo-real
for $Sp(N)$, there is no integration over $\mQt$.  It is not an
independent variable, but related to $\mQ$ by Eq.\
(\ref{eq:def_Qtilde}).  We will write the symmetry properties of
the tensor field $\mPhi$ as
\begin{eqnarray}
\mPhi^T = 
 \begin{cases}
  \sigma \mPhi & SO(N), \\
  \sigma J\mPhi J^{-1} & Sp(N). 
 \end{cases}
\end{eqnarray}
where $\sigma = \pm 1$.  The matrix $m(\mPhi)$ has symmetry properties as
given in Eq.\ (\ref{eq:sym_prop_m}).

Now we perform two independent transformations 
\begin{eqnarray}
\label{eq: transformations: mm}
\delta \mPhi = B P {1 \over z - \mPhi}
,\quad
\delta \mQ_f = \lambda_{f f'} {1 \over z - \mPhi} \mQ_{f'}
\end{eqnarray}
where $B$ (number) and $\lambda$ (matrix) are independent and
infinitesimal.  
To make sure that $\delta \mPhi$ has the same 
symmetry properties as $\mPhi$ itself, we have introduced 
the appropriate projector $P$ in (\ref{eq: transformations: mm}), 
see Eq.\ (\ref{eq:def_projector}).  
The measure in
(\ref{eq: mm partition function}) changes as
\begin{eqnarray}
d \mPhi &\to& d\mPhi\, J_\mPhi = d\mPhi\, (1 + \Delta_\mPhi),
\nonumber\\
d \mQ &\to& d\mQ\, J_\mQ = d\mQ\, (1 + \Delta_\mQ)
\end{eqnarray}
to first order in $B$ and $\lambda$, 
where the corresponding changes in the Jacobians are 
\begin{eqnarray}
\Delta_\mPhi &=& 
B 
P_{ij,ab} 
\left( {1 \over z - \mPhi} \right)_{ia} 
\left( {1 \over z - \mPhi} \right)_{bj}
\nonumber\\
\Delta_\mQ &=& 
\left( \Tr {1 \over z - \mPhi} \right)
\left( \tr \lambda \right),
\end{eqnarray}
where $\tr$ is a trace over the flavor indices.
The classical pieces change by 
\begin{eqnarray}
\delta \, \Tr[W(\mPhi) ]
= B \, \Tr\left[W'(\mPhi) P {1 \over z - \mPhi}\right]
= B \, \Tr {W'(\mPhi) \over z - \mPhi}
.
\end{eqnarray}
One can show the second equality using symmetry properties of 
$W$: since it only enters $Z$ in the form of the trace, we should 
take $W(\mPhi^T) = W(\mPhi)$ for $SO$ and 
$W(\mPhi^T) = J W(\mPhi) J^{-1}$ for $Sp$. 
Similarly, 
\begin{eqnarray}
\delta ( \mQt m \mQ )
&=& 
\mQt \left( {\lambda^T m \over z - \sigma \mPhi} 
+ {m \lambda \over z - \mPhi} \right)\mQ 
+ B \mQt m' \left( P {1 \over z - \mPhi} \right) \mQ 
\nonumber\\
&=&
2 \mQt {m \lambda \over z - \mPhi} \mQ 
+ B \mQt {m' \over z - \mPhi} \mQ 
,
\end{eqnarray}
where we used a similar symmetry property of the matrix $m$. 
Finally, with the explicit form of the projectors 
(\ref{eq:def_projector}) we find that in all four cases the 
statement $\delta Z = 0$ gives two independent loop equations 
(one for $B$, and one for $\lambda$): 
\begin{align}
\label{eq: exact mm loop equations}
{1\over 2}
\left\langle
\left(\mg
\Tr {1 \over z - \mPhi}
\right)^2
\right\rangle
&\pm
{\sigma\over 2} {\mg}
\left\langle
{\mg} \Tr {1 \over (z - \mPhi)(z - \sigma \mPhi)}
\right\rangle\notag\\
&\qquad\qquad
 =
\left\langle
{\mg} \Tr {W'(\mPhi) \over z - \mPhi}
\right\rangle
+ {\mg}
\left\langle
\mQt {m'(\mPhi) \over z - \mPhi} \mQ
\right\rangle,
\nonumber\\
\left\langle
{\mg} \Tr {1 \over z - \mPhi}
\right\rangle
\delta_{ff'}
&=
2
\left\langle
\mQt_{\ft} {m_{\ft f}(\mPhi)  \over z - \mPhi} \mQ_{f'}
\right\rangle,
\end{align}
for $SO$ and 
$Sp$,, respectively.
This is Eq.\ (\ref{eq:exact_mm_loop_eqs}) 
quoted in Section \ref{section: mm loop equations}.

As it is written, equation (\ref{eq: exact mm loop equations}) 
includes all orders in $\mg$. The anomaly term in the first 
equation (\ref{eq: exact mm loop equations}) factorizes as 
\begin{eqnarray}
 \left\langle
\left(\mg
\Tr {1 \over z - \mPhi}
\right)^2
\right\rangle
= 
\left\langle
\mg
\Tr {1 \over z - \mPhi}
\right\rangle^2
\times \left[ 1 + \CO(\mg^2) \right]
\end{eqnarray}
as can be seen from a diagram expansion. 
With this and the definition of matrix model resolvents 
(\ref{eq:def_mm_resolvents}) and (\ref{eq:mm_resolvents_expansion}), 
we obtain the loop equations (\ref{eq:loop_eq_mm}).


\EndAppendix

\chapter{The string theory prescription}
\label{IKRSV}

We consider ${\CN}=1$ supersymmetric $U(N)$, $SO(N)$, and $Sp(N)$ gauge
theories, with two-index tensor matter and added tree-level
superpotential, for general breaking patterns of the gauge group.  By
considering the string theory realization and geometric transitions, we
clarify when glueball superfields should be included and extremized, or
rather set to zero; this issue arises for unbroken group factors of low
rank.  The string theory results, which are equivalent to those of the
matrix model, refer to a particular UV completion of the gauge theory,
which could differ from conventional gauge theory results by residual
instanton effects.  Often, however, these effects exhibit miraculous
cancellations, and the string theory or matrix model results end up
agreeing with standard gauge theory.
In particular, these string theory considerations explain and remove
some apparent discrepancies between gauge theories and matrix models in
the literature.

\section{Introduction}

Large $N$ topological string duality \cite{Gopakumar:1999ki} embedded in
superstrings \cite{Vafa:2001wi, Cachazo:2001jy} has led to a new
perspective on ${\CN}=1$ supersymmetric gauge theories: that the exact
effective superpotential can be efficiently computed by including
glueball fields. For example, in a theory with gauge group $G$, with
tree-level superpotential leading to a breaking pattern
\begin{align}
 G(N)\rightarrow \prod _{i=1}^K G_i(N_i)~,
 \label{eq:higgsbp}
\end{align}
the dynamics is
efficiently encoded in a superpotential $W_{\textrm{eff}}(S_1, \dots S_K; g_j,
\Lambda )$ ($g_j$ are the parameters in $W_{\textrm{tree}}$ and $\Lambda$ is the
dynamical scale).  Further, string theory implies \cite{Cachazo:2001jy}
\begin{align}
   W_{\textrm{eff}}(S_i; g_j, \Lambda)=
  \sum _{i=1}^K
  \left(
    h_i{\partial \CF(S_i) \over \partial S_i}-2\pi i \tau _i S_i
  \right),
 \label{eq:weffgen}
\end{align}
with $h_i$ and $\tau _i$ the fluxes through $A_i$ and $B_i$ three-cycles
in the geometry, as will be reviewed in sect.\
\ref{IKRSV:geo_trans_U(N)_SO/Sp(N)}.  The prepotential $\CF(S_i)$ in
\eqref{eq:weffgen} is computable in terms of geometric period integrals,
which yields \cite{Cachazo:2001jy}
\begin{align}
   {\partial \CF(S_i)\over \partial S_i}
  =
  S_i \left(\log \left({\Lambda _i^3\over S_i}\right)+1\right)
  +{\partial \over \partial S_i}
  \sum_{i_1, \dots , i_K\geq 0}c_{i_1\dots i_K}S_1^{i_1}\cdots S_K^{i_k},
 \label{eq:prepotgen}
\end{align}
with coefficients $c_{i_1\dots i_K}$ depending on the $g_j$ (but not on
the gauge theory scale $\Lambda$).  In \cite{Dijkgraaf:2002fc} it was shown how
planar diagrams of an associated matrix model can also be used to
compute \eqref{eq:weffgen} and \eqref{eq:prepotgen}.  Based on the
stringy examples, this was generalized in \cite{Dijkgraaf:2002dh} to a more
general principle to gain non-perturbative information about the strong
coupling dynamics of gauge theories, by extremizing the perturbatively
computed glueball superpotential.

There are two aspects to the above statements: first that the glueball
fields $S_i$ are the `right' variables to describe the IR physics, and
second that perturbative gauge theory techniques suffice to compute the
glueball superpotential.  The latter statement has now been proven in
two different approaches for low powers of the glueball fields $S_i$ in
\eqref{eq:prepotgen} \cite{Dijkgraaf:2002xd, Cachazo:2002ry}.  For
powers of the glueball fields $S_i$ larger than the dual Coxeter number
of the group, an ambiguity sets in for the glueball computation of the
coefficients $c_{i_1\dots i_K}$ in both of these approaches.  The matrix
model provides a natural prescription for how to resolve this ambiguity,
essentially by continuing from large $N_i$.  It was argued in
\cite{Dijkgraaf:2003xk, Aganagic:2003xq} that the string geometry /
matrix model result (since the string geometry and matrix model results
are identical, we refer to them synonymously) has the following meaning:
it computes the $F$-terms for different supersymmetric gauge theories,
which can be expressed in terms of $G(N+k|k)$ supergroups.  The
$W_{\textrm{eff}}(S_i)$ is independent of $k$, and the above ambiguity
can be eliminated by taking $k$, and hence the dual Coxeter number,
arbitrarily large.  The $G(N)$ theory of interest is obtained from the
$G(N+k|k)$ theory by Higgsing; but there can be residual instanton
contributions to $W_{\textrm{eff}}$ \cite{Aganagic:2003xq}, which can
lead to apparent discrepancies between the matrix model and gauge theory
results.  We will somewhat clarify here when such residual instanton
effects do, or do not, lead to discrepancies with standard gauge theory
results.

There is another, more non-trivial assumption in \cite{Dijkgraaf:2002dh}
: the statement that the glueball fields $S_i$ are the `right' variables
in the IR.  This assumption was motivated from the string dualities
\cite{Gopakumar:1999ki} - \cite{Cachazo:2001jy}, where geometric
transition provide the explanation of why the glueball fields are the
natural IR variables: heuristically, $\ev{S_i}$ corresponds to
confinement.  However, this is not quite correct: it also applies to
abelian theories, as had been noted in \cite{Cachazo:2001jy} . So the
deep explanation of why we should choose certain dynamical $S$ variables
remains mysterious.

In this paper, we will uncover the precise prescription for the correct
choice of IR variables.  This will be done from the string theory
perspective, by arguing in which cases there is a geometric transition
in string theory.  For the general breaking pattern \eqref{eq:higgsbp},
our prescription for treating the glueball field $S_i$, corresponding to
the factor $G_i$ in \eqref{eq:higgsbp}, is as follows: \textit{If
$h(G_i)>0$ we include $S_i$ and extremize $W_{\textrm{eff}}(S_i)$ with
respect to it.  On the other hand, if $h(G_i)\leq 0$ we do not include
or extremize $S_i$, instead we just set $S_i\rightarrow 0$.}  Here we
define the generalized dual Coxeter numbers\footnote{Our convention for
symplectic group is such that $Sp(N)\subset SU(N)$ with even $N$, and
hence $Sp(2)\cong SU(2)$.}
\begin{equation}
 \begin{split}
   h(U(N))  &= N,\\
  h(Sp(N)) &= {N\over 2}+1,\\
  h(SO(N)) &= N-2,
\end{split}
  \label{eq:wlowuiv}
\end{equation}
which are generalized in that \eqref{eq:wlowuiv} applies for all $N\geq
0$.  In particular $h(U(1))= h(Sp(0))=1$, so when some $G_i$ factor in
\eqref{eq:higgsbp} is $U(1)$ or $Sp(0)$, our prescription is to include
the corresponding $S_i$ and extremize with respect to it. On the other
hand, $h(U(0))=0$ and $h(SO(2))=0$, so when some $G_i$ factor in
\eqref{eq:higgsbp} is $U(0)$ or $SO(2)$, our prescription is to just set
the corresponding $S_i=0$ from the outset.  (Note that $U(1)$ and
$SO(2)$ are treated differently here.)

This investigation was motivated by trying to understand the
discrepancies found in \cite{Kraus:2003jf} for $Sp(N)$ theory with
antisymmetric tensor matter, where the superpotentials from the matrix
model and gauge theory were found to differ at order $h$ in perturbation
theory and beyond.  The analysis considered the trivial breaking pattern
$Sp(N)\rightarrow Sp(N)$ and a single glueball was introduced
corresponding to the single unbroken gauge group factor.
In \cite{Aganagic:2003xq}, various gauge theories including this example
were studied, and an explanation for the discrepancies was proposed in
terms of the conjecture, mentioned above, that the string theory /
matrix model actually computes the superpotential of the large $k$
$G(N+k|k)$ supergroup theories, rather than the ordinary $G(N)$ theory.
In this context, the trivial breaking pattern considered in \cite{Kraus:2003jf}
should be understood as $Sp(N)\to Sp(N)\times Sp(0)$, which is completed
to $Sp(N+k|k)\to Sp(N+k_1|k_1)\times Sp(k_2|k_2)$.  In particular,
$Sp(0)$ factors, while trivial in standard gauge theory, are non-trivial
in the string theory geometry / matrix model context:
there can be a residual instanton contribution to the superpotential
when one Higgses $Sp(k_2|k_2)$ down to $Sp(0|0)=Sp(0)$, as explicitly
seen in \cite{Aganagic:2003xq}\ for the case of breaking
$Sp(0)\rightarrow Sp(0)$ with quadratic $W_{\textrm{tree}}$\footnote{ It
was suggested in the original version of \cite{Aganagic:2003xq}\ that
such $Sp(0)$ residual instanton contributions could also play a role for
the case of cubic and higher order $W_{\textrm{tree}}$ (where they had
not yet been fully computed) and could explain the apparent matrix model
vs.\ standard gauge theory discrepancies found in \cite{Kraus:2003jf}.
As we will discuss, we now know that this last speculation was not
correct.  The corrected proposal of \cite{Aganagic:2003xq}\ is still
that the matrix model computes the superpotential of the $G(N+k|k)$
theory, but where the matrix model side of the computation should be
corrected, as we discuss in this paper, to include glueball fields for
the $Sp(0)$ factors.}.  Related aspects of ``$Sp(0)$'' being non-trivial
in the string / matrix model context were subsequently discussed in
\cite{Cachazo:2003kx, Ahn:2003ui, Landsteiner:2003ph}.

However, it turns out that one also needs to modify the matrix model
side of the computation to take into account the $Sp(0)$ factors. 
This was found by Cachazo \cite{Cachazo:2003kx},
who showed that the loop equations determining $T(z)\equiv \Tr({1\over
z-\Phi})$ and $R(z)\equiv -{1\over 32\pi ^2}\Tr({W_\alpha W^\alpha \over
z-\Phi})$ for the $Sp(N)$ theory with antisymmetric tensor matter
\cite{Alday:2003dk, Kraus:2003jv} could be related to those of a
$U(N+2K)$ gauge theory with adjoint matter, with $Sp(N) \rightarrow
Sp(N)\times Sp(0)^{K-1}$ mapped to $U(N+2K)\rightarrow U(N+2) \times
U(2)^{K-1}$.
It was thus shown in \cite{Cachazo:2003kx} that vanishing period of $T(z)dz$
through a given cut, corresponding to an $Sp(0)$ factor, does not imply
that the cut closes up on shell (aspects of the periods in this theory
were also discussed in \cite{Matone:2003bx}).
This fits with our above prescription that the $Sp(0)$ glueballs should
be included and extremized in the string theory / matrix model picture,
as would be done for $U(2)$, rather than set to zero, as was originally
done in \cite{Kraus:2003jf}.  We stress that we are not yet even discussing whether
or not the string theory / matrix model result agrees with standard
gauge theory.  Irrespective of any comparison with standard gauge
theory, the prescription to obtain the actual string theory / matrix
model result is as described above \eqref{eq:wlowuiv}.  Having obtained
that result, we can now discuss comparisons with standard gauge theory
results.  As seen in \cite{Cachazo:2003kx}, by solving the $U(N+2K)$ loop
equations for the present case, this corrected matrix model result now
agrees perfectly with standard gauge theory!  This will be discussed
further here, with all glueball fields $S_i$ included.

This agreement, between the matrix model result and standard gauge
theory, is in a sense surprising for this particular theory, in light of
the $Sp(k|k)$ description of \cite{Aganagic:2003xq}\ for the unbroken
$Sp(0)$ factors, with the resulting residual instanton contributions to
the superpotential.  As we will explain later in this paper, the
agreement here between matrix models and standard gauge theory is thanks
to a remarkable cancellation of the residual instanton effect terms,
which could have spoiled the agreement.  The cancellation occurs upon
summing over the $i$ in \eqref{eq:weffgen} from $i=1\dots K$.

There are similar remarkable cancellations of the ``residual instanton
contributions'' to the superpotential in many other examples, which we
will also discuss.  In fact, in all cases that we know of, the only
cases where the residual instantons do not cancel is when the gauge
theory clearly has some ambiguity, requiring a choice of how to define
the theory in the UV; string theory / matrix model gives a particular
such choice.  Examples of such cases is when the LHS of
\eqref{eq:higgsbp} is itself $U(1)$ or $Sp(0)$ super Yang--Mills, as
discussed in \cite{Aganagic:2003xq}.  Other examples where the residual
instanton contributions do not cancel, is when the superpotential is of
high enough order such that not all operators appearing in it are
independent, e.g.\ terms like $\Tr\, \Phi ^n$, for a $U(N)$ adjoint
$\Phi$, when $n>N$.  In standard gauge theory, there are then potential
ambiguities involved in reducing such composite operators to the
independent operators, since classical operator identities can receive
quantum corrections.  The residual instanton contributions, which do not
cancel generally in these cases, imply specific quantum relations for
these operators, corresponding to the specific UV completion.  See
\cite{Dorey:2002tj, Balasubramanian:2002tm} for related issues.

The organization of this paper is as follows: In section
\ref{IKRSV:GT_exmpl} we summarize the gauge theories under
consideration. In section \ref{IKRSV:geo_trans_U(N)_SO/Sp(N)} we review
the type IIB string theory construction of these gauge theories.  We
also discuss maps of the exact superpotentials of $Sp$ and $SO$ theories
to those of $U$ theories, generalizing observations of \cite{Ita:2002kx,
Ashok:2002bi, Janik:2002nz, Cachazo:2003kx, Landsteiner:2003ph}.  In
section \ref{IKRSV:ST_prscrptn} we explain, from the string theory
perspective in which cases we have a geometric transition. In section
\ref{IKRSV:exmpls} we consider examples, where the glueball fields $S_i$ of
all group factors are correctly accounted for on the matrix model side.
The results thereby obtained via matrix models are found to agree with
those of standard gauge theory.  In many of these examples, this
agreement relies on a remarkable interplay of different residual
instanton contributions, which sometimes fully cancel.  Residual
instantons are discussed further in section \ref{IKRSV:resdl_instntns}, with
examples illustrating cases where they do, or do not, cancel.  In
appendix \ref{IKRSV:MM_calc_spot}, a proof of a general relation between the $S^2$
and $\RPtwo$ contributions to the matrix model free energy is given, and
also the matrix model computation of superpotential is presented.  In
appendix \ref{IKRSV:GT_calc_spot} the gauge theory computation of the
superpotential is discussed.

\section{The gauge theory examples}
\label{IKRSV:GT_exmpl}

The specific examples of $\CN=1$ supersymmetric gauge theories
which we consider, with breaking patterns as in \eqref{eq:higgsbp}, are
as follows:  
\begin{align}
 \begin{array}{r@{}l@{~~~~}r@{~}c@{~}l}
 U(N) &\text{ with adjoint $\Phi$:}       &U(N) &\rightarrow &\textstyle\prod _{i=1}^KU(N_i),\\
 SO(N)&\text{ with adjoint $\Phi$:}       &SO(N)&\rightarrow &SO(N_0)\times \textstyle\prod _{i=1}^{K}U(N_i), \\
 Sp(N)&\text{ with adjoint $\Phi$:}       &Sp(N)&\rightarrow &Sp(N_0)\times \textstyle\prod _{i=1}^{K}U(N_i),\\
 SO(N)&\text{ with symmetric $S$:}        &SO(N)&\rightarrow &\textstyle\prod_{i=1}^K SO(N_i),\\
 Sp(N)&\text{ with antisymmetric $A$:}    &Sp(N)&\rightarrow &\textstyle\prod_{i=1}^K Sp(N_i),\\
 U(N) &\text{ with $\Phi+S+\widetilde S$:}&U(N)&\rightarrow &SO(N_0)\times \textstyle\prod _{i=1}^{K} U(N_i),\\
 U(N) &\text{ with $\Phi+A+\widetilde A$:}&U(N)&\rightarrow &Sp(N_0)\times \textstyle\prod _{i=1}^{K} U(N_i).
\end{array}
 \label{eq:higgscases}
\end{align}
For $U(N)$ with adjoint $\Phi$, the tree-level superpotential is taken to be 
\begin{align}
   W_{\textrm{tree}} &= \Tr[W(\Phi)],\qquad
  W(x) = \sum_{j=1}^{K+1} {g_j\over j}x^j,
 \label{ump22Mar04}
\end{align}
with $K$ potential wells.  In the classical vacua, with breaking pattern
as in \eqref{eq:higgscases}, $\Phi$ has $N_i$ eigenvalues equal to the
root $a_i$ of
\begin{align}
  W'(x)=\sum _{j=1}^{K+1}g_{j}x^{j-1}\equiv g_{K+1}\prod_{i=1}^K(x-a_i),
 \label{eq:Wprimei}
\end{align}
with $\sum _{i=1}^K N_i=N$.  For $SO(N)$ with symmetric tensor or
$Sp(N)$ with antisymmetric tensor we take
\begin{align}
   W_{\textrm{tree}}=\half \Tr W(S), 
  \qquad \hbox{or} \qquad
  W_{\textrm{tree}}=\half \Tr W(A),
\label{eq:bsospx}
\end{align}
respectively, where $W(x)$ is as in \eqref{ump22Mar04}, the factor of
$\half$ is for convenience, because the eigenvalues of $S$ or $A$ appear
in pairs, and the indices are contracted with $\delta ^a_b$ for $SO(N)$
or $J^a_b$ for $Sp(N)$.  For $SO/Sp(N)$\footnote{We often write $SO(N)$
and $Sp(N)$ simply as $SO/Sp(N)$.  When we use a ``$\pm$'' sign for
$SO/Sp(N)$, it means ``$+$'' for $SO(N)$ and ``$-$'' for $Sp(N)$.
Similarly ``$\mp$'' means ``$-$'' for $SO(N)$ and ``$+$'' for $Sp(N)$.}
with adjoint matter, the tree-level superpotential is
\begin{align}
   W_{\textrm{tree}} &= \half \Tr[W(\Phi)],\qquad
  W(x) = \sum_{j=1}^{K+1} {g_{2j}\over 2j}x^{2j},
 \label{eq:basp}
\end{align}
since all Casimirs of the adjoint $\Phi$ are even, and the $\half$ is
again for convenience because the eigenvalues appear in pairs.  $\Phi$'s
eigenvalues sit at the zeros of
\begin{align}
   W'(x)=\sum _{j=1}^{K+1}g_{2j}x^{2j-1}\equiv
  g_{2K+2}x\prod _{i=1}^{K}(x^2-a_i^2).
 \label{eq:Wprimeisp}
\end{align}
The breaking pattern in \eqref{eq:higgscases} has $N_0$ eigenvalues of
$\Phi$ equal to zero, and $N_i$ pairs at $\pm a_i$, so $N=N_0+2\sum
_{i=1}^{K}N_i$ for $SO/Sp(N)\rightarrow SO/Sp(N_0)\times \prod
_{i=1}^{K}U(N_i)$ (with the convention $Sp(2)\cong SU(2)$).

The next to last example in \eqref{eq:higgscases} is the $\CN=2$ $U(N)$
theory with a matter hypermultiplet in the two-index symmetric tensor
representation, breaking $\CN=2$ to $\CN=1$ by a
superpotential as in \eqref{ump22Mar04}:
\begin{align}
 W=\sum _{j=1}^{K+1}{g_j\over j}\Tr \Phi ^j +\sqrt{2}\,\Tr \widetilde S \Phi S.
\end{align}
In addition to the possibility of $\Phi$'s eigenvalues sitting in any of
the $K$ critical points $W'(x)$ analogous to \eqref{eq:Wprimei}, there
is a vacuum where $N_0$ eigenvalues sits at $\phi =0$, with
$\ev{S\widetilde S}\neq 0$, breaking $U(N_0)\rightarrow SO(N_0)$.  The
last example in \eqref{eq:higgscases} is the similar theory where the
$\CN=2$ hypermultiplet is instead in the antisymmetric tensor
representation $A$, rather than the symmetric tensor $S$.  These last
two classes of examples were considered in
\cite{Klemm:2003cy, Naculich:2003cz, Naculich:2003ka}.

In all of these theories, the low energy superpotential is of the
general form
\begin{align}
 W_{\textrm{low}}(g_j, \Lambda)=W_{\textrm{cl}}(g_j)+W_{\textrm{gc}}(\Lambda _i)+W_H(g_j,
 \Lambda).
 \label{eq:wlowg}
\end{align}
$W_{\textrm{cl}}(g_j)$ is the classical contribution (evaluating
$W_{\textrm{tree}}$ in the appropriate minima).
$W_{\textrm{gc}}(\Lambda _i)$ is the gaugino condensation contribution
in the unbroken gauge groups of \eqref{eq:higgsbp},
\begin{align}
 W_{\textrm{gc}}(\Lambda _j)=\sum _{i=1}^K h_ie^{2\pi i n_i/h_i}\Lambda _i^3,
 \label{eq:wgcisi}
\end{align}
where $h_i=C_2(G_i)$ is the dual Coxeter number of the group $G_i$ in
\eqref{eq:higgsbp}, with the phase factors associated with the ${\mathbb
Z}_{2h_i}\rightarrow {\mathbb Z}_2$ chiral symmetry breaking of the
low-energy $G_i$ gaugino condensation.  The scales $\Lambda _i$ are
related to $\Lambda$ by threshold matching for the fields which got a
mass from $W_{\textrm{tree}}$ and the breaking \eqref{eq:higgsbp}; some
examples, with breaking patterns as in \eqref{eq:higgscases}, are as
follows.  For $U(N)$ with adjoint $\Phi$:
\begin{align}
 \Lambda _i^{3N_i}=\Lambda ^{2N} W''(a_i)^{N_i}
 \prod _{j\neq i}m_{W_{ij}}^{-2N_j}
 =g_{K+1}^{N_i} \Lambda ^{2N}
 \prod _{j\neq i}(a_j-a_i)^{N_i-2N_j}.
 \label{eq:lamin}
\end{align}
For $SO/Sp(N)$ with adjoint, breaking $SO/Sp(N)\rightarrow SO/Sp(N_0)\times \prod
_{i=1}^{K}U(N_i)$,
\begin{equation}
\begin{split}
  \Lambda _0^{3(N_0\mp 2)}&=g_{2K+2}^{N_0\mp 2}\Lambda ^{2(N\mp 2)}
 \prod _{i=1}^{K}a_i^{2(N_0\mp 2)-4N_i}\cr
 \Lambda _i^{3N_i}&=2^{-N_i}g_{2K+2}^{N_i}\Lambda ^{2(N\mp 2)}a_i^{-2(N_0\mp 2)}\prod _{ j\neq i}^{K}
 (a_i^2-a_j^2)^{N_i-2N_j}.
\end{split} 
\label{eq:lamsospa}
\end{equation}
Note that, although this relation looks very similar for $SO(N)$ and
$Sp(N)$, its meaning is different for two groups in gauge theory due to
the index of the embedding of $U(N_i)$; for $SO(N)$, the $U(N_i)$
one-instanton factor $\Lambda_i^{b(U(N_i))}$, $b(U(N_i))=3N_i$, is
related to the $SO(N)$ one-instanton factor $\Lambda^{b(SO(N))}$,
$b(SO(N))=2(N-2)$.  On the other hand, for $Sp(N)$, it is related to the
$Sp(N)$ {\em two\/}-instanton factor $\Lambda^{2b(Sp(N))}$,
$b(Sp(N))=N+2$.
For $SO(N)$ with symmetric tensor,
\begin{align}
 \Lambda _i ^{3(N_i-2)}=g_{K+1}^{N_i+2}\Lambda ^{2N-8}\prod _{j\neq i=1}^K
 (a_i-a_j)^{N_i+2-2N_j}.
 \label{eq:lamsos}
\end{align}
For $Sp(N)$ with antisymmetric $A$:
\begin{align}
 \Lambda _i^{3(N_i+1)}=g_{K+1}^{N_i-1}\Lambda ^{2N+4}\prod _{j\neq
 i}^K(a_i-a_j)^{N_i-1-2N_j}.
 \label{eq:laminsp}
\end{align}

Finally, the term $W_H(g_j, \Lambda)$ in \eqref{eq:wlowg} are additional
non-perturbative contributions, which can be regarded as coming from the
massive, broken parts of the gauge group.

In the description with the glueballs $S_i$ integrated in, as in
\eqref{eq:weffgen}, the gaugino condensation contribution comes from the
first term in \eqref{eq:prepotgen}:
\begin{align}
 W_{\textrm{gc}}(S_i, \Lambda)=\sum _{i=1}^K h_i S_i \left(\log
 \left({\Lambda _i^3 \over S_i}\right)+1\right)~,
 \label{eq:wgca}
\end{align}
and $W_H(g_i, \Lambda)$ comes from the last terms in
\eqref{eq:prepotgen}, upon integrating out the $S_i$.  When the minima
$a_i$ of the superpotential are widely separated, the contributions
$W_H$ from these last terms are subleading as compared with
$W_{\textrm{gc}}$.  As in \eqref{eq:weffgen}, the full glueball
superpotential \eqref{eq:weffgen} can be computed via the string theory
geometric transition, in terms of certain period integrals
\cite{Vafa:2001wi, Cachazo:2001jy}, as will be reviewed in the next
section, or via the matrix models.  In that context, the term
$W_{\textrm{gc}}(S_i, \Lambda _i)$ comes from the integration measure
(as is also natural in field theory, since it incorporates the $U(1)_R$
anomaly) and $W_H(S_i, g_j)$ can be computed perturbatively
\cite{Dijkgraaf:2002fc, Dijkgraaf:2002dh}.  The perturbative computation of
$W_H(S_i, g_j)$ can also be understood directly in the gauge theory
\cite{Dijkgraaf:2002xd}, up to ambiguities in terms $S^n$ with $n>h=C_2(G)$.  The
string theory/ matrix model constructions yield a specific way of
resolving these ambiguities, which correspond to a particular UV
completion of the gauge theory \cite{Dijkgraaf:2003xk, Aganagic:2003xq}.

As discussed in the introduction, our present interest will be in
analyzing this circle of ideas when some of the gauge group factors in
\eqref{eq:higgsbp} are of low rank, or would naively appear to be trivial, e.g.\
$U(1)$, $U(0)$, $SO(2)$, $SO(0)$, and $Sp(0)$.

\section{Geometric transition of $U(N)$ and $SO/Sp(N)$ theories}
\label{IKRSV:geo_trans_U(N)_SO/Sp(N)}

In this section we briefly review the type IIB geometric engineering of
relevant $U(N)$ and $SO/Sp(N)$ theories, and their geometric transition.

%
\bigskip
\subsection{$U(N)$ with adjoint and $W_{\textrm{tree}}=\Tr \sum_{j=1}^{K+1}{g_i\over j}\Phi ^j$}

The Calabi--Yau 3-fold relevant to this theory \cite{Cachazo:2001jy,Cachazo:2001gh}
is the non-compact $A_1$ fibration
\begin{align}
   W'(x)^2+y^2+u^2+v^2=0~.
 \label{xcw22Mar04}
\end{align}
This fibration has $K$ conifold singularities at the critical points of
$W(x)$, i.e.\ at $W'(x)=0$.  Near each of the singularities, the
geometry \eqref{xcw22Mar04} is the same as the usual conifold
$x'{}^2+y^2+u^2+v^2=0$, which is topologically a cone with base
$S^2\times S^3$.

The singularities can be resolved by blowing up a 2-sphere $S^2=\Pone$
at each singularity.  We can realize the $U(N)$ gauge theory with
adjoint matter and superpotential \eqref{ump22Mar04} in type IIB
superstring theory compactified on this resolved geometry, with $N$
D5-branes partially wrapping the $K$ $\Pone$'s; two dimensions of the
D5-brane worldvolume wrap the $\Pone$'s and the remaining four
dimensions fill the flat Minkowski space.  The gauge theory degrees of
freedom correspond to the open strings living on these D5-branes.  The
classical supersymmetric vacuum is obtained by distributing the $N_i$
D5-branes over the $i$-th 2-sphere $\Pone_i$ with $i=1,\cdots,K$.  The
corresponding breaking pattern of the gauge group is as in
\eqref{eq:higgscases}: $U(N)\to \prod _{i=1}^KU(N_i)$.

At low energy, the gauge theory confines (when $N_i>1$), each $U(N_i)$
factor developing nonzero vev of the glueball superfield $S_i$.  In
string theory this is described by the geometric transition
\cite{Gopakumar:1999ki, Vafa:2001wi, Cachazo:2001jy} in which the resolved
conifold geometry with $\Pone$'s wrapped by D5-branes is replaced by a
deformed conifold geometry
\begin{align}
 W'(x)^2+f_{K-1}(x)+y^2+u^2+v^2=0~,
 \label{xea22Mar04}
\end{align}
where $f_{K-1}(x)$ is a polynomial of degree $K-1$ in $x$ and
parametrizes the deformation.  After the geometric transition, each
2-sphere $\Pone_i$ wrapped by $N_i$ D5-branes is replaced by a 3-sphere
$A_i$ with 3-form RR flux through it:
\begin{align}
 \oint_{A_i} H = N_i~,
 \label{xgz22Mar04}
\end{align}
where $H = H_{RR} + \tau H_{NS}$ and $\tau=C^{(0)}+ie^{-\Phi}$ is the
complexified coupling constant of type IIB theory.
We define the periods of the Calabi--Yau geometry \eqref{xea22Mar04} by
\begin{align}
   S_i \equiv  {1\over 2\pi i}\oint_{A_i} \Omega,
  \qquad
  \Pi_i \equiv \int_{B_i}^{\Lambda_b} \Omega,
 \label{xjv22Mar04}
\end{align}
where $\Omega$ is the holomorphic 3-form, $B_i$ is the noncompact
3-cycle dual to the 3-cycle $A_i$, and $\Lambda_b$ is a cutoff needed to
regulate the divergent $B_i$ integrals. The IR cutoff $\Lambda_b$ is to
be identified with the UV cutoff of the 4d gauge theory.  The set of
variables $S_i$ measure the size of the blown up 3-spheres, and can be
used to parametrize the deformation in place of the $K$ coefficients of
the polynomial $f_{K-1}(x)$.

The dual theory after the geometric transition is described by a 4d,
$\CN=1$ $U(1)^K$ gauge theory, with $K$ $U(1)$ vector superfields $V_i$
and $K$ chiral superfields $S_i$.  If not for the fluxes, this theory
would be $\CN=2$ supersymmetric $U(1)^K$, with $(V_i, S_i)$ the
$\CN=2$ vector super-multiplets, $S_i$ being the Coulomb branch
moduli.  This low-energy $U(1)^K$ theory has a non-trivial prepotential
$\CF(S_i)$ and the dual periods $\Pi _i$ in \eqref{xjv22Mar04} can be written as
\begin{align}
 \Pi _i (S)={\partial \CF\over \partial S_i}.  
 \label{eq:Piper}
\end{align}
Without
the fluxes, this prepotential can be understood as coming {}from
integrating out D3 branes which wrap the $A_i$ cycles, and are charged
under the low-energy $U(1)$'s \cite{Strominger:1995cz}.

The effect of the added fluxes is to break $\CN=2$ supersymmetry to
$\CN=1$ by the added superpotential \cite{Taylor:2000ii, Mayr:2001hh}
\begin{align}
   W_{\textrm{flux}}
  =  \int H\wedge \Omega
  =  \sum_i
  \left(
    \oint_{A_i} H \int_{B_i}^{\Lambda_b} \Omega
    -\int_{B_i}^{\Lambda_b} H \oint_{A_i} \Omega
  \right)~.
 \label{xop22Mar04}
\end{align}
In the present $U(N)$ case, \eqref{xgz22Mar04} and \eqref{xjv22Mar04}
gives
\begin{align}
   W_{\textrm{flux}}
  =
  \sum_{i=1}^K (N_i \Pi_i  - 2\pi i \alpha S_i)~,
 \label{bdqh22Mar04}
\end{align}
where
\begin{align}
   \int_{B_i}^{\Lambda_b} H = \alpha\label{epny1Apr04}
\end{align}
is the 3-form NS flux through the 3-cycle $B_i$ and identified
with the bare coupling constant of the gauge theory by
\begin{align}
   2\pi i \alpha  =  {8\pi^2 \over g_b^2}=V.\label{bdto22Mar04}
\end{align}
where $V$ is the complexified volume of the $\Pone$'s.  The
${\CN}=1$ $U(1)^K$ vector multiplets $V_i$ remain massless, but the
$S_i$ now have a superpotential, which fixes them to sit at discrete
vacuum expectation values, where they are massive.  The fields $S_i$ are
identified with the glueballs on the gauge theory side.

The superpotential $W_{\textrm{flux}}(S_i)$ is the full, exact,
effective superpotential in \eqref{eq:weffgen}.  As can be verified by
explicit calculations \cite{Vafa:2001wi, Cachazo:2001jy}, the leading
contribution to \eqref{bdqh22Mar04} is always of the form
\begin{align}
   W_{\textrm{flux}} \sim \sum_{i=1}^K [N_i S_i(1-\ln(S_i/\Lambda_i^3)) - 2
 \pi
 i \alpha S_i],\label{bdvb22Mar04}
\end{align}
where $\Lambda_i$ is related to the scale $\Lambda_b$ via precisely the
relation \eqref{eq:lamin}.  This leading term \eqref{bdvb22Mar04} is the gaugino condensation
part of the superpotential, as in \eqref{eq:prepotgen}.

%
\bigskip
\subsection{$SO$ and $Sp$ theories}

The string theory construction of $SO/Sp(N)$ theories can be obtained
from the above $U(N)$ construction, by orientifolding the geometry
before and after the geometric transition by a certain $\Ztwo$ action.
The geometric construction of $SO/Sp(N)$ theory with adjoint was
discussed in \cite{Sinha:2000ap, Edelstein:2001mw, Ashok:2002bi, Landsteiner:2003rh}, and in that case the
invariance of the geometry \eqref{xea22Mar04} under the $\Ztwo$ action
requires that the polynomial $W(x)$ be even.  The geometric construction
of $SO/Sp(N)$ theory with symmetric/antisymmetric tensor was studied in
\cite{Csaki:1998mx, Landsteiner:2003xe, Landsteiner:2003ph}.

In the classical vacuum of the ``parent'' $U(2N)$ theory, the gauge
group is broken into a product of $U(N_i)$ groups.  When a $U(N_i)$
factor is identified with another $U(N_i)$ by the $\Ztwo$ orientifold
action, they lead to a single $U(N_i)$ factor.  When a $U(N_i)$ factor
is mapped to itself by the $\Ztwo$ orientifold action, it becomes an
$SO(N_i)$ or $Sp(N_i)$, depending on the charge of the orientifold
hyperplane.  As a result, the classical vacuum of the ``daughter''
$SO/Sp(N)$ theory has gauge group broken as in \eqref{eq:higgscases},
depending on whether the theory is $SO$ or $Sp$ with adjoint, or $SO$
with symmetric tensor, or $Sp$ with antisymmetric tensor:
\begin{align}
   SO/Sp(N)\to \prod_i G_i(N_i),
  \qquad
  G_i=U, SO, {\textrm{~or~}} Sp.
\end{align}

The 3-form RR fluxes from orientifold hyperplanes makes an additional
contribution to the superpotential \eqref{xop22Mar04}, and the flux superpotential can be
written as
\begin{align}
  W_{\textrm{flux}}  =  \sum_i [\Nh_i \Pi _i (S_i)- 2 \pi i \eta_i \alpha S_i],
 \label{bdyq22Mar04}
\end{align}
where
\begin{equation}
 \begin{split}
 \Nh_i&=
 \begin{cases}
  N_i                & \qquad G_i=U(N_i), \\
  N_i/2 \mp 1 & \qquad G_i=SO/Sp(N_i), 
 \end{cases}\\
 \eta_i &=
 \begin{cases}
  1     & \qquad G_i=U, \\
  1/2 & \qquad G_i=SO/Sp. 
 \end{cases}
 \label{beex22Mar04}
\end{split}
\end{equation}
$\Nh_i$ is the net 3-form RR flux through the $A_i$ cycle.  For $U(N_i)$
and $Sp(N_i)$, $\Nh_i$ in \eqref{beex22Mar04} is the dual Coxeter number \eqref{eq:wlowuiv},
while for $SO(N_i)$ it is half\footnote{So we get $h$ replaced with $h/2$
for $SO$ groups in \eqref{eq:wgcisi}.  While one could absorb the overall factor
of 2 into the definition of $\Lambda$, the number of vacua should be $h$
whereas here we apparently get $h/2$ for $SO$ groups.  This is because
the we don't see spinors or the ${\mathbb Z}_2$ part of the center which
acts on them; it's analogous to $U(2N)$ being restricted to vacua with
confinement index 2.}  the dual Coxeter number \eqref{eq:wlowuiv}.  The $1/2$ in
\eqref{beex22Mar04} is because the integration over the $A_i$ cycles should be halved
due to the $\Ztwo$ identification.

\subsection{Relations between $SO/Sp$ theories and $U(N)$ theories}
\label{IKRSV:rel_SO/Sp_and_U(N)}

The result \eqref{bdyq22Mar04}, with \eqref{beex22Mar04}, gives the exact superpotential of the $SO/Sp$
theories in terms of the same periods $S_i$ and $\Pi (S_i)$ as an
auxiliary $U$ theory.  This was first noted in \cite{Cachazo:2003kx} at the level
of the Konishi anomaly equation as a map between the resolvents of $Sp$
theory with antisymmetric matter and $U$ theory with adjoint matter. In
\cite{Landsteiner:2003ph}, it was generalized to the map between the resolvents of
$SO/Sp$ theories with two-index tensor matter and $U$ theory with
adjoint matter, and string theory interpretation was discussed.
In this subsection we will derive this map from the string theory
perspective using the flux superpotential \eqref{bdyq22Mar04}.
Furthermore, we will clarify the relation of the superpotential and the
scale of the $SO/Sp$ theories to those of the $U$ theory.  The map
between resolvents can be derived from these results.  For $Sp(N)$
theory with an antisymmetric tensor, the scale relation was obtained in
a different way in \cite{Cachazo:2003kx}.

As a first example, consider $SO(N)$ with an adjoint, with the breaking
pattern as in \eqref{eq:higgscases}.  The geometric transition result
\eqref{bdyq22Mar04} and \eqref{beex22Mar04} implies that the exact
superpotential is the same as for the $U(N-2)$ theory with adjoint, with
breaking pattern map
\begin{gather}
 SO(N)\rightarrow SO(N_0)\times \prod _{i=1}^K U(N_i)
 ~~\iff~~
 U(N-2)\rightarrow U(N_0-2)\times \prod_{i=1}^K U(N_i)^2.
 \label{eq:soaaux}
\end{gather}
The map between the superpotential is
\begin{align}
  W^{SO(N)}_{\textrm{exact}}=\half W^{U(N-2)}_{\textrm{exact}}.
 \label{eq:soaauxsp}
\end{align}
%
The $SO(N)$ scale matching relation \eqref{eq:lamsospa} is compatible with
the map \eqref{eq:soaaux}, since \eqref{eq:lamin} for the theory
\eqref{eq:soaaux} reproduces \eqref{eq:lamsospa}.

Likewise, $Sp(N)$ with adjoint has the same exact superpotential
as for the $U(N+2)$ theory with adjoint, with
\begin{gather}
 Sp(N)\to Sp(N_0)\times \prod _{i=1}^K U(N_i)
  ~~\iff~~
  U(N+2)\rightarrow U(N_0+2)\times \prod _{i=1}^K U(N_i)^2,\cr
  W^{Sp(N)}_{\textrm{exact}}=\half W^{U(N+2)}_{\textrm{exact}}.
 \label{eq:spaaux}
\end{gather}
The $Sp(N)$ scale matching relations \eqref{eq:lamsospa} follow from the
$U(N)$ matching relations \eqref{eq:lamin} with the replacement
\eqref{eq:spaaux}, with the understanding that the $U(N+2)$ and
$U(N_0+2)$ one-instanton factors correspond to the $Sp(N)$ and
$Sp(N_0)$ two-instanton factors; this is related to the index of the
embedding mentioned after \eqref{eq:lamsospa}, and is accounted for by
dividing the $U(N+2)$ superpotential by two, as above.

Next consider $Sp(N)$ with antisymmetric tensor $A$ and breaking pattern
as in \eqref{eq:higgscases}.  The geometric transition result \eqref{bdyq22Mar04} and
\eqref{beex22Mar04} implies that the exact superpotential is the same as for the
$U(N+2K)$ theory with adjoint and breaking pattern
\begin{align}
 Sp(N)\rightarrow \prod _{i=1}^K Sp(N_i)
 ~~\iff~~
 U(N+2K)\rightarrow \prod _{i=1}^K U(N_i+2).
 \label{eq:spasaux}
\end{align}
In the present case, comparing the matching relations \eqref{eq:laminsp} for
$Sp(N)$ with symmetric tensor with the matching relations \eqref{eq:lamin} for
$U(N+2K)$ with adjoint, for the mapping as in \eqref{eq:spasaux} requires that
the scales of the original $Sp(N)$ on the LHS of \eqref{eq:spasaux} and the
$U(N+2K)$ on the RHS of \eqref{eq:spasaux} be related as
\begin{align}
 \Lambda _{U(N+2K)}^{N+2K}=g_{K+1}^{-2}\Lambda _{Sp(N)}^{N+4}.
 \label{eq:spausr}
\end{align}
Then the $\Lambda _i$ of the unbroken groups on both sides of
\eqref{eq:spasaux} coincide, with the understanding that the $U(N_i+2)$
one-instanton factors correspond to the $Sp(N_i)$ two-instanton factors,
as above.  The map between the superpotential is
\begin{gather}
   W^{Sp(N)}_{\textrm{exact}}=\half [W^{U(N+2K)}_{\textrm{exact}}-\Delta W_{\textrm{cl}}],
  \qquad
  \Delta W_{\textrm{cl}}=2\sum_{i=1}^K W(a_i),
  \cr
 \hbox{i.e.\ writing}\quad W_{\textrm{exact}}=W_{\textrm{cl}}+W_{\textrm{quant}},
 \qquad  W_{\textrm{quant}}^{Sp(N)}=
 \half W_{\textrm{quant}}^{U(N+2K)}.
\label{eq:spausrsp}
\end{gather}
Note that, in order for the superpotentials on the two sides of
\eqref{eq:spasaux} to fully coincide, one must compensate the classical
mismatch $\Delta W_{\textrm{cl}}$, since each well is occupied by two
additional eigenvalues in the theory on the RHS of \eqref{eq:spasaux}.
In the string theory geometric transition realization, this constant
shift, which is independent of $N$, $\Lambda$, and the glueball fields
$S_i$, is most naturally interpreted as an additive shift of the
superpotential on the $Sp$ side, which can be regarded as coming from
the orientifold planes both before and after the transitions.  The
classical shift of $\Delta W_{\textrm{cl}}$ leads to slightly different
operator expectation values (as computed via $W_{\textrm{eff}}(g_p,
\Lambda)$ as the generating function) between the $Sp$ and $U$ theory,
as was seen in the example of \cite{Cachazo:2003kx}.  Also, writing the map as in
\eqref{eq:spasaux}, we want the vacuum with confinement index 2
\cite{Cachazo:2003kx}.  We could equivalently replace the RHS of
\eqref{eq:spasaux} with $U(N/2+K)\rightarrow \prod _{i=1}^KU(N_i/2+1)$, in
which case we would not have to divide by 2 in \eqref{eq:spausrsp}.

Likewise, $SO(N)$ with symmetric tensor $S$ has exact superpotential
related to that of a $U(N-2K)$ theory with adjoint as
\begin{gather}
 SO(N)\rightarrow \prod _{i=1}^K SO(N_i)
 ~~\iff~~
 U(N-2K)\rightarrow \prod _{i=1}^K U(N_i-2),\cr
  W^{SO(N)}_{\textrm{exact}}=\half [W^{U(N-2K)}_{\textrm{exact }}+\Delta W_{\textrm{cl}}],
  \qquad
  \Delta W_{\textrm{cl}}=2\sum_{i=1}^K W(a_i).
 \label{eq:sosaux}
\end{gather}
Comparing the matching relations \eqref{eq:lamsos} for $SO(N)$ with symmetric
tensor with those of \eqref{eq:lamin} for $U(N-2K)$ with adjoint requires that
the scales of the original $SO(N)$ on the LHS of \eqref{eq:sosaux} and those of
the $U(N-2K)$ theory on the RHS be related as
\begin{align}
 \Lambda_{U(N-2K)}^{2(N-2K)}=g_{K+1}^4\Lambda _{SO(N)}^{2N-8}.
 \label{eq:sosusr}
\end{align}
Then the $\Lambda _i$ of the unbroken groups on both sides of
\eqref{eq:sosaux} coincide.  Again, in order for the superpotentials on
the two sides of \eqref{eq:sosaux} to fully coincide, one must correct
for the classical mismatch $\Delta W_{\textrm{cl}}$ coming from the fact
that the $U(N-2K)$ theory has two fewer eigenvalues in each well.

In appendix \ref{IKRSV:MM_calc_spot} we will discuss these relations from the matrix model
viewpoint.  In this context, the relation relevant for \eqref{eq:soaaux}
and \eqref{eq:spaaux} was conjectured in
\cite{Ita:2002kx, Ashok:2002bi} based on explicit diagrammatic
calculations, and it was proven for the case of unbroken gauge group
$N_0=N$ in \cite{Janik:2002nz}.  This will be generalized in Appendix
\ref{IKRSV:MM_calc_spot} to all breaking patterns.  Likewise, the matrix model relation
relevant for \eqref{eq:spasaux} and \eqref{eq:sosaux} will be proven in
the appendix; this is a generalization of the connection found in
\cite{Cachazo:2003kx} for the theories in \eqref{eq:spasaux} and
\eqref{eq:sosaux}.

\section{String theory prescription for low rank}
\label{IKRSV:ST_prscrptn}

The discussion of the previous section applies for all $N_i\geq 0$.  We
now discuss under which circumstances one expects a transition in string
theory, where $S_i^3$'s grows, and therefore an effective glueball field
$S_i$ should be included in the superpotential.  Whether or not there is
a geometric transition in string theory is a local question, so each
$S_i^3$ can be studied independently.  Near any $S_i^3$ the local
physics is just a conifold singularity, so we only need to consider the
case of a conifold singularity.

\subsection{Physics near a conifold singularity}

As we saw, $U(N)$, $SO/Sp(N)$ gauge theory can be realized in type IIB
theory as the open string theory living on the D5-branes partially
wrapped on the exceptional $\Pone$ of a resolved conifold geometry.
There is a $\Pone$ associated with each critical point of the polynomial
$W(x)$.  By the geometric transition duality \cite{Vafa:2001wi,
Cachazo:2001jy}, this gauge theory is dual to the closed string theory
in the deformed conifold geometry where the $\Pone$'s have been blown
down and $S^3$'s are blown up instead.

Let us focus on one $\Pone$ with $N\ge 0$ D5-branes wrapping it.  This
corresponds to focusing on one critical point on the gauge theory side.
We allow $N=0$, which corresponds to an unoccupied critical point.  In
the neighborhood, the geometry after the geometric transition is
approximately a deformed conifold $x^2+y^2+z^2+w^2=\mu$ with a blown up
$S^3$.  The low energy degrees of freedom in the four-dimensional theory
are the $\CN=1$ $U(1)$ photon vector superfield $V$ and the $\CN=1$
chiral superfield $S$.  The chiral superfield $S$ is neutral under the
$U(1)$, and can be thought of as in the adjoint representation of the
$U(1)$.  The bosonic component of $S$ is proportional to $\mu$ and
measures the size of the $S^3$.

%

First, consider the case {\textit{without\/}}
fluxes.  Then the closed string theory has $\CN=2$ and there is one
$\CN=2$ $U(1)$ vector multiplet $(V,S)$.
It is known that as the size of the $S^3$ goes to zero there appears an
extra massless degree of freedom \cite{Strominger:1995cz}, which corresponds to the
D3-brane wrapping the $S^3$.  The mass of the wrapped BPS D3-brane is
proportional to the area, $S$, of the $S^3$, so the mass becomes zero as
the $S^3$ shrinks to zero, i.e.\ as $S\to0$.
This extra degree of freedom is described as an $\CN=2$ hypermultiplet
charged under the $U(1)$ (of $V$).  Let us write this hypermultiplet in
$\CN=1$ language as $(Q,\Qt)$, where $Q$ and $\Qt$ are both $\CN=1$
chiral superfields with opposite $U(1)$ charges.  The $\CN=2$
supersymmetry requires the superpotential
\begin{align}
 W_Q = \sqrt{2}\,Q\Qt S,\label{bffs22Mar04}
\end{align}
which indeed incorporates the above situation that the $Q,\Qt$ become
massless as $S\to 0$.  The $D$-flatness is
\begin{align}
 |Q|^2 - |\Qt|^2 = 0 ,
\end{align}
and the $F$-flatness is
\begin{align}
 QS=\Qt S=Q\Qt=0.
\end{align}
The only solution to these is
\begin{align}
 Q=\Qt=0, \qquad S:{\textrm{any}},
\end{align}
which just means that $S\sim \mu$ is a modulus.
%

Now let us come back to the case with the fluxes.  As reviewed in the
last section, the fluxes give rise to a superpotential
\eqref{xop22Mar04} which breaks $\CN=2$ to $\CN=1$ \cite{Taylor:2000ii, Mayr:2001hh}.
As in \eqref{bdyq22Mar04}, the local flux superpotential contribution is
\begin{align}
   W_{\textrm{flux}}(S)\simeq \Nh S[1-\ln(S/\Lambda^3)] - 2\pi i\eta \alpha S,\label{bfjk22Mar04}
\end{align}
where we just keep the leading order term in \eqref{bdyq22Mar04}, as in \eqref{bdvb22Mar04}, with 
\begin{align}
   \Nh=
 \begin{cases}
    N         & U(N), \\
    N/2 \mp 1 & SO/Sp(N), 
 \end{cases}
  \qquad
  \eta =
 \begin{cases}
    1   & U(N), \\
    1/2 & SO/Sp(N). 
 \end{cases}
\end{align}
The scale $\Lambda$ is written in terms of the bare coupling $\Lambda_b$
and the coupling constants in the problem, as before, and $2\pi i
\alpha$ is related to the bare gauge coupling by \eqref{bdto22Mar04}.

In the following, we discuss the cases with $\Nh=0$, $\Nh>0$ and $\Nh<0$
in order.
\begin{itemize}
 \item
      %
      %
       \textbf{$\boldsymbol{\hat N=0}$ case}\\
      In this case, the total superpotential is simply the sum of
      \eqref{bffs22Mar04} and \eqref{bfjk22Mar04}:
      \begin{align}
       W=\sqrt{2}\,Q\Qt S-2\pi i\eta \alpha S.
      \end{align}
      The only solution to the equation of motion is
      \begin{align}
       |Q|^2=|\Qt|^2,\qquad Q\Qt={2\pi i\eta \alpha\over \sqrt{2}},\qquad S=0.
      \end{align}
      This is consistent with the fact that $\alpha$ is proportional to the
      volume of the $\Pone$, and the D3-brane condensation $\langle Q\Qt
      \rangle $ corresponds to the size of the $\Pone$.  Furthermore, since
      $\ev{S}=0$, the superpotential vanishes: $W=0$.  Therefore, for $\Nh=0$,
      i.e.\ for $U(0)$ and $SO(2)$, geometric transition does not take place
      and we should set the corresponding glueball field $S\rightarrow 0$ from
      the beginning.
      
 \item
      %
      \textbf{$\boldsymbol{\hat N>0}$ case}\\
      In this case, there is a net RR flux through the $A$-cycle:
      $\oint_A H=\Nh$.  This means that the D3-brane hypermultiplet
      $(Q,\Qt)$ is infinitely massive, because the RR flux will induce
      $\Nh$ units of fundamental charge on the D3-brane.  Since the
      D3-brane is wrapping a compact space $S^3$, the fundamental charge
      on it should be canceled by $\Nh$ fundamental strings attached to
      it.  Those fundamental strings extend to infinity and thus cost
      infinite energy\footnote{This phenomenon is the same as that
      observed in the context of AdS/CFT \cite{Witten:1998xy,
      Gross:1998gk, Gubser:1998fp}.}. Therefore, we can forget about
      $Q,\Qt$ in this case, and the full superpotential is given just by
      the flux contribution \eqref{bfjk22Mar04}.  The equation of motion
      gives
      \begin{align}
       S^{\Nh} \simeq \Lambda^{3\Nh} e^{-2\pi i\eta \alpha},
      \end{align}
      which corresponds to the confining vacua of the gauge theory. Note that
      this case includes $U(1)$ and $Sp(0)$; these theories have a dual
      confining description.  This may sound a little paradoxical, but is
      related to the fact that the string theory computes not for the standard
      $G(N)$ gauge theory but the associated $G(N+k|k)$ higher rank gauge
      theory, which is confining and differs from standard $U(1)$ and $Sp(0)$
      due to residual instanton effects \cite{Aganagic:2003xq}.

 \item 
       %
       \textbf{$\boldsymbol{\hat N<0}$ case}\\
       In this case, the same argument as the $\Nh>0$ case tells us that we
       should not include the D3-brane fields $Q,\Qt$.  Hence the
       superpotential is just the flux part \eqref{bfjk22Mar04}, which again leads to
       \begin{align}
	S^{\Nh} \simeq \Lambda^{3\Nh} e^{-2\pi i\eta \alpha}.\label{bfnu22Mar04}
       \end{align}
       However, now \eqref{bfnu22Mar04} is physically unacceptable,
      since $S$ diverges in the weak coupling limit where the bare
      volume of $\Pone$ becomes $V=2\pi i \alpha\to \infty$ ($g_b\to 0$)
      --- i.e.\ taking $\Pone$ large would lead to $S^3$ also being
      large, which does not make sense geometrically.  The resolution is
      that $S$ cannot be a good variable: the $S^3$ does not actually
      blow up, and $S$ should be set to zero, $S \rightarrow 0$, also
      for this case.  Though $S$ is set to zero, the non-zero flux can
      lead to a non-zero superpotential contribution
      $W_{\textrm{flux}}=\Nh \Pi (S\rightarrow 0)$.

       \bigskip Note that the above result concerning the sign of $\Nh$ does
       not mean the gauge theory prefers D5-branes to anti-D5-branes; it just
       means that one should choose the sign of the NS flux (i.e.\ the sign of
       $2\pi i \alpha$) appropriately.  If one wraps the $\Pone$ with
       anti-D5-branes, one should flip the sign of the NS flux in order to have
       a blown up $S^3$ (which can be viewed as a generalization of Seiberg
       duality to $N_f=0$).
\end{itemize}

\subsection{General prescription}

Although we focused on the physics around just one $\Pone$ in the above,
the result is applicable to general cases where we have multiple $\Pone$
wrapped with D5-branes, because the geometry near each $\Pone$ is
identical to the conifold geometry considered above.  Therefore, if we
replace $\Nh$ with $\Nh_i$, all of the above conclusions carry over.

Once we have understood the physics, we can forget about the D3-brane
hypermultiplet $(Q,\Qt)$ and state the result as a general prescription
for how string theory treats $U(0)$, $SO(0)$, $SO(2)$, and $Sp(0),U(1)$
groups in the geometric dual description: 
\begin{itemize}
\item
\textbf{$\boldsymbol{U(0)}$, $\boldsymbol{SO(0)}$, $\boldsymbol{SO(2)}$:}\\
There are no glueball variables associated
with these gauge groups, so we should take the corresponding
 $S\rightarrow 0$.

\item
\textbf{All other groups, including $\boldsymbol{Sp(0)}$ and $\boldsymbol{U(1)}$:}\\
We should consider and extremize the corresponding glueball field $S$.
\end{itemize}

This prescription should also be applied when using the matrix
model \cite{Dijkgraaf:2002fc, Dijkgraaf:2002vw, Dijkgraaf:2002dh} to compute the glueball
superpotentials.

\section{Examples}
\label{IKRSV:exmpls}

Let us scan over all of the examples of \eqref{eq:higgscases},
considering the vacuum where the gauge group is unbroken, and ask when
glueball fields $S_i$ for the apparently trivial groups in
\eqref{eq:higgscases} should be set to zero, or included and extremized.
For the first three cases in \eqref{eq:higgscases}, $U(N)$, $SO(N)$, and
$Sp(N)$ with adjoint, the breaking \eqref{eq:higgscases} is
$G\rightarrow G\times U(0)^{K-1}$, and the glueball fields $S_i$ for the
$U(0)$ factors are to be set to zero.  This justifies the analysis of
these theories in the unbroken vacua in \cite{Ferrari:2002jp, Janik:2002nz}.
The next case is $SO(N)$ with a symmetric tensor $S$, where the vacuum
with unbroken gauge group is to be understood as $SO(N)\rightarrow
SO(N)\times SO(0)^{K-1}$, and again the glueball fields $S_i$ for the
$SO(0)$ factors are set to zero.  This eliminates the
Veneziano--Yankielowicz part of the superpotential for $SO(0)$, but the
$-1$ unit of flux associated with each $SO(0)$ does contribute to flux
terms $\Nh_i \Pi _i=-\Pi_i$ in \eqref{bdyq22Mar04}, even though this
does not contain $SO(0)$ glueballs any more.

The next case is $Sp(N)$ with an antisymmetric tensor, where the vacuum
with unbroken gauge group is to be understood as $Sp(N)\rightarrow
Sp(N)\times Sp(0)^{K-1}$.  Unlike the above cases, here we must keep and
extremize the $S_i$ for the $Sp(0)$ factors, as will be further
discussed shortly.

For the next to last example in \eqref{eq:higgscases}, $U(N)$ with $\Phi
+S+\widetilde S$, the vacuum with unbroken gauge group is to be
understood as $U(N)\rightarrow SO(0)\times U(N)\times U(0)^{K-1}$, and
the glueball fields $S_i$ for $SO(0)$ and $U(0)$ are to be set to zero.
Finally, for the last example in \eqref{eq:higgscases}, $U(N)$ with
$\Phi +A+\widetilde A$, the vacuum with unbroken gauge group is to be
understood as $U(N)\rightarrow Sp(0)\times U(N)\times U(0)^{K-1}$.
Though the string engineering of these examples differs somewhat from
those discussed in sect.\ 4 (it was obtained in \cite{Landsteiner:2003rh}),
the general prescription of sect.\ 4 is expected to carry over in
general: the glueball field $S_0$ for the $Sp(0)$ factor should be
included and extremized, rather than set to zero.  On the other hand,
the $S_i$ for the $U(0)$ factors are set to zero.  These latter two
theories in \eqref{eq:higgscases} were considered in \cite{Naculich:2003ka}
and it was noted there that for the case with antisymmetric one expands
on the matrix model side around a different vacuum than would be naively
expected; this indeed corresponds to keeping and extremizing the
glueball field $S_0$ for the $Sp(0)$ factor, as we have discussed.

We now illustrate some other breaking patterns in the examples of
\eqref{eq:higgscases}, from the matrix model perspective, for the case
of $K=2$.  We also compare with standard gauge theory results and
generally find agreement, even in cases where there was room for
disagreement because of the possibility of residual instanton effects
along the lines of \cite{Aganagic:2003xq}.  As will be discussed in more detail in
the following section, the agreement is thanks to a remarkable interplay
of different residual instanton contributions.

\subsection{$SO/Sp(N)$ theory with adjoint}
Consider $\CN=2$ $SO/Sp(N)$ theory broken to $\CN=1$ by a tree level
superpotential for the adjoint chiral superfield $\Phi$:
\begin{align}
   W_{\textrm{tree}}=\half \Tr[W(\Phi)],\qquad
  W(x)={m\over 2}x^2+{g\over 4}x^4.\label{bfpc22Mar04}
\end{align}
In the $SO$ case, we can skew-diagonalize $\Phi$ as
\begin{align}
 \Phi
 &\sim{\textrm{diag}}[\lambda_1,\cdots,\lambda_{N/2}]\otimes i\sigma^2.
\end{align}
The superpotential \eqref{bfpc22Mar04} has critical points at
$\lambda=0$ and $\lambda=\pm\sqrt{m/ g}$.  The classical supersymmetric
vacuum of the theory is given by distributing $N_0$ of the $N$
``eigenvalues'' $\lambda_i$ at the critical point $\lambda=0$ and $N_1$
``eigenvalue'' pairs at $\lambda=\pm \sqrt{m/g}$, with $N_0+2N_1=N$.  In
this vacuum, the gauge group breaks as $SO(N)\to SO(N_0)\times U(N_1)$.
Similarly in the $Sp$ case we have $Sp(N)\to Sp(N_0)\times U(N_1)$.

In the matrix model prescription, the effective superpotential in these
vacua is calculated by matrix model as
\begin{align}
   W_{\textrm{DV}}(S_0,N_0;S_1,N_1)
  &=
  \left({N_0\over2} \mp 1\right)S_0\left[1-\ln\left({S_0\over\Lambda_0^3}\right)\right]
 \cr
 &\qquad\qquad\qquad
 +N_1S_1\left[1-\ln\left(\frac{S_1}{\Lambda_1^3}\right)\right]+W_{\textrm{pert}},
  \cr
  W_{\textrm{pert}}
  &=
  N_0{\partial \CF_{S^2}\over \partial S_0}
  + N_1{\partial \CF_{S^2}\over \partial S_1} + 4\CF_{\RPtwo}.\label{bfsa22Mar04}
\end{align}
where $\CF_{S^2}$ and $\CF_{\RPtwo}$ are the $S^2$ and $\RPtwo$
contributions, respectively, to the free energy of the associated
$SO/Sp(N)$ matrix model, as defined in Appendix \ref{IKRSV:MM_calc_spot}.  The scales
$\Lambda_0$, $\Lambda_1$ in \eqref{bfsa22Mar04} are the energy scales of the low energy
$SO/Sp(N_0)$, $U(N_1)$ theories with the $\Phi$ field integrated
out, respectively.  They are related to the high energy scale $\Lambda$
by the matching conditions as in \eqref{eq:lamsospa}, which yields
\begin{align}
   (\Lambda_0)^{3(N_0/2 \mp 1)}
  &=
  m^{N_0/2 -N_1\mp 1} g^{N_1}
  \Lambda^{N \mp 2}
  ,\cr
  (\Lambda_1)^{3N_1}
  &=2^{-N_1}
  m^{-N_0\pm 2} g^{N_0+N_1\mp 2} \Lambda^{2(N \mp 2)} .\label{bfvc22Mar04}
\end{align}

The matrix model free energy is computed in Appendix
\ref{IKRSV:MM_calc_spot}, and the result is
\begin{align}
   W_{\textrm{pert}}
  =&
  \left({N_0\over2} \mp 1\right)
  \biggl[
    \left( {3\over 2}S_0^2-8S_0S_1+2S_1^2\right) \alpha\cr
 &\qquad\qquad\qquad
 + \left( -{9\over 2}S_0^3 + 42S_0^2S_1 - 36S_0S_1^2 + 4S_1^3\right)
 \alpha^2
    \cr
    &
    \qquad\qquad\qquad
    +\left(
      {45\over 2}S_0^4 - {932\over 3}S_0^3S_1 + 523S_0^2S_1^2
      -{608\over 3}S_0S_1^3 + {40\over 3}S_1^4
    \right) \alpha^3
  \biggr]
  \cr
  &+{N_1}
  \biggl[
    \left( -2S_0^2+2S_0S_1 \right) \alpha
    + \left( 7S_0^3 -18S_0^2S_1+6S_0S_1^2 \right) \alpha^2
  \cr
  &
  \qquad\qquad\qquad
  + \left(
      -{233\over 6}S_0^4 + {524\over 3}S_0^3S_1
      - 152S_0^2S_1^2 + {80\over 3}S_0S_1^3
    \right) \alpha^3
  \biggr]\cr
  &+\CO(\alpha^4)~,
 \label{bfwj22Mar04}
\end{align}
where $\alpha\equiv g/m^2$.  The result \eqref{bfwj22Mar04} agrees with
the one obtained in \cite{Fuji:2003vv}, where the glueball
superpotential was calculated by evaluating the periods
\eqref{xop22Mar04}.  The full result, \eqref{bfsa22Mar04} and
\eqref{bfwj22Mar04}, has the expected general form \eqref{bdyq22Mar04}:
\begin{align}
 W_{\textrm{eff}}=\left({N_0\over2}\mp 1\right) \Pi _0 (S_0, S_1)+N_1\Pi _1(S_0, S_1)-2\pi i 
 \alpha \left(\half S_0+S_1\right).
\end{align}
(Here $\alpha$ is the flux defined by \eqref{epny1Apr04}, not to be
confused with the expansion parameter $\alpha$ in \eqref{bfwj22Mar04}.)

The general prescription in section \ref{IKRSV:ST_prscrptn} reads in the
present case as follows:
\begin{itemize}
 \item 
       \textbf{$\boldsymbol{SO/Sp(N)\to SO/Sp(N)\times U(0)}$ (unbroken $\boldsymbol{SO/Sp}$):}\\
       Set $N_1=0$, $S_1=0$.  Then the superpotential is
       \begin{align}
	W_{\textrm{eff}}(S_0,N_0)
	&=
	\left({N\over2} \mp 1\right)S_0[1-\ln(S_0/\Lambda_0^3)]
	+2\left({N\over2} \mp 1\right)
	\left.{\partial \CF_{S^2}\over \partial S_0}\right|_{S_1=0}.\label{bfzm22Mar04}
       \end{align}
       This superpotential coincides with that of $U(N\mp 2)$ with
       adjoint and breaking pattern $U(N\pm 2)\rightarrow U(N\mp
       2)\times U(0)\times U(0)$, as expected from the map
       \eqref{eq:soaaux} or \eqref{eq:spaaux}.  As shown in
       \cite{Janik:2002nz} for $SO(N)$, this matrix model result agrees
       with that of standard gauge theory, via using the corresponding
       Seiberg--Witten curve.

 \item \textbf{$\boldsymbol{SO(N)\to SO(0)\times U(N/2)}$:}\\
       Set $N_0=0$ and take $S_0\rightarrow 0$, which eliminates the
       Veneziano--Yankielowicz part for the $SO(0)$.  Then the superpotential
       is
       \begin{align}
	W_{\textrm{eff}}(S_1)
	&=
	{N\over 2}S_1[1-\ln(S_1/\Lambda_1^3)]
	+4\left.\CF_{\RPtwo}\right|_{S_0=0}.\cr
	&=-\Pi _0(S_0, S_1)|_{S_0=0}+{N\over2}\Pi _1(S_0, S_1)|_{S_0=0}-2\pi i \alpha S_1.\label{bgaf22Mar04}
       \end{align}
       Note that $\left.{\partial \CF_{S^2}\over \partial
       S_1}\right|_{S_0=0}=0$ because $\CF_{S^2}$ does not contain terms
       with $S_1$ only (all terms are of the form $S_0^nS_1^m$ with
       $n>0$).  And though the Veneziano--Yankielowicz part of the
       superpotential for $SO(0)$ is eliminated via $S_0\rightarrow 0$,
       the $-1$ units of flux associated with $SO(0)$ does make a
       contribution in \eqref{bgaf22Mar04}, with the non-zero terms in
       $-\Pi _0(S_0, S_1)|_{S_0=0}$ in the second line of
       \eqref{bgaf22Mar04} coming from the term $4\CF
       _{\RPtwo}|_{S_0=0}$.

 \item
       \textbf{$\boldsymbol{SO(N)\to SO(2)\times U(N/2-1)}$:}\\
       Set $N_0=2$, $S_0=0$ and remove the Veneziano--Yankielowicz part
       for the $SO(2)$.  Then the superpotential is
       \begin{align}
	W_{\textrm{DV}}(S_1) &= \left({N\over2}-1\right)S_1[1-\ln(S_1/\Lambda_1^3)],
       \end{align}
       where we used $\left.{\partial \CF_{S^2}\over \partial
       S_1}\right|_{S_0=0}=0$ again.  Integrating out $S_1$ gives
       \begin{align}
	W_{\textrm{low}}
	=\left({N\over2}-1\right)\Lambda_1^3 
	=\half \left({N\over2}-1\right)g\Lambda^4\label{bgdf22Mar04} 
       \end{align}
 \item
       \textbf{$\boldsymbol{Sp(N)\to Sp(0)\times U(N/2)}$:}\\
       Set $N_0=0$ in the equation and keep both $S_0$ and $S_1$.
       Then the superpotential is
       \begin{align}
	W_{\textrm{DV}}(S_0,N_0,S_1,N_1)
	&=
	S_0[1-\ln(S_0/\Lambda_0^3)] + NS_1[1-\ln(S_1/\Lambda_1^3)]+W_{\textrm{pert}},\cr
	W_{\textrm{pert}}
	&=
	{N\over 2}{\partial \CF_{S^2}\over \partial S_1} + 4\CF_{\RPtwo}
	=
	{N\over 2}{\partial \CF_{S^2}\over \partial S_1}
	+2{\partial \CF_{S^2}\over \partial S_0}.\label{bgeb22Mar04}
       \end{align}
\end{itemize}

%
%
%
%
%
\begin{table}
  \begin{center}
  \epsfxsize=14cm \epsfbox{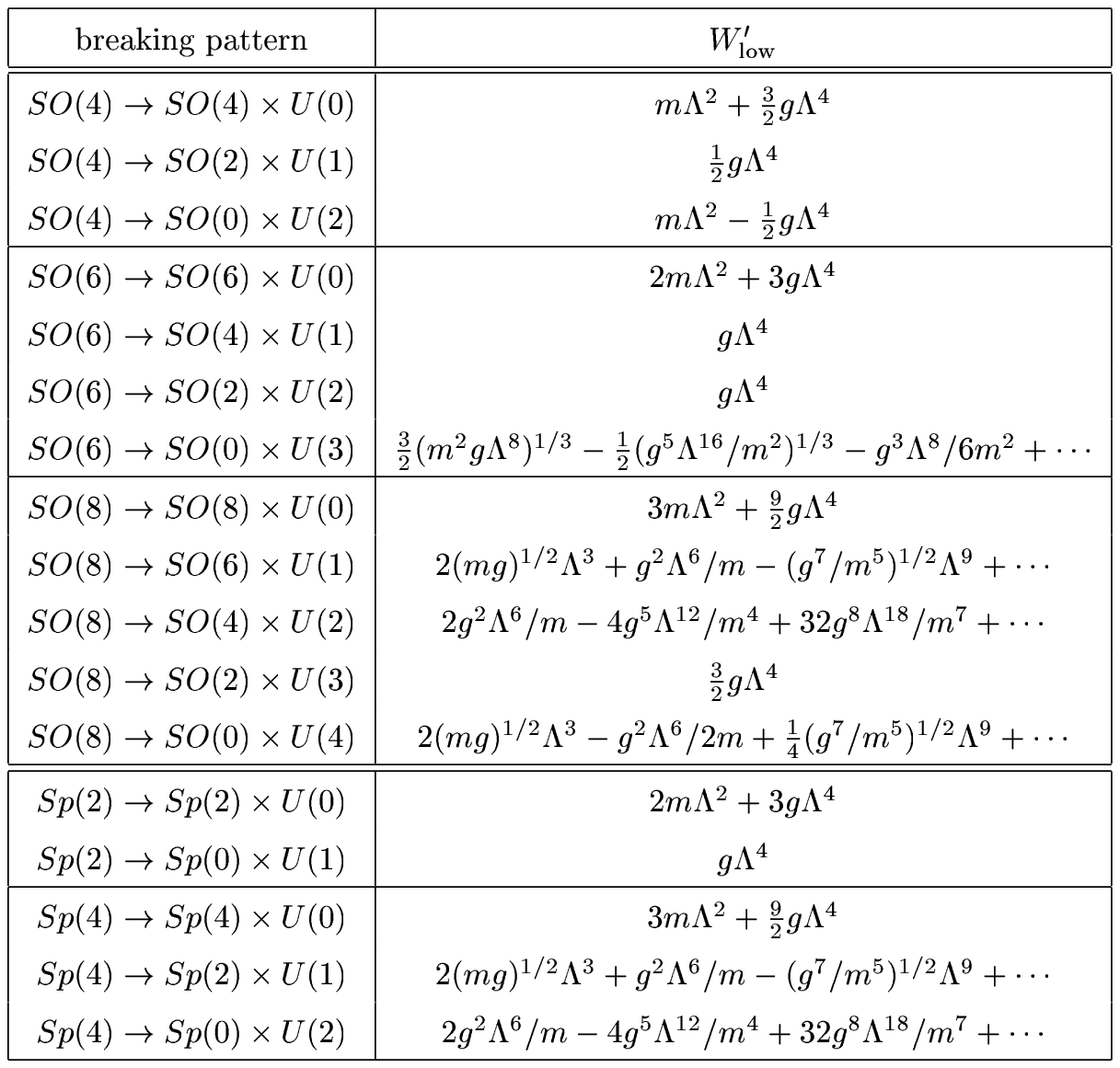}
  \caption
   [Low energy superpotential calculated from gauge theory
   and matrix model]
   {The low energy superpotential calculated from the factorization of the
   Seiberg--Witten curve and from matrix model.  In the above, the 
   classical contribution has been subtracted:
   $W_{\textrm{low}}=-{N_1 m^2/ 4g}+W'_{\textrm{low}}$.
   \label{superpottable}}
 \end{center}
\end{table}
For various breaking patterns, we integrated out the glueball
superfield(s) from the glueball superpotential
\eqref{bfzm22Mar04}--\eqref{bgeb22Mar04}, and calculated the low energy
superpotential $W_{\textrm{low}}$ as a function of coupling constants
$m$, $g$, and the scale $\Lambda$.  Having obtained the actual matrix
model results, we can compare to the superpotential as computed via
standard gauge theory methods, such as via factorizing of the
Seiberg--Witten curve.  This method is reviewed in Appendix \ref{IKRSV:GT_calc_spot}, and the
results are found to agree with the matrix model results completely.
The resulting $W_{\textrm{low}}$ is shown in Table \ref{superpottable}.

We have been considering $SO(N)$ theory with even $N$.  The $SO(2N+1)$
theory with adjoint in the $SO(2N+1)\to U(1)^N$ vacuum was studied
diagrammatically in \cite{Abbaspur:2003hf}.

\subsection{$Sp(N)$ theory with antisymmetric tensor}

Consider $Sp(N)$ theory with an antisymmetric tensor chiral superfield
$A=-A^T$.  Take cubic tree level superpotential
\begin{align}
   W_{\textrm{tree}}=\half \Tr[W(\Phi)],\qquad
  W(x)={m\over 2}x^2+{g\over 3}x^3,\label{bggr22Mar04}
\end{align}
where $\Phi=AJ$, and $J$ is the invariant antisymmetric tensor
$J={\unit}_{N/2} \otimes i\sigma^2$.  We do not require $A$ to be
traceless, i.e.\ $\Tr[\Phi]=\Tr[AJ]\neq 0$.  By a complexified $Sp(N)$
gauge rotation, $\Phi$ can be diagonalized as \cite{Cho:1996bi}
\begin{align}
   \Phi
  &\cong  {\textrm{diag}}[\lambda_1,\cdots,\lambda_{N/2}]\otimes {\unit}_2,
 \qquad
  \lambda_i\in {\mathbb C}.
\end{align}
The superpotential \eqref{bggr22Mar04} has critical points at
$\lambda=0,-m/g$.  The classical supersymmetric vacuum of the theory is
given by distributing $N_1$ and $N_2$ eigenvalues at the critical point
$\lambda=0$ and $\lambda=-{m/g}$, respectively, with $N_1+N_2=N$,
breaking $Sp(N)\to Sp(N_1)\times Sp(N_2)$.

The glueball superpotential is calculated from the associated $Sp(N)$
matrix model as
\begin{align}
  W_{\textrm{DV}}(S_1,N_1;S_2,N_2)
  =&
  \left({N_1\over 2} + 1\right)S_1[1-\ln(S_1/\Lambda_1^3)]\cr
  &\qquad\qquad
  +\left({N_2\over 2} + 1\right)S_2[1-\ln(S_2/\Lambda_2^3)]
  +W_{\textrm{pert}},
  \cr
  W_{\textrm{pert}}
  &=
  N_1{\partial \CF_{S^2}\over \partial S_1}
  + N_2{\partial \CF_{S^2}\over \partial S_2} + 4\CF_{\RPtwo}.\label{bgkm22Mar04}
\end{align}
The scales $\Lambda_1$, $\Lambda_2$ in \eqref{bfsa22Mar04} respectively
are the energy scales of the low energy $Sp(N_1)$, $Sp(N_2)$ theories
with the $\Phi$ field integrated out.  They are related to the high
energy scale $\Lambda$ by the matching conditions as in
\eqref{eq:laminsp}, which yields
\begin{align}
  (\Lambda_1)^{3({N_1/2}+1)} &= m^{N_1/2-N_2-1} g^{N_2} \Lambda^{N+4} ,\cr
  (\Lambda_2)^{3({N_2/2}+1)} &= (-1)^{N_2/2-1}   m^{-N_1+N_2/2-1} g^{N_1} \Lambda^{N+4}.
 \label{bglg22Mar04}
\end{align}

The matrix model free energy is computed in Appendix \ref{IKRSV:MM_calc_spot}, and the
result is
\begin{align}
  W_{\textrm{pert}}
 =&\,
 (N_1+2)
 \biggl[
   \left( -{S_1^2} + 5 S_1 S_2 - {5\over 2}S_2^2 \right)\alpha\cr
   &\qquad\qquad\qquad
   +
   \left(
    -{16\over 3}S_1^3 + {91\over 2}S_1^2 S_2 - {59}S_1 S_2^2
    + {91\over 6} S_2^3
   \right)\alpha^2
   \cr
   &\qquad\qquad\qquad
 +
   \biggl(
    -{140\over 3}S_1^4
    + {1742\over 3}S_1^3 S_2
    - {1318}S_1^2 S_2^2\cr
 &\qquad\qquad\qquad\qquad\qquad
    + {2636\over 3}S_1 S_2^3
    - {871\over 6} S_2^4
   \biggr)\alpha^3
 \biggr]
 \cr
 &+(N_2+2)\biggl[S_1\leftrightarrow S_2,\, \alpha\to -\alpha\biggr],\label{bgmi22Mar04}
\end{align} 
where $\alpha\equiv g^2/m^3$.  This is as expected from the map \eqref{eq:spasaux} and
\cite{Cachazo:2003kx}.

In particular, let us concentrate on the unbroken case, $N_2=0$.  Unlike 
\cite{Kraus:2003jf, Kraus:2003jv}, we do not set $S_2=0$, but rather 
keep $S_2$ non-zero and extremize with respect to it, according to our general
prescription, to obtain the actual matrix model result.  
After integrating out $S_1$ and $S_2$ from \eqref{bgmi22Mar04}, we
obtain the superpotential as a power series in $\Lambda_1$ and
$\Lambda_2$ as
\begin{align}
   W_{\textrm{low}}= \left( {N\over2}+1 \right) \Lambda_1^3+\Lambda_2^3
  +({\textrm{higher~order~terms~in~}}\Lambda_{1,2}).\label{bgqk22Mar04}
\end{align}
The matching relation \eqref{bglg22Mar04} gives
\begin{align}
   \Lambda_2^3=-(\Lambda_1^3)^{{N/2}+1}\alpha^{{N/2}},\label{bgta22Mar04}
\end{align}
so the terms containing $\Lambda_2$ in \eqref{bgqk22Mar04} starts to
contribute to the superpotential at order $(\Lambda_1^3)^{{N/2}+1}$, i.e.\
like $Sp(N)$ instantons.  If we use the relation \eqref{bgta22Mar04} and
write out all the terms in \eqref{bgqk22Mar04}, we obtain
\begin{equation}
\begin{split}
  N=0: & \quad W_{\textrm{low}}=\CO(\alpha^4), \cr
  N=2: & \quad W_{\textrm{low}}=2\Lambda_1^3 + \CO(\alpha^4),\cr
  N=4: & \quad W_{\textrm{low}}=3\Lambda_1^3 - \Lambda_1^6 \alpha
          -2 \Lambda_1^9 \alpha^2 -{187\over 27} \Lambda_1^{12}\alpha^3
    +\CO(\alpha^4), \cr
  N=6: & \quad W_{\textrm{low}}=4\Lambda_1^3 - 3\Lambda_1^6 \alpha
          -{47\over 6} \Lambda_1^9 \alpha^2
    -{75\over 2} \Lambda_1^{12}\alpha^3
    +\CO(\alpha^4), \cr
  N=8: & \quad W_{\textrm{low}}=5\Lambda_1^3 - 5\Lambda_1^6 \alpha
          -{13} \Lambda_1^9 \alpha^2
    -{65} \Lambda_1^{12}\alpha^3
    +\CO(\alpha^4). 
\end{split}\label{bgvy22Mar04}
\end{equation}
Thus properly accounting for $S_2$, it turns out that these matrix model
results agree perfectly, up to the order presented, with the standard
gauge theory results (Eq.\ (4.13) of \cite{Kraus:2003jf}).  (The discrepancies
found in \cite{Kraus:2003jf} set in at order $\Lambda_1^{3(N/2+1)}$, and are
canceled e.g.\ by \eqref{bgta22Mar04}.)  In \eqref{bgvy22Mar04}, for
$N=0,2$, there were rather remarkable cancellations between the
instanton contributions from $Sp(N_1)$ and $Sp(N_2)$.  This will be
further discussed and generalized in the next section.

The matrix model prediction for the superpotential of the $SO(N)$
theory with symmetric tensor can similarly be obtained by simply
changing the $N_i/2+1$ in \eqref{bgmi22Mar04} to $N_i/2-1$.  It should be
possible to compute the superpotential from gauge theory using the
duality for this theory \cite{Intriligator:1995ff}.  The result is expected
to be compatible with the map \eqref{eq:sosaux} to the superpotential
computed for the $U(N-2K)$ theory with adjoint.

\section{Residual instantons:  string theory (matrix model) versus
gauge theory}
\label{IKRSV:resdl_instntns}

A remarkable aspect of the string theory (matrix model) computation of
the effective superpotential is that \eqref{eq:prepotgen} can be
obtained purely in terms of the dynamics of the low-energy $\prod
_{i=1}^K G(N_i)$ theory on the RHS of \eqref{eq:higgsbp}.  The only
information needed about the high-energy $G(N)$ gauge theory is the
perturbative contribution of the $G(N)/\prod _i G_i(N_i)$ ghosts to the
glueball superpotential \eqref{eq:prepotgen}, as discussed in
\cite{Dijkgraaf:2002pp}, along with the matching relations connecting the scales
$\Lambda _i$ of the low-energy $G_i(N_i)$ factors to the scale $\Lambda$
of the high-energy $G(N)$ theory.  This is very different from the
conventional description of standard gauge theory, where there can be
non-perturbative contributions to $W_{\textrm{low}}$ which are not
readily seen in terms of the low-energy theory on the RHS of
\eqref{eq:higgsbp}.  An example of such an effect is instantons in the
broken part of the group when $\pi _3 (G(N)/\prod _i G_i(N_i))\neq 0$
(see e.g.\ \cite{Csaki:1998vv}).  Nevertheless, the string theory/ matrix
model does properly reproduce such effects, via a low-energy
description.

A gauge theory interpretation for the string theory/ matrix model
results was given in \cite{Dijkgraaf:2003xk, Aganagic:2003xq}: the string theory
/ matrix model results actually refer to a particularly natural UV
completion of the original $G(N)$ theory, where it is embedded in the
supergroup $G(N+k|k)$ with $k$ large.  This latter theory has a Higgs
branch, where $k$ can be reduced successively, eventually Higgsing the
theory down to the original $G(N)$ theory.  More generally, the theory
with breaking pattern \eqref{eq:higgsbp} is replaced with
\begin{align}
 G(N+k|k)\rightarrow \prod _{i=1}^KG_i(N_i+k_i|k_i),
 \label{eq:higgsbpsg}
\end{align}
which has a Higgs branch flat direction connecting it to
\eqref{eq:higgsbp}.  Consideration of the particular matter content of
the $G(N+k|k)$ theories along the Higgs branch, which often has extended
supersymmetry, suggests that no dynamically generated superpotential
ever lifts this Higgs branch moduli space, i.e.\ that the
superpotentials of these particular theories are always independent of
the location of the theory on this Higgs branch \cite{Aganagic:2003xq}.  Moving
along the Higgs branch has the effect of reducing $k$, and this expected
independence of the superpotential of the position on the Higgs branch
fits with the fact that the $G(N+k|k)$ matrix model results are $k$
independent, because all $k$ dependence cancels in the supertraces.

Because of the expected independence of the superpotential on the Higgs
branch, and because we Higgs back to the original $G(N)$ theory, in most
cases, this ``F-completion'' of the original $G(N)$ theory into the
$G(N+k|k)$ theory is of no consequence.  There are, however, a few rare
exceptions, where the superpotential of the Higgsed $G(N+k|k)$ theory
differs from that of the standard $G(N)$ theory.  This difference comes
from residual instantons in $G(N+k|k)/G(N)$, which need not decouple
even if $G(N+k|k)$ is Higgsed to $G(N)$ far in the UV.  As verified in
\cite{Aganagic:2003xq}, these residual instanton contributions precisely account
for the few differences between the string theory (matrix model) results
and standard gauge theory, for example the glueball superpotentials,
with coefficient $h=1$, for $U(1)$ and $Sp(0)$, e.g.\ with an adjoint
and quadratic superpotential.

In many cases, however, these residual instanton contributions sum up to
yield precisely the result expected from standard gauge theory,
including superpotential contributions which in standard gauge theory
would not have had a known low-energy description. In particular,
residual instanton contributions which could have lead to potential
discrepancies with standard gauge theory often completely cancel.  The
cancellation occurs once one sums over the different terms $i$ in
\eqref{eq:weffgen}, upon using the precise matching relation between the
low-energy scales $\Lambda _i$, and the original high-energy scale
$\Lambda$.

As an example, consider $U(K)$ with adjoint matter and breaking pattern
$U(K)\rightarrow U(1)^K$. For n$K=1$ the string theory (matrix model)
description includes a residual instanton effect, yielding
$W_{\textrm{low}}=\Lambda _L^3$ rather than the standard gauge theory
answer $W_{\textrm{sgt}}=0$ \cite{Aganagic:2003xq}.  But for all $K>1$
the string theory/ matrix model result is $W_{\textrm{low}}=0$, in
agreement with the standard gauge theory expectation for
$U(K)\rightarrow U(1)^K$.  The result $W_{\textrm{low}}=0$ looks like a
remarkable cancellation because the glueball superpotential
$W_{\textrm{eff}}(S_1, \dots S_K)$ is quite non-trivial.  Nevertheless,
upon solving for the $\ev{S_i}$ and plugging back in, the exact result
for $W_{\textrm{low}}=W_{\textrm{eff}}(\ev{S_i})$ is zero, as was proven
in \cite{Cachazo:2002pr}.

To illustrate this cancellation, consider the leading order gaugino
condensation contribution to $W_{\textrm{eff}}(S_i)$ in the string
theory (matrix model) constructions, where the unbroken $U(1)$ factors
in $U(K)\rightarrow U(1)^K$ contribute as in \eqref{eq:wgca}, with
$h_i=1$, unlike in standard gauge theory:
\begin{align}
 W_{\textrm{gc}}(S_i)=\sum _{i=1}^kS_i \left(\log({\Lambda _i^3\over
 S_i})+1\right),
 \label{eq:wgcaa}
\end{align}
with $\Lambda _i^3=g_{K+1}\Lambda ^{2N}\prod _{j\neq i} (a_j-a_i)^{-1}$
by using \eqref{eq:lamin} with all $N_i=1$.  Though this is a
non-trivial superpotential, it vanishes upon integrating out the $S_i$:
\begin{align}
 W_{\textrm{gc}}(\ev{S_i})=\sum _{i=1}^K \Lambda _i ^3=
 g_{K+1}\Lambda ^{2N}\sum _{i=1}^K \prod _{j\neq i}
 (a_j-a_i)^{-1}=g_{k+1}^2\Lambda ^{2N}\oint {dx\over 2\pi i}{1\over
 W'(x)}=0.
 \label{eq:wgcasv}
\end{align} 
The contour in \eqref{eq:wgcasv} encloses all the zeros of $W'(x)$, and
we get zero for all $K>1$ by pulling the contour off to infinity.  We
see here why $K=1$ is different: we then get a residue at infinity,
leading to the low energy superpotential $W_{U(1)}=\Lambda _L^3$, as in
\eqref{eq:wgcisi}, with $h_{U(1)}=1$ as in \eqref{eq:wlowuiv}.

To give another example of such a cancellation of residual instanton
effects, consider the string theory (matrix model) result for $Sp(N)$
with an antisymmetric tensor $A$, with $W_{\textrm{tree}}$ having $K$
critical points, for the case $N=0$.  For the case of $K=1$, the
superpotential is just a mass term for $A$ and the low-energy
superpotential is the $Sp(0)$ gaugino condensation superpotential, with
$h(Sp(0))=1$: $W=\Lambda ^3$, unlike standard gauge theory.  Again, this
can be understood as a residual instanton effect in the F-completion of
$Sp(N)$ to $Sp(N+k|k)$, which is present precisely for the case $N=0$
\cite{Aganagic:2003xq}.  For a higher order superpotential, $K>1$, we would write
the breaking pattern as $Sp(0)\rightarrow Sp(0)^K$.  For all $K>1$, the
residual instanton effects all cancel, precisely as in the
$U(K)\rightarrow U(1)^K$ example discussed above; in fact, the two
theories have the same effective superpotential $W_{\textrm{eff}}$
(aside from the classical difference), as discussed in \cite{Cachazo:2003kx}
and sect.\ 3.3.  Thus, for example, \eqref{eq:wgcasv} can also be
interpreted as the leading gaugino condensation contributions from the
$Sp(0)^K$ factors, and where we now use the matching relation
\eqref{eq:laminsp} to relate the $\Lambda _i^3$ to
$g_{K+1}\Lambda^{2K}\prod _{j\neq i}(a_j-a_i)^{-1}$.  Again, there is
complete cancellation in $W_{\textrm{eff}}$ here, except for the case
$K=1$.

More generally, for $Sp(N)$ with antisymmetric, breaking as\break
$Sp(N)\rightarrow \prod _{i=1}^KSp(N_i)$, the results obtained via the
string theory / matrix model glueball potential $W_{\textrm{eff}}(S_1,
\dots S_K)$, upon integrating out the $S_i$, appears to always agree
with standard gauge theory results for the superpotential
\cite{Cho:1996bi,Csaki:1996eu}, as seen in the examples of
\cite{Cachazo:2003kx} and \eqref{bgvy22Mar04}.  This agreement comes about
via a remarkable interplay between the different terms $i$ in
\eqref{bdyq22Mar04}.  If we treated the scales $\Lambda _i$ of the
$Sp(N_i)$ factors as if they were initially independent, each term
$\Nh_i \Pi (\ev{S_i}) -2\pi i \eta _i \ev{S_i}$ in \eqref{bdyq22Mar04}
would be a complicated function of $\Lambda _i$, which does not have a
known, conventional, interpretation in terms of standard gauge theory
for the low-energy $Sp(N_i)$ factor.  But upon adding the different $i$
terms and using the matching relations relating $\Lambda _i$ to
$\Lambda$, e.g.\ \eqref{eq:laminsp}, one nevertheless obtains the
standard gauge theory results, thanks to an intricate interplay between
the different terms $i$.

By the map of \eqref{eq:spasaux} \cite{Cachazo:2003kx}, the agreement between
string theory / matrix models and standard gauge theory for $Sp(N)$ with
antisymmetric can be phrased as such an agreement for $U(N/2+K)$ with
adjoint and breaking pattern $U(N/2+K)\rightarrow \prod _{i=1}^KU(N_i/2+1)$.

As another example, consider $U(N)$ with adjoint $\Phi$ and
superpotential having $K=N-1$, in the vacuum where $U(N)\rightarrow
U(2)\times U(1)^{N-2}$.  Factorizing the Seiberg--Witten curve yields
for the exact superpotential \cite{Elitzur:1996gk}
\begin{align}
 W_{\textrm{exact}}=W_{\textrm{cl}}(g)\pm 2 g_{N}\Lambda ^N.  
 \label{eq:wscph}
\end{align}
The map of \cite{Cachazo:2003kx} and subsection \ref{IKRSV:rel_SO/Sp_and_U(N)}
relates this to $Sp(2)\rightarrow Sp(2)\times Sp(0)^{N-2}$, where the
exact gauge theory result agrees with \eqref{eq:wscph}, up to the
classical shift, upon using the relation \eqref{eq:spausr}.  A priori,
one might expect the string theory / matrix model result to disagree
with \eqref{eq:wscph}, due to residual instanton contributions from the
$U(1)^{N-2}$ or the $Sp(0)^{N-2}$ in $U(N)\rightarrow U(2)\times
U(1)^{N-2}$ and $Sp(2)\rightarrow Sp(2)\times Sp(0)^{N-2}$ respectively.
But the string theory / matrix model result nevertheless agrees with
\eqref{eq:wscph}, thanks to the interplay between the different terms.
Consider, in particular, the case $U(3)\rightarrow U(2)\times U(1)$.
The fact that \eqref{eq:wscph} will only hold if remarkable
cancellations occur upon integrating out $S_1$ and $S_2$ from the
non-trivial $W(S_1, S_2)$, was discussed in \cite{Cachazo:2001jy}, where
the cancellations were verified to indeed occur, up to order $\alpha
^3$.  This is checked to one higher order in \eqref{bgvy22Mar04}, since
it is related to $Sp(2)\rightarrow Sp(2)\times Sp(0)$ by the map of
\cite{Cachazo:2003kx} and subsection \ref{IKRSV:rel_SO/Sp_and_U(N)}.  The
leading order cancellation, say in terms of $U(3)\rightarrow U(2)\times
U(1)$, is between $U(1)$ gaugino condensation, $\Lambda _2^3$, and a
higher order term coming from integrating out $S_1$ from
$W_{\textrm{pert}}(S_i)$.

The residual instanton contributions associated with the UV completion\break
\eqref{eq:higgsbpsg}, as opposed to the standard gauge theory results
for \eqref{eq:higgsbp} do not always cancel, however.  The cases where
we find non-cancellations are when the degree of the superpotential is
sufficiently large, so that it contains terms which are not independent
moduli.  As an example, consider $U(1)$ with $W_{\textrm{tree}}$ as in
\eqref{ump22Mar04} having $K$ minima, breaking $U(1)\rightarrow
U(1)\times U(0)^{K-1}$.  The gaugino condensation contribution to the
superpotential, according to the string theory (matrix model)
construction, is given by \eqref{eq:wgca} with $h_1=1$ and all other
$h_i=0$ and their $S_i$ set to zero.  Upon integrating out $S_1$, we
thus obtain the superpotential
\begin{align}
 W_{\textrm{gc}}=\Lambda _1^3=\Lambda ^2W''(a_1)=g_{K+1}\Lambda
 ^2\prod _{j\neq 1} (a_j-a_1),
 \label{eq:wgczz}
\end{align}
where we used the matching relation \eqref{eq:lamin} with $N_1=1$ and
all other $N_j=0$.

The full low-energy effective superpotential $W_{\textrm{low}}(g_i,
\Lambda)$ can be regarded as the generating function for the operator
expectation values:
\begin{align}
 \ev{u_j}={\partial W_{\textrm{low}}(g_i, \Lambda)\over \partial g_j}
 \qquad u_i\equiv {1\over j}
 \Tr \Phi ^j.
 \label{eq:wlowgf}
\end{align}
In the $U(1)$ theory, we have classical relations $u_j={1\over j}u_1^j$.
But the quantum contribution \eqref{eq:wgczz} (along with additional,
higher order contributions) imply quantum deformation of these classical
relations, due to the residual instanton effects in the
$U(1+k|k)\rightarrow U(1+k_1|k_1)\times U(k_2|k_2)\dots U(k_K|k_K)$
F-completion.  For the simplest such example, consider $U(1)$ with
$W_{\textrm{tree}}=\half m \Phi ^2 +\lambda \Phi$.  The low-energy
superpotential is
\begin{align}
 W_{\textrm{low}}=-{\lambda ^2\over 2m}+m\Lambda ^2,
 \label{eq:wlowexq}
\end{align}
with the first term the classical contribution and the second the
residual instanton.  Using \eqref{eq:wlowgf} we then get
\begin{align}
 \ev{u_1}=-{\lambda \over m}, \quad, \ev{u_2}={\lambda ^2\over
 2m^2}+\Lambda ^2
 \qquad\hbox{i.e.}\quad \ev{u_2}=\half \ev{u_1^2}+\Lambda ^2,
 \label{eq:evqis}
\end{align}
which can be regarded as an instanton correction to the composite
operator $u_2$.

As another such example, consider $U(2)$ with an adjoint and
$W_{\textrm{tree}}$ having $K$ minima, in the vacuum where the gauge
group is broken as $U(2) \rightarrow U(1)\times U(1)\times U(0)^{K-2}$.
The gaugino condensation contribution to $W_{\textrm{low}}$ is
\begin{align}
 W_{\textrm{gc}}=\Lambda _1^3+\Lambda _2^3=g_{K+1}\Lambda
 ^4\left({\prod _{j=3}^K
 (a_j-a_1)-\prod _{j=3}^K (a_j-a_2)\over a_2-a_1}\right).
 \label{eq:wlowggf}
\end{align}
For example, for $U(2)$ with $W_{\textrm{tree}}$ having $K=3$ critical
points, we break $U(2)\rightarrow U(1)\times U(1)\times U(0)$ and
\eqref{eq:wlowggf} leads to
\begin{align}
 W_{\textrm{gc}}=g_4\Lambda ^4.
 \label{eq:wlowcgexx}
\end{align}
Computing expectation values as in \eqref{eq:wlowgf} this leads to
\begin{align}
 \ev{u_4}=\ev{u_4}_{\textrm{cl}}+\Lambda ^4,
 \label{eq:uivc}
\end{align}
which can be interpreted as an instanton contribution to the composite
operator $u_4=\Tr \Phi ^4$ in $U(2)$ gauge theory.  More generally, for
$U(N)$ gauge theory, the independent basis of operators $u_j={1\over
j}\Tr \Phi ^j$ are only those with $j\leq N$, those with $j>N$ can be
expressed as products of these basis operators via classical relations.
But these relations can be affected by instantons.  In particular, for
$U(N)$ with an adjoint, the instanton factor is $\Lambda ^{2N}$, so
operators $u_j$ with $j\geq 2N$ can be affected.  The above residual
instanton contributions of the $U(N+k|k)$ UV completion can be
interpreted as implying specific such instanton corrections to the
higher Casimirs $u_j$.

A similar situation arises in the $\CN=1^*$ $U(N)$ theory, where the
effective superpotential of the matrix model and conventional gauge
theory differ by a contribution $N^2m^3 E_2(N \tau )$ \cite{Dorey:2002tj};
this was interpreted in \cite{Dorey:2002tj} as differing operator definitions
of $\Tr \Phi ^2$ between gauge theory and the matrix model at the level
of instantons.  Related issues for multi-trace operators were seen in
\cite{Balasubramanian:2002tm}.

\section{Conclusions}

To compute the correct string theory / matrix model results, we should
include or not include the glueball fields $S_i$ according to the
prescription of this paper.  Upon doing so, in all examples that we know
of, the string theory / matrix model results agree with the results of
standard gauge theory, at least in those cases where the relevant gauge
theory does not suffer from UV ambiguities.  In the case where such
ambiguities are present, for example in defining composite operators
appearing in $W_{\textrm{tree}}$, the string theory / matrix model
results correspond to a particular UV definition of the theory.  The
agreement with standard gauge theory results is often due to a
remarkable interplay between the different low-energy terms, found upon
integrating out the glueball fields $S_i$, and connecting their scales
$\Lambda _i$ via the appropriate matching relation to the scale
$\Lambda$ of the original theory.  In some cases, this interplay leads
to complete cancellations of the residual instanton contributions to
$W_{\textrm{low}}$ coming {}from the $G(N+k|k)$ completion
\cite{Dijkgraaf:2003xk, Aganagic:2003xq}.  Perhaps there is some
additional structure governing the glueball superpotentials, which would
make these remarkable cancellations more manifest.

%
%


\BeginAppendix

\section{Matrix model calculation of superpotential}
\label{IKRSV:MM_calc_spot}

In this appendix, after giving a proof for a general relation that
relates $S^2$ and $\RPtwo$ contributions to the $SO/Sp$ matrix model free
energy, we compute explicitly the free energy of the matrix models
associated with $SO/Sp$ gauge theory with adjoint and $Sp$ gauge theory
with antisymmetric tensor.  These matrix model results are used in
section \ref{IKRSV:exmpls} to evaluate the glueball superpotential of the
corresponding gauge theories.

\subsection{Proof for relation between $\CF_{S^2}$ and $\CF_{\RPtwo}$}

Here we prove a general relation between the $S^2$ and $\RPtwo$
contributions to the $SO/Sp(N)$ matrix model free energy:
\begin{align}
  \CF_{\RPtwo}= 
 \begin{cases}
 \displaystyle
 \mp{1\over2}{\partial \CF_{S^2}\over \partial S_0}
   & \text{$SO/Sp$ with adjoint,}\\
 \displaystyle
  \mp{1\over2}\sum_{i=1}^K {\partial \CF_{S^2}\over \partial S_i}
   & 
  \text{$SO/Sp$ with symmetric/antisymmetric tensor.}
 \end{cases}
  \label{bhuq22Mar04}
\end{align}
The first equation was conjectured in \cite{Ita:2002kx, Ashok:2002bi}
based on explicit diagrammatic calculations, and proven in
\cite{Janik:2002nz} for the case of unbroken vacua.  Here we will give a
general matrix model proof for arbitrary breaking pattern.  These
relations are equivalent to the maps \eqref{eq:soaaux},
\eqref{eq:spaaux}, \eqref{eq:spasaux}, and \eqref{eq:sosaux}, which we
obtained in subsect.\ \ref{IKRSV:rel_SO/Sp_and_U(N)}
 immediately from the string theory geometric
transition construction, accounting for the orientifold contributions to
the fluxes.

\bigskip
Consider $U(\mN)$ and $SO/Sp(\mN)$ matrix models which correspond to
$U(N)$ and $SO/Sp(N)$ gauge theories with a two-index tensor matter
field.  The partition function is
\begin{align}
  \mZ
 =
 e^{-\frac{1}{\mg^2}\mF(\mS_i)}
 =
 \int d\mPhi \,
   e^{-\frac{1}{\mg}W_{\textrm{tree}}(\mPhi)}.\label{bhxy22Mar04}
\end{align}
We denote matrix model quantities by boldface letters, following the
notation of \cite{Kraus:2003jv}.  $\mPhi$ is an $\mN\times\mN$ matrix
corresponding to the $\Phi$ field in gauge theory, and the ``action''
$W_{\textrm{tree}}$ is defined in \eqref{ump22Mar04}, \eqref{eq:bsospx}
and \eqref{eq:basp}.
The matrix integral \eqref{bhxy22Mar04} is evaluated perturbatively
around the general broken vacua of \eqref{eq:higgscases}, with $N_i$
replaced by $\mN_i$.  We take the large $\mN$ limit $\mN_i\to\infty$,
$\mg\to 0$ with $\mg \mN_i\equiv\mS_i$ kept finite.
The dependence of the free energy $\mF(\mS_i)$ on $\mN_i$ are eliminated
in favor of $\mS_i$, and $\mF(\mS_i)$ is expanded in the 't Hooft
expansion as
\begin{align}
 \mF(\mS_i)
 =\sum_{\CM} \mg^{2-\chi(\CM)} \CF_{\CM}(\mS_i)
 =\CF_{S^2}+\mg\CF_{\RPtwo}+\cdots\label{bhzb22Mar04}
\end{align}
where the sum is over all compact topologies $\CM$ of the matrix model
diagrams written in the 't Hooft double-line notation, and
$\chi(\CM)$ is the Euler number of $\CM$.

The matrix model resolvent is defined as follows:
\begin{align}
   \mR(z)
  \equiv
  \mg\Bracket{\Tr\left[\frac{1}{z-\mPhi}\right]}
  =
  \mR_{S^2}(z)+\mg\mR_{\RPtwo}(z)+\cdots.
\end{align}
For $U(N)$ theory with adjoint, the expansion parameter is $\mg^2$
instead of $\mg$, and in particular, $\mR_{\RPtwo}(z)\equiv 0$.
The resolvent and the free energy are related as
\begin{align}
  \mR_{\CM}(z)
 ={S\over z}\delta_{\chi(\CM),2}
 +{1\over z^2}{\partial\CF_{\CM}\over\partial g_1}
 +{2\over z^3}{\partial\CF_{\CM}\over\partial g_2}
 +{3\over z^4}{\partial\CF_{\CM}\over\partial g_3}
 +\cdots,\label{bibz22Mar04}
\end{align}
where $S=\sum_{i=1}^K S_i$. The resolvents can be determined uniquely by
solving the matrix model loop equations (the loop
equations for the relevant matrix models are summarized in \cite{Kraus:2003jv}),
under the condition
\begin{align}
  \oint_{A_i}{dz\over 2\pi i}\mR_{S^2}(z)=\mS_i,\qquad
 \oint_{A_i}{dz\over 2\pi i}\mR_{\RPtwo}(z)=0.\label{biew22Mar04} 
\end{align}
$A_i$ is the contour around the $i$-th critical point of $W(z)$.  In
general, $\mR(z)$ develops a cut around each critical point
in the large $\mN_i$ limit, and $A_i$ is taken to encircle the $i$-th
cut. %
Note that the expression \eqref{bibz22Mar04} should be understood as a Laurent
expansion around $z=\infty$, and converges only if $|z|$ is
larger than $r$ such that all the singularities (cuts) of the resolvent
are inside the circle $C:~|z|=r$.

On the other hand, gauge theory resolvents $R(z)$, $T(z)$ (see e.g.\
\cite{Cachazo:2002ry}) are determined uniquely by solving the Konishi
anomaly equations (the Konishi anomaly equations for the relevant gauge
theories are summarized in \cite{Kraus:2003jv}), under the condition
\begin{align}
   \oint_{A_i}{dz\over 2\pi i}R(z)=S_i,
  \qquad
  \oint_{A_i}{dz\over 2\pi i}T(z)=N_i.
 \label{bigl22Mar04}
\end{align}

As was shown in \cite{Kraus:2003jv}, the matrix model resolvents
$\mR_{S^2}(z)$, $\mR_{\RPtwo}(z)$ are related to the gauge theory
resolvents $R(z)$, $T(z)$ as
\begin{align}
  R(z)=\mR_{S^2}(z),\qquad
 T(z)=
 \!\!\sum_{U,\,SO,\,Sp} \!\! 
   N_i{\partial\over\partial S_i}\mR_{S^2}(z)
 +4\mR_{\RPtwo}(z)\label{bije22Mar04}
\end{align}
with $S_i$ and $\mS_i$ identified\footnote{The relation
\eqref{bije22Mar04} is an obvious generalization of the formula in
\cite{Kraus:2003jv}, which was for unbroken vacua, to an arbitrary
breaking pattern.  The gauge theory resolvents $R(z)$, $T(z)$ given in
\eqref{bije22Mar04} clearly satisfy the condition \eqref{bigl22Mar04}
provided that the matrix model resolvents $\mR_{S^2}(z)$,
$\mR_{\RPtwo}(z)$ satisfy the condition \eqref{biew22Mar04}.}.

\bigskip
First, consider $SO(2N)/Sp(N)$ theory with adjoint.  The general
breaking pattern is $SO/Sp(2N)\to SO/Sp(2N_0)\times U(N_1)\times \cdots
\times U(N_K)$ (Eq.\ \eqref{eq:higgscases}), where $N = N_0 +
2\sum_{i=1}^K N_i$. Note that the eigenvalues are distributed in a
symmetric manner under $z\leftrightarrow -z$, and hence
\eqref{bigl22Mar04} is
\begin{align}
  \oint_{A_{0}}{dz\over 2\pi i}R(z)&=S_0,&
  \oint_{A_{i}}{dz\over 2\pi i}R(z)
  &=\oint_{A_{-i}}{dz\over 2\pi i}R(z)=S_i,
  \cr
  \oint_{A_{0}}{dz\over 2\pi i}T(z)&=N_0,&
  \oint_{A_{i}}{dz\over 2\pi i}T(z)
  &=\oint_{A_{-i}}{dz\over 2\pi i}T(z)=N_i, 
\end{align}
where $i=1,\dots,K$. The contours $A_i$ and $A_{-i}$ encircle
counterclockwise the cuts around $z=a_i$ and $z=-a_i$, respectively. The
relation \eqref{bije22Mar04} holds as it is, with the summation understood as
over $SO/Sp(N_0)$ and $U(N_i)$, $i=1,\dots,K$.

It was shown in \cite{Landsteiner:2003ph} that the resolvents of this
$SO/Sp(N)$ theory are related to the resolvents $\tilde R(z)$ and
$\Tt(z)$ of $U(\Nt\!\equiv\!N\mp 2)$ theory with adjoint as follows:
\begin{align}
   R(z)=\tilde R(z),\qquad
  T(z)=\Tt(z)\pm{2\over z}.\label{bink22Mar04}
\end{align}
The tree level superpotential of the $U(\Nt)$ theory is related to the
one for the $SO/Sp(N)$ theory as $W^{U}(z)=W^{SO/Sp}(z)$ (see \eqref{ump22Mar04} and
\eqref{eq:basp}), and the breaking pattern is $U(\Nt)\to U(N_{-K})\times \cdots
\times U(N_{-1}) \times U(N_0\mp 2) \times U(N_1)\times \cdots \times
U(N_K)$ with $N_{-i}=N_i$, $1\le i \le K$.  Note that since there is no
$z\leftrightarrow -z$ symmetry in the $U(\Nt)$ theory, the $U(N_{-i})$
factors that are ``images'' for $SO/Sp(N)$ are ``real'' for
$U(\Nt)$.  In addition, the glueball $\St$ of the $U(\Nt)$ theory is
related to the glueball $S$ of the $SO/Sp(N)$ theory as $\St_0=S_0$,
$\St_{i}=\St_{-i}=S_i$, $1\le i \le K$. Therefore, e.g.\ the first
equation in \eqref{bink22Mar04} is more precisely
\begin{align}
  R(z,S_j)
 =
 \Rt(z,\St_{j})\bigr|_{\St_0=S_0,\,\St_{i}=\St_{-i}=S_i}\,.\label{biqa22Mar04}
\end{align}
Differentiating \eqref{biqa22Mar04} with respect to $S_j$, we obtain
\begin{align}
   {\partial R\over\partial S_0} =
  {\partial \Rt\over \partial \St_0}
  \biggr|_{\St_0=S_0,\,~~~~\atop\St_i=\St_{-i}=S_i}
  ,\quad
  {\partial R\over\partial S_j} =
  \biggl(
  {\partial \Rt\over \partial \St_j}
  + {\partial \Rt\over \partial \St_{-j}}
  \biggr)
  \biggr|_{\St_0=S_0,\,~~~~\atop\St_i=\St_{-i}=S_i}
  ,\label{biqz22Mar04}
\end{align}
where $1\le j\le K$.

Now, using \eqref{bije22Mar04}, let us translate the relation
\eqref{bink22Mar04} among gauge theory resolvents into a relation among
matrix model resolvents:
\begin{align}
  \mR_{S^2}&=\tilde\mR_{S^2},\\
 N_0{\partial\mR_{S^2}\over\partial S_0}
 + & \sum_{i=1}^K N_i {\partial\mR_{S^2}\over\partial S_i}
 + 4\mR_{\RPtwo}\cr
 &=
 (N_0\mp 2){\partial\tilde\mR_{S^2}\over\partial S_0}
 + \sum_{i=1}^K N_i
 \biggl(
  {\partial\tilde\mR_{S^2}\over \partial \St_i}
  + {\partial\tilde\mR_{S^2}\over \partial \St_{-i}}
 \biggr)
 \pm {2\over z}.
\end{align}
Here $\tilde\mR_{S^2}$ is the matrix model resolvent associated with the
$U(\Nt)$ theory.  Using \eqref{biqz22Mar04} and the relations
$\mR_{S^2}=R$, $\tilde\mR_{S^2}=\Rt$ (Eq.\ \eqref{bije22Mar04}), we obtain
\begin{align}
  \mR_{\RPtwo}(z)
 =
 \mp {1\over 2} {\partial\over\partial S_0}\mR_{S^2}(z)
 \pm {1\over 2z}.
\end{align}
By expanding the resolvents around $z=\infty$ using \eqref{bibz22Mar04} and comparing
the coefficients, we obtain a relation between matrix
model free energies:
\begin{align}
  j{\partial \CF_{\RPtwo} \over \partial g_j}
 =
 \pm {1\over 2} {\partial\over \partial S_0}
 \left( S\delta_{j0}+  j{\partial \CF_{S^2}\over \partial g_j} \right)
 \mp {1\over 2}\delta_{j0}.
\end{align}
where $j=0,2,\cdots,2(K+1)$.  The $j=0$ case is trivially satisfied
since $S=S_0+2\sum_{i=1}^K S_i$ here, while the
$j=2,4,\dots,2(K+1)$ cases lead to the first equation of \eqref{bhuq22Mar04}, which we
wanted to prove.

\bigskip
Next, consider $SO/Sp(N)$ theory with symmetric/antisymmetric
tensor.  The breaking pattern is $SO(N)\to \prod_{i=1}^K SO(N_i)$ or
$Sp(N)\to \prod_{i=1}^K Sp(N_i)$ (Eq.\ \eqref{eq:higgscases}), where
$N=\sum_{i=1}^K N_i$.  It was shown in \cite{Cachazo:2003kx} that the
resolvents of this $SO/Sp$ theory is related to the resolvents $\tilde
R(z)$ and $\Tt(z)$ of $U(\Nt\!\equiv\! N\mp 2K)$ theory with adjoint as
follows:
\begin{align}
   R(z)=\tilde R(z),\qquad
  T(z)=\Tt(z)\pm{d\over dz}\ln[W'(z)^2+f_{K-1}(z)].\label{biup22Mar04}
\end{align}
The tree level superpotential of the $U(\Nt)$ theory is related to the
one for the $SO/Sp(N)$ theory as $W^{U}(z)=W^{SO/Sp}(z)$ (see
\eqref{ump22Mar04} and \eqref{eq:bsospx}), and the breaking pattern is
$U(\Nt)\to \prod_{i=1}^{K} U(N_i\mp 2)$.  The glueball $S_i$ of the
$U(\Nt)$ theory is taken to be the same as the glueball of the
$SO/Sp(N)$ theory.  In \eqref{biup22Mar04}, $f_{K-1}(z)$ is a
polynomial of degree $K-1$.  Using \eqref{bije22Mar04}, we can translate
the relation \eqref{biup22Mar04} among gauge theory resolvents into a
relation among matrix model resolvents:
\begin{align}
  \mR_{\RPtwo}(z)
 =
 \mp {1\over 2} \sum_{i=1}^K {\partial\over\partial S_i}\mR_{S^2}(z) \pm
 {1\over 4}{d\over dz}\ln[W'(z)^2+f_{K-1}(z)].\label{bivq22Mar04}
\end{align}
In order to extract the relation between matrix model free energies, let
us multiply \eqref{bivq22Mar04} by $z^j$ ($0\le j\le K+1$) and integrate over $z$
along the contour $C$, introduced under \eqref{biew22Mar04}, which
encloses all the cuts around the critical points of $W(z)$.  Taking
\begin{align}
  W'(z)^2+f_{K-1}(z)
 =g_{K+1}^2\prod_{i=1}^K (z-a_i^+)(z-a_i^-),
\end{align}
the branching points of the cuts are at $z=a_i^\pm$.  The second term on
the right hand side in \eqref{bivq22Mar04} does not contribute to the contour integral
unless $j=0$:
\begin{align}
  \mp &{1\over 4}\oint_C {dz\over 2\pi i}
 \sum_{i=1}^K z^j\left( {1\over z-a_i^+}+{1\over z-a_i^-}\right)\cr &=
 \pm {1\over 4}\oint_C {dw\over 2\pi i}
 \sum_{i=1}^K {1\over w^{j+1}}
 \left( {1\over 1-a_i^+w}+{1\over 1-a_i^-w}\right)
 =
 \mp {K\over 2}\delta_{j0},
\end{align}
where $w=1/z$, because all the poles $w=1/a_i^{\pm}$ are outside of the
contour $C$ (on the $w$-plane).  On the contour $C$, we can use the
Laurent expansion \eqref{bibz22Mar04} to evaluate the contribution from the other
terms, and the final result is
\begin{align}
 j{\partial \CF_{\RPtwo} \over \partial g_j}
 =
 \pm {1\over 2}\sum_{i=1}^K {\partial\over \partial S_i}
 \left( S\delta_{j0}+  j{\partial \CF_{S^2}\over \partial g_j} \right)
 \mp {K\over 2}\delta_{j0}.
\end{align}
The $j=0$ case is trivially satisfied, while the $1\le j\le K+1$ cases
lead to the second equation of \eqref{bhuq22Mar04}, which we wanted to
prove.

\subsection{Computation of matrix model free energy: $SO/Sp(N)$ theory
with adjoint}

Let us consider $SO(\mN)$ matrix model which corresponds to $SO(N)$
gauge theory with adjoint.  The tree level superpotential is taken to be
quartic \eqref{bfpc22Mar04}.  The matrix variable $\mPhi$ in
\eqref{bhxy22Mar04} is a real antisymmetric matrix and can be
skew-diagonalized as
\begin{align}
   \mPhi \cong
  {\textrm{diag}}[\lambda_1,\cdots,\lambda_{\mN/2}]\otimes i\sigma^2.
\end{align}
By changing the integration variables from $\mPhi$ to $\lambda_i$, we
obtain %
\begin{align}
   \mZ\sim\int
  \prod_{i=1}^{\mN/2}d\lambda_i
  \prod_{i<j}^{\mN/2}(\lambda_i^2-\lambda_j^2)^2\,\,
  e^{-{1\over \mg}\sum_{i=1}^{\mN/2}
  \left(-{m\over 2}\lambda_i^2+{g\over 4}\lambda_i^4\right)},\label{bizx22Mar04}
\end{align}
where $\prod_{i<j}^{\mN/2}(\lambda_i^2-\lambda_j^2)^2$ is the Jacobian for
this change of variables \cite{Mehta:1991, Ashok:2002bi}.  The
polynomial $-{m\over 2}\lambda^2+{g\over 4}\lambda^4$ has critical
points at $\lambda=0,\pm\sqrt{m/g}$, around which we would like to do
perturbative expansion.  For this purpose, we separate $\lambda$'s into
two groups as
\begin{align}
   \lambda_i =
  \begin{cases}
    \lambda_{i_0}^{(0)}            & i_0=1,\dots,\mN_0/2,\\
    \sqrt{m/g}+\lambda_{i_1}^{(1)} & i_1=1,\dots,\mN_1,
  \end{cases}
\end{align}
with $\mN_0+2\mN_1=\mN$, corresponding to the classical supersymmetric
vacuum with breaking pattern $SO(\mN ) \to SO(\mN _0)\times U(\mN _1)$.
%
We would like to evaluate the matrix integral \eqref{bizx22Mar04} perturbatively around
$\lambda^{(0,1)}=0$.
If we expand the matrix model free energy in the coupling constant $g$
as %
\begin{align}
  \mF =g f_1(\mN_0,\mN_1)+ g^2 f_2(\mN_0,\mN_1)+\cdots,\label{bjcy22Mar04}
\end{align} 
the loop expansion tells us that $f_n(\mN_0,\mN_1)$ is a polynomial of
degree $n+2$.  Therefore, by performing the matrix integral by computer
for small values of $\mN_0$ and $\mN_1$, one can determine the
polynomial $f_n$.  If we rewrite $\mN_{0,1}$ in favor of $S_{0,1}=\mg
\mN_{0,1}$, the expansion \eqref{bjcy22Mar04} arranges itself into the
't Hooft expansion \eqref{bhzb22Mar04}, from which one can read off
$\CF_{S^2}$, $\CF_{\RPtwo}$, etc.

Following the procedure sketched above, we computed the matrix model
free energy as
\begin{align}
   \CF_{S^2}=&
  \left( {1\over 4}S_0^3-2 S_0^2 S_1+ S_0 S_1^2 \right)\alpha
  +
  \left(
    -{9\over 16}S_0^4 + 7S_0^3 S_1 - {9}S_0^2 S_1^2+ {2}S_0 S_1^3
  \right)\alpha^2
  \cr
  &
  +
  \left(
    {9\over 4} S_0^5
    -{233\over 6} S_0^4 S_1
    +{262\over 3} S_0^3 S_1^2
    -{152\over 3} S_0^2 S_1^3
    +{20\over 3} S_0 S_1^4
  \right)\alpha^3
  +\CO(\alpha^4),
 \label{bjfr22Mar04}
\end{align}
where we defined $\alpha\equiv{g/m^2}$.  We also checked explicitly that
the relation \eqref{bhuq22Mar04} holds.  Substituting
\eqref{bjfr22Mar04} into the DV relation \eqref{bfsa22Mar04}, we obtain
the superpotential \eqref{bfwj22Mar04}.

The $Sp(N)$ result is obtained similarly, with the result as in
\eqref{bhuq22Mar04}.

\subsection{Computation of matrix model free energy: $Sp(N)$ theory with
antisymmetric tensor}

Consider the $Sp(\mN)$ matrix model which corresponds to $Sp(N)$ gauge
theory with an antisymmetric tensor.  The superpotential is taken to be
quartic \eqref{bggr22Mar04}.  The matrix variable $\mPhi$ satisfies
$\mPhi=\mA J$, $\mA^T=-\mA$.  The ``action'' $W_{\textrm{tree}}$ is
given in \eqref{bggr22Mar04}.  By a complexified $Sp(\mN)$ gauge
rotation, the matrix $\mPhi$ can be brought to the form
\cite{Cho:1996bi}
\begin{align}
   \Phi \cong {\textrm{diag}}[\lambda_1,\dots, \lambda_{\mN/2}]
  \otimes{\unit}_2, \qquad
  \lambda_i\in {\mathbb C}.
\end{align}
By changing the integration variables from $\Phi$ to $\lambda_i$, we
obtain %
\begin{align}
   Z\sim
  \int
    \prod_{i=1}^{\mN/2}d\lambda_i
    \prod_{i<j}^{\mN/2}(\lambda_i-\lambda_j)^4\,\,
    e^{
        -{1\over \mg}\sum_{i=1}^{\mN/2}
        \left( {m\over 2}\lambda_i^2+{g\over 3}\lambda_i^3 \right)
      }.
\end{align}
where $\prod_{i<j}^{\mN/2}(\lambda_i-\lambda_j)^4$ comes from the
Jacobian for this change of variables.  The polynomial ${m\over
2}\lambda^2+{g\over 3}\lambda^3$ has two critical points $z=0,-{m\over
g}$, around which we would like to do perturbative expansion.  For this
purpose, we separate $\lambda$'s into two groups as
\begin{align}
   \lambda_i =
  \begin{cases}
    \lambda_{i_0}^{(1)}      & i_1=1,\dots,\mN_1/2,\\
    -m/g+\lambda_{i_1}^{(2)} & i_2=1,\dots,\mN_2/2,
  \end{cases}
\end{align}
with $\mN_1+\mN_2=\mN$.  This corresponds to the classical
supersymmetric vacuum with breaking pattern $Sp(\mN ) \to Sp(\mN_1)\times
Sp(\mN_2)$.

The matrix integral can be performed just the same way as for the
$SO/Sp(N)$ theory with adjoint, as described in the last
subsection. After substitution $S_{1,2}=\mg \mN_{1,2}$, we obtain
\begin{align}
  \CF_{S^2}=&
 \left(
  -{S_1^3\over 3}+{S_2^3\over 3}
  +{5\over 2}S_1^2 S_2 -{5\over 2}S_1 S_2^2
 \right)\alpha
 \cr &+
 \left(
  -{4\over 3}S_1^4 + {91\over 6}S_1^3 S_2 - {59\over 2}S_1^2 S_2^2 +
 {91\over 6}S_1 S_2^3-{4\over 3}S_2^4
 \right)\alpha^2
 \cr &+
 \left(
  -{28\over 3}S_1^5
  + {871\over 6}S_1^4 S_2
  - {1318\over 3}S_1^3 S_2^2
  + {1318\over 3}S_1^2 S_2^3
  - {871\over 6}S_1 S_2^4
  +{28\over 3}S_2^5
 \right)\alpha^3\cr
 &+\CO(\alpha^4),\cr
 \label{bjkw22Mar04}
\end{align}
where $\alpha\equiv g^2/m^3$.  We also checked explicitly that the
relation \eqref{bhuq22Mar04} holds. Substituting \eqref{bjkw22Mar04}
into \eqref{bgkm22Mar04}, we obtain the glueball superpotential\break
\eqref{bgmi22Mar04}.

\section{Gauge theory calculation of superpotential}
\label{IKRSV:GT_calc_spot}

In this appendix, we compute the exact superpotential of the $\CN=1$
$SO/Sp(N)$ theory with adjoint in various vacua by considering
factorization of the $\CN=2$ curve.
This factorization method was developed in \cite{Cachazo:2001jy} for
$U(N)$, and generalized in \cite{Cachazo:2003zk} to the case with
unoccupied critical points (in other words, the $n<K$ case
below). Inclusion of fundamentals was considered in
\cite{Ookouchi:2002be, Balasubramanian:2003tv}.  The generalization to
$SO/Sp$ gauge group, which discuss below, was given in
\cite{Edelstein:2001mw, Fuji:2003vv, Feng:2003gb, Ahn:2003cq,
Ahn:2003vh, Ahn:2003ui}.

First consider $\CN=2$ $SO(N)$ theory broken to $\CN=1$ by the following
polynomial tree level superpotential for the adjoint chiral superfield
$\Phi$:
\begin{align}
   W_{\textrm{tree}}&=\half\Tr[W(\Phi)], \cr 
  W(x) &= \sum_{j=1}^{K+1}{g_{2j}\over 2j}x^{2j},\qquad
  W'(x)=g_{2K+2}\,x\prod_{i=1}^{K}(x^2-a_i^2). 
\end{align}
The classical supersymmetric vacua are obtained by putting $N_0$
eigenvalues of $\Phi$ at $x=0$ and $N_i$ pairs of eigenvalues at $x=\pm
a_i$, where $i=1,\cdots,K$ and $N_0+2\sum_{i=1}^K N_i=N$.  In this vacuum
the gauge group breaks as $SO(N)\to SO(N_0)\times \prod_{i=1}^K
U(N_i)$.  We allow some of $N_i$ to vanish, i.e.\ we allow
``unoccupied'' critical points.  Let the number of nonzero $N_{i\ge 1}$
be $n$.  Then, the $\CN=2$ curve governing this $SO(N)$ theory
factorizes as \cite{Edelstein:2001mw, Fuji:2003vv, Feng:2003gb}:
\begin{align}
   y^2=P_{N}^2(x)-4x^4\Lambda^{2N-4}
  = \begin{cases}
      [xH_{N-2n-2}(x)]^2 F_{2(2n+1)}(x) & N_0>0,\\
      H_{N-2n}(x)^2 F_{4n}(x) & N_0=0. 
    \end{cases}
     \label{bjob22Mar04} 
\end{align}
Here $P$, $H$ and $F$ are polynomials in $x$ of the subscripted degree, 
which are invariant under $x\to -x$, i.e.\ they are
actually polynomials in $x^2$.  This factorization is required to have the 
appropriate number of independent, massless, monopoles and dyons, which
must condense to eliminate some of the low-energy photons.  The polynomial $F$ 
is related to the
tree level superpotential as
\begin{align}
 \begin{cases} 
  F_{2(2n+1)}(x)={1\over g_{2K+2}^2} W'_{2K+1}(x)^2+f_{2K}(x)  &   n=K, \\
  F_{2(2n+1)}(x)Q_{2K-2n}(x)^2={1\over g_{2K+2}^2}W'_{2K+1}(x)^2+f_{2K}(x)
  & n<K
 \end{cases}
 \label{bjri22Mar04} 
\end{align}
with some polynomial $Q_{2K-2n}(x)$, $f_{2K}(x)$ of the subscripted
degrees, and\break $W'_{2K+1}(x)$ is as in \eqref{eq:Wprimeisp}.
Equation \eqref{bjri22Mar04} is for $N_0>0$, and for $N_0=0$ one must use the second
equation with $Q_{2K-2n}$ replaced by $Q_{2K-2n+2}$.

For $N_0>0$ we can write the solution of \eqref{bjob22Mar04} in terms of
that of the corresponding $U(N-2)$ breaking pattern, via:
\begin{align}
 P^{SO(N)}_{N}(x)=x^2P_{N-2}^{U(N-2)}(x).
 \label{eq:psopu}
\end{align}

The low energy superpotential is given by 
\begin{align}
  W_{\textrm{low}} = \half\sum_{j=1}^{K+1}g_{2j}\ev{u_{2j}},\label{bjvt22Mar04} 
\end{align}
where the $\ev{u_{2j}}$ are constrained to satisfy \eqref{bjob22Mar04}.  Implementing this 
leads to the result that  \cite{Ahn:2003cq}:
\begin{align}
   \left\langle\Tr{1\over x-\Phi}\right\rangle 
  = {d\over dx}\ln 
  \left[P_{N}(x)+\sqrt{P_{N}(x)^2-4x^4\Lambda^{2N-4}}\right].\label{bjzh22Mar04}
\end{align}
Plugging back into \eqref{bjvt22Mar04} gives $W_{\textrm{low}}$.  Note
that the superpotential takes this simple form \eqref{bjvt22Mar04} only
after one integrates out the monopoles and dyons, whose equation of
motion led to the factorization constraint \eqref{bjob22Mar04}
\cite{Seiberg:1994rs, Seiberg:1994aj, Janik:2002nz}.

The $Sp(N)$ theory can be solved similarly.  The $\CN=2$ curve
factorizes in the vacuum with breaking pattern $Sp(N)\to Sp(N_0)\times
\prod_{i=1}^K U(N_i)$ as \cite{Edelstein:2001mw, Fuji:2003vv,
Feng:2003gb}
\begin{align}
   y^2&=B_{N+2}(x)^2-4\Lambda^{2N+4}= x^2H_{N-2n}(x)^2  F_{2(2n+1)}(x), \cr 
  B_{N+2}(x)&\equiv x^2P_{N}(x)+2\Lambda^{N+2}.\label{bkah22Mar04} 
\end{align}
The polynomial $F_{2(2n+1)}(x)$ is related to $W'(x)$ by \eqref{bjri22Mar04}.
The mapping of the $Sp(N)$ theory to a $U(N+2)$ theory, as in
\eqref{eq:spaaux}, can be written as a solution of \eqref{bkah22Mar04}
in terms of solutions of the corresponding $U(N+2)$ factorization
problem:
\begin{align}
 B_{N+2}(x)\equiv x^2P_{N}^{Sp(N)}(x)+2\Lambda^{N+2} = P_{N+2}^{U(N+2)}(x).  
 \label{eq:psppu}
\end{align}
The $\Lambda ^{2N+2}$ shift in \eqref{eq:psppu} is an $Sp(N)$ residual
instanton effect, associated with the index of the embedding of the
$U(N_i)$ factors in $Sp(N)$ \cite{Intriligator:1995id, Csaki:1998vv}.

Again, the superpotential is given as in \eqref{bjvt22Mar04}, subject to
the constraint that $\ev{u_j}$ satisfy \eqref{bkah22Mar04}.
Implementing this, the $\ev{u_j}$ can be obtained from $B_{N+2}(x)$
by \cite{Ahn:2003cq}
\begin{align}
   \left\langle\Tr{1\over x-\Phi}\right\rangle 
  = {d\over dx}\ln  \left[B_{N+2}(x)+\sqrt{B_{N+2}(x)^2-4\Lambda^{2N+4}}\right].\label{bkdy22Mar04} 
\end{align}

We now consider the exact $W_{\textrm{eff}}$ for a few $SO/Sp(N)$ cases,
to illustrate and clarify the general features\footnote{More $SO/Sp$
examples can be found in \cite{Ahn:2003cq}.}.  We take quartic tree
level superpotential
\begin{align}
 W_{\textrm{tree}}=\half \Tr W(\Phi), \qquad  W(x)={m\over 2}x^2+{g\over 4}x^4, 
\end{align}
which corresponds to $K=1$.

\subsection{$SO(N)$ unbroken}

By the map \eqref{eq:soaaux}, this maps to $U(N-2)$ unbroken, for which
$P^{U(N-2)}(x)= 2\Lambda^{N-2} T_{N-2}(x/2\Lambda)$, with
$T_N(x=\half(t+t^{-1}))=\half (t^N+t^{-N})$ a Chebyshev polynomial
\cite{Douglas:1995nw}.  Thus, using \eqref{eq:psopu},
$P^{SO(N)}(x)=2\Lambda^{N-2} x^2 T_{N-2}(x/2\Lambda)$, as found in
\cite{Janik:2002nz}.  This then leads to \cite{Janik:2002nz}
\begin{align}
 \ev{u_{2p}}
 \equiv 
 {1\over 2p} \ev{\Tr \Phi ^{2p}} = {N-2\over 2p} {2p\choose p}
 \Lambda ^{2p}.  
 \label{eq:sounbrv}
\end{align}
In particular,
\begin{align}
 \ev{u_2}=(N-2)\Lambda^2,\quad
 \ev{u_4}=3\left({N\over 2}-1\right)\Lambda^4,
\end{align}
and the low-energy superpotential is $W_{\textrm{low}}=\half (m\ev{u_2}+g\ev{u_4})$:
\begin{align}
  W_{\textrm{low}}=\left({N\over2}-1\right)\left( m\Lambda^2+{3\over 2}g\Lambda^4\right). 
\end{align}

\subsection{$SO(N)\to SO(2)\times U(N/2-1)$}

By the map \eqref{eq:soaaux}, this maps to $U(N-2)\rightarrow
U(0)\times U(N/2-1)\times U(N/2-1)$.  Using \eqref{eq:soaaux}, the
multiplication map of \cite{Cachazo:2001jy} for the $U(N-2)$ theory leads to
a similar multiplication map for the $SO(N)$ theory, which was
discussed in \cite{Ahn:2003cq}.  Using this, we can construct the solution to
the factorization problem for general $N$ in terms of that of say $N=4$,
i.e.\ $SO(4)\rightarrow SO(2)\times U(1)$.  In this case, equation
\eqref{bjob22Mar04} is
\begin{align}
  y^2=P_4^2-4x^4 \Lambda^{4} = x^2 F_6. 
\end{align}
The solution to this factorization problem is
\begin{align}
  P_4=x^2(x^2-a^2),\quad
 F_6=x^2[(x^2-a^2)^2-4\Lambda^2],  
\end{align}
from which we can see the breaking pattern $SO(4)\to SO(2)\times U(1)$.  Using
\eqref{bjzh22Mar04} gives
\begin{align}
  u_2=a^2,\quad u_4={a^4\over 2}+\Lambda^4. 
\end{align}
Further, the condition \eqref{bjri22Mar04}
\begin{align}
  F_6={1\over g^2}W_3'{}^2+f_2 
\end{align}
leads to
\begin{align}
  a^2=-{m\over g},\quad f_2=-4\Lambda^2 x^2. 
\end{align}
The solution for general $SO(N)\rightarrow SO(2)\times U(N/2-1)$, the
multiplication map gives the solution to the factorization problem as
$P_{N}(x)=2x^2 \Lambda^{N-2} T_{N/2-1}((x^2-a^2)/2\Lambda ^2)$, with
$T_{N/2-1}$ the Chebyshev polynomial defined above.  The effect is to
rescale $u_2$, $u_4$, and hence $W_{\textrm{low}}$ by an overall factor
of $N/2-1$:
\begin{align}
  W_{\textrm{low}}=\left({N\over 2}-1\right)\left(-{m^2\over 4g}+{1\over 2}g\Lambda^4\right). 
\end{align}
This agrees with the result \eqref{bgdf22Mar04}.  
\subsection{$SO(4)\to U(2)$}

More generally, we could consider the breaking pattern
$SO(N)\rightarrow U(N/2)$.  The map of \eqref{eq:soaaux} is less useful
here, when $N_0=0$, since it suggests mapping to $U(N-2)\rightarrow
U(-2)\times U(N/2)\times U(N/2)$ and the $U(-2)$ needs to be interpreted.
In general, this breaking pattern leads to a complicated
$W_{\textrm{low}}(\Lambda)$.  We will here illustrate the case $SO(4)\to
U(2)$, corresponding to $N=4,~N_0=0,~n=1,~K=1$. Equation
\eqref{bjob22Mar04} is
\begin{align}
  y^2=P_4^2-4x^4 \Lambda^{4} = H_2^2 F_4. 
\end{align}
The solution to this factorization problem is
\begin{align}
  P_4=(x^2-a^2)^2+2\Lambda^2x^2,\quad H_2=x^2-a^2,\quad
 F_4=(x^2-a^2)^2+4\Lambda^2x^2. 
\end{align}
In the classical $\Lambda\to 0$ limit, this shows $P_4\to (x^2-a^2)^2$,
implying the breaking pattern $SO(4)\to U(2)$.  \eqref{bjzh22Mar04} gives
\begin{align}
  u_2=2(a^2-\Lambda^2),\quad
 u_4=(a^4-2\Lambda^2)^2-\Lambda^4. 
\end{align}
Further, the condition \eqref{bjri22Mar04}
\begin{align}
  F_4x^2={1\over g^2}W_3'{}^2+f_2 
\end{align}
leads to
\begin{align}
 a^2=-{m\over g}+2\Lambda^2,\quad
 f_2=4\Lambda^2x^2\Bigl(-{m\over g}+\Lambda^2\Bigr). 
\end{align}
Therefore the exact superpotential is
\begin{align}
  W_{\textrm{low}} = -{m^2\over 2g}+m\Lambda^2+{1\over 2}g\Lambda^4. 
\end{align}

\subsection{$Sp(4)\to U(2)$, $Sp(2)\times U(1)$} 

This corresponds to $N=4,~n=1,~K=1$. Equations \eqref{bkah22Mar04} and
\eqref{bjri22Mar04} are
\begin{align}
  y^2=B_6^2-4\Lambda^{12} = x^2H_2^2 F_6,\qquad
 F_6={1\over g^2}W_3'{}^2+f_2. 
\end{align}
This factorization problem is solved by \cite{Ahn:2003cq}:
\begin{align}
  P_4=(x^2-a^2)^2+{4\Lambda^6\over a^4}(x^2-2a^2),
 \qquad {m\over g}=-a^2+{4\Lambda^6\over a^4}\label{bksq22Mar04} 
\end{align}
From \eqref{bkdy22Mar04}, we obtain
\begin{align}
  u_2=2a^2-{4\Lambda^6\over a^4},\qquad
 u_4=a^4+{8\Lambda^{12}\over a^8}.\label{bktq22Mar04} 
\end{align}
This solution continuously connects two classically different vacua with
breaking pattern $Sp(4)\to U(2)$ and $Sp(4)\to
Sp(2)\times U(1)$. Correspondingly there are two ways to take the
classical limit: i) $\Lambda\to 0$ with $a$ fixed, or ii)
$\Lambda,a\to 0$ with $w=2\Lambda^3/a^2$ fixed.  In these limits,
$P_4(x)$ goes to i) $(x^2-a^2)^2$ or ii) $x^2(x^2+w^2)$,
showing the aforementioned breaking pattern.

In the $Sp(4)\to U(2)$ case, we solve the second equation of
\eqref{bksq22Mar04} with the condition $a^2\to -m/g$ as $\Lambda\to 0$.
The solution is
\begin{align}
   a^2=-{m \over g}+{4g^2\Lambda^6\over
 m^2}+{32g^5\Lambda^{12}\over m^5}+ {448g^8\Lambda^{18}\over m^8}+\cdots.
\end{align}
From \eqref{bktq22Mar04} and \eqref{bjvt22Mar04}, one obtains the exact superpotential:
\begin{align}
   W_{\textrm{low}}
  =
  -{m^2\over 2g}
  -{2g^2\Lambda^6\over m}
  -{4g^5\Lambda^{12}\over m^4} 
  -{32g^8\Lambda^{18}\over m^7}
  +\cdots. 
\end{align}

In the $Sp(4)\to Sp(2)\times U(1)$ case, we solve the second equation of
\ref{bksq22Mar04}  with the condition $w^2\to m/g$ as $\Lambda\to 0$.
It is
\begin{align}
  w={m^{1/2}\over g^{1/2}}+{g\Lambda^3\over m}
 -{3g^{5/2}\Lambda^6\over 2m^{5/2}}+{4g^4\Lambda^9\over
 m^4}+\cdots. 
\end{align}
From \eqref{bktq22Mar04} and \eqref{bjvt22Mar04}, one obtains the exact
superpotential:
\begin{align}
   W_{\textrm{low}}
  =
  -{m^2\over 4g}
  +2m^{1/2}g^{1/2}\Lambda^3
  +{g^2\Lambda^6\over m} 
  -{g^{7/2}\Lambda^9\over m^{5/2}}
  +\cdots. 
\end{align}
%

\makeatletter
\def\eqnarray{%
  \stepcounter{equation}\def\@currentlabel{\p@equation\theequation}%
  \global\@eqnswtrue \m@th \global\@eqcnt\z@ \tabskip\@centering
  \let\\\@eqncr
  $$\everycr{}\halign to\displaywidth\bgroup
    \hskip\@centering$\displaystyle\tabskip\z@skip{##}$\@eqnsel
   &\global\@eqcnt\@ne \hfil$\;{##}\;$\hfil
   &\global\@eqcnt\tw@ $\displaystyle{##}$\hfil\tabskip\@centering
   &\global\@eqcnt\thr@@ \hb@xt@\z@\bgroup\hss##\egroup
    \tabskip\z@skip
    \cr}
\makeatother

\def\ap{{\alpha'}}

\newcommand{\nn}{\nonumber}
\def\La{\Lambda}
\def\nonu{\nonumber}
\newcommand{\bea}{\begin{eqnarray}}
\newcommand{\eea}{\end{eqnarray}}
 \def\Si{\Sigma}
\def\W #1{\widetilde{#1}}
\def\braket#1{\left\langle #1 \right\rangle}
\newcommand{\bean}{\begin{eqnarray*}}
\newcommand{\eean}{\end{eqnarray*}}
\def\a{{\alpha}}
\def\WH #1{\widehat{#1}}
\def\O #1{\overline{#1}}

\EndAppendix

\chapter{Adding flavors}
\label{AFOS}

We present two results concerning the relation between poles and cuts by
using the example of $\CN=1$ $U(N_c)$ gauge theories with matter fields
in the adjoint, fundamental and anti-fundamental representations.
The first result is the on-shell possibility of poles, which are
associated with flavors and on the second sheet of the Riemann surface,
passing through the branch cut and getting to the first sheet.
The second result is the generalization of hep-th/0311181 (Intriligator,
Kraus, Ryzhov, Shigemori, and Vafa) to include flavors.  We clarify when
there are closed cuts and how to reproduce the results of the strong
coupling analysis by matrix model, by setting the glueball field to zero
from the beginning.
%
%
We also make remarks on the possible stringy explanations of the results
and on generalization to $SO(N_c)$ and $USp(2N_c)$ gauge groups.

\section{Introduction}
\setcounter{equation}{0}
                                                                                                           

String theory can be a powerful tool to understand four dimensional
supersymmetric gauge theory which exhibits rich dynamics and allows an
exact analysis.  In \cite{Cachazo:2003yc}, using the generalized Konishi
anomaly and matrix model \cite{Dijkgraaf:2002fc, Dijkgraaf:2002vw,
Dijkgraaf:2002dh}, ${\mathcal N}=1$ supersymmetric $U(N_c)$ gauge theory
with matter fields in the adjoint, fundamental and anti-fundamental
representations was studied.  The resolvents in the quantum theory live
on the two-sheeted Riemann surface defined by the matrix model curve.
Their quantum behavior is characterized by the structure around the
branch cuts and poles, which are related to the RR flux contributions in
the Calabi--Yau geometry and flavor fields, respectively.  A pole
associated with flavor on the first sheet is related to the Higgs vacua
(corresponding to classical nonzero vacuum expectation value of the
fundamental) while a pole on the second sheet is related to the
pseudo-confining vacua where the classically vanishing vacuum
expectation value of the fundamental gets nonzero values due to quantum
correction.

It is known \cite{Cachazo:2003yc} that Higgs vacua and pseudo-confining vacua,
which are distinct in the classical theory, are smoothly transformed
into each other in the quantum theory. This transition is realized on
the Riemann surface by moving poles located on the second sheet to pass
the branch cuts and enter the first sheet. This process was analyzed in
\cite{Cachazo:2003yc} at the off-shell level by fixing the value of glueball fields during
the whole process.  However, in an on-shell process, the position of
poles and the width and position of branch cuts are correlated (when the
flavor poles are moved, the glueball field is also changed).  It was
conjectured in \cite{Cachazo:2003yc} that for a given branch cut,
there is an upper bound for the number of poles (the number of flavors)
which can pass through the cut from the second sheet to the first sheet.

Our first aim of this paper is to confirm this conjecture and give the
corresponding upper bound for various gauge groups (in particular, we
will concentrate on the $U(N_c)$ gauge group). The main result is that
if $N_f \geq N_c $, the poles will not be able to pass through the cut
to the first sheet where $N_c$ is the effective fluxes associated with
the cut (and can be generalized to other gauge groups).

Another important development was made in \cite{Intriligator:2003xs}, which was
inspired by \cite{Kraus:2003jf}.  In \cite{Intriligator:2003xs}, which we will refer
to as IKRSV, it was shown that, to correctly compute the prediction of
string theory (matrix model), it is crucial to determine whether the
glueball is really a good variable or not.  A prescription was given,
regarding when a glueball field corresponding to a given branch cut
should be set to zero before extremizing the off-shell glueball superpotential.
The discussion of IKRSV was restricted to $\CN=1$ gauge theories with an
adjoint and no flavors, so the generalization to the case with
fundamental flavors is obviously the next task.

Our second aim of this paper is to carry out this task. The main result
is the following.  Assuming $N_f$ poles around a cut associated with
gauge group $U(N_{c,i})$, when $N_f\geq N_{c,i}$ there are situations in
which we should set $S_i=0$ in matrix model computations.  More
concretely, situations with $S_i=0$ belong to either of the following
two branches: the baryonic branch for $N_{c,i}\leq N_f< 2N_{c,i}$, or
the $r=N_{c,i}$ non-baryonic branch for $N_f\ge 2N_{c,i}$. Moreover,
when $S_i=0$, the gauge group is completely broken and there should
exist some extra, charged massless field which is not incorporated in
matrix model.

In section \ref{sec:bg}, as background, we review basic materials for
${\mathcal N}=1$ supersymmetric $U(N_c)$ gauge theory with an adjoint chiral
superfield, and $N_f$ flavors of quarks and anti-quarks.  The chiral
operators and the exact effective glueball superpotential are given. We
study the vacuum structure of the gauge theory at classical and quantum
levels.  We review also the main results of IKRSV.\ \ In addition to all
these reviews, we present our main motivations of this paper.


In section \ref{sec:1cut}, we apply the formula for the off-shell
superpotential obtained in \cite{Cachazo:2003yc} to the case with quadratic tree
level superpotential, and solve the equation of motion derived from it.
We consider what happens if one moves $N_f$ poles associated with
flavors on the second sheet through the cut onto the first sheet,
on-shell.
Also, in subsection \ref{subsec:general'n_IKRSV}, we briefly touch the
matter of generalizing IKRSV in the one cut model.

In section \ref{sec:2cut}, we consider cubic tree level superpotential.
On the gauge theory side, the factorization of the Seiberg--Witten curve
provides an exact superpotential.  We reproduce this superpotential by
matrix model, by extremizing the effective glueball superpotential with
respect to glueball fields after setting the glueball field to zero when
necessary.  We present explicit results for $U(3)$ theory with all
possible breaking patterns and different number of flavors
($N_f=1,2,3,4$, and 5).

In section \ref{sec:conclusion}, after giving concluding remarks, we
repeat the procedure we did in previous sections for $SO(N_c)/USp(2N_c)$
theories, briefly.

In the appendix, we present some proofs and detailed calculations which
are necessary for the analysis in section \ref{sec:2cut}.

Since string theory results in the dual Calabi--Yau geometry are
equivalent to the matrix model results, we refer to them synonymously
through the paper.  There exist many related works to the present paper.
For a list of references, we refer the reader to \cite{Argurio:2003ym}.

\section{Background}
\setcounter{equation}{0}
\label{sec:bg}

In this section, we will summarize the relevant background needed for
the study of ${\mathcal N}=1$ supersymmetric gauge theory with matter
fields.

\subsection{The general picture of matrix model with flavors}

The generalized Konishi anomaly interpretation to the matrix model
approach for $\CN=1$ supersymmetric gauge theory with flavors was given
in \cite{Seiberg:2002jq,Cachazo:2003yc}. Here we make only a brief summary on some
points we will need.

Let us consider $\CN=1$ supersymmetric $U(N_c)$ gauge theory, coupled to
an adjoint chiral superfield $\Phi$, $N_f$ fundamentals $Q^f$, and $N_f$
anti-fundamentals $\Qt_\ft$.  The tree level superpotential is taken to
be
\bea 
\label{csw2-2.1}
W_{\text{tree}} & = & \Tr\, W(\Phi)+\sum_{f,\ft}\W Q_{\W f}\, m^{\W f}{}_{f}(\Phi) Q^f~,
\eea
where the function $W(z)$ and the matrix $m_f^\ft(z)$ 
are polynomials 
\bea 
W(z) & = & \sum_{k=0}^{n} \frac{g_k z^{k+1}}{k+1}, \qquad
m^{\W f}{}_{f}(z)  =  \sum_{k=1}^{l+1} (m_k)^{\W f}{}_{f} z^{k-1}~.
\nonu
\eea
Classically we can have the ``pseudo-confining vacua'' where
the vacuum expectation values of  $Q$, $\W Q$  are zero, or the
``Higgs vacua'' where the vacuum expectation values of $Q$, $\W Q$ are 
nonzero so that the total rank of the remaining gauge groups is
reduced. 
These two vacua, which seem to have a big difference classically, are
not fundamentally distinguishable from each other in the quantum theory
and in fact can be continuously transformed into each other, as we will
review shortly, in the presence of flavors \cite{Cachazo:2003yc}.

Supersymmetric vacua of gauge theory are characterized by the vacuum
expectation values of chiral operators \cite{Svrcek:2003az}.  They are
nicely packaged into the following functions called resolvents
\cite{Cachazo:2003yc,Seiberg:2002jq}: \footnote{We set 
$w_\alpha(z)\equiv \frac{1}{
4\pi}\Ev{\Tr \frac{W_\a}{ z-\Phi}}$ to zero because in supersymmetric
vacua $w_\alpha(z)=0$.}
\begin{align}
 T(z) & =  \Ev{\Tr \frac{1}{ z-\Phi}}, \label{csw2-2.10-a}\\
 R(z) & =  -\frac{1}{ 32\pi^2} \Ev{\Tr 
\frac{W_\a W^\a}{ z-\Phi}}, 
\label{csw2-2.10-c}
\\
 M(z)^f{}_{\W f} & =  \Ev{\W Q_{\W f} \frac{1}{ z-\Phi} Q^f} 
\label{csw2-2.10-d}
\end{align}
where $W_\a$ is (the lowest component of) the field strength superfield.
Classically, $R(z)$ vanishes while $T(z)$, $M(z)$ have simple poles on
the complex $z$-plane at infinity and at the eigenvalues of $\Phi$.
Each eigenvalue of $\Phi$ is equal to one of zeros of $W'(z)$ or $B(z)$,
where
\bea  
\label{csw2-2.2-1}
W'(z)= g_n \prod_{i=1}^n (z-a_i),\qquad
B(z) \equiv \det m(z)=B_L \prod_{I=1}^L (z-z_I).
\eea
In the pseudo-confining vacuum, every eigenvalue of $\Phi$ is equal to
$a_i$ for some $i$.  On the other hand, in the Higgs vacuum, some
eigenvalues of $\Phi$ are equal to $z_I$ for some $I$.

In the quantum theory, the resolvents
\eqref{csw2-2.10-a}, \eqref{csw2-2.10-c}
and \eqref{csw2-2.10-d} are determined by the
generalized Konishi anomaly equations \cite{Cachazo:2003zk, Seiberg:2002jq,Cachazo:2003yc}:
\begin{equation}
\begin{split}
%
~[W'(z) T(z)]_-+ \Tr [m'(z) M(z)]_-  & =  2 R(z) T(z),  \\
~[W'(z) R(z)]_- & =  R(z)^2, \\
~[(M(z) m(z))^{f'}_{~f}]_- & =  R(z) \delta^{f'}_{f}, \\
~[( m(z) M(z))^{\W f'}_{~ \W f}]_- & =  R(z) \delta^{\W f'}_{\W f},
\end{split}
\label{csw2-2.14} 
\end{equation}
where the notation $[~]_-$ means to drop the nonnegative powers in a
Laurent expansion in $z$.   From the second 
equation of (\ref{csw2-2.14}),
one obtains \cite{Cachazo:2002ry}
\bea 
R(z) = \frac{1}{ 2} \left( W'(z)-\sqrt{W'(z)^2+f(z)}\right),
\nonu
\eea 
where $f(z)$ is a polynomial of degree $(n-1)$ in $z$.  This implies
that in the quantum theory the zeros $z=a_i$ ($i=1,2,\dots, n$) of
$W'(z)$ are blown up into cuts $A_i$ along
intervals\footnote{\label{phys_mean_cut} $a_i^{\pm}$ are generally
complex and in such cases we take $A_i$ to be a straight line connecting
$a_i^-$ and $a_i^+$.  Note that there is no physical meaning to the
choice of the cut; it can be any path connecting $a_i^-$ and $a_i^+$.}
$[a_i^-,a_i^+]$ by the quantum effect represented by $f(z)$, and the
resolvents \eqref{csw2-2.10-a}--\eqref{csw2-2.10-d} are defined on a
double cover of the complex $z$-plane branched at the roots $a_i^\pm$ of
$ W'(z)^2+f(z)$.  This double cover of the $z$-plane can be thought of
as a Riemann surface $\Sigma$ described by the matrix model curve
\bea
\label{csw2-2.18}
\Sigma: \quad y_m^2= W'(z)^2+f(z).
\eea
This curve is closely related to the factorization form of 
$\CN=2$ curve in the strong coupling analysis.

Every point $z$ on the $z$-plane is lifted to two points on the
Riemann surface $\Sigma$ which we denote by $q$ and $\qt$ respectively.
For example, $z_I$ is lifted to $q_I$ on the first sheet and $\W q_I$ on
the second sheet.  We write the projection from $\Sigma$ to the
$z$-plane as $z_I=z(q_I)=z(\widetilde q_I)$, following the notation of
\cite{Cachazo:2003yc}.
 
The classical singularities of the resolvents $T(z)$, $M(z)$ are
modified in the quantum theory to the singularities on $\Sigma$, as
follows.
For $T(z)$, the classical poles at
$z_I$ are lifted to poles at $q_I$ or $\W q_I$, depending on which
vacuum the theory is in, while the classical poles at $a_i$ with residue
$N_{c,i}$ are replaced by cuts with periods $\frac{1}{ 2\pi i}\oint_{A_i}
T(z) dz=N_{c,i}$.  For $M(z)$, the classical poles at $z_I$ are also lifted
to poles at $q_I$ or $\W q_I$.
More specifically, by solving the last two equations of (\ref{csw2-2.14}),
one can show \cite{Cachazo:2003yc}
\begin{align}
M(z) = R(z) \frac{1}{m(z)}
 &-\sum_{I=1}^L \frac{ (1-r_I)R(q_I) }{ (z-z_I)} \frac{1}{ 2\pi i}\oint_{q_I} \frac{dx}{ m(x)} \notag\\
 &-\sum_{I=1}^L \frac{ r_I R(\W q_I)}{(z-z_I)} \frac{1}{ 2\pi i} \oint_{\W q_I} \frac{dx}{ m(x)},
\label{Higgs-M} 
\end{align}
where $(q_I,\W q_I)$ are the lift of $z_I$ to the first sheet and to the
second sheet of $\Sigma$, and $r_I=0$ for  poles on the second sheet
and $r_I=1$ for  poles on the first sheet.
Furthermore, for $T(z)$, by solving the first equation of
(\ref{csw2-2.14}),
\bea \label{Higgs-T-z} T(z) & = & \frac{ B'(z)}{ 2
B(z)}- \sum_{I=1}^L \frac{ (1-2 r_I) y(q_I)}{ 2y(z) (z-z_I)}+
\frac{c(z)}{ y(z)},
\eea
where
\bea \label{c-z} c(z) & = & \braket{ \Tr
\frac{W'(z)-W'(\Phi) }{ z-\Phi}} -\frac{1}{ 2} \sum_{I=1}^{L} \frac{
W'(z)-W'(z_I)}{ z-z_I}. 
\eea
Practically it is hard to use (\ref{c-z}) to obtain $c(z)$ and we use
the following condition instead:
\bea
\label{off-shell} \frac{1}{ 2\pi i}\oint_{A_i} T(z) dz=N_{c,i}. 
\eea 
Finally, the exact, effective glueball superpotential is given by \cite{Cachazo:2003yc}
\begin{equation}
 \label{Wexact-y} 
\begin{split}
  W_{\text{eff}}  =  
 &-\frac{1}{ 2} \sum_{i=1}^n N_{c,i} \int_{\widehat{B}^r_i} y(z) dz\\
 &-\frac{1}{ 2} \sum_{I=1}^L (1-r_I)\int_{\widetilde{q}_I} ^{\widetilde{\Lambda}_0} y(z) dz
 -\frac{1}{ 2} \sum_{I=1}^L r_I \int_{q_I} ^{\widetilde{\Lambda}_0} y(z) dz 
 \\
 &+\frac{1}{ 2}(2N_c-L) W(\Lambda_0) 
 +\frac{1}{ 2} \sum_{I=1}^L W(z_I) \\
 &-\pi i(2N_c-L)S+ 2\pi i \tau_0 S +2\pi i \sum_{i=1}^{n-1} b_i S_i,
\end{split}
\end{equation}
where
\bea \label{scale} 
2\pi i \tau_0 =\log\left(
\frac{B_L \Lambda^{2N_c-N_f} }{ \Lambda_0^{2N_c-L}} \right).
\eea
Here, $\Lambda_0$ is the cut-off of the contour integrals, $\Lambda$ is
the dynamical scale, $S\equiv \sum_{i=1}^n S_i$, and $b_i\in
\mathbb{Z}$.  $\widehat{B}_i^r$ is the regularized contour from
$\widetilde{\La}_0$ to $\La_0$ through the $i$-th cut and $\La_0$ and
$\widetilde{\La}_0$ are the points on the first sheet and on the second
sheet, respectively.
The glueball field is defined as 
\bea 
S_i =\frac{1}{2\pi i} \oint_{A_i} R(z) dz.
\nonu
\eea

In the above general solutions \eqref{Higgs-M}, \eqref{Higgs-T-z}, we
have $r_I=1$ or $r_I=0$, depending on whether the pole is on the first
sheet or on the second sheet of $\Sigma$, respectively.  The relation
between these choices of $r_I$ and the phase of the system is as
follows.  Let us start with all $r_I=0$, i.e., all the poles are on the
second sheet. This choice corresponds to the pseudo-confining vacua
where the gauge group is broken as $U(N_c)\to \prod_{i=1}^{n} U(N_{c,i})$
with $\sum_{i=1}^{n} N_{c,i}=N_c$. Now let us move a single pole through,
for example, the $n$-th cut to the first sheet. This will break the
gauge group as $\prod_{i=1}^{n} U(N_{c,i})\to \prod_{i=1}^{n-1} U(N_{c,i})
\times U(N_{c,n}-1)$.  
Note that the rank of the last factor is now $(N_{c,n}-1)$ so
that $\sum_{i=1}^{n} N_{c,i}=N_c-1<N_c$.  Namely, the gauge group is Higgsed
down. In this way, by passing poles through cuts, one can go
continuously from the pseudo-confining phase to the Higgs phase, as
advocated before.

However, if we consider this process of passing poles through a cut to
the first sheet {\em on-shell\/}, then there should be an obstacle at a
certain point. For example, if initially we have $N_{c,n}=1$, 
after passing
a pole we would end up with an $U(0)$.  This sudden jump of the number
of $U(1)$'s in the low energy gauge theory is not a smooth physical
process, because the number of massless particles (photons) changes
discontinuously.  So we expect some modifications to the above
picture. In \cite{Cachazo:2003yc}, it was suggested that in an on-shell process,
the $n$-th cut will close up in such a situation so that the pole cannot
pass through. It is one of our motivations to show that this is indeed
true.  More precisely, the cut does not close up completely and the pole
can go through a little bit further to the first sheet and then will be
bounced back to the second sheet.

\subsection{The vacuum structure}

In the last subsection we saw that different distributions of poles over
the first and the second sheets correspond to different phases of the
theory.  In this subsection we will try to understand this vacuum
structure of the gauge theory at both classical and quantum levels for a
specific model (for more details, see 
\cite{Argyres:1996eh,Carlino:2000uk,Carlino:2001ya,Balasubramanian:2003tv,Ahn:2003vh,Ahn:2003ui}).
For simplicity we will focus on $U(N_c)$ theory with $N_f$ flavors and
the following tree level superpotential \footnote{We used the convention
of \cite{Demasure:2002jb} for the normalization of the second term. Different choices
are related to each other by redefinition of $\W Q$ and $Q$.}
%
%
%
%
\bea \label{quadraticpot}
W_{\text{tree}} & = & \frac{1}{2} m_A \Tr\,\Phi^2
-
\sum_{I=1}^{N_f} \widetilde{Q}_{I}
 (\Phi + m_f) Q^{I}.
\eea
%
This corresponds to taking polynomials in \eqref{csw2-2.1} as
\begin{align}
 W(z)&=\frac{m_A}{ 2}z^2,\qquad 
 m^{\widetilde I}_{~I}(z)=-(z+m_f)\delta_{I}^{\widetilde I}.
\nonu
\end{align}
All $N_f$ flavors have the same mass $m_f$, and the mass function
defined in \eqref{csw2-2.2-1} is given by
\begin{align}
 B(z)&=(-1)^{N_f}(z+m_f)^{N_f}.
\nonu
\end{align}
Therefore, poles associated with  flavors are located at
\begin{align}
 z_I=-m_f\equiv z_f, \qquad I=1,2,\dots, N_f~.
 \label{def_zf}
\end{align}
In the quantum theory, some of these poles are lifted to $q_f$ on the
first sheet and others are lifted to $\qt_f$ on the second sheet.


The $D$- and $F$-flatness for the superpotential (\ref{quadraticpot})
is given by
\begin{align*}
0  &=  [ \Phi, \Phi^{\dagger}],   \quad 
0  =  Q Q^{\dagger}- \W Q^{\dagger} \W Q, \\
0  &=  m_A \Phi- Q \W Q , \quad
0  =  (\Phi + m_f) Q= \W Q  (\Phi + m_f).
\end{align*}
Solutions are a little different for $m_f\neq 0$ and $m_f=0$,
because $m_f=0$ is the root of $W'(z)=z$. The case of $W'(-m_f)=0$
was discussed in \cite{Argyres:1996eh,Balasubramanian:2003tv} which we will refer to as the
classically massless case.

In the $m_f\neq 0$ case, the solution is given by
\begin{equation}\label{sol2DF}
\begin{split}
\Phi & =  \left[ \begin{array}{cc}  -m_f I_{K\times K} & 0 \\
0 & 0_{(N_c-K)\times (N_c-K)} \end{array} \right], \\
Q   & =   \left[ \begin{array}{cc}  A_{K\times K} & 0 \\
0 & 0_{(N_c-K)\times (N_f-K)} \end{array} \right], ~~~
 ^t \W Q  =  \left[ \begin{array}{cc}  ^t\! \W A_{K\times K} &
\W B_{K\times (N_f-K)} \\
0 & 0_{(N_c-K)\times (N_f-K)} \end{array} \right]
\end{split}
\end{equation}
with 
\bean
- m_f m_A I_{K\times K}= A \,^t\!\W A, ~~~~~~
A A^{\dagger}= \W A^{\dagger} \W A+ \W B^* \,^t\!\W B.
\eean
The gauge group is Higgsed down to $U(N_c-K)$ where 
\bea 
K_{m_f\neq 0} \leq \min\left(N_c, N_f\right).
\nonu
\eea
To understand the range of $K_{m_f\neq 0} $, first note that the
$\Phi$ breaks the gauge group as $U(N_c)\to U(K)\times U(N_c-K)$. Now
the $U(K)$ factor has effectively $N_f$ massless flavors and because
$\ev{\W Q Q} \neq 0$, $U(K)$ is further Higgsed down to 
$U(0)$.

For $m_f=0$, we have $\Phi=0$ and $Q, \W Q$ are still of the above form
\eqref{sol2DF} with one special requirement: $\W A=0$. Because of this
we have
\bea 
K_{m_f=0} \leq \min\left(N_c, \left[\frac{N_f}{ 2}\right]\right),
\nonu
\eea
where $[~]$ means the integer part.  The integer $K_{m_f=0}$ precisely
corresponds to the $r$-th branch discussed in \cite{Argyres:1996eh,Balasubramanian:2003tv}.  The
$m_f=0$ case is different from the $m_f\neq 0$ case as follows. First,
$\Phi$ does not break the $U(N_c)$ gauge group, i.e., $U(N_c)\to
U(N_c)$. Secondly, the $r$-th branch is the intersection of the Coulomb
branch in which $\ev{\W Q Q}=0$ and the Higgs branch in which $\ev{\W Q
Q}\neq 0$, whereas for $m_f\neq 0$ the vacuum expectation value $\ev{\W Q Q}$ must be nonzero
and the gauge group must be Higgsed down. For these reasons, $K_{m_f\neq
0}$ and $K_{m_f=0}$ have different ranges.

The above classical classification of  $r$-th branches is also valid
in the quantum theory (including the baryonic branch). 

The quantum $r$-th branch can also be discussed by using the
Seiberg--Witten curve.  In the $r$-th branch, the curve factorizes as
\begin{align*}
 y_{\CN=2}^2
 &= P_{N_c}(x)^2 -4 \Lambda^{2N_c-N_f} (x+m_f)^{N_f}\\
 &=(x+m_f)^{2r} \left[ P_{N_c-r}^2(x) - 4 \La^{2N_c-N_f} (x+m_f)^{N_f-2r} \right].
\end{align*}
Because $N_c-r \geq 0$ (coming from $P_{N_c-r}(x)$) and $N_f -2r \geq 0$
(coming from the last term), we have $r \leq N_c$ and 
$r \leq N_f/2 $,
which leads to the range
\bea 
r \leq \min\left(N_c, \left[\frac{N_f}{ 2}\right]\right).
\label{range_r-vac}
\eea
The relation between this classification of the Seiberg--Witten curve
and the above classification of  
$r$-branches, in the $m_f\neq 0$ and
$m_f=0$ cases, is as follows.  In the $m_f=0$ case, we have one-to-one
correspondence where the $r$ is identified with $K_{m_f=0}$. In the
$m_f\neq 0$ case, for a given $r$ of the curve, there exist two cases:
either $K_{m_f\neq 0}=r$ for $K_{m_f\neq 0}\leq [N_f/2]$, or $K_{m_f\neq
0}=N_f-r$ for $K_{m_f\neq 0}\geq [N_f/2]$.

\subsection{The work of IKRSV}

Now we discuss another aspect of the model. In the above, we saw that
there is a period condition (\ref{off-shell}) for $T(z)$.  So, if for
the $i$-th cut we have $\oint_{A_i} T(z) dz=
N_{c,i}=0$, then it seems that,
in the string theory realization of the gauge theory, there is no RR
flux provided by D5-branes through this cut and the cut is
closed. Because of this, it seems that we should set the corresponding
glueball field $S_i=0$.  Based on this naive expectation, Ref.\
\cite{Kraus:2003jf} calculated the effective superpotential of
$USp(2N_c)$ theory with an 
antisymmetric tensor by the matrix model, which
turned out to be different from the known results obtained by holomorphy
and symmetry arguments (later Refs. \cite{Kraus:2003jv,Alday:2003dk} confirmed this
discrepancy).

This puzzle intrigued several papers \cite{Aganagic:2003xq,
Cachazo:2003kx, Intriligator:2003xs, Ahn:2003ui, Matone:2003bx,
Landsteiner:2003ph, Naculich:2003ka, Gomez-Reino:2004rd}.  In
particular, in \cite{Cachazo:2003kx}, it was found that although
$N_{c,i}=0$, we cannot set $S_i=0$. The reason became clear by later
studies. Whether a cut closes or not is related to the total RR flux
which comes from both D5-branes and orientifolds. For $USp(2N_c)$ theory
with antisymmetric tensor, although the RR flux from D5-branes is zero,
there exists RR flux coming from the orientifold with positive RR
charges, thus the cut does not close. That the cut does not close can
also be observed from the Seiberg--Witten curve \cite{Ahn:2003ui} where
for such a cut, we have two single roots in the curve, instead of a
double root.  All these results were integrated in
\cite{Intriligator:2003xs} for $\CN=1$ gauge theory with adjoint. Let us
define $\WH N_c=N_c$ for $U(N_c)$, $N_c/2-1$ for $SO(N_c)$ and $2N_c+2$
for $USp(2N_c)$. Then the conclusion of \cite{Intriligator:2003xs} can
be stated as
\begin{quote}
 If $\WH N_c>0$, we should include $S_i$ and extremize $W_{\text{eff}}(S_i)$ with respect to it. On the other hand, if $\WH N_c\leq 0$, 
 we just set $S_i= 0$ instead.
\end{quote}
In \cite{Intriligator:2003xs}, it was argued that this prescription of setting $S_i=0$
can be explained in string theory realization by considering an extra
degree of freedom which corresponds to the D3-brane wrapping the blown
up $S^3$ and becomes massless in the $S\to 0$ limit
\cite{Strominger:1995cz}.  Our second motivation of this paper is to
generalize this conclusion to the case with flavors.
We will discuss the precise condition when one should set $S_i=0$ in
order to get agreement with the gauge theory result, in the case with
flavors.

\subsection{Prospects from the strong coupling analysis}
\label{subsec:prospects_strong_cpl}

Before delving into detailed calculations, let us try to get some
general pictures from the viewpoint of factorization of the
Seiberg--Witten curve.  
Since we hope to  generalize IKRSV, we are interested in the case
where some $S_i$ vanish.  Because $S_i$ is related to the size of a
cut in the matrix model curve, which is essentially the same as the
Seiberg--Witten curve, we want some cuts to be closed in the
Seiberg--Witten curve.  Namely, we want a double root in the
factorization of the curve, instead of two single roots.

%

For $U(N_c)$ theory with $N_f$ flavors of the same mass $m_f=-z_f$, tree
level superpotential \eqref{csw2-2.1}, 
and breaking pattern
$U(N_c)\to \prod_{i=1}^n U(N_{c,i})$, the factorization form of the
curve is \cite{Balasubramanian:2003tv}
\begin{equation}
  \label{factor-UN}
\begin{split}
 P_{N_c}(z)^2-4\La^{2N_c-N_f} (z-z_f)^{N_f} &= H_{N-n}(z)^2 F_{2n}(z),\\
 F_{2n}(z) &= W'(z)^2+f_{n-1}(z),
\end{split}
\end{equation}
where the degree $2n$ polynomial $F_{2n}(z)= W'(z)^2+f_{n-1}(z)$
generically has $2n$ single roots.  How can we have a double root
instead of two single roots?

For a given fixed mass, for example $z_f=a_1$, there are three cases where
we have a double root, as follows.  (a) There is no $U(N_{c,1})$ 
group factor
associated with the root $a_1$, namely $N_{c,1}=0$. (b) The $U(N_{c,1})$
factor is in the baryonic branch. This can happen for $N_{c,1}\leq N_f<
2N_{c,1}$. (c) The $U(N_{c,1})$ factor is in the $r$-th non-baryonic
branch with $r=N_{c,1}$. This can happen only for $N_f\geq
2N_{c,1}$. Among these three cases, (b) and (c) \cite{Balasubramanian:2003tv} are new for
theories with flavors, and will be the focus of this paper.
However, it is worth pointing
out that the factorization form in the
cases (b) and (c) are not the one given in (\ref{factor-UN}) for a fixed
mass, but the one given in (\ref{Nocorrection}). 

Can we keep the factorization form (\ref{factor-UN}) while having an
extra double root? We can, but instead of a fixed mass we must let the
mass  ``floating,'' which means the following. There will be multiple
solutions to the factorization form (\ref{factor-UN}), and for any given
solution the $2n$ single roots of $F_{2n}(x)$, denoted by $a_i^\pm$,
$i=1,\dots,n$, are functions of $z_f$.  Now, we tune $z_f$ so that
$a_i^+(z_f)=a_i^-(z_f)$, i.e., so that two single roots combine into one
double root. Since for different solutions this procedure will lead to
different values of $z_f$, we call this situation the ``floating'' mass.

Now we have two ways to obtain extra double roots: one is with a
fixed mass, but to go to the baryonic or the $r=N_{c,1}$ 
non-baryonic branch,
while the other is to start with a general non-baryonic branch but using
a floating mass. In fact it can be shown that these two methods are
equivalent to each other when the double root is produced.
In the calculations in section \ref{sec:2cut} we will use the floating
mass to check our proposal.

\section{One cut model---quadratic tree level superpotential }
\setcounter{equation}{0}
\label{sec:1cut}

In this section we will study whether a cut closes up if one tries to
pass too many poles through it.  If the poles are near the cut, the
precise form of the tree level superpotential (namely the polynomials
$W(z)$, $m^{\tilde{f}}{}_f(z)$) is inessential and we can simplify the
problem to the quadratic tree level superpotential given by
(\ref{quadraticpot}).
For this superpotential, we will compute the effective glueball
superpotential using the formalism reviewed in the previous section.
Then, by solving the equation of motion, we study the {\em on-shell\/}
process of sending poles through the cut, 
and see whether the poles can pass  or not.

Also, on the way, we make an observation on the relation between the
exact superpotential and the vacuum expectation value of the tree level
superpotential.

\subsection{The off-shell $W_{\text{eff}}$, $M(z)$ and $T(z)$}

First, let us compute the effective glueball superpotential for the
quadratic superpotential \eqref{quadraticpot}.  The matrix model curve
\eqref{csw2-2.18} is related to $W'(z)$ in this case as
\bea \label{matrix-curve}
y_m^2 & = & W'(z)^2+f_{0}(z)= m_A^2 z^2-\mu \equiv
m_A^2 (z^2-\W \mu),~~~~~~~\W \mu=\frac{\mu }{ m_A^2}.
\eea

Let us consider the case with $K$ poles on the first sheet at $q_I=q_f$,
for which $r_I=1$, and with $(N_f-K)$ poles on the second sheet at $\W
q_I=\W q_f$, for which $r_I=0$ (recall that $(q_f,\W q_f)$ is the lift
of $z_f$
defined in \eqref{def_zf}).
Using the curve (\ref{matrix-curve}) and various formulas summarized in
the previous section, one can compute
\bean
S & = &  \frac{1}{ 2\pi i}\oint_{A} R(z) dz=
\frac{m_A\W \mu }{ 4} =\frac{  \mu}{ 4 m_A},
\nonu \\
\Pi & = &2\int_{\sqrt{\W \mu}}^{\Lambda_0} y(z) dz
 =   m_A\Lambda_0^2 - 2 S - 2 S \log \frac{ \Lambda_0^2 m_A }{ S},\notag \\
\Pi_{f,I}^{r_I=0} 
& = & \int_{\W q_I}^{\W \Lambda_0} y(z)d z
 = -\int_{q_I}^{\Lambda_0} y(z)d z\notag\\
&=& \frac{- m_A\Lambda_0^2}{ 2}- 2S \log \frac{ z_I}{ \Lambda_0}\\
&&\quad +2 S\left[ -\log\left( \frac{1}{ 2}+ \frac{1}{ 2}\sqrt{ 1- 
\frac{4 S}{ m_A z_I^2}}\right)
+\frac{m_A z_I^2 }{ 4 S}\sqrt{ 1- \frac{4 S}{ m_A z_I^2}}+
\frac{1}{ 2}\right],
\notag\\
\Pi_{f,I}^{r_I=1} & = & \int_{q_I}^{\W\Lambda_0} y(z) dz
=-\int_{\W q_I}^{\Lambda_0} y(z) dz \notag\\
&=& \frac{- m_A\Lambda_0^2}{ 2}
- 2S \log \frac{ z_I}{ \Lambda_0}\\
&&\quad+2 S\left[ -\log\left( \frac{1}{ 2}- 
\frac{1}{ 2}\sqrt{ 1- \frac{4S}{ m_A z_I^2}}\right)
-\frac{m_A z_I^2 }{ 4 S}
\sqrt{ 1- \frac{4S}{ m_A z_I^2}}+\frac{1}{ 2}\right],
\notag
\eean
where we dropped $\CO( {1/ \Lambda_0})$ terms.  
We have traded $q_I$, $\W q_I$ for $z_I$ in the square roots,
so that the sign convention is such that $\sqrt{1- \frac{4 S}{ m_A
z_I^2}}\sim 1-\frac{2 S}{ m_A z_I^2}$ and $\sqrt{z_I^2- 
\frac{4 S}{ m_A
}}\sim z_I$ for $|z_I|$ very large. 
Substituting this into (\ref{Wexact-y}), we obtain
\begin{align}
 &W_{\text{eff}}(S) = S \left[ N_c+ \log \left( \frac{m_A^{N_c}
 \Lambda^{2N_c-N_f} \prod_I z_I }{ S^{N_c}} \right) \right]\nonumber\\
  &\quad
 -\sum_{I,r_I=0} S
 \left[ -\log\left( \frac{1}{ 2}+ \frac{1}{ 2}
 \sqrt{ 1- \frac{4S}{ m_A z_I^2}}
 \right)
 +\frac{m_A z_I ^2 }{ 4 S}
 \left(\sqrt{ 1- \frac{4 S}{ m_A z_I ^2}} -1\right)+
 \frac{1}{ 2}\right]
 \nonumber \\
 &\quad
 -\sum_{I,r_I=1} S
 \left[ -\log\left( \frac{1}{ 2}- \frac{1}{ 2}
 \sqrt{ 1- \frac{4 S}{ m_A z_I^2}}
 \right)
 +\frac{m_A z_I ^2}{ 4 S}
 \left(-\sqrt{ 1- \frac{4S}{ m_A z_I ^2}} -1\right)+
 \frac{1}{ 2}\right],
 \label{quadratic-exact}
\end{align}
where $\Lambda$ is the dynamical scale of the corresponding ${\mathcal N}=2$
gauge theory defined in \eqref{scale}.

Let us compute resolvents also.
The resolvent $M(z)$ is an $N_f\times N_f$ matrix.  Using
(\ref{Higgs-M}), we find that for the $I$-th eigenvalue
\bea 
M_{I}(z) & = & \frac{ -R(z) }{  (z- q_I)}
+\frac{(1-r_I) R(q_I) }{ (z- q_I)}+
\frac{r_I R(\W q_I) }{  (z- q_I)}.
\nonu
\eea
Expanding $M_I(z)$ around $z=\infty$ we can read off the
following vacuum expectation values
\begin{align*}
 \ev{\W Q Q}_{I} 
    &=  \frac{m_A}{ 2} 
{ \Bigl[z_I +(2 r_I-1)\sqrt{z_I^2-\W \mu}\Bigr]}, \\
 \ev{ \W Q \Phi Q}_{I} 
    &= \frac{m_A}{ 4} 
{ \Bigl[ 2 z_I^2 +2(2r_I-1) z_I\sqrt{z_I^2-\W \mu}-\W \mu\Bigr]},\\
 \Ev{\sum_I -( \W Q \Phi Q- z_I \W Q Q)}  &=  
\frac{ N_f m_A \W \mu}{ 4} = N_f S.
\end{align*}
There is something worth noting here. One might naively expect that the
exact superpotential is simply the vacuum expectation value of the tree
level superpotential (\ref{quadraticpot}), as is the case without
flavors.  However, this naive expectation is wrong!  Although we have
$\ev{W_{\text{tree,fund}}} = \ev{-\sum_I ( \W Q \Phi Q- z_I \W Q Q)}\neq 0$
if $S\neq 0$, we still have
\bea \label{promise} 
W_{\text{eff,on-shell}} = \braket{ \frac{m_A}{ 2} \Tr\, \Phi^2}~,
\eea 
as we will see shortly. The reason for (\ref{promise}) can be explained
by symmetry arguments \cite{Intriligator:1994sm,Elitzur:1996gk}. Although the tree level part
$\ev{W_{\text{tree,fund}}}$ for fundamentals is generically nonzero and
also contributes to $W_{\text{low}}$, the contribution is precisely
canceled by the dynamically generated superpotential $W_{\text{dyn}}$,
leaving only the $\Phi$ part of $W_{\text{tree}}$
\footnote{We would like
to thank K. Intriligator for explaining this point to us.}.

Let us calculate the resolvent $T(z)$ also.  For the present example,
$c(z)=m_A(N_c-\frac{N_f}{ 2})$ from (\ref{c-z}).  Therefore, using
(\ref{Higgs-T-z}), we obtain the expansion of the resolvent $T(z)$:
%
\begin{align}
 T(z) & =  \frac{N_c}{ z}+\sum_{I}\left[ \frac{ z_I}{ 2}+
 \frac{ 2 r_I-1}{ 2} \sqrt{ z_I^2- \W \mu} \right] \frac{1}{ z^2} \nonumber \\
 &  + \left[ \frac{ \W \mu(2N_c-N_f) }{ 4}+ \sum_{I}\left(
 \frac{ z_I^2}{ 2}+ \frac{2 r_I-1}{ 2} z_I  
 \sqrt{ z_I^2- \W \mu} \right)
 \right]\frac{1}{ z^3}+\cdots 
 \label{exact-T(z)}
\end{align}
From this we can read off $\braket{\Tr\,\Phi^n}$.  For example, for
$K=0$ we have
\begin{align*}
 \braket{\Tr\, \Phi} &=  \frac{N_f}{ 2} 
 \left(z_f-\sqrt{ z_f^2-\W \mu}\right) , \\
 \braket{\Tr\, \Phi^2}  &=   \frac{N_f}{ 2} 
 z_f\left(z_f-\sqrt{ z_f^2-\W \mu}\right)    + 
 \frac{(2N_c-N_f) \W \mu }{ 4}.
\end{align*}
%

\subsection{The on-shell solution}

Now we can use the above off-shell expressions to find the on-shell
solution. First we rewrite the superpotential as
\begin{align}
 &W_{\text{eff}} = S \left[ N_c+ \log \left(\frac{
 \Lambda_1^{3N_c-N_f}}{ S^{N_c} }\right) \right]\notag\\
 &- (N_f-K)S
 \left[ -\log\left( \frac{z_f}{ 2}+ \frac{1}{ 2}
 \sqrt{ z_f^2- \frac{4S}{ m_A }}
 \right)
 +\frac{m_A z_f}{ 4 S}
 \left(\sqrt{ z_f^2- \frac{4 S}{ m_A}} -z_f\right)+
 \frac{1}{ 2}\right]\notag\\
 &-K S \left[ -\log\left( \frac{z_f}{ 2}- 
 \frac{1}{ 2}\sqrt{ z_f^2- \frac{4 S}{
 m_A }} \right) +\frac{m_A z_f }{ 4 S} \left(-\sqrt{ z_f^2- 
 \frac{4S}{
 m_A }} -z_f\right)+\frac{1}{ 2}\right],
 \label{superpot_quad}
\end{align}
where $\Lambda_1^{3N_c-N_f}\equiv m_A^{N_c} \Lambda^{2N_c-N_f}$.  We
have set all $z_I$ to be $z_f$, i.e., all masses are the same, as in
\eqref{def_zf}.  From this we obtain%
\footnote{
It
is easy to check that the equation (\ref{equ-W-1}) with parameters
$(N_c,N_f,K)$ is the same as the one with parameters $
(N_c-r,N_f-2r,K-r)$.  Also from the expression (\ref{exact-T(z)}) it
is straightforward to see we
have $\braket{\Tr \Phi^2}_{N_c,N_f,K}=\braket{\Tr
\Phi^2}_{N_c-r,N_f-2r,K-r} +r z_I^2$.  
All of these facts are the result
of the ``addition map'' observed in \cite{Balasubramanian:2003tv}. Furthermore, one can
show that both (\ref{equ-W-1}) and (\ref{exact-T(z)}) for $K=0$ are
exactly the same as the one given by the strong coupling analysis in
\cite{Demasure:2002jb} and the weak coupling analysis in \cite{Demasure:2003sk, Erlich:2003aw}.
}
\begin{align}
 0  =   
    \log  \biggl(\!\frac{\Lambda_1^{3N_c-N_f}}{ S^{N_c}}\!\biggr)
 &+ K \log 
 \left( \frac{ z_f- \!\sqrt{ z_f^2- \frac{4 S}{ m_A }}  }{ 2} \right)\notag \\
 &+(N_f-K) \log \left( 
 \frac{ z_f+ \!\sqrt{ z_f^2- \frac{4 S}{ m_A }}  }{ 2} \right),
 \label{equ-W-1}
\end{align}
or
\footnote{As we mentioned before, for $m_f\neq 0$ the allowed Higgs branch
requires $K\leq N_f$ but the strong coupling analysis gives
$K\leq N_f/2$. The resolution for that puzzle is that
if $K> N_f/2$, it is given by $(N_f-K)$-th branch of the curve.
Since the same $r$-th branch of the curve gives both
$r$ and $(N_f-r)$ Higgs branches, we expect that
$r$ and $(N_f-r)$ Higgs branches are  related. This 
relation is given by $\W S= S b^2$, $\W z_f= z_f b$ with
$b = \frac{m_A \Lambda^2 }{ S}$. It can be shown that with the above 
relation, the equation of motion of $S$ for the 
$K$-th branch is changed to the equation of motion of $\W S$ 
for the $(N_f-K)$-th branch.}
\begin{align}
 0  = \log \left(\Sh^{-N_c} \right) 
 &+ K \log \left(\frac{\zh_f- \sqrt{\zh_f^2- {4\Sh }}  }{ 2}\right)\notag \\
 &+(N_f-K) \log \left( \frac{ \zh_f+ \sqrt{\zh_f^2- {4 \Sh }}  }{ 2}\right),
 \label{equ-W-4}
\end{align}
where we have defined dimensionless quantities 
$\Sh=\frac{S}{ m_A
\Lambda^2}= \frac{\W \mu}{ 4 \Lambda^2}$ 
and $\zh_f=\frac{z_f}{ \Lambda}$.
Note that using these massless quantities the cut is from along the
interval $[-2\sqrt{\Sh}, 2\sqrt{\Sh}]$.

Using (\ref{equ-W-1}) or (\ref{equ-W-4}) it is easy to show that
\bea 
W_{\text{eff,on-shell}} & = & \left(N_c-
\frac{N_f}{ 2}\right) S+ \frac{ m_A z_f^2}{ 4}
\left[ N_f +(2K-N_f) \sqrt{ 1-\frac{4 S}{ m_A z_f^2}} \right].
\nonu
\eea
Also by expanding $T(z)$ in the present case, just as we did in
\eqref{exact-T(z)}, we can read off
\bean
\braket{ \frac{m_A }{ 2} \Tr\, \Phi^2} 
 & = &
 \left(N_c-\frac{N_f}{ 2}\right) S + \frac{ m_A z_f^2}{ 4}
\left[ N_f+(2K-N_f) \sqrt{1-\frac{4 S}{ m_A z_f^2}} \right],
\eean
which gives us the relation
\bean
\braket{ \frac{m_A}{ 2} \Tr\, \Phi^2  }
={W_{\text{eff,on-shell}}}
\eean
as we promised in (\ref{promise}).

Equation (\ref{equ-W-1}) is hard to solve. But if we want just to
discuss whether the cut closes up when we bring $z_f\to 0$, we can set
$K=0$ \footnote{If $K\neq 0$, we will have 
$U(N_c)\to U(K)\times
U(N_c-K)$ and the problem reduces to of 
$U(N_c-K)$ with $(N_f-2K)$
flavors in the 0-th branch.}, for which \eqref{equ-W-1} reduces to
\bea \label{eom_quad}
 z_f= \omega^r_{N_f}\Sh^{\frac{N_c}{ N_f}} +
 \omega^{-r}_{N_f} \Sh^{\frac{N_f-N_c}{ N_f}}.
\eea
Here, $\omega_{N_f}$ is the $N_f$-th root of unity,
$\omega_{N_f}=e^{2\pi i/N_f}$, and $r=0,1,\dots,(N_f-1)$ 
corresponds to
different branches of solutions.  It is also amusing to note that
above solution has the Seiberg duality \cite{Seiberg:1994pq} where
electric theory with $(N_c,N_f)$ is mapped to 
a magnetic theory with $(N_f-N_c,N_f)$.

With these preparations, we can start to discuss the on-shell process of
passing poles through the cut from the second sheet.

\subsection{Passing poles through a branch cut}

Consider moving $N_f$ poles on top of each other at infinity on the
second sheet toward the cut along a line passing through the origin and
making an angle of $\theta$ with the real axis.  Namely,
take\footnote{Throughout this subsection, we will use the dimensionless
quantities ${\zh}_f$, $\Sh$, etc.\ and omit the hats on them to avoid
clutter, unless otherwise mentioned.}
\begin{align}
 z(\widetilde q_f)=z(q_f)=  p e^{i\theta}, \qquad p,\theta\in\mathbb{R}, 
 \label{move_pole_z(q)}
\end{align}
and change $p$ from $p=\infty$ to $p=-\infty$ (see Fig.\
\ref{fig:poles_coming_in}).  This equation \eqref{move_pole_z(q)} needs
some explanation.  Remember that $z_f=z(q_f)=z(\qt_f)$ denotes the
projection from the Riemann surface $\Sigma$ (Eq.\ \eqref{matrix-curve})
to the $z$-plane. For each point $z$ on the $z$-plane, there are two
corresponding points: $q$ on the first sheet and $\qt$ on the second
sheet.  Although we are starting with poles at $\qt_f$ on the second
sheet, we do not know in advance if the poles will pass through the cut
and end up on the first sheet, or it will remain on the second sheet.
Therefore we cannot specify which sheet the poles are on, and that is
why we used $z(q_f)$, $z(\qt_f)$ in \eqref{move_pole_z(q)}, instead of
$q_f$ or $\qt_f$.

\begin{figure}[h]
\begin{center}
  \epsfxsize=7cm \epsfbox{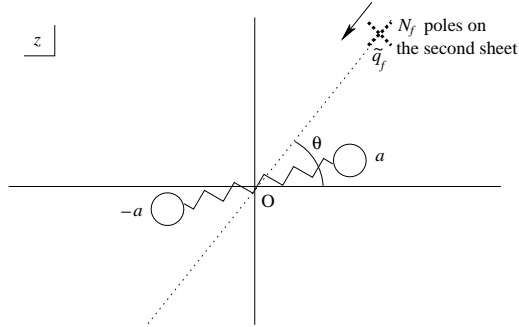}
\end{center}
 \vspace{-.5cm}
\caption[A process in which $N_f$ poles on the second sheet approach the
branch cut]{\sl A process in which $N_f$ poles at $\qt_f$ on the second
sheet far away from the cut approach the branch cut on the double
sheeted $z$-plane, along a line which goes through the origin and makes
an angle $\theta$ with the real $z$ axis.  The ``$\times$'' with dotted
lines denotes the poles on the second sheet, moving in the direction of
the arrow.  The two branch points $\pm a$ are connected by the branch
cut, which is denoted by a zigzag.  } \label{fig:poles_coming_in}
\end{figure}

Below, we study the solution to the equation of motion \eqref{eom_quad},
changing $p\in \mathbb{R}$ from $p=\infty$ to $p=-\infty$.  By
redefining $z_f$, $S$ by $z_f\to z_f e^{2\pi i r/(N_f-2N_c)}$, $S\to S
e^{4\pi i r/(N_f-2N_c)}$, we can bring \eqref{eom_quad} to the following
form:
\begin{align}
 z_f = p e^{i\theta}= S^{t}+S^{1-t},
\label{eom_quad(2)}
\end{align}
where
\begin{align}
 \frac{N_c}{ N_f}\equiv t.
\nonu
\end{align}
Henceforth we will use \eqref{eom_quad(2)}.  
Because $z_f$ as well as $S$ is complex, the position of the branch
points (namely, the ends of the cut), $\pm a$, where
\begin{align}
 a\equiv \sqrt{4S}
\nonu
\end{align}
is also complex, which means that in general the cut makes some finite
angle with the real axis, as shown in Fig.\ \ref{fig:poles_coming_in}.

\subsubsection*{\bf{$\bullet $ $\pmb{N_f=N_c}$}}

As the simplest example, let us first consider the $N_f=N_c$ (i.e.,
$t=1$) case.  We will see that the poles barely pass through the cut
but get soon bounced back to the second sheet.


The equation of motion \eqref{eom_quad(2)} is, in this case,
\begin{align}
 z_f=p e^{i \theta} = S + 1.
\label{eom_quad_Nf=N}
\end{align}
Therefore, as we change $p$, the position of the branch points changes
according to
\begin{align}
 a=\sqrt{4S}
 = 2\sqrt{z_f - 1} = 2\sqrt{pe^{i\theta} - 1}.
\label{pos_brch_pt}
\end{align}

Let us look closely at the process, step by step.  The point is that
transition between the first and the second sheet can happen only when
the cut becomes parallel to the incident direction of the poles, or when
the poles pass through the origin.
\begin{itemize}
 \item[(1)] $p\simeq +\infty$, on the second sheet:\\ 
            In this case, we can approximate the
            right hand side of \eqref{pos_brch_pt} as
            \begin{align*}
             a&
             =2p^{\half}e^{\frac{i\theta}{ 2}}(1-p^{-1}e^{-i\theta})^{\half}\\
             &\simeq 2p^{\half}e^{\frac{i\theta}{ 2}} e^{-\half p^{-1}e^{-i\theta}}
             =2p^{\half}e^{-\half p^{-1}\cos{\theta}}\;
             e^{i(\frac{\theta}{ 2}+\half p^{-1}\sin\theta)}.
            \end{align*} 
            Therefore, when the poles are far away, the angle between
            the cut and the real axis is approximately $\frac{\theta}{
            2}>0$ (we assume $0<\theta<\frac{\pi}{ 2}$).  Furthermore,
            as the poles approach ($p$ becomes smaller), the cut
            shrinks (because of $p^{\half}$) and rotates
            counterclockwise (because of $e^{\frac{i}{
            2}p^{-1}\sin\theta}$).  This corresponds to Fig.\
            \ref{fig:cfg_cut_poles}a.
            
 \item[(2)] Because the cut is rotating counterclockwise, as the poles
            approach, the cut will eventually become parallel to the
            incident direction, at some point.  This
            happens when
            \begin{align*}
             a^2&=4(pe^{i\theta}-1)=4\left[(p\cos\theta-1)+ip\sin\theta\right] 
             \propto e^{2i\theta} = \cos 2\theta+i\sin 2\theta.
            \end{align*}
            By simple algebra, one
            obtains
            \begin{align}
             p=2\cos\theta, \qquad a=2e^{i\theta}.\label{horzntl_pt}
            \end{align}
            Note that this is the only solution; the cut becomes
            parallel to the incident direction only once. Because
            $0<p<|a|=2$ (we are assuming $0 < \theta < \pi/2$), by the
            time the cut becomes parallel to the incident direction, the
            poles have come inside of the interval $[-2e^{i\theta},
            2e^{i\theta}]$, along which the cut extends when it is
            parallel to the incident direction.
	    This implies that the poles cross \footnote{As mentioned in
	    footnote \ref{phys_mean_cut}, there is no real physical
	    meaning to the position of the cut itself.} the cut at this
	    point, and enter into the first sheet.
            Fig.\ \ref{fig:cfg_cut_poles}b shows the situation when this
            transition is about to happen.
            In Fig.\ \ref{fig:cfg_cut_poles}c,  the poles are
            just crossing the cut.
            Fig.\ \ref{fig:cfg_cut_poles}d corresponds to the situation just
            after the transition happened; the poles have passed the cut
            and are now proceeding on the first sheet.
            
            Here we implicitly assumed that the cut is still rotating
            counterclockwise with a finite angular velocity, but this
            can be shown by expanding $a$ around \eqref{horzntl_pt} as
            $p=2\cos\theta+\Delta p$.  A short computation shows
            \begin{align*}
             a&\simeq 2  e^{\half \Delta p \, \cos\theta}
             e^{i\theta-\frac{i}{ 2} \Delta p \, \sin\theta}
            \end{align*}
            which implies that the cut shrinks and rotates
            counterclockwise if we move the poles to the left ($\Delta p$
            decreases).
            
\begin{figure}[htbp]
\begin{center}
  \epsfxsize=13.5cm \epsfbox{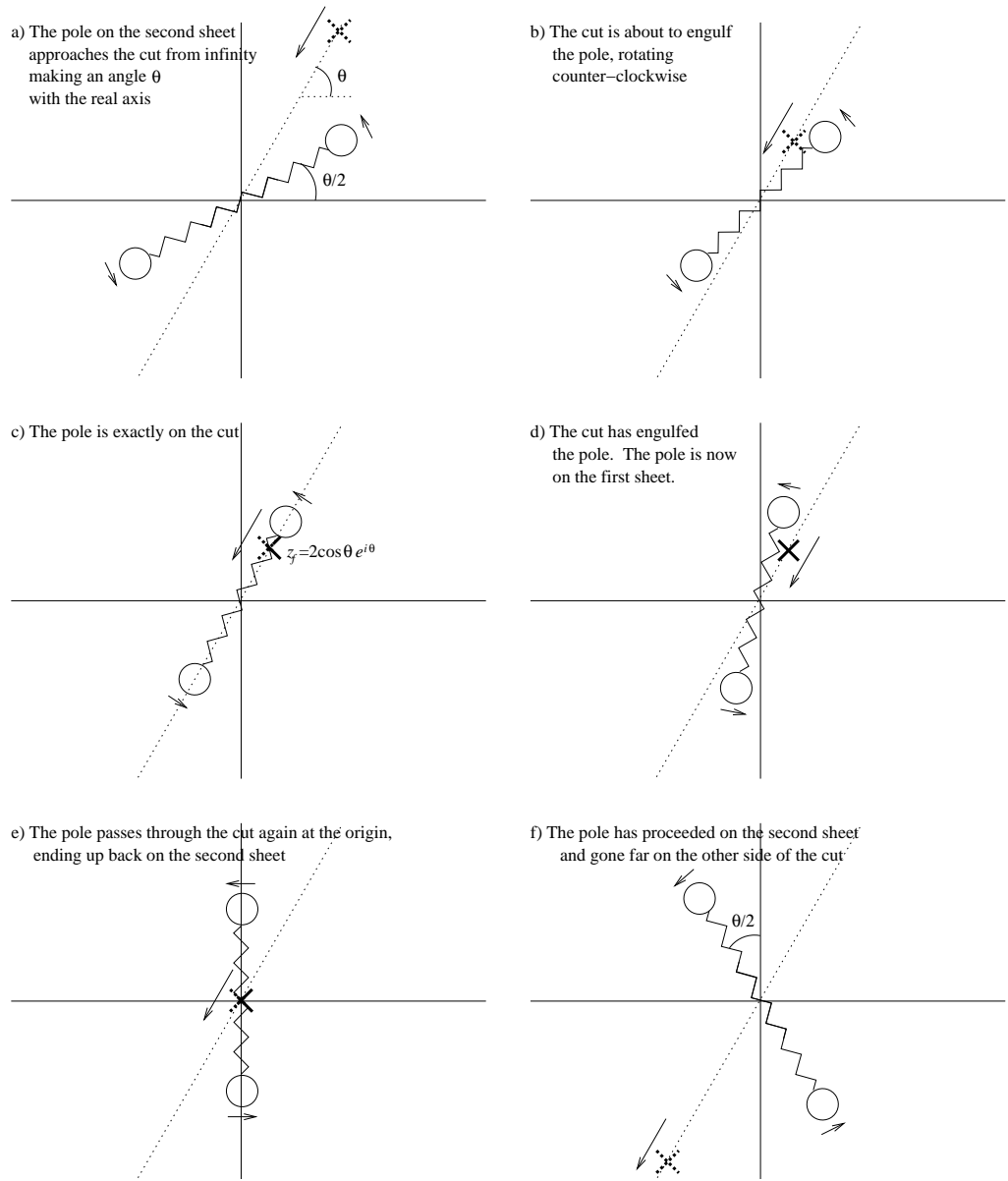}
\end{center}
\begin{center}
\caption[Six configurations of the branch cut and the poles]{\sl Six configurations of the branch cut and the poles.  The
poles are depicted by ``$\times$'' and moving along a line at an angle
$\theta$ with the real axis, as the arrow on it indicates.  The
``$\times$'' in solid (dotted) lines denotes poles on the first (second)
sheet.  The branch cut is rotating counterclockwise (as the arrows on
its sides indicate), changing its length.  }
\label{fig:cfg_cut_poles}
\end{center}
\end{figure}

 \item[(3)] $p\simeq 0$:\\ If the poles proceed on the real line
            further, it eventually reaches the origin $p=0$.
            By expanding \eqref{pos_brch_pt} around $p=0$, one obtains
            \begin{align}
             a&= 2(e^{\pi i}+pe^{i\theta})^{\half}
             \simeq 2e^{-\half p \cos\theta} e^{i(
             \frac{\pi}{ 2}-\half p \sin\theta)}.
             \label{Nf=Nc,zf=0}
            \end{align}
            Therefore the cut has a finite size ($|a|=2$) at $p=0$ and
            along the imaginary axis, still rotating counterclockwise, but now
            expanding. Because the cut goes through the origin, the poles
            pass through the cut again and comes back onto the second
            sheet (Fig.\ \ref{fig:cfg_cut_poles}e).

 \item [(4)] $p\simeq -\infty$, on the second sheet:\\ 
            If the poles have gone far past the cut so that $p<0$, $|p|\gg
            1$, we can approximate \eqref{pos_brch_pt}, as before, as 
            \begin{align*}
             a
             &= 2|p|^{\half}e^{\frac{i}{2}(\theta+\pi)}
             (1-p^{-1}e^{-i\theta})^{\half}
             \simeq 
             2|p|^{\half}e^{\half |p|^{-1}\cos{\theta}}\;
             e^{i(\frac{\pi}{ 2}+\frac{\theta}{2}-
             \half |p|^{-1}\sin\theta)}.
            \end{align*}
            Therefore, as the poles go away, the cut expands and
            rotates counterclockwise.  The angle between the cut and the
            real axis asymptotes to $(\frac{\pi}{2}+
            \frac{\theta}{2})$
            (Fig.\ \ref{fig:cfg_cut_poles}f).

\end{itemize}

In the above we assumed that $0<\theta<\frac{\pi}{2}$.  If
$\frac{\pi}{2}<\theta<\pi$, 
the only difference is that the order of steps
(2) and (3) are exchanged.  If $\theta<0$, the cut rotates clockwise
instead of counterclockwise.

When is the cut shortest in this whole process?
From \eqref{pos_brch_pt}, one easily obtains 
\begin{align}
 & |a|=2[(p-\cos\theta)^2+\sin^2\theta]^{1/4}\ge 2|\sin\theta|^{1/2}.
 \label{cut_length}
\end{align}
Therefore, when the poles are at $z_f=pe^{i\theta}=\cos\theta\,
e^{i\theta}$, which is between the steps (2) and (3) above, the cut
becomes shortest.  In particular, in the limit $\theta\to 0$ or
$\theta\to \pm\pi$, the cut completely closes up instantaneously.  These
correspond to configurations with either a horizontal cut with poles
colliding sideways, or a vertical cut with poles colliding from right
above or from right below.  Actually the existence of the $S=0$ solution
is easy to see in \eqref{eom_quad_Nf=N}: it is just $z_f=1,S=0$.

Summary: for $N_f=N_c$, when one moves poles on the second sheet from
infinity along a line toward a cut, poles pass through the cut onto
the first sheet and move away from the cut by a short distance. 
Then 
poles are bounced back to the second sheet again.  Therefore, one can
never move poles far away from the cut on the first sheet.  During
the process, in certain situations, the cut completely closes up.

\subsubsection*{\bf{$\bullet $ $\pmb{N_f\neq N_c}$}}

Now let us consider a more general case with $N_f\neq N_c$.  We again
consider a situation where poles on the second sheet approach a cut.
This time we will be brief and sketchy, because a detailed analysis such
as the one we did for the $N_f=N_c$ case would be rather lengthy due to
the existence of multiple branches, and would not be very illuminating.

First, let us ask how we can see whether poles are on the first sheet or
on the second sheet, from the behavior of $S$ versus $p$.  
Because this
is not apparent in the equation of motion of the form
\eqref{eom_quad(2)}, let us go back to
\begin{align}
 0=\frac{\partial W_{\text{eff}}}{ \partial  S}
 \propto
 \log S^{-N_c}+{N_f}\log 
 \frac{z_f\mp \sqrt{z_f^2-4S}}{ 2}
 ~~\Longrightarrow~~
 S^{t}
 =\frac{z_f\mp \sqrt{z_f^2-4S}}{ 2}.
 \label{eom_quad_pre}
\end{align}
which led to the equation \eqref{eom_quad(2)}.  Here the ``$-$''
(``$+$'') sign corresponds to $q_f$ on the first (second) sheet. For
$|z_f|^2\gg |4S|$, the square root can be approximated as
$\sqrt{z_f^2-4S}=z_f(1-4S/z_f^2)^{1/2}\simeq z_f(1-2S/z_f^2)$ (our sign
convention was discussed above \eqref{quadratic-exact}).  Therefore
\eqref{eom_quad_pre} is, on the first sheet,
\begin{align}
 S^{t} \simeq  \frac{z_f-z_f(1-2S/z_f^2)}{ 2}
 = \frac{S}{ z_f}
 ~~~~~~\Longrightarrow~~~~~~
 z_f\simeq S^{1-t}, 
 \label{S-q_sht1}
\end{align}
while on the second sheet
\begin{align}
 S^t \simeq \frac{z_f+z_f(1-2S/z_f^2)}{ 2}
 \simeq  {z_f}
 ~~~~~~\Longrightarrow~~~~~~
 z_f\simeq S^t.
 \label{S-q_sht2}
\end{align}

Now let us solve \eqref{eom_quad(2)} for $|z_f|\gg 1$.  By carefully
comparing the magnitude of the two terms in \eqref{eom_quad(2)}, one
obtains
\begin{align}
 \begin{array}{ccccl@{}c}
 N_f<N_c&(1<t) &\rightarrow\!\!&
  \left\{
  \begin{array}{l}
   z_f\simeq S^t \\[1ex]
   z_f\simeq S^{1-t}
  \end{array}
  \right.
  &
  \begin{array}{@{}l}
   \rightarrow~~ |S|\simeq |p|^{\frac{1}{ t}},    \\[1ex]
    \rightarrow~~ |S|\simeq |p|^{-\frac{1}{ t-1}},
  \end{array}
  &
  \begin{array}{c}
    |S|\gg 1    \\[1ex]
    |S|\ll 1
  \end{array}
  \\[4ex]
 N_c<N_f<2N_c&(\thalf<t<1) &\rightarrow\!\!&
 z_f\simeq S^t 
 & \rightarrow~~ |S|\simeq |p|^{\frac{1}{ t}}, &|S|\gg 1 \\[2ex]
 2N_c<N_f&(0<t<\thalf) &\rightarrow\!\!& z_f\simeq S^{1-t} 
 & \rightarrow~~ |S|\simeq |p|^{\frac{1}{ 1-t}}, & |S|\gg 1
\end{array}
 \label{large-q_behavior}
\end{align}
It is easy to show that $|z_f|^2\gg |4S|$ in all cases. So by using
\eqref{S-q_sht1}, \eqref{S-q_sht2}, we conclude that the first and the
third lines in \eqref{large-q_behavior} correspond to poles on the
second sheet, while the second and the last lines correspond to poles
on the first sheet.  This implies that, only for $N_f<N_c$, poles on
the second sheet can pass through the cut 
all the way and go infinitely
far away on the first sheet from a cut, 
as we will see explicitly in the
examples below.  
For $N_c<N_f<2N_c$, if one tries to pass poles through a cut, then
either poles will be bounced back to the second sheet, or the cut
closes up before the poles reach it.  
For $N_f>2N_c$, there is no solution corresponding to poles moving
toward a cut from infinity on the second sheet.  This should be related
to the fact that in this case glueball $S$ is not a good IR field.
Therefore, this one cut model is not applicable for $N_f>2N_c$.

To argue that these statements are true, rather than doing an analysis
similar to the one we did for the $N_f=N_c$ case, we will present some
explicit solutions for some specific values of $t=\frac{N_c}{ N_f}$ and
$\theta$, and argue general features.  Before looking at explicit
solutions, note that there are multiple solutions to Eq.\
\eqref{eom_quad(2)} which can be written as
\begin{align}
 z_f=(S^{\frac{1}{ N_f}})^{N_c}+(S^{\frac{1}{ N_f}})^{N_f-N_c}.
\label{eom_quad_ito_T}
\end{align}
From the degree of this equation, one sees that the  number of
the solutions to \eqref{eom_quad_ito_T} is:
\begin{align}
\begin{array}{ccl}
 N_f\le N_c     &\Longrightarrow & \text{$2N_c-N_f$ solutions,}\\
 N_c\le N_f\le 2N_c&\Longrightarrow & \text{$N_c$ solutions,}\\
 2N_c\le N_f    &\Longrightarrow & \text{$N_f-N_c$ solutions.}
\end{array}
\nonu
\end{align}
%
Therefore, if we solve the equation of motion \eqref{eom_quad(2)}, in
general we expect multiple branches of solutions.  We do not have to
consider the $N_f$ branches of the root $S^{\frac{1}{ N_f}}$, because it
is taken care of by the phase rotation we did above \eqref{eom_quad(2)}.

Now let us look at  explicit solutions for $U(2)$ example.
For ${N_f}=\half N_c$ ($t=2$), the solution to the equation of motion
\eqref{eom_quad(2)} is
\begin{align}
 S&=
 -\frac{1}{ 3}
 \left(
\frac{27+\sqrt{729-108z_f^3}}{ 2}\right)^{\!\!\!1/3}\!\!\!
 - \left(\frac{27+\sqrt{729-108z_f^3}}{ 2}\right)^{\!\!\!-1/3},
\nonu
\end{align}
where three branches of the cubic root are implied.
In Fig.\ \ref{fig:S-q_rf=1/2} we plotted $|S|$ versus $p$ for these
branches, for a randomly chosen value of the angle of incidence,
$\theta=\pi/6$.  Even if one changes $\theta$, there are always three
branches whose general shapes are similar to the ones in Fig.\
\ref{fig:S-q_rf=1/2}.  These three branches changes into one another
when $\theta$ is changed by $2\pi/3$.
One can easily see which
branch corresponds to what kind of processes, 
by the fact that on the
first sheet $|S|\ll 1$ as $|p|\to \infty$, while on the second sheet
$|S|\gg 1$ as $|p|\to \infty$.
The three branches correspond to the process in which: 
i)  poles go from the second sheet to the first sheet through the
cut, without any obstruction,
ii)  poles go from the first sheet to the second sheet (this is not
the process we are interested in), and
iii)  poles coming from the second sheet get reflected back to the
second sheet.
Note that the cut has never closed 
in all cases, because $|S|$ is always
nonvanishing.

%
%
%
%
\begin{figure}[ht]
 \vspace{.5cm}
 \begin{center}
  \begin{tabular}{ccc}
   \epsfxsize=4cm \epsfbox{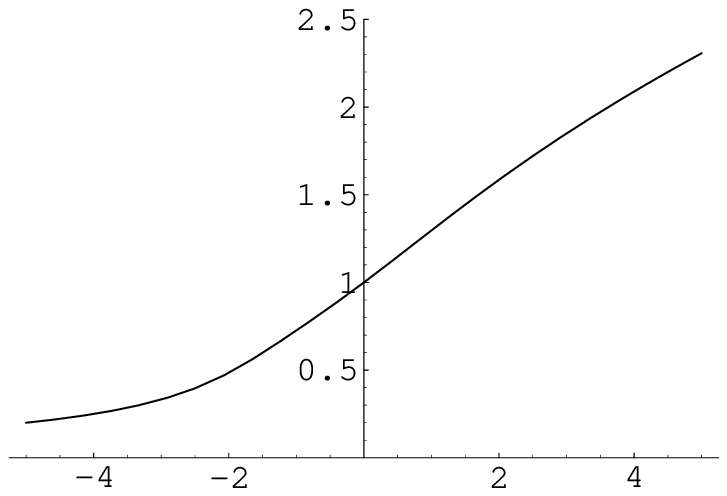}&
   \epsfxsize=4cm \epsfbox{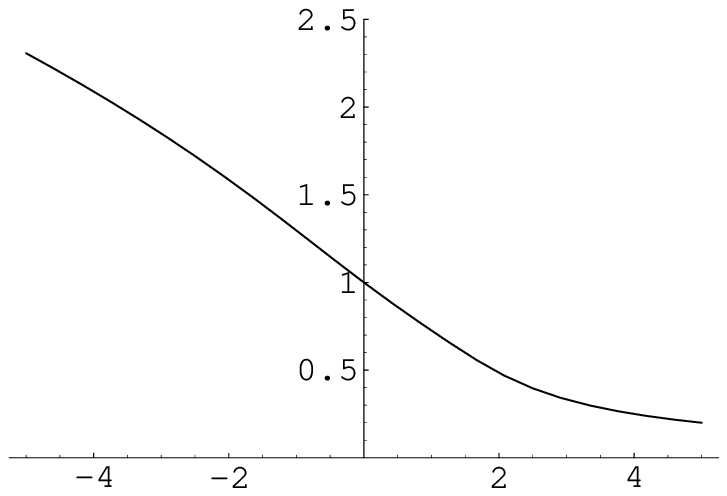}&
   \epsfxsize=4cm \epsfbox{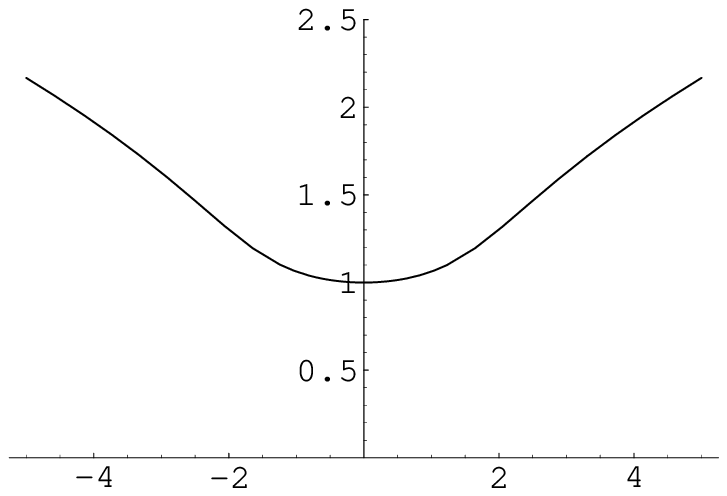}\\
   branch i) & branch ii) &branch iii) \\
  \end{tabular}
  \caption[The graph of $|S|$ versus $p$ for ${N_f}=\half N_c$ ($t=2$),
  $\theta=\pi/6$]{\sl The graph of $|S|$ versus $p$ for ${N_f}=\half
  N_c$ ($t=2$), $\theta=\pi/6$.  The vertical axis is $|S|$ and the
  horizontal axis is $p$.
  Although we are showing just the $\theta=\pi/6$ case, there are
  similar looking three branches for any value of $\theta$, which change
  into one another when $\theta$ is changed by $2\pi/3$.
  }
  \label{fig:S-q_rf=1/2}
 \end{center}
\end{figure}

Similarly, for ${N_f}=\frac{3}{ 2}N_c$ ($t=\frac{2}{ 3}$), 
the solution to the equation of motion
\eqref{eom_quad(2)} is
\begin{align}
 S&=\frac{1}{ 2}\left[-3z_f-1\pm(z_f+1)\sqrt{4z_f+1}\right].
\nonu
\end{align}
This time there are two branches, which change into each other when the
$\theta$ is changed by $\pi$.  We plotted $|S|$ versus $p$ for
$\theta=\pi/2$ in Fig.\ \ref{fig:S-q_rf=3/2}.  It shows two
possibilities: i) poles coming from the second sheet get reflected
back to the second sheet, for which $|S|\neq 0$ as $p\to 0$, ii) the cut
closes up before poles passes through it, for which $|S|\to 0$ as $p\to
0$.

%
%
%
%
\begin{figure}[ht]
 \vspace{.5cm}
 \begin{center}
  \begin{tabular}{cc}
   \epsfxsize=4cm \epsfbox{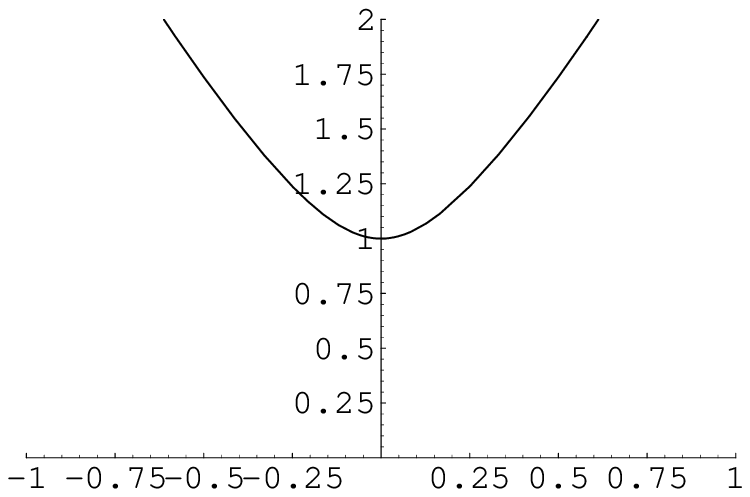}&
   \epsfxsize=4cm \epsfbox{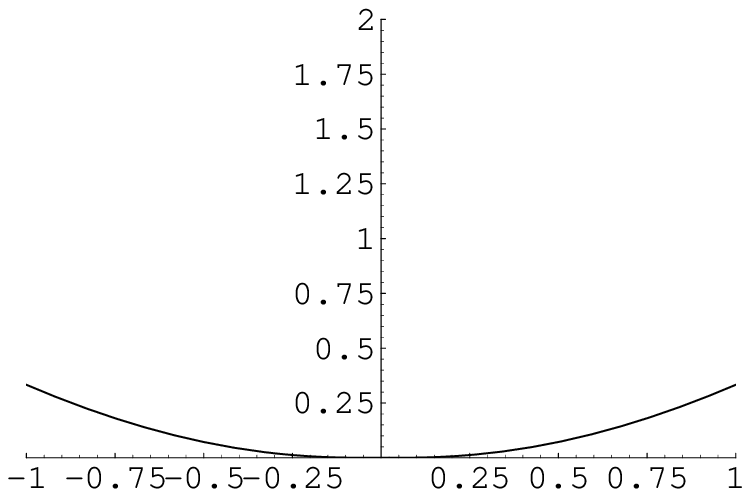}\\
   branch i) & branch ii)
  \end{tabular}
 \caption[The graph of $|S|$ versus $p$ for ${N_f}=\frac{3}{ 2}N_c$
  ($t=\frac{2}{ 3}$) or $N_f=3N_c$ ($t=\frac{1}{ 3}$), for
  $\theta=\pi/2$]{\sl The graph of $|S|$ versus $p$ for ${N_f}=\frac{3}{ 2}N_c$
  ($t=\frac{2}{ 3}$) or $N_f=3N_c$ ($t=\frac{1}{ 3}$), for
  $\theta=\pi/2$.  The two values of $t$ give the same graph.  The
  vertical axis is $|S|$ and the horizontal axis is $p$.  Although we
  are showing just the $\theta=\pi/2$ case, there are similar looking
  two branches for any value of $\theta$, which change into each other
  when $\theta$ is changed by $\pi$,
  } \label{fig:S-q_rf=3/2}
 \end{center}
\end{figure}


The $N_f=3N_c$ ($t=\frac{1}{ 3}$) case is also described by the same
Fig.\ \ref{fig:S-q_rf=3/2}.  However, as we discussed below
\eqref{large-q_behavior}, it does not correspond to a process of poles
approaching the cut from infinity on the second sheet; it corresponds to
poles on the first sheet and we cannot give any physical interpretation
to it.

\bigskip
These demonstrate the following general features: 
\begin{itemize}
 \item For $N_f< N_c$, one can move poles at infinity on the second
       sheet through a cut all the way to infinity on the first sheet
       without obstruction, if one chooses the incident angle
       appropriately.  If the angle is not chosen appropriately, the
       poles will be bounced back to the second sheet.

 \item For $N_c\le N_f<2N_c$, one cannot move poles at infinity on the second
       sheet through a cut all the way to infinity on first sheet.
       If one tries to, either i) the cut rotates and sends the poles
       back to the second sheet, or ii) the cut closes up before the
       poles reach it.

 \item For $2N_c<N_f$, the one cut model does not apply directly.
\end{itemize}

%

$N_f=2N_c$ is an exceptional case, 
for which the equation of
motion \eqref{eom_quad(2)} becomes
\begin{align}
 z_f=2 S^{1/2}.
\label{sol_Nf=2Nc}
\end{align}
Therefore $|S|\to 0$ as $z_f\to 0$, and the cut always closes before the
poles reach it.  

\subsubsection*{Subtlety in $S=0$ solutions}

If $p=0$, or equivalently if $z_f=0$, there is a subtle, but important
point we overlooked in the above arguments.  For $z_f=0$, the
superpotential \eqref{superpot_quad} becomes
\begin{align}
 W_{\text{eff}}
 &=
 S\left[\left( N_c-\frac{N_f}{ 2}\right)
 +\log\left(
\frac{(-1)^{N_f/2}m_A^{N_c-N_f/2}
\Lambda^{2N_c-N_f}}{ S^{N_c-N_f/2}}\right)
 \right]
 \notag\\
 &=
 S\left[\left( N_c-\frac{N_f}{ 2}\right)+
 \log\left(
\frac{(-1)^{-N_f/2}m_A^{N_c-N_f/2}
\Lambda_0^{2N_c-N_f}}{ S^{N_c-N_f/2}}\right)
 \right]
 +2\pi i \tau_0 S\notag\\
 &=
 \left( N_c-\frac{N_f}{ 2}\right)S
\left[1+ \log \left(\frac{\widetilde\Lambda_0^3}{ S} \right)
 \right]
 +2\pi i \tau_0 S
 \label{Weff_zf=0}
\end{align}
with $\widetilde\Lambda_0^{3(N_c-N_f/2)}\equiv
(-1)^{-N_f/2}m_A^{N_c-N_f/2}\Lambda_0^{2N_c-N_f}$.  Only from here to
\eqref{q=0_exclusion}, $S$ means the dimensionful quantity
($S=m_A\Lambda^2 \Sh$; see below \eqref{equ-W-4}).  In addition, in the
second line of \eqref{Weff_zf=0}, we rewrite the renormalized scale
$\Lambda$ in terms of the bare scale $\Lambda_0$ and the bare coupling
$\tau_0$ using the relation \eqref{scale}.  In our case,
$B_L=(-1)^{N_f}$.
The equation of motion derived from \eqref{Weff_zf=0}
is
\begin{align}
 \left( N_c-\frac{N_f}{ 2}\right)
\log \left(\frac{\widetilde\Lambda_0^3}{ S} \right) 
+ 2\pi i \tau_0
 =0
\nonu
\end{align}
and the solution is
\begin{align}
 S=
 \begin{cases}
 \widetilde\Lambda_0^3 \,e^{\frac{2\pi i \tau_0}{ N_c-N_f/2}}   
& N_f\neq 2N_c~, \\
 \text{no solution}  & N_f=2N_c~.
 \end{cases}
 \label{q=0_exclusion}
\end{align}

For $N_f<N_c$, \eqref{q=0_exclusion} is consistent with the fact that
the $|S|$ versus $p$ graphs in Fig.\ \ref{fig:S-q_rf=1/2} all go through
the point $(p,|S|)=(0,1)$ (now $S$ means the dimensionless quantity).
Also for $N_f=N_c$, \eqref{q=0_exclusion} is consistent with the result
\eqref{Nf=Nc,zf=0} ($|a|=2$, so $|S|=1$).  On the other hand, for
$N_f>N_c$, \eqref{q=0_exclusion} implies that we should exclude the
origin $(p,|S|)=(0,0)$ from the $|S|$-$p$ graphs in Fig.\
\ref{fig:S-q_rf=3/2}, which is the only $S=0$ solution (this includes
the $N_f=2N_c$ case \eqref{sol_Nf=2Nc}).  \footnote{One may think that
if one uses the first line of \eqref{Weff_zf=0}, then for $N_f=2N_c$,
$W_{\text{eff}}=S\log[(-1)^{N_f/2}]$ and there are solutions for some
$N_f$.  However, the glueball superpotential that string theory predicts
\cite{Cachazo:2001jy,Ookouchi:2002be} is the third line of \eqref{Weff_zf=0} which is in
terms of the bare quantities $\Lambda_0$ and $\tau_0$.  If $N_f=2N_c$,
then the log term vanishes and one cannot define a new scale $\Lambda$
as we did in \eqref{scale} to absorb the linear term $2\pi i \tau_0 S$.}

Therefore, the above analysis seems to indicate that, for $N_f>N_c$, the
$S=0$ solution at $p=0$, or equivalently $z_f=0$ is an exceptional case
and should be excluded. On the other hand, as can be checked easily,
gauge theory analysis based on the factorization method shows that there
is an $S=0$ solution in the baryonic branch. Thus we face the problem of
whether the baryonic $S=0$ branch for $N_f>N_c$ can be described in
matrix model, as alluded to in the previous discussions.

Note that, there is also an $S=0$ solution for $N_c=N_f$ in certain
situations, as discussed below \eqref{cut_length}.  For this solution,
which is in the non-baryonic branch, there is no subtlety in the
equation of motion such as \eqref{q=0_exclusion}, and it appears to be a
real on-shell solution.  This will be discussed further below.

\subsection{Generalization of IKRSV}
\label{subsec:general'n_IKRSV}

In the above and in subsection \ref{subsec:prospects_strong_cpl}, we
argued that for $N_f\geq N_c$ the $S=0$ solutions are real, on-shell
solutions based on the factorization analysis.  More accurately, there
are two cases with $S=0$: the one in the maximal non-baryonic branch
with $N_f\geq 2N_c$ and the other one in the baryonic branch with
$N_c\le N_f < 2N_c$. The case of non-baryonic branch cannot be discussed
in the one cut model, which is applicable only to $N_f<2N_c$.  On the
other hand, the baryonic one did show up in the previous subsection, but
we just saw above that those solutions should be excluded by the matrix
model analysis.  What is happening?
Is it impossible to describe the baryonic branch in matrix model?

Recall that the glueball field $S$ has to do with the strongly coupled
dynamics of $U(N_c)$ theory.  That $S=0$ in those solutions means that
there is no strongly coupled dynamics any more, namely the $U(N_c)$
group has broken down completely.  The only mechanism for that to happen
is by condensation of a massless charged particle which makes the $U(1)$
photon of the $U(N_c)$ group massive.
Therefore, in order to make $S=0$ a solution, we should incorporate such
an extra massless degree of freedom, which is clearly missing in the
description of the system in terms only of the glueball $S$.
This extra degree of freedom should exist even in the $N_f=N_c$ case
where $S=0$ really is an on-shell solution as discussed below
\eqref{cut_length}; we just could not directly see the degree of freedom
in this case.

The analysis of \cite{Intriligator:2003xs} hints on what this extra massless degree of
freedom should be in the matrix model / string theory context.  Note
that, the superpotential \eqref{Weff_zf=0} is of exactly the same form
as Eq.\ (4.5) of \cite{Intriligator:2003xs}, if we interpret $N_c-N_f/2\equiv \Nh$ as
the amount of the net RR 3-form fluxes.
In \cite{Intriligator:2003xs} it was argued that, if the net RR flux $\Nh$ vanishes,
one should take into account an extra degree of freedom corresponding to
D3-branes wrapping the blown up $S^3$ in the Calabi--Yau geometry
\cite{Strominger:1995cz}, and condensation of this extra degree of
freedom indeed makes $S=0$ a solution to the equation of motion.  The
form of the superpotential \eqref{Weff_zf=0} strongly suggests that the
same mechanism is at work for $N_f=2N_c$ in the $r=N_c$ non-baryonic
branch; condensation of the D3-brane makes $S=0$ a solution.
Furthermore, as discussed in \cite{Intriligator:2003xs}, for $N_f>2N_c$ the glueball
$S$ is not a good variable and should be set to zero.  A concise way of
summarizing this conclusion is: if the generalized dual Coxeter number
$h=N_c-N_f/2$ is zero or negative, we should set $S$ to zero in the
$r=N_c$ non-baryonic branch.

However, this is not the whole story, as we have discussed in subsection
\ref{subsec:prospects_strong_cpl}.  As we saw above, we need some extra
physics also for $N_c\le N_f<2N_c$ in order to explain the matrix model
result in the baryonic branch.
We argue below that this extra degree of freedom at least in the $N_c <
N_f<2N_c$ case should also be the D3-brane wrapping $S^3$ which shrinks
to zero when the glueball goes to zero: $S\to 0$.

The original argument of \cite{Intriligator:2003xs} is not directly applicable for
$N_f<2N_c$ because there are nonzero RR fluxes penetrating such a
D3-brane ($\Nh\neq 0$).  These RR fluxes induce fundamental string
charge on the D3-brane.  Because the D3-brane is compact, there is no
place for the flux to end on (note that this flux is not the RR one but
the one associated with the fundamental string charge).  Hence it should
emanate some number of fundamental strings.  If there are no flavors,
there is no place for such fundamental strings to end on, so they should
extend to infinity.  This fact led to the conclusion of \cite{Intriligator:2003xs}
that the D3-brane wrapping $S^3$ is infinitely massive and not relevant
unless $\Nh=0$.

However, in our situation, there are places for the fundamental strings
to end on --- noncompact D5-branes which give rise to flavors
\cite{Cachazo:2001jy,Ookouchi:2002be}.  In particular, precisely in the $z_f=0$ case,
where we have $S=0$ solutions for $N_c<N_f<2N_c$, the D3-brane wrapping
$S^3$ intersects the noncompact D5-branes in the $S\to 0$ limit, hence
the 3-5 strings stretching between them are massless.  Therefore the
D3-brane with these fundamental strings on it is massless and should be
included in the low energy description.
It is well known \cite{Witten:1998xy} that such a D-brane with
fundamental strings ending on it can be interpreted as baryons in gauge
theory.\footnote{That the D3-brane wrapping $S^3$ cannot exist for
$N_f<N_c$ can probably be explained along the same line as
\cite{Witten:1998xy}, by showing that those 3-5 strings are fermionic.
Also, note that the gauge group here is $U(N_c)$, not $SU(N_c)$ as 
in \cite{Witten:1998xy}, hence the ``baryon'' is charged under the $U(1)$.
  }
Condensation of this baryon degree of freedom should make $S=0$ a
solution, making the photon massive and breaking the $U(N_c)$ down to
$U(0)$.  The precise form of the superpotential for this extra degree of
freedom must be more complicated than the one proposed in \cite{Intriligator:2003xs}
for the case without flavors.
%

%
%

All these analyses tell us the following prescription: 
\begin{align}
 \label{prescription}
 \quad
 \begin{minipage}{4.5in}
  \em
  Using the floating mass condition that all $N_f$ poles are on top of
  one branch point\/%
  \addtocounter{footnote}{1}%
  \protect\footnotemark[\arabic{footnote}] 
  on the Riemann surface, we will have 
  an $S=0$ solution for $N_f\geq 2N_c$. For $N_c< N_f<2N_c$ there are two
  solutions: one with $S=0$ in the baryonic branch and one with
  $S\neq 0$ in the non-baryonic branch. In multi-cut cases, this 
  applies to each cut by replacing $N_c,$ $S$ with the corresponding
  $N_{c,i}$, $S_i$ for the cut.
 \end{minipage}
\end{align}
\footnotetext[\arabic{footnote}]{This condition will not work for the
$N_f=N_{c,i}$ case.}

In the next section we will discuss the condition we have used
in above prescription. Also by explicit examples, we
will demonstrate that when the gauge theory has a solution with closed
cuts ($S_i=0$), one can reproduce its superpotential in matrix model by
setting the corresponding glueballs $S_i$ to zero by hand.

  %
  %


\section{Two cut model---cubic tree level superpotential} 
\setcounter{equation}{0}
\label{sec:2cut}

Now, let us move on to $U(N_c)$ theory with cubic tree level
superpotential, where we have two cuts.  We will demonstrate that 
for each closed cut we can set $S=0$ by hand to 
reproduce the correct gauge theory superpotential using matrix model.

Specifically, we take the tree level superpotential to be
\begin{equation}
 \begin{split}
 W_{\text{tree}}&=\Tr[W(\Phi)]-\sum_{I=1}^{N_f} \widetilde{Q}_I (\Phi-z_f) Q^I,\\
 W(z)&=\frac{g}{ 3}z^3+\frac{m}{ 2}z^2,\qquad
 W'(z)=gz\left(z+\frac{m}{ g}\right)\equiv g(z-a_1)(z-a_2). 
\end{split}
\label{hnhn8Apr04}
\end{equation}
Here we wrote down $W(z)$ in terms of $g_2=g$, $g_1=m$ for definiteness,
but mostly we will work with the last expression in terms of $g$,
$a_{1,2}$.
The general breaking pattern in the pseudo-confining phase is $U(N_c)\to
U(N_{c,1})\times U(N_{c,2})$, $N_{c,1}+N_{c,2}=N_c$, $N_{c,i}>0$.  In the quantum theory, the
critical points at $a_1$ and $a_2$ blow up into cuts along the intervals
$[a_1^-,a_1^+]$ and $[a_2^-,a_2^+]$, respectively.  Namely, we end up
with the matrix model curve \eqref{csw2-2.18}, which in this case is
\begin{align}
 y_m^2= W'(z)^2+f_1(z)= g^2(z-a_1^-)(z-a_1^+)(z-a_2^-)(z-a_2^+).
\label{hpni8Apr04}
\end{align}
We will call the cuts along $[a_1^-,a_1^+]$ and $[a_2^-,a_2^+]$
respectively the ``first cut'' and the ``second cut'' henceforth.  One
important difference from the quadratic case is that, we can study a
process where $N_f\ge 2N_{c,i}$ flavor poles are near the $i$-th cut in
the cubic case. 

%

As we have mentioned, our concern is whether the cut is closed or not.
Also from the experiences in the factorization it can be seen that for
$N_f>N_{c,i}$, when closed cut is produced, the closed cut and the poles are
on top of each other \footnote{We do not discuss the $N_f=N_{c,i}$ case
where closed cut and poles are not at the same point.  However because the
$S=0$ solution in this case is an on-shell solution, we can reproduce
the gauge theory result in matrix model without setting $S=0$ by hand.}.
With all these considerations we take the following condition to
constrain the position of the poles:\footnote{We could choose
$z_f=a_1^+$ or $z_f=a_2^\pm$ instead of \eqref{zf=a1-}, but the result
should be all the same, so we take \eqref{zf=a1-} without loss of
generality.}
\begin{align}
 z_f=a_1^-.
\label{zf=a1-}
\end{align}
If there are $S_1=0$ solutions in which the closed cut and the poles are
on top of each other, then all such solutions can be found by solving
the factorization problem under the constraint \eqref{zf=a1-}, since for
such solutions $z=a_1^-=a_1^+$ obviously.  One could impose a further
condition $S_1=0$, or equivalently $a_1^-=a_1^+$ if one wants just
closed cut solutions, but we would like to know that there also are
solutions with $S_1\neq 0$ for $N_f<2N_{c,1}$, so we do not do that.

To summarize, what we are going to do below is: first we  explicitly solve
the factorization problem under the constraint \eqref{zf=a1-}, and
confirm that the $S_1=0$ solution exists  when $N_f>N_{c,i}$. 
 Then, we reproduce the gauge theory
superpotential in matrix model by setting $S_1=0$ by hand.

\bigskip Before plunging into that, we must discuss one aspect of the
constraint \eqref{zf=a1-} and the $r$-branches, in order to understand
the result of the factorization method.  If one solves the factorization
equation for a given flavor mass $m_f=-z_f$ (without imposing the
constraint \eqref{zf=a1-}), then in general one will find multiple
$r$-branches labeled by an integer $K$ with range $0\le K\le \text{min}(N_c,[\frac{N_f}{2}])$ (see Eq.\ \eqref{range_r-vac}).  This is
related to the fact that the factorization method cannot distinguish
between the poles on the first sheet and the ones on the second sheet.
The $r$-branch labeled by $K$ corresponds to distributing $N_f-K$ poles
on the second sheet and $K$ poles on the first sheet.  We are not
interested in such configurations; we want to put $N_f$ poles at the
same point on the same sheet.  However, as we discuss now, we actually
do not have to worry about the $r$-branches under the constraint
\eqref{zf=a1-}.

The $r$-branches with different $K$ are different vacua in
general. However, under the constraint \eqref{zf=a1-} these $r$-branches
become all identical because at the branch point $z=a_i^\pm$ there is no
distinction between the first and second sheets.  This can be easily
seen in the matrix model approach.  From the equation (\ref{Wexact-y}),
the effective glueball superpotential for the two cut model with $N_f-K$
poles at $\qt_f$ on the second sheet and $K$ poles at $q_f$ on the first
sheet is
\begin{align*}
 W_{\text{eff}}& = -\frac{1}{ 2} ( N_{c,1} \Pi_1+ N_{c,2} \Pi_2)
 -\frac{1}{ 2} (N_f-K) \Pi_f^{(2)}
 -\frac{1}{ 2} K \Pi_{f} ^{(1)}\\
 &\quad +\frac{1}{ 2}(2N_c-N_f) W(\Lambda_0)
 +\frac{1}{ 2} N_f W(q)  \\
 &\quad -\pi i(2N_c-N_f)S+ 2\pi i \tau_0 S
 +2\pi i  b_1 S_1,
\end{align*}
where the periods are defined by
\begin{align*}
 S_i  &=  \frac{1}{ 2\pi i }\int_{A_{i}} R(z) dz,\qquad
 \Pi_i  =  2\int_{a_{i}^-}^{\Lambda_0} y(z) dz,\\
 \Pi_f^{(2)}  &=  \int_{\qt_f}^{\W\Lambda_0} y(z) dz, \\
 \Pi_{f}^{(1)} 
 &=  \int_{q_f}^{\W \Lambda_0} y(z) dz
 =\left[\int_{q_f}^{\W q_f} + \int_{\W q_f}^{\W\Lambda_0}\right] y(z) dz
 \equiv \Delta \Pi_f +\Pi_f^{(2)} 
\end{align*}
with $i=1,2$.  The periods $\Pi_f^{(1)}$, $\Pi_f^{(2)}$ are associated
with the poles on the first sheet and the ones on the second sheet,
respectively.  The contour $C_2$ for $\Pi_f^{(2)}$ is totally on the
second sheet, while the contour $C_1$ for $\Pi_f^{(1)}$ is from $q_f$
on the first sheet, through a cut, to $\W\Lambda_0$ on the second sheet.
These contours are shown in Fig.\ \ref{fig:contours}.
This $r$-branch with $K$ poles on the first sheet can be reached by
first starting from the pseudo-confining phase with all $N_f$ poles at
$\qt_f$ ($K=0$) and then moving $K$ poles through the cut to $q_f$.  The
path along which the poles are moved in this process is the difference
in the contours, $C_1-C_2\equiv \Delta C$ \footnote{There is ambiguity
in taking $\Delta C$; for example we can take $\Delta C$ to go around
$a_1^+$ in Fig.\ \ref{fig:contours}.  However, the difference in
$\int_{\Delta C} y(z)dz$ for such different choices of $\Delta C$ is
$2\pi i n S_1$, $n\in \mathbb{Z}$, which can be absorbed in redefinition
of the theta angle and is immaterial.}.

\begin{figure}[h]
\begin{center}
  \epsfxsize=7cm \epsfbox{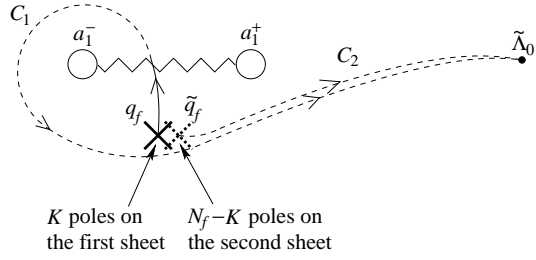}
\end{center}
\vspace{-.5cm}
\caption[Contours $C_1$ and $C_2$ defining $\Pi_f^{(1)}$ and
 $\Pi_f^{(2)}$]{\sl Contours $C_1$ and $C_2$ defining $\Pi_f^{(1)}$ and
 $\Pi_f^{(2)}$, respectively.  The part of a contour on the first sheet
 is drawn in a solid line, while the part on the second sheet is drawn
 in a dashed line.  The $N_f-K$ poles on the second sheet and the $K$
 poles on the first sheet are actually on top of each other (more
 precisely, their projections to the $z$-plane are.)}
 \label{fig:contours}
\end{figure}

When we impose the constraint \eqref{zf=a1-}, then the difference
$\Delta C$ vanishes (Fig.\ \ref{fig:contours_deg}).  Therefore there is
no distinction between $C_1$, $C_2$ and hence $\Pi_f^{(1)}=\Pi_f^{(2)}$
for any $K$.  In other words, all $K$-th branches collapse\footnote{In
fact this collapse was observed in \cite{Ahn:2003vh,Ahn:2003ui} for $SO(N_c)$ and
$USp(2N_c)$ gauge groups with massless flavors. We have seen that there
are only two branches, i.e., Special branch and Chebyshev branch, which
correspond to the baryonic branch and the non-baryonic branch in
$U(N_c)$ case.}  to the same branch under the constraint \eqref{zf=a1-}.

\begin{figure}[h]
\begin{center}
  \epsfxsize=7cm \epsfbox{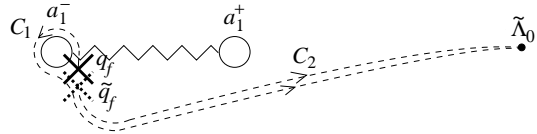}
\end{center}
 \vspace{-.5cm}
\caption[Under the constraint \eqref{zf=a1-}, contours $C_1$ and
 $C_2$ become degenerate: $C_1=C_2$]{\sl Under the constraint \eqref{zf=a1-}, contours $C_1$ and
 $C_2$ become degenerate: $C_1=C_2$.}  \label{fig:contours_deg}
\end{figure}

\bigskip Now, let us explicitly solve the factorization problem under
the constraint \eqref{zf=a1-}, and check that the $S=0$ solutions exist
as advertised before.  In solving the factorization problem, we do not
have to worry about the $r$-branches because there is no distinction
among them under the constraint \eqref{zf=a1-}.  Then, we compute the
exact superpotential using the data from the factorization and reproduce
it in matrix model by setting $S=0$ by hand when $S=0$ on the gauge
theory side.

For simplicity and definiteness, we consider the case of $U(3)$ gauge
group henceforth.  We consider $N_f<2N_c$ flavors, namely $1\le N_f\le 5$,
because $N_f\ge 2N_c$ cases are not asymptotically free and cannot be
treated in the framework of the Seiberg--Witten theory.

\subsection{Gauge theory computation of superpotential}
\label{QuantumMassless}

In this subsection, we solve the factorization equation under the
constraint \eqref{zf=a1-} and compute the exact superpotential, for the
system (\ref{hnhn8Apr04}) with $U(3)$ gauge group and with various
breaking patterns.  

\subsubsection{Setup}

The factorization equation for $U(3)$ theory with $N_f$ flavors with mass
$m_f=-z_f$ is given by\footnote{From the result of Appendix
(\ref{Nocorrection}), the matrix model curve with flavors does not
change even for $N_f > N_c$,
contrary to [7].}  \cite{Balasubramanian:2003tv}
\begin{align}
  \widetilde{P}_3(\W z)^2-4\La^{6-N_f}(\W z-z_f)^{N_f}
 &=  \W H_{1}(\W z)^2\left[W^{\prime}(\W z)^2+f_{1}(\W z) \right]\nonumber\\
 &= \W H_{1}(\W z)^2 (\W z-a_1^-)(\W z-a_1^+)(\W z-a_2^-)(\W z-a_2^+) ,\label{cvn11Apr04}
\end{align}
where we set $g=1$ for simplicity and $W^{\prime}(\W z)$ is given by
(\ref{hnhn8Apr04}).  The breaking pattern is assumed to be $U(3)\to
U(N_{c,1})\times U(N_{c,2})$ with $N_{c,i}>0$.
Here we used new notations to clarify the shift of the coordinate below.
For quantities after the shift, we use letters without tildes. Enforcing
the constraint \eqref{zf=a1-}
and shifting
$\zt$ as $\W z=z+a_1^-$, we can rewrite this relation as follows:
\begin{align}
 P_3(z)^2-4\La^{6-N_f}z^{N_f}&= 
 H_{1}(z)^2\left[ \W W^{\prime}(z)^2+ \W f_1(z)\right]\notag\\
 &=
 H_{1}(z)^2\left[z(z-\W a_1^+)(z-\W a_2^-)(z-\W a_2^+) \right]\nonumber\\
 &\equiv  H_{1}(z)^2 \left[z(z^3+Bz^2+Cz+D) \right].
 \label{factorization} 
\end{align}
Because of the shift, the polynomials $P_3(z)$ and $H_1(z)$ are
different in form from $\W P_3(\W z)$ and $\W H_1(\W z)$ in
(\ref{cvn11Apr04}).  We parametrize the polynomials $P_3(z)$ and
$H_1(z)$ as
\begin{eqnarray}
P_3(z)=z^3+az^2+bz+c,\qquad  H_1(z)=z-A. 
\label{p3h1}
\end{eqnarray}
The parameters $B, C$ can be written in terms of the parameters in
\bea
\W W^{\prime}(z)=  W^{\prime}(\W z)=\W z(\W z+m_A)=
(z+a_1^-)(z+a_1^-+m_A)\equiv (z-a_1) (z-a_2) 
\nonu
\eea
by comparing the coefficients in
(\ref{hpni8Apr04}):
\bea
 a_1&= &\frac{-B \mp {\sqrt{3\,B^2 - 8\,C}}}{4},\ \  
a_2=\frac{-B \pm {\sqrt{3\,B^2 - 8\,C}}}{4}, \nonumber \\
\Delta^2 &\equiv & (\W a_2-\W a_1)^2=(a_2-a_1)^2= \frac{{{3\,B^2 - 8\,C}}}{4}.
 \label{constraint}
\eea
The ambiguity in signs in front of the square roots can be fixed by
assuming $a_1<a_2$. Finally we undo the shift by noting that 
\bea
W(\W z) & = & \frac{1}{ 3} \W z^3+\frac{m_A}{ 2} \W z^2
 =  
 \frac{z^3}{ 3}-(a_1+a_2)\frac{z^2}{ 2}+a_1 a_2 z
 + \frac{1}{ 6}(a_1^3-3a_1^2 a_2). \label{field-form}
\eea

With all this setup, we can compute the superpotential as follows.
First we factorize the curve according to (\ref{factorization}).  Then
we find the Casimirs $U_1, U_2, U_3$ from $P_3(z)$ \footnote{From the
coefficients of $P_3(z)$, namely $a$, $b$ and $c$, one can compute the
Casimirs $U_k=\frac{1}{ k}\braket{\Tr\, \Phi^k}$ using the quantum
modified Newton relation Appendix (\ref{PN2}) as explained in Appendix
\ref{comp_spot-gt}.}  and solve for $a_1, a_2$ using the last equation
of (\ref{constraint}). Finally we put all these quantities into
(\ref{field-form}) to get the effective action as
\bea
 W_{\text{low}}
 =\braket{\Tr\, W(\Phi) }
 &= & 
 U_3 - (a_1+a_2)U_2 + a_1 a_2  U_1 
 + \frac{N_c}{ 6}(a_1^3-3a_1^2 a_2).
 \label{W_low_U(3)}
\eea
Here, for $a_{1}$, $a_2$, one can use the first two equations of
\eqref{constraint}.  We also want to know whether the first cut (the one
along the interval $[a_1^-,a_1^+]$) is closed or not; we expect that the
cut closes if we try to bring too many poles near the cut.  As is
obvious from \eqref{factorization}, this can be seen from the value of
$D$. If $D=0$, the cut is closed, while if $D\neq 0$, the cut is open.

\subsubsection{The result of factorization problem}

\begin{table}
\begin{center}\footnotesize
\renewcommand{\arraystretch}{1.2}
 \begin{tabular}{||c||@{\hspace{-0ex}}c@{\hspace{-0ex}}||c|c|c||@{\hspace{-1ex}}c@{\hspace{-1ex}}||} 
  \hline
  $N_f$  & 
  \begin{minipage}{1.25in}\begin{center}breaking pattern\\$\widehat{U(N_{c,1})}\times U(N_{c,2})$\end{center}\end{minipage}  & 
      $N_f\le N_{c,1}$ & $N_{c,1}<N_f<2N_{c,1}$ & $2N_{c,1}\le N_f$&
  \begin{minipage}{.75in}\begin{center}the first\\cut is\end{center}\end{minipage} \\  \hline\hline
 $1$ &  $\widehat{U(2)}\times U(1)$ & $\bigcirc$ & -          & -         & open    \\ \cline{2-6}
     &  $\widehat{U(1)}\times U(2)$ & $\bigcirc$ & -          & -         & open    \\ \hline
 $2$ &  $\widehat{U(1)}\times U(2)$ & -          & -          & $\bigcirc$& closed  \\ \cline{2-6}
     &  $\widehat{U(2)}\times U(1)$ & $\bigcirc$ & -          & -         & open    \\ \hline
 $3$ &  $\widehat{U(1)}\times U(2)$ & -          & -          & $\bigcirc$& closed  \\ \cline{2-6}
     &  $\widehat{U(2)}\times U(1)$ & -          & $\bigcirc$ & -         & closed  \\ \cline{2-6}
     &  $\widehat{U(2)}\times U(1)$ & -          & $\bigcirc$ & -         & open    \\ \hline
 $4$ &  $\widehat{U(1)}\times U(2)$ & -          & -          & $\bigcirc$& closed  \\ \cline{2-6}
     &  $\widehat{U(2)}\times U(1)$ & -          & -          & $\bigcirc$& closed  \\ \hline
 $5$ &  $\widehat{U(1)}\times U(2)$ & -          & -          & $\bigcirc$& closed  \\ \cline{2-6}
     &  $\widehat{U(2)}\times U(1)$ & -          & -          & $\bigcirc$& closed  \\ \hline
\end{tabular} 
\begin{center}
\caption[The result of factorization of curves for $U(3)$ with up to
$N_f=5$ flavors]{\sl The result of factorization of curves for $U(3)$ with up to
$N_f=5$ flavors.  ``$\,\bigcirc$'' denotes which inequality $N_f$ and $N_{c,1}$ satisfy.
}
\end{center}
\end{center}
\end{table}

We explicitly solved the factorization problem for $U(3)$ gauge theories
with $N_f=1,2,\dots, 5$ and summarized the result in Table 1.  Let us
explain about the table.  
$\widehat{U(N_{c,1})}\times U(N_{c,2})$ denotes the breaking pattern of the
$U(3)$ gauge group.  The hat on the first factor means that the pole is
at one of the branch points of the first cut (Eq.\ \eqref{zf=a1-}) which
is associated with the first factor $U(N_{c,1})$. 
Of course this choice is arbitrary and
we may as well choose $U(N_{c,2})$, ending up with the same result.
Finally, whether the cut is closed or not depends on whether $D=0$ or
not, as explained below \eqref{W_low_U(3)}.

%
%
%

Now let us look carefully at Table 1, comparing it with the prescription
\eqref{prescription} based on the analysis of the one cut model.

First of all, for $N_f\ge 2N_{c,1}$, the first cut is always closed.  We
will see below that in these cases with a closed cut the superpotential
can be reproduced by setting $S_1=0$ by hand in the corresponding matrix
model, confirming the prescription \eqref{prescription} for
$N_f>2N_{c,1}$.


Secondly, the lines for $N_f=3$ and $\widehat{U(2)}\times U(1)$
correspond to the $N_{c,1}<N_f<2N_{c,1}$ part of the prescription
\eqref{prescription}.  There indeed are both an open cut solution and a
closed cut solution.
We will see below that the superpotential of the closed cut solution can
be reproduced by setting $S_1=0$ by hand in the corresponding matrix
model.  On the other hand, the superpotential of the open cut solution
can be reproduced by not setting $S_1=0$, namely by treating $S_1$ a
dynamical variable and extremizing $W_{\text{eff}}$ with respect to it.
In fact these two solutions are baryonic branch for a closed cut and
non-baryonic branch for an open cut.

Finally, for $N_f\le N_{c,1}$, the cut is always open, which is also
consistent with the prescription \eqref{prescription}.  In this case,
the superpotential of the open cut solution can be reproduced by
extremizing $W_{\text{eff}}$ with respect to it, as we will see below.
For $N_f=N_{c,1}$ there should be an $S_1=0$ ($a_1^-=a_1^+$) solution
for some $z_f$ (corresponding to $U(2)$ theory with $N_f=2$ in the $r=0$
branch) in the quadratic case, but under the constraint \eqref{zf=a1-}
we cannot obtain that solution.

%
%
%


\bigskip
Below we present  resulting exact superpotentials, for all possible
breaking patterns.  For simplicity, we do not take care of phase factor
of $\La$ which gives rise to the whole number of vacua.  For details of
the calculation, see Appendix \ref{comp_spot-gt}.

\subsubsection*{Results}

Definitions: $ W_{\text{cl}}= -\frac{1}{3}$ for $\widehat{U(1)}\times
U(2)$ and $ W_{\text{cl}}= -\frac{1}{6}$ for $\widehat{U(2)}\times
U(1)$.  For simplicity we set $g=1$ and $\Delta=a_2-a_1=-m/g=1$.

\begin{itemize}
 \item $\boldsymbol{\widehat{U(1)}\times U(2)}$ {\bf with } $\boldsymbol{N_f=1}$
       \begin{align*}
        W_{\text{low}}
         &= W_{\text{cl}} - 2T - \frac{5T^2}{2} + 
         \frac{115 T^3}{12}\\ 
	&\qquad\qquad
	- \frac{245T^4}{4} 
	+ 
         \frac{30501 T^5}{64} - \frac{12349 T^6}{3}+\cdots,
         \quad 
         T\equiv  \La^{\frac{5}{2}}.
          \nonu
       \end{align*}

 \item  $\boldsymbol{\widehat{U(2)}\times U(1)}$ {\bf with } $\boldsymbol{N_f=1}$ 
       \begin{multline*}
        W_{\text{low}}
         = W_{\text{cl}} - \frac{5T^2}{2} + \frac{5T^3}{3} - 
         \frac{11T^4}{3} + 11T^5 - \frac{235T^6}{6}+\cdots,
         \quad T\equiv \Lambda ^{\frac{5}{3}}.  \nonu
       \end{multline*}

 \item $\boldsymbol{\widehat{U(1)}\times U(2)}$ {\bf with } $\boldsymbol{N_f=2}$ 
       \begin{align*}
        W_{\text{low}}
         &= W_{\text{cl}} + 2T^2 - 6T^4 - \frac{32T^6}{3} \\
	&\qquad\qquad
	- 
         40T^8 - 192T^{10} - \frac{3136T^{12}}{3}+\cdots,
         \quad T \equiv  \La. \nonu
       \end{align*}

 \item $\boldsymbol{\widehat{U(2)}\times U(1)}$ {\bf with } $\boldsymbol{N_f=2}$
       \begin{multline*}
        W_{\text{low}}
         = W_{\text{cl}} - 2T^4 - \frac{16T^6}{3} - 24T^8 - 
         128T^{10} - \frac{2240T^{12}}{3}+\cdots,\quad 
         T\equiv  \Lambda.  \nonu
       \end{multline*}

 \item  $\boldsymbol{\widehat{U(1)}\times U(2)}$ {\bf with } $\boldsymbol{N_f=3}$ 
       \begin{align*}
        W_{\text{low}}
         &= W_{\text{cl}} + 2T - \frac{19T^2}{2} + 
         \frac{51T^3}{4} \\
	&\qquad\qquad
	+ \frac{157T^4}{4} + 
         \frac{5619T^5}{64} + \frac{33T^6}{2}+\cdots,\quad 
         T\equiv  \Lambda^{\frac{3}{2}}.  \nonu
       \end{align*}

 \item  $\boldsymbol{\widehat{U(2)}\times U(1)}$ {\bf with } $\boldsymbol{N_f=3}${\bf : two solutions}
       \begin{align*}
        W_{\text{low,baryonic}}
         &=W_{\text{cl}} + T - \frac{5T^2}{2} - 33T^3 \\
	&\qquad\qquad
	- 
         543T^4 - 10019T^5 - \frac{396591T^6}{2}+\cdots,\quad
         T\equiv  \La^3.  \nonu
         \\
        W_{\text{low}}
         &= W_{\text{cl}} + {\Lambda }^3.\nonu
       \end{align*}

 \item  $\boldsymbol{\widehat{U(1)}\times U(2)}$ {\bf with } $\boldsymbol{N_f=4}$ 
       \begin{align*}
        W_{\text{low}}
         &= W_{\text{cl}} + 2T - 13T^2 + \frac{176T^3}{3} - 
         138T^4 + 792T^6 \nonu \\
	&\qquad\qquad- 9288T^8 + 
         137376T^{10} 
        - 2286144T^{12}+\cdots,\quad
         T\equiv  \Lambda.  \nonu
       \end{align*}

 \item $\boldsymbol{\widehat{U(2)}\times U(1)}$ {\bf with } $\boldsymbol{N_f=4}$ 
       \begin{align*}
        W_{\text{low}}
         &= W_{\text{cl}} + T - 6T^2 - \frac{40T^3}{3} - 
         56T^4 - 288T^5 \\
	&\qquad\qquad 
	- \frac{4928T^6}{3} -9984T^7 - 63360T^8  - \frac{1244672T^9}{3} \\
	&\qquad\qquad
	-          2782208T^{10} - 19009536T^{11}\cdots,\quad
         T\equiv \Lambda^2 . 
       \end{align*}

 \item  $\boldsymbol{\widehat{U(1)}\times U(2)}$ {\bf with } $\boldsymbol{N_f=5}$ 
       \begin{align*}
        W_{\text{low}}
         & = W_{\text{cl}} - 2T - \frac{33T^2}{2}  - 
         \frac{1525T^3}{12}\\&\qquad\qquad
	- \frac{3387T^4}{4} 
	- \frac{314955T^5}{64} 
          - \frac{74767T^6}{3}+\cdots,
         \quad T\equiv \Lambda ^{\frac{1}{2}}. \nonu
       \end{align*}
 \item $\boldsymbol{\widehat{U(2)}\times U(1)}$ {\bf with } $\boldsymbol{N_f=5}$ 
\begin{eqnarray}
W_{\text{low}}
= W_{\text{cl}}+ T - \frac{19T^2}{2} + 
  \frac{154T^3}{3} - 132T^4 + 828T^6+\cdots,\quad 
  T\equiv \Lambda.  \nonu
\end{eqnarray}
\end{itemize}

\subsection{Matrix model computation of superpotential}

In this subsection we compute the superpotential of the system
\eqref{hnhn8Apr04} in the framework of \cite{Cachazo:2003yc}.  If all the $N_f$
flavors have the same mass $m_f=-z_f$, the effective glueball
superpotential $W_{\text{eff}}(S_j)$ for the pseudo-confining phase with
breaking pattern $U(N_c)\to \prod_{i=1}^n U(N_{c,i})$, $\sum_{i=1}^n
N_{c,i}=N_c$ is, from \eqref{Wexact-y},
\begin{align}
 W_{\text{eff}}(S_j)
 &=
 -\frac{1}{ 2}\sum_{i=1}^n N_{c,i} \Pi_i - \frac{N_f}{ 2} \Pi_f
 +\left(N_c-\frac{N_f}{ 2}\right)W(\Lambda_0) 
 +\frac{N_f}{ 2}W(z_f)
 \cr
 &\qquad
 -2\pi i\left(N_c-\frac{N_f}{ 2}\right)S
 +2\pi i\tau_0 S
 +2\pi i\sum_{i=1}^{n-1} b_i S_i,\label{jziq11Apr04}
\end{align}
where the periods associated with adjoint and fundamentals are defined
by
\begin{gather*}
 \Pi_i(S_j)\equiv 2 \int_{a_i^-}^{\Lambda_0} y(z) dz,\qquad
 \Pi_f(S_j)\equiv \int_{\W z_f}^{\widetilde\Lambda_0} y(z) dz 
  = -  \int_{z_f}^{\Lambda_0} y(z) dz,\\
 y(z)=\sqrt{W'(z)^2+f_1(z)}~.
\end{gather*}

For cubic tree level superpotential \eqref{hnhn8Apr04},
the periods $\Pi_{1,2}(S_j)$ were computed by explicitly evaluating the
period integrals by power expansion in \cite{Cachazo:2001jy}, as
\begin{align}
 \frac{\Pi_1}{ 2g\Delta^3}=\,&
 \frac{1}{ g\Delta^3}[W(\Lambda_0)-W(a_1)]
 +s_1\left[1+\log\left(\frac{\lambda_0^2}{s_1}\right)\right]
 +2s_2\log \lambda_0\cr
 &+(
 -2s_1^2
 +10s_1s_2
 -5s_2^2
 )
 +\left( 
   -\frac{32}{ 3} s_1^3
   +91 s_1^2 s_2
   -118 s_1 s_2^2
   + \frac{91}{ 3}s_2^3
 \right)
 \cr
 &+\left( 
   -\frac{280}{3}s_1^4
   +\frac{3484}{3}s_1^3s_2
   - 2636s_1^2 s_2^2
   + \frac{5272}{3}s_1 s_2^3
   - \frac{871}{3}s_2^4 
 \right)
 +\cdots,
 \cr
 \frac{\Pi_2}{ 2g\Delta^3}=\,&
 \frac{1}{ g\Delta^3}[W(\Lambda_0)-W(a_2)]
 +s_2\left[1+\log\left(\frac{\lambda_0^2}{-s_2}\right)\right]
 +2s_1\log \lambda_0\notag\\
 &
 +(2s_2^2-10s_1s_2+5s_1^2)
 +\left(
 -\frac{32}{ 3}s_2^3
 +91s_1s_2^2
 -118s_1^2s_2
 +\frac{91}{ 3}s_1^3
 \right)
 \notag\\
 &
 +\left(
 \frac{280}{3}s_2^4
 -\frac{3484}{3}s_1s_2^3
 +2636s_1^2s_2^2
 -\frac{5272}{3}s_1^3s_2
 +\frac{871}{3}s_1^4
 \right)
 +\cdots,\label{jwhq8Apr04}
\end{align}
where $\Delta\equiv a_2-a_1$, $s_i\equiv S_i/g\Delta^3$, and
$\lambda_0\equiv \Lambda_0/\Delta$.


Under the constraint \eqref{zf=a1-}, the contours defining $\Pi_1$ and
$\Pi_f$ coincide, so
\begin{align}
 \frac{1}{ 2}\Pi_1=
-\Pi_f=\int_{z_f=a_1^-}^{\Lambda_0}y(z)dz.\label{mqes8Apr04}
\end{align} 
Using this, we can rewrite \eqref{jziq11Apr04} as
\begin{align}
 W_{\text{eff}}(S_1,S_2)
 =\,&
 \Bigl[N_{c,1} W(a_1)+N_{c,2} W(a_2)\Bigr]
  -\frac{N_f}{ 2} \Bigl[W(a_1)-W(z_f)\Bigr]
 \cr 
 &
 -{\Nt_{c,1}} \Bigl[\half\Pi_1-W(\Lambda_0)+W(a_1)\Bigr]
 -N_{c,2} \Bigl[\half\Pi_2-W(\Lambda_0)+W(a_2)\Bigr]
 \cr
 &
 -2\pi i(\Nt_{c,1}+N_{c,2})S+2\pi i\tau_0 S  +2\pi ib_1 S_1.
 \label{juho8Apr04}
\end{align}
Here we rearranged the terms taking into account the fact that the
periods take the form $\half\Pi_i=W(\Lambda_0)-W(a_i)+\text{(quantum
correction of order $\CO(S_i)$)}$, and also the fact that we are
considering $z_f=a_1^-\simeq a_1$ (thus the second term).  The first line
corresponds to the classical contribution, while the second and third
lines correspond to quantum correction.  Furthermore, we defined
$\Nt_{c,1}\equiv N_{c,1}-N_f/2$.

We would like to extremize this $W_{\text{eff}}$ \eqref{juho8Apr04} with
respect to $S_{1,2}$, and compute the low energy superpotential that can
be compared with the $W_{\text{low}}$ obtained in the previous subsection
using gauge theory methods.
In doing that, one should be careful to the fact that one should treat
the mass $z_f$ as an external parameter which is independent of
$S_{1,2}$ although we are imposing the constraint \eqref{zf=a1-},
$z_f=a_1^-=a_1^{-}(S_1,S_2)$.  Where is the $z_f$ dependence in
\eqref{juho8Apr04}?  Firstly, $z_f$ appears explicitly in the second
term in \eqref{juho8Apr04}.  Therefore, when we differentiate $W_\text{eff}$ with respect to $S_{1,2}$, we should exclude this term.  Secondly,
there is a more implicit dependence on $z_f$ in $\Pi_f =
-\int_{z_f}^{\Lambda_0} y(z)dz$, which we replaced with $-\Pi_1/2$ 
using
\eqref{mqes8Apr04}.  If we forget to treat $z_f$ as independent of
$S_i$, then we get an apparently unwanted, extra contribution as
$\frac{\partial \Pi_f }{ \partial S_i} = -\int_{z_f}^{\Lambda_0}
\frac{\partial y(z) }{\partial S_i} dz -\frac{\partial z_f}{ \partial
S_i}\cdot y(z)|_{z=z_f} $.  However, this last term actually does not
make difference because
\begin{align*}
y(z)|_{z=z_f}=g\left.\sqrt{(z-z_f)(z-a_1^+)(z-a_2^-)(z-a_2^+)}\right|_{z=z_f}=0.
\end{align*}
Therefore what one should do is: i) plug the expression
\eqref{jwhq8Apr04} into \eqref{juho8Apr04}, ii) solve the equation of
motion for $S_{1,2}$ using \eqref{juho8Apr04} without the second term,
and then iii) substitute back the value of $S_{1,2}$ into
\eqref{juho8Apr04}, now with the second term included.

Solving the equation of motion can be done by first writing the
Veneziano--Yankielowicz term (log and linear terms) as
\begin{align*}
 & -{\Nt_{c,1}} \Bigl[\half\Pi_1-W(\Lambda_0)+W(a_1)\Bigr]
 -N_{c,2} \Bigl[\half\Pi_2-W(\Lambda_0)+W(a_2)\Bigr]
 \\
 &\qquad\qquad\qquad\qquad\qquad\qquad
 -2\pi i(\Nt_{c,1}+N_{c,2})S+2\pi i\tau_0 S  +2\pi ib_1 S_1
  \\
 &\qquad\qquad
 = 
 g\Delta^3\biggl\{ \Nt_{c,1} s_1 [ 1-\log(s_1/\lambda_1^3)] +  N_{c,2} s_2 [ 1-\log(s_2/\lambda_2^3)] 
 + \CO(s_i^2)
 \biggr\},
\end{align*}
where
\bea
 \lambda_1^{3\Nt_{c,1}} 
 & =& 
\lambda_0^{2(\Nt_{c,1}+2N_{c,2})} e^{2\pi i (\Nt_{c,1}+N_{c,2})-2\pi i \tau_0-2\pi i b_1},\nonu \\
 \lambda_2^{3N_{c,2}} & = & 
(-1)^{N_{c,2}}\lambda_0^{2\Nt_{c,1}+2N_{c,2}}
 e^{2\pi i (\Nt_{c,1}+N_{c,2})-2\pi i \tau_0},
\nonu
\eea
and then solving the equation of motion perturbatively in
$\lambda_{1,2}$.
In this way, one can straightforwardly reproduce the results obtained in
the previous section in the case with the first cut {\em open\/}.  In the
case with the first cut {\em closed\/}, in order to reproduce the results
in the previous section, one should first set $S_1=0$ by hand, and then
extremize $W_{\text{eff}}$ with respect to the remaining dynamical variable
$S_2$.
%

Following the procedure above, we checked explicitly that extremizing\break
$W_{\text{eff}}(S_1,S_2)$ (open cut) or $W_{\text{eff}}(S_1\!=\!0,S_2)$
(closed cut) reproduces the $W_{\text{low}}$ up the order presented in
the previous section, for all breaking patterns for $U(3)$ theory.

\bigskip 
In the above, we concentrated the explicit calculations of effective
superpotentials in $U(3)$ theory with cubic tree level superpotential.
These explicit examples are useful to see that the prescription
\eqref{prescription} really works; one can first determine using
factorization method when we should set $S_i=0$ by hand, and then
explicitly check that the superpotential obtained by gauge theory can be
reproduced by matrix model.

However, if one wants only to show the equality of the two effective
superpotentials on the gauge theory and matrix model sides, one can
actually prove it in general cases.  In Appendix \ref{CVmethod}, we
prove this equivalence for $U(N_c)$ gauge theory with an degree $(k+1)$
tree level superpotential where $k+1<N_c$.  There, we show the
following: if there are solutions to the factorization problem with some
cuts closed, then the superpotential $W_{\text{low}}$ of the gauge theory
can be reproduced by extremizing the glueball superpotential $W_{\text{eff}}(S_i)$ on the matrix model side, after setting the corresponding
glueball fields $S_i$ to zero by hand.  Note that, on the matrix side we
do not know when we should set $S_i$ to zero {\it a priori\/}; we can
always set $S_i$ to zero in matrix model, but that does not necessarily
correspond to a physical solution on the gauge theory side that solves
the factorization constraint.

\section{Conclusion and some remarks}
\setcounter{equation}{0}
\label{sec:conclusion}

In this paper, taking $\CN=1$ $U(N_c)$ gauge theory with an adjoint and
flavors, we studied the on-shell process of passing $N_f$ flavor poles
on top of each other on the second sheet through a cut onto the first
sheet.  This corresponds to a continuous transition from the
pseudo-confining phase with $U(N_c)$ unbroken to the Higgs phase with
$U(N_c-N_f)$ unbroken (we are focusing on one cut).  We confirmed the
conjecture of \cite{Cachazo:2003yc} that for $N_f<N_c$ the poles can go all the
way to infinity on the first sheet, while for $N_f\ge N_c$ there is
obstruction.  There are two types of obstructions: the first one is that
the cut rotates, catches  poles and sends them back to the first
sheet, while the second one is that the cut closes up before  poles
reach it.  The first obstruction occurs for $N_c\le N_f<2N_c$ whereas
the second one occurs for $N_c<N_f $.

If a cut closes up, the corresponding glueball $S$ vanishes, which means
that the $U(N_c)$ group is completely broken down.  This can happen only
by condensation of a charged massless degree of freedom, which is
missing in the matrix model description of the system.  With a massless
degree of freedom missing in the description, the $S=0$ solution should
be singular in matrix model in some sense.  Indeed, we found that the
$S=0$ solution of the gauge theory does not satisfy the equation of
motion in matrix model (with an exception of the $N_f=N_c$ case, where
the $S=0$ solution does satisfy the equation of motion).  How to cure
this defect of matrix model is simple --- the only thing the missing
massless degree of freedom does is to make $S=0$ a solution, so we just
set $S=0$ by hand in matrix model.  We gave a precise prescription
\eqref{prescription} when we should do this, i.e.,
{\em in the baryonic branch for $N_{c,i}\leq N_f< 2N_{c,i}$ and in the
$r=N_{c,i}$ non-baryonic branch\/}, and checked it with specific
examples.

The string theory origin of the massless degree of freedom can be
conjectured by generalizing the argument in \cite{Intriligator:2003xs}.  We argued
that it should be the D3-brane wrapping the blown up $S^3$, along with
fundamental strings emanating from it and ending on the noncompact
D5-branes in the Calabi--Yau geometry.

Although we checked that the prescription works, the string theory
picture of the $S=0$ solution needs further refinement, which we leave
for future research.
For example, although we argued that some extra degree of freedom makes
$S=0$ a solution, we do not have the precise form of the superpotential
including that extra field.  It is desirable to derive it and show that
$S=0$ is indeed a solution, as was done in \cite{Intriligator:2003xs} in the case
without flavors.
%
%
Furthermore, we saw that there is an on-shell $S=0$ solution for
$N_f=N_c$.  Although this solution solves the equation of motion in
matrix model, there should be a massless field behind the scene.  It is
interesting to look for the nature of this degree of freedom.  It cannot
be the D3-branes with fundamental strings emanating from it,
since for this solution the noncompact D5-branes are at finite distance
from the collapsed $S^3$ and the 3-5 strings are massive.
Finally, we found that the $S=0$ solution is in the baryonic branch.  It
would be interesting to ask if one can describe the baryonic branch in
the matrix model framework by adding some extra degrees of freedom.

\bigskip
In the following, we study some aspects of the theory, which we could
not discuss so far.  
We will discuss generalization to $SO(N_c)$ and $USp(2N_c)$ gauge groups
by computing the effective superpotentials with quadratic tree level
superpotential.

\subsection{$SO(N_c)$ theory with flavors}
\setcounter{equation}{0}

Here we consider the one cut model for $SO(N_c)$ gauge theory with
$N_f$ flavors.
The tree level superpotential of the theory 
is obtained from ${\mathcal N}=2$ SQCD  by adding the mass 
$m_A$ for the adjoint scalar $\Phi$
\bea
W_{\text{tree}} = \frac{m_A}{2} \Tr\, \Phi^2 + Q^{f} \Phi Q^{f'}
J_{ff'} + 
Q^f \widetilde{m}_{ff'} Q^{f'}.
\label{tree}
\eea
where $f=1, 2, \cdots, 2N_f$ and the 
symplectic metric $J_{ff'}$ and 
mass matrix for quark $\widetilde{m}_{ff'}$
are given by 
\bea
J=\left(\begin{array}{cc} 0 & 1 \\ -1 & 0 \end{array} \right)
\otimes {\unit}_{N_f\times N_f}, \qquad
\widetilde{m}=\left(\begin{array}{cc} 0 & 1 \\ 1 & 0 \end{array} 
\right)
\otimes \mbox{diag} (m_1, \cdots, m_{N_f})~.
\nonu
\eea 
For this simple case the matrix model curve is given by
\bea
y(z)^2= m_A^2 \left( z^2 - 4 \mu^2 \right).  
\label{soy}
\eea
This Riemann surface  is a double cover of the 
complex $z$-plane branched at the roots of $y_m^2$ (that is 
$z=\pm 2 \mu$).

The effective superpotential receives 
contributions from both the sphere and the disk amplitudes 
in the matrix model \cite{Seiberg:2002jq}  
and the explicit form was given in \cite{Cachazo:2003yc} for $U(N_c)$
gauge theory with flavors. 
Now we apply this procedure to our $SO(N_c)$ gauge theory
with flavors and it turns out the following expression  
\begin{align*}
 W_{\text{eff}} & =  -\frac{1}{2} \left( \sum_{i=-n}^{n} N_{c,i} -
 2 \right) \int_{\widehat{B}_i^r} 
 y(z) dz -\frac{1}{4}
 \sum_{I=1}^{2N_f}
 \int_{\widetilde{q}_I}^{\widetilde{\La}_0} 
 y(z) dz \\
 &\qquad\qquad
 + \frac{1}{2} \left(2N_c-4-2N_f \right) W(\La_0)
  + \frac{1}{2} \sum_{I=1}^{2N_f} W(z_I) \\
 &\qquad\qquad
 - \pi i \left(  2N_c-4-2N_f \right) S + 2\pi i \tau_0 S + 2\pi i 
 \sum_{i=1}^{n} b_i S_i
\end{align*}
where
$S=S_0 + 2 \sum_{i=1}^{n} S_i $ and $z_I$ is the root of
\bea
B(z) = \mbox{det} \; m(z)=
\prod_{I=1}^{N_f}\left(z^2-z_I^2 \right)~.
\nonu
\eea

Since the curve (\ref{soy}) is same as the one 
(\ref{matrix-curve}) of $U(N_c)$ gauge
theory, we can use the integral results given there to
write down the effective superpotential as
\begin{align*}
 &W_{\text{eff}} = 
 S  \left[  \frac{(N_c-2)}{2}  + \log 
 \left( \frac{2^{\frac{N_c-2}{2}} m_A^{\frac{N_c-2}{2}} \La^{N_c-2-N_f}
    \mbox{det} z}{S^{\frac{N_c-2}{2}}} \right)
 \right]
 \nonu \\
 & -S \sum_{I=1, r_I=0}^{N_f} \left[- \log 
 \left( \frac{1}{2} + \frac{1}{2} \sqrt{1-
 \frac{2S}{m_A z_I^2}} \right) + \frac{m_A z_I^2}{2S} \left( \sqrt{1-
 \frac{2S}{m_A z_I^2}} -1 \right) + \frac{1}{2} \right] 
 \nonu \\
 & -S \sum_{I=1, r_I=1}^{N_f} \left[- \log 
 \left( \frac{1}{2} - \frac{1}{2} \sqrt{1-
 \frac{2S}{m_A z_I^2}} \right) + \frac{m_A z_I^2}{2S} \left( -\sqrt{1-
 \frac{2S}{m_A z_I^2}} -1 \right) + \frac{1}{2} \right] .
 \nonu 
\end{align*}

We can solve $M(z)$ and $T(z)$ as did for $U(N_c)$ gauge theory.
For simplicity we take all $r_I=0$, i.e., all poles at the 
second sheet. 
For the $I$-th block diagonal matrix element of $M(z)$
($I=1, \cdots, N_f$)  it is given by  
\bea
M_I(z) 
& =&  \left(\begin{array}{cc} 0 
& -\frac{R(z) -R(q_I=m_I)}{z-m_I} \\ \frac{R(z) - R(q_I=-m_I)}{z+m_I} & 0 
\end{array} \right)
\nonu 
\eea
where 
\bea
R(z) =m_A  \left(z -  \sqrt{z^2- 4\mu^2} \right). 
\nonu
\eea
Expanding $M_I(z)$ in the series of $z$ we can find 
$
\braket  { Q^f \Phi Q^{f'}
J_{ff'} + 
Q^f \widetilde{m}_{ff'} Q^{f'}} = 2 N_f S$.

The gauge invariant operator $T(z)$ can be constructed similarly
as follows:
\bea
T(z) = \frac{B^{\prime}(z)}{2B(z)}-\sum_{I=1}^{N_f} 
\frac{y(q_I) z_I }{ y(z) \left(z^2-z_I^2 \right)} +
\frac{c(z)}{y(z)} -\frac{2}{z} 
\frac{R(z)}{y(z)}
\label{sotz}
\eea
where
\bea
c(z) = \left< \Tr \frac{W^{\prime}(z) -W^{\prime}(\Phi)}{z-\Phi} 
\right> - \sum_{I=1}^{N_f} \frac{z W^{\prime}(z)-z_I W^{\prime}(z_I)}{
\left(z^2-z_I^2\right)}.
\nonu
\eea
For the theory without the quarks, 
the Konishi anomaly was derived in \cite{Alday:2003dk,Kraus:2003jv,Ahn:2003vu}.
The last term in (\ref{sotz}) reflects the action of orientifold.
For our example we have 
\bea
c(z)= m_A \left( N_c-N_f \right). 
\nonu
\eea
and 
\bean
T(z) & = &  
\frac{1}{z} N_c +
\frac{1}{z^3}
\left[ \sum_{I=1}^{N_f}  z_I \left( z_I-  
\sqrt{z_I^2 - 4 \mu^2} \right)
+   2\mu^2 \left( N_c-2-N_f \right)   \right] \nonu \\
& & +
\frac{1}{z^5}
\left[ \sum_{I=1}^{N_f} z_I^4 -\sum_{I=1}^{N_f} \sqrt{z_I^2-4\mu^2}
  z_I 
\left( z_I^2 + 2\mu^2 \right) +
6\mu^4 \left( N_c-N_f \right) -12 \mu^4  \right] \\
  &&+ {\mathcal O} \left(1/z^7 \right),
\nonu 
\eean
where for equal mass of flavor, we get
\begin{align*}
 \ev{\Tr\, \Phi^2 } &= N_f  q \left( q-  \sqrt{q^2 - 4 \mu^2} \right)
 +   2\mu^2 \left( N_c-2-N_f \right)  \\
 \ev{\Tr\, \Phi^4 } &= N_f q^4 -N_f \sqrt{q^2-4\mu^2} q 
 \left( q^2 + 2\mu^2 \right) +
 6\mu^4 \left( N_c-2-N_f \right).
\end{align*}


Let us assume the mass of flavors are the same and $K$ of them
(in this case, $r_I=1$)
locate at the first sheet while the remainder $(N_f-K)$ where
$r_I=0$ are
at the second sheet. Then from the effective superpotential,  
it is ready to extremize this with respect to 
the glueball field $S$ 
\footnote{
One can easily check that this equation with parameters 
$(N_c, N_f, K)$ is equivalent to the one with 
parameters $(N_c-2r, N_f-2r, K-r)$. The equation of 
motion for glueball field is the same.
Since the equation of motion for both $r$-th Higgs branch and
$\left(N_f-r\right)$-th Higgs branch is the same, one expects that
both branches have some relation.
By redefinition of
$
S \rightarrow \frac{4m_A^2 \La^4}{S} \equiv \widetilde{S}, 
z_f \rightarrow \frac{2m_A
  \La^2}{S} z_f \equiv \widetilde{z_f}$
we get the final relation between $K$ Higgs branch and 
$\left(N_f-K \right)$ 
Higgs branch.}
\begin{align}
 0&= \log \left( \frac{\La_1^{\frac{3}{2}(N_c-2)-N_f}}
 {S^{\frac{N_c-2}{2}}} \right) \nonu\\
 &\quad
 + K \log\left(\frac{z_f-\sqrt{{z_f}^2-\frac{2S}{m_A}}}{2} \right) +
 \left(N_f-K \right) \log \left( \frac{z_f+\sqrt{{z_f}^2-\frac{2S}{m_A}}}{2} \right).
 \label{minimum}
\end{align}
Or by rescaling the fields 
$\Sh = \frac{S}{2m_A \La^2}$, $\zh_f = \frac{z_f}{\La} $
one gets the solution and consider for $K=0$ case  
\bea
1= \Sh^{-\frac{N_c-2}{2}} 
\left(  \frac{\zh_f+ \sqrt{\zh_f^2-4\Sh}}{2}  \right)^{N_f}.
\nonu
\eea
This equation is the same as the one 
in $U(N_c)$ case with $N_c\to \frac{N_c-2}{ 2}$, 
so the discussion of passing poles will go through without
modification and the result is  when $N_f\geq \frac{N_c-2}{ 2}$,
the on-shell 
poles at the second plane cannot pass the cut to
reach the first sheet far away from the cut.

By using the condition (\ref{minimum}), one gets the on-shell 
effective
superpotential
\begin{align*}
 W_{\text{eff,on-shell}}  &=     \frac{1}{2} \left(N_c-2-N_f \right) S  \\
 &\quad
 + \frac{1}{2}  m_A {z_f}^2 \left( N_f  + \left( 2 K-N_f \right) 
 \sqrt{1- \frac{2S}{m_A {z_f}^2}}  \right).  
\end{align*}
It can be checked that this is the same as 
$ \frac{1}{2} m_A \left< \Tr\,  \Phi^2 \right>$.

\subsection{$USp(2N_c)$ theory with flavors}

For the $USp(2N_c)$ gauge theory with $N_f$ flavors we will sketch
the discussion because most of them 
is similar to $SO(N_c)$ gauge theory.
The tree level superpotential is given by (\ref{tree}) but with
 $J$ the symplectic metric, and $\widetilde{m}_{ff'}$
 the quark mass  given by 
\bea
J=\begin{pmatrix} 0 & 1 \\ -1 & 0 \end{pmatrix} 
\otimes {\unit}_{N_c \times N_c}, \qquad
\widetilde{m}=\begin{pmatrix} 0 & -1 \\ 1 & 0 \end{pmatrix} 
\otimes \mbox{diag} (m_1, \cdots, m_{N_f})~.
\nonu
\eea 
We parametrize   the matrix model curve and the resolvent $R(z)$ 
as
\bea
y_m^2 & = &  W^{\prime}(z)^2 + f(z) = m_A^2 \left( z^2 + 4 \mu^2 
\right), 
\nonu \\
R(z)& = & m_A  \left(z -  \sqrt{z^2+ 4\mu^2} \right). 
\nonu
\eea
The effective superpotential is given by
\begin{align*}
 &W_{\text{eff}}  =  S   \left(N_c+1 \right) \left[ 1 + \log 
 \left( \frac{\tilde{\La}^3}{S} \right)
 \right] \nonu \\
 & -S \sum_{I=1, r_I=0}^{N_f} \left[- \log 
 \left( \frac{1}{2} + \frac{1}{2} \sqrt{1+
 \frac{2S}{m_A z_I^2}} \right) - \frac{m_A z_I^2}{2S} \left( \sqrt{1+
 \frac{2S}{m_A z_I^2}} -1 \right) + \frac{1}{2} \right] 
 \nonu \\
 & -S \sum_{I=1, r_I=1}^{N_f} \left[- \log 
 \left( \frac{1}{2} - \frac{1}{2} \sqrt{1+
 \frac{2S}{m_A z_I^2}} \right) - \frac{m_A z_I^2}{2S} \left( -\sqrt{1+
 \frac{2S}{m_A z_I^2}} -1 \right) + \frac{1}{2} \right] 
\nonu 
\end{align*}
and $T(z)$ is given by 
\bea
T(z) = \frac{B^{\prime}(z)}{2B(z)}-\sum_{I=1}^{N_f} 
\frac{y(q_I) z_I }{ y(z) \left(z^2-z_I^2 \right)} +
\frac{c(z)}{y(z)} +\frac{2}{z} 
\frac{R(z)}{y(z)}.
\nonu
\eea
Note the last term (different sign) compared with the $SO(N_c)$
gauge theory. From the solution of $M(z)$, we can show that
although
$$
\left<  Q^f (\Phi J)_{ff'} Q^{f'}
+ 
Q^f \widetilde{m}_{ff'} Q^{f'} J 
 \right> = 2 N_f S\neq 0,$$ we still have
on-shell relation  
$ \frac{1}{2} m_A \left< \Tr\,  \Phi^2 \right> 
= W_{\text{eff}}$.

The equation of motion is given by \footnote{
One can easily check that this equation with parameters 
$(2N_c, N_f, K)$ is equivalent to the one with 
parameters $(2N_c-2r, N_f-2r, K-r)$. In other words, the equation of 
motion for glueball field is the same.
Since the equation of motion for both $r$-th Higgs branch and
$\left(K-r\right)$-th Higgs branch is equivalent to each other, 
one expects that
both branches have some relation.} 
\begin{align*}
 0= \log \Sh^{-N_c-1} 
 &+ K \log  \left(\frac{\zh_f-\sqrt{\zh_f^2+4\Sh}}{2} \right)\\
 &+\left( N_f-K \right) \log  \left( \frac{\zh_f+\sqrt{\zh_f^2+4\Sh}}{2} \right)
\end{align*}
where $\Sh = \frac{S}{2m_A \La^2}$, $ \zh_f = \frac{z_f}{\La} 
$. From this we can read out the following result:
when $N_f\geq N_c+1$,
the on-shell poles at the second sheet cannot pass through 
the cut  to reach 
the first sheet far away from the cut.

\BeginAppendix

\section{On matrix model curve with $N_f(>N_c)$  
flavors
}
\setcounter{equation}{0}

In this Appendix we prove by strong coupling analysis that matrix model
curve corresponding to $U(N_c)$ supersymmetric gauge theory with $N_f$
flavors is exactly the same as the one without flavors when the degree
$(k+1)$ of tree level superpotential $W_{\text{tree}}$ is less than 
$N_c$
\footnote{The generalized Konishi anomaly equation of $R(z)$ given in
(\ref{csw2-2.14}) is same with or without flavors, so the form of the
solution is the same for gauge theory with or without flavor.  In this
Appendix we use another method to prove this result.}.  This was first
proved in \cite{Ookouchi:2002be} but the derivation was valid only for the
range $N_f<N_c$. Then in \cite{Balasubramanian:2003tv}, the proof was extended to the
cases with the range $2N_c > N_f\ge N_c$.  However, in \cite{Balasubramanian:2003tv}, the
characteristic function $P_{N_c}(x)$ was defined by $P_{N_c}(x)=\det
(x-\Phi)$, without taking into account the possible quantum corrections
due to flavors.  In consequence, it appeared that the matrix model
curve is changed by addition of flavors. In this Appendix, we use the
definition of $P_{N_c}(x)$ proposed in Eq.\ (C.2) of \cite{Cachazo:2003yc}:
\begin{eqnarray}
P_{N_c}(x)=x^{N_c} \mbox{exp}\left(-\sum_{i=1}^{\infty}\frac{U_i}{x^i} 
\right)+\La^{2N_c-N_f}\frac{\widehat{B}(x)}{x^{N_c}}\mbox{exp}\left(
\sum_{i=1}^{\infty}\frac{U_i}{x^i} \right), \label{PN2}
\end{eqnarray}
which incorporates quantum corrections and reduces to $P_{N_c}(x)=\det
(x-\Phi)$ for $N_f=0$, and see clearly that the matrix model curve is
not changed, even when the number of flavors is more than $N_c$.  
Since
$P_{N_c}(x)$ is a polynomial in $x$, (\ref{PN2}) can be used to express
$U_r$ with $r>N_c$ in terms of $U_r$ with $r\le N_c$ by imposing the
vanishing of the negative power terms in $x$.

Assuming that the unbroken gauge group at low energy is $U(1)^n$ with
$n\le k$, the factorization form of Seiberg--Witten curve can be written
as,
\begin{eqnarray}
P_{N_c}^2(x)-4\La^{2N_c-N_f}\widehat{B}(x)=H_{N_c-n}^2(x)F_{2n}(x).
\nonu
\end{eqnarray}
The effective superpotential with this double root constraint can 
be written as follows
\footnote{If we want to generalize this proof 
to more general cases in which $k+1$ is greater than and equals 
to $N_c$, we have to take care more constraints like Appendix A in
\cite{Cachazo:2003zk}, which should be straightforward.}. 
\begin{align*}
W_{\text{eff}} =  \sum_{r=0}^{k}g_rU_{r+1}
 &+  
\sum_{i=1}^{N_c-n}\Biggl[L_i \oint \frac{P_{N_c}(x)-2\epsilon_i 
\La^{N_c-\frac{N_f}{2}}\sqrt{\widehat{B}(x)}}{x-p_i}dx\\
 &\qquad\qquad\qquad
 +B_i \oint \frac{P_{N_c}(x)-
 2\epsilon_i \La^{N_c-\frac{N_f}{2}}\sqrt{\widehat{B} (x)}}{(x-p_i)^2}dx \Biggr].
\end{align*}
The equations of motion for $B_i$ and $p_i$ are given as follows
respectively:
\begin{align*}
0&=\oint \frac{P_{N_c}(x)-2\epsilon_i \La^{N_c-\frac{N_f}{2}}\sqrt
{\widehat{B}(x)}}{(x-p_i)^2}dx,\\
 0&=2B_i \oint \frac
{P_{N_c}(x)-2\epsilon_i \La^{N_c-\frac{N_f}{2}}\sqrt{\widehat{B}(x)}}
{(x-p_i)^3}dx. 
\end{align*}
Assuming that the factorization form does not have any triple or higher
roots, we obtain $B_i=0$ at the level of equation of motion.  Next we
consider the equation of motion for $U_r$:
\begin{eqnarray}
0=g_{r-1}+\sum_{i=1}^{N_c-n}\oint \left[\frac{P_{N_c}}{x^r}-2
\frac{x^{N_c}}{x^r}\mbox{exp}\left( -\sum_{i=1}^{\infty} 
\frac{U_i}{x^i} \right) \right]\frac{L_i}{x-p_i}dx
\nonu
\end{eqnarray}
where we used $B_i=0$ and (\ref{PN2}) to evaluate $\frac{\partial
P_{N_c}}{\partial U_r}$. Now, as in \cite{Cachazo:2003zk}, we multiply this by
$z^{r-1}$ and sum over $r$.
\begin{eqnarray}
W^{\prime}(z)=-\oint \frac{P_{N_c}}{x-z}\sum_{i=1}^{N_c-n} \frac{L_i}
{x-p_i}dx+\oint \frac{2x^{N_c}}{x-z}\mbox{exp} \left(-\sum_{k=1}^{
\infty}\frac{U_k}{x^k} \right)\sum_{i=1}^{N_c-n}\frac{L_i}{x-p_i}dx. 
\nonu
\end{eqnarray}
Defining the polynomial $Q(x)$ in terms of 
\begin{eqnarray}
\sum_{i=1}^{N_c-n}\frac{L_i}{x-p_i}=\frac{Q(x)}{H_{N_c-n}(x)},
\label{Qpolynomial}
\end{eqnarray}
and also using (\ref{PN2}) and factorization form, we obtain
\begin{eqnarray}
W^{\prime}(z)&=&-\oint \frac{P_{N_c}}{x-z}\frac{Q(x)}{H_{N_c-n}(x)} 
dx+\oint \frac{P_{N_c}}{x-z}\frac{Q(x)}{H_{N_c-n}(x)} dx +\oint 
\frac{Q(x)\sqrt{F_{2n}(x)}}{x-z} dx \nonu \\
&=& \oint \frac{Q(x)\sqrt{F_{2n}(x)}}{x-z} dx.
\nonu
\end{eqnarray}
This is nothing but $(2.37)$ in \cite{Cachazo:2003zk}. Since 
$W^{\prime}(z)$ is a polynomial of degree $k$, the $Q$ 
should be a polynomial of degree $(k-n)$. Therefore, we 
conclude that the matrix model curve is not changed by 
addition of flavors:
\begin{eqnarray}
y_m^2=F_{2n}(x)Q^2_{k-n}(x)=W_k^{\prime}(x)^2+{\mathcal O}(x^{k-1}). 
\label{Nocorrection}
\end{eqnarray}

\section{Equivalence between $W_{\text{low}}$ and 
$W_{\text{eff}}(\ev{S_i})$ with flavors \label{CVmethod}} 
\setcounter{equation}{0}

In this Appendix we prove the equivalence $W_{\text{low}}$ in $U(N_c)$
gauge theory with $W_{\text{eff}}(\langle S_i \rangle)$ in corresponding
dual geometry when some of the branch cuts on the Riemann surface are
closed and the degree $(k+1)$ of the tree level superpotential $W_{\text{tree}}$ is less than $N_c$. This was first proved in \cite{Ookouchi:2002be},
however, the proof was only applicable in the $N_f<N_c$ cases.
Especially, the field theory analysis in \cite{Ookouchi:2002be} did not work
for $N_c \le N_f< 2N_c$ cases. 
Furthermore, as we saw in the main text, for some particular choices of
$z_I$ (position of the flavor poles), extra double roots appear in the
factorization problem.  In section \ref{sec:2cut}, we dealt with $U(3)$
with cubic tree level superpotential and saw the equivalence of two
effective superpotentials for such special situations. To include these
cases we are interested in the Riemann surface that has some closed
branch cuts. Therefore our proof is applicable for $U(N_c)$ gauge
theories with $W_{\text{tree}}$ of degree $k+1$ $(<N_c)$ in which some of
branch cuts are closed and number of flavors is in the range $N_c\le N_f
< 2N_c$. In addition, we restrict our discussion to the Coulomb phase.

In the discussion below, we follow the strategy developed by 
Cachazo and Vafa in \cite{Cachazo:2002pr} and use (\ref{PN2}) as the 
definition of $P_{N_c}(x)$. We have only to show the two
relations:
\begin{eqnarray}
W_{\text{low}}(g_r,z_I,\La)\big|_{\La \to 0} &=&W_{\text{eff}}
(\langle S_i \rangle)\big|_{\La \to 0}, \label{proof1} \\
\frac{\partial W_{\text{low}}(g_r,z_I,\La )}{\partial \La}&=&
\frac{\partial W_{\text{eff}}(\langle S_i \rangle )}{\partial 
\La}, \label{proof2}
\end{eqnarray}
the equivalence of two effective superpotentials in the classical limit
and that of the derivatives of the superpotentials with respect to
$\La$.

\subsection{Field theory analysis }

Let $k$ be the order of $W_{\text{tree}}^{\prime}$ and 
$n$$(\le k)$ be the number of
$U(1)$ at low energy. Since we are interested in cases with 
degenerate branch cuts, let us consider the following factorization 
form
\footnote{In the computation below, we will use relation 
(\ref{Qpolynomial}) and put $g_{k+1}=1$.}:
\begin{align}
P_{N_c}(x)^2-4\La^{2N_c-N_f}\widehat{B}(x)&=
F_{2n}(x)\left[Q_{k-n}(x)
 \widetilde{H}_{N_c-k}(x)\right]^2 \nonumber\\
 &\equiv F_{2n}(x)\left[H_{N_c-n}(x)\right]^2, \label{FactoForm} \\
W^{\prime}(x)^2+f_{k-1}(x)&=F_{2n}(x)Q_{k-n}(x)^2 .
\nonu
\end{align}
If $k$ equals to $n$, all the branch cuts in $F_{2k}(x)$ are open. 
The low energy effective superpotential is given by
\begin{align*}
 W_{\text{low}}  = 
\sum_{r=1}^{n+1}g_r U_r 
&+\sum_{i=1}^l 
\Biggl[L_i\left(P_{N_c}(p_i)-2\epsilon_i 
\La^{N_c-\frac{N_f}{2}}\sqrt{\widehat{B}(p_i)} \right)  \\
 &\qquad\qquad\qquad
+
Q_i \frac{\partial}{\partial p_i}
\left(P_{N_c}(p_i)-2\epsilon_i 
\La^{N_c-\frac{N_f}{2}}\sqrt{\widehat{B}(p_i)} \right) \Biggr], 
\end{align*}
where $l\equiv N-n$ and $P_{N_c}(x)$ is defined by (C.3) or (C.4)
in \cite{Cachazo:2003yc},
\begin{eqnarray}
P_{N_c}(x)=\ev{ \det (x-\Phi) } +\left[ \La^{2N_c-N_f} 
\frac{\widehat{B}(x)}{x^{N_c}} \mbox{exp} \left( \sum_{i=1}^{\infty} 
\frac{U_i}{x^i} \right)  \right]_+. \label{rr1}
\end{eqnarray}
The second term is specific to the $N_f \ge N_c$ case, 
representing quantum correction.  Define
\begin{eqnarray}
K(x)\equiv  \left[ \La^{2N_c-N_f} \frac{\widehat{B}(x)}{x^{N_c}} 
\mbox{exp} \left( \sum_{i=1}^{\infty} \frac{U_i}{x^i} \right)  
\right]_+. 
\nonu
\end{eqnarray}
The first term in (\ref{rr1}) can be represented as  $\langle 
\det (x-\Phi)   \rangle \equiv \sum_{k=0}^{N_c}x^{N_c-k}s_k$. The 
relation between $U_i$'s and $s_k$'s are given by the ordinary 
Newton
relation, $ks_k+\sum_{r=1}^krU_rs_{k-s}=0$. 
From the variations of $W_{\text{low}}$ with respect to $p_i$ and 
$Q_i$, we
conclude that $Q_i=0$ at the level of the equation of motion. 
In
addition, the variation of $W_{\text{low}}$ with respect to $U_r$ 
leads to
\begin{eqnarray}
g_r&=&-\sum_{i=1}^lL_i \frac{\partial P_{N_c}(p_i)}{\partial U_r} 
=\sum_{i=1}^l\sum_{j=0}^{N_c} L_i p_i^{N_c-j}s_{j-r}-\sum_{i=1}^l 
L_i \frac{\partial K(p_i)}{\partial U_r} \nonu \\
&=&\sum_{i=1}^l\sum_{j=0}^{N_c} L_i p_i^{{N_c}-j}s_{j-r}-\sum_{i=1}^l 
L_i \left[ \La^{2{N_c}-N_f}\frac{\widehat{B}(p_i)}{p_i^{{N_c}+r}} 
\mbox{exp} \left(\sum_{k=1}^{\infty} \frac{U_k}{p_i^k} \right) 
\right]_+ .
\nonu
\end{eqnarray}
Let us define
\begin{eqnarray}
{\mathcal G}_r(p_i) \equiv \left[ \La^{2{N_c}-N_f}\frac{
\widehat{B}(p_i)}{p_i^{{N_c}+r}} \mbox{exp} \left(\sum_{k=1}^
{\infty} \frac{U_k}{p_i^k} \right) \right]_+.
\nonu
\end{eqnarray}
By using these relations, let us compute $W_{\text{cl}}^{\prime}$
\begin{align}
W_{\text{cl}}^{\prime}
 &=\sum_{r=1}^{N_c} g_r x^{r-1}\nonu \\
&= \sum_{r=-\infty}^{N_c} \sum_{i=1}^l \sum_{j=0}^{N_c} x^{r-1}p_i^
{{N_c}-j}s_{j-r}L_i \notag\\
 &\qquad\qquad
 -\frac{1}{x} \sum_{i=1}^l L_i \det (p_i-\Phi)-
\sum_{r=1}^{{N_c}}\sum_{i=1}^{l}L_i {\mathcal G}_{r}(p_i)x^{r-1}
\nonu \\
&= \sum_{i=1}^l \frac{\det (x-\Phi)}{x-p_i}L_i -\frac{1}{x} 
\sum_{i=1}^l L_i \det (p_i-\Phi)-\sum_{r=1}^{{N_c}}\sum_{i=1}^{l}L_i 
{\mathcal G}_{r}(p_i)x^{r-1}
\nonu \\
&= \sum_{i=1}^l \frac{P_{N_c}(x)}{x-p_i}L_i-\sum_{i=1}^l \frac{K(x)}
{x-p_i}L_i-\frac{1}{x}\sum_{i=1}^lL_i P_{N_c}(p_i)+\frac{1}{x}
\sum_{i=1}^l L_i K(p_i)
\nonu \\
&  -\sum_{r=1}^{N_c}\sum_{i=1}^l L_i 
{\mathcal G}_{r}(p_i)x^{r-1} 
\label{rr2} 
\end{align}
where we dropped $ {\mathcal O}(x^{-2})$.
The fifth term above can be written as
\begin{eqnarray}
-\sum_{r=1}^{N_c}\sum_{i=1}^l L_i {\mathcal G}_{r}(p_i)x^{r-1}= -
\sum_{r=-\infty}^{N_c}\sum_{i=1}^l L_i {\mathcal G}_{r}(p_i)x^{r-1}+
\sum_{i=1}^l \frac{1}{x} L_i K(p_i)+{\mathcal O}(x^{-2}).
\nonu
\end{eqnarray}
After some manipulation with the factorization form 
(\ref{FactoForm}), we obtain a relation
\begin{eqnarray}
Q_{k-n}(x)\sqrt{F_{2n}(x)} &=& \frac{P_{N_c}(x) }{\widetilde{H}_{{N_c}-k}}- 
\frac{2\La^{2{N_c}-N_f}\widehat{B}(x)}{\widetilde{H}_{{N_c}-k}x^N}\mbox{exp} 
\left(\sum_{i=1}^{\infty} \frac{U_i}{x^i} \right)\nonu \\
&=& \frac{P_{N_c}(x) }{\widetilde{H}_{{N_c}-k}}-\frac{2K(x)}{
\widetilde{H}_{{N_c}-k}}+{\mathcal O}(x^{-2}). 
\nonu
\end{eqnarray}
Substituting this relation into (\ref{rr2}) we obtain 
\begin{eqnarray}
W_{\text{cl}}^{\prime} & = & 
Q_{k-n}(x)\sqrt{F_{2n}(x)}-\sum_{i=1}^l \frac{1}{x} 
\left[L_iP_{N_c}(p_i)-2L_i K(p_i) \right]+\sum_{i=1}^l \frac{K(x)}
{x-p_i}L_i\nonu \\
&&-\sum_{r=-\infty}^{N_c} \sum_{i=1}^l L_i {\mathcal G}_r(p_i)x^{r-1}+
{\mathcal O}(x^{-2}).  
\nonu
\end{eqnarray}
Finally let us use the following relation
\footnote{
For simplicity, we ignore $\sum_{i=1}^{l}L_i$. To prove the 
relation, let us consider the circle integral over $C$. Until 
now, since we assumed $p_i<x$, the point $p_i$ should be 
included in the contour $C$. Multiplying $\frac{1}{x^k}$ $k\ge 0$ 
and taking the circle integral, we can pick up the coefficient 
$c_{k-1}$ of $x^{k-1}$. In addition, if we denote a polynomial 
$M\equiv \La^{2{N_c}-N_f}\frac{\widehat{B}(x)}{x^{N_c}}\mbox{exp}
\left(\sum_{i=1}^{\infty}\frac{U_i}{x^i} \right)\equiv \sum_
{j=-\infty}^{\infty}a_jx^j$
we can obtain following relation,
\begin{eqnarray}
\frac{\left[M \right]_+}{x^k}=\left[ \frac{M}{x^k} \right]_+ 
+\sum_{j=0}^{k-1}a_j x^{-(k-j)} \iff \frac{K(x)}{x^k}={\mathcal 
G}_k(x) +\sum_{j=0}^{k-1}a_j x^{-(k-j)}\nonu
\end{eqnarray}
where right hand side means circle integral of left hand side. 
Thus, we obtain the $c_{k-1}$ as
\begin{align*}
c_{k-1}&= \sum_{j=0}^{k-1} \oint_{x=0}\frac{a_j x^{j-k}}{x-p_i}dx 
+\sum_{j=0}^{k-1} \oint_{x=p_i}\frac{a_j x^{j-k}}{x-p_i}dx \\
&= -\sum_{j=0}^{k-1}\oint_{x=0} a_j x^{j-k}\sum_{n=1}^{\infty}
\frac{x^n}{p_i^{n+1}} dx +\sum_{j=0}^{k-1} a_j p_i^{j-k}=0 \nonu
\end{align*}
where we used that around $x=0$,
$\frac{1}{x-p_i}=-\sum_{n=1}^{\infty}\frac{x^n}{p_i^{n+1}}$. 
Therefore
the left hand side of (\ref{mar67}) can be written as 
$\sum_{j=-\infty}^{\infty}c_{j}x^j=\sum_{j=-\infty}^{-2}c_jx^j=
{\mathcal O}(x^{-2})$.
}
;
\begin{eqnarray}
\sum_{i=1}^l \frac{K(x)}{x-p_i}L_i-\sum_{r=-\infty}^{N_c} 
\sum_{i=1}^l L_i {\mathcal G}_r(p_i)x^{r-1}={\mathcal O}(x^{-2}). 
\label{mar67}
\end{eqnarray}

After all, by squaring $W_{\text{cl}}^{\prime}$, we have 
\begin{eqnarray}
Q_{k-n}(x)^2F_{2n}(x)&=&W_{\text{cl}}^{\prime 2}(x)+ 2
\sum_{i=1}^l \frac{1}{x} \left[L_iP_{N_c}(p_i)-2L_i K(p_i) 
\right]x^{k-1} +{\mathcal O}(x^{k-2}), \nonu \\
b_{k-1}&=& 2\sum_{i=1}^l \frac{1}{x} \left[L_iP_{N_c}(p_i)-
2L_i K(p_i) \right].
\nonu
\end{eqnarray}

On the other hand the variation of $W_{\text{low }}$ with respect to $\La$
is given by
\begin{eqnarray}
\frac{\partial W_{\text{low}}}{\partial \log \La^{2{N_c}-N_f} }=
\sum_{i=1}^lL_iK(p_i)-\frac{1}{2}\sum_{i=1}^lL_i P_{N_c}(p_i)=-
\frac{b_{k-1}}{4} \label{db_{n-1}}.
\end{eqnarray}
This is one of the main results for our proof.  In the dual geometry
analysis below, we will see the similar relation.

In the classical limit, we have only to consider the expectation value
of $\Phi$. In our assumption, gauge symmetry breaks as $U({N_c})\to
\prod_i^{n} U(N_{c,i})$ we have $\Tr\,\Phi=\sum_{i=1}^n
N_{c,i}a_i$. Therefore in the classical limit $W_{\text{low}}$ behaves as
\begin{eqnarray}
W_{\text{cl}}=\sum_{i}^n N_{c,i} W(a_i). \label{Apr1}
\end{eqnarray}
By comparing (\ref{db_{n-1}}) and (\ref{Apr1}) and the 
similar result which we will see in the dual geometry 
analysis below, we will show (\ref{proof1}) and (\ref{proof2}). 
Let us move to the dual geometry analysis.

\subsection{Dual geometry analysis with some closed branch cuts}

As we have already seen in the main text, a solution with $\langle S_i
\rangle=0$ appears for some special choice of $z_I$, the position of
flavor poles.
In our present proof, however, we put some of $S_i$ to be zero from the
beginning, {\em without specifying $z_I$\/}. More precisely, what we
prove in this Appendix is as follows: for a given choice of $z_I$, if
there exists a solution to the factorization problem with some of
$\langle S_i \rangle$ vanishing, then we can construct a dual geometry
which gives the same low energy effective superpotential as the one
given by the solution to the factorization problem, by setting some of
$S_i$ to zero from the beginning. Therefore, this analysis {\em does
not\/} tell us when a solution with $\langle S_i \rangle=0$ appears. To
know that within the matrix model formalism, we have to go back to
string theory and consider an explanation such as the one given in
\cite{Intriligator:2003xs}.

Now let us start our proof. Again, let $k$ be the degree of tree level
superpotential $W_{\text{tree}}^{\prime}(x)$ and $n$ be the number of
$U(1)$ at low energy.  To realize this situation, we need to consider
that $(k-n)$ branch cuts on the Riemann surface should be closed, which
corresponds to $\langle S_i \rangle=0$. Here, there is one important
thing: As we know from the expansion of $W_{\text{eff}}$ in terms of $\La$
(e.g. see (\ref{juho8Apr04})), we cannot obtain any solutions with
$\langle S_i \rangle=0$ if we assume that $S_i$ is dynamical and solve
its equation of motion.  Therefore to realize the situation with
vanishing $\langle S_i \rangle$, we must put $S_i=0$ at the off-shell
level by hand. With this in mind, let us study dual geometry which
corresponds to the gauge theories above. In the field theory, we assumed
that the Riemann surface had $(k-n)$ closed branch cuts. Thus, in this
dual geometry analysis we must assume that at off-shell level, $(k-n)$
$S_i$'s must be zero. For convenience, we assume that first $n$ $S_i$'s
are non-zero and the remaining $(k-n)$ vanish,
\begin{eqnarray}
S_i\neq 0 ,\ \ i=1,\cdots n,\ \ \ S_i=0,\ \  i=n+1,\cdots k.
\nonu
\end{eqnarray}
Therefore the Riemann surface can be written as
\begin{eqnarray}
y^2=F_{2n}(x)Q_{k-n}^2 =W^{\prime}(x)^2+b_{k-1}x^{k-1}+\cdots
\nonu
\end{eqnarray}
The effective superpotential in dual geometry corresponding to
$U({N_c})$ gauge theory with $N_f$ flavors was given in \cite{Cachazo:2003yc}
(See also (\ref{Wexact-y})) and in the classical limit it behaves as
\footnote{Remember that in this Appendix we are assuming only Coulomb
branch. For the Higgs branch, see (7.11) and (7.12) in \cite{Cachazo:2003yc}.}
\begin{eqnarray}
W_{\text{eff}} \big|_{\text{cl}}= \sum_{i=1}^n N_{c,i}W(a_i). \label{sqb}
\end{eqnarray}

As discussed in the previous section, existence of flavors does not
change the Riemann surface $y(x)$. In other words, Riemann surface is
not singular at $x_I$ (roots of $B(x)$),
\begin{eqnarray}
\oint_{x_I} y(x) dx =0,\qquad y(x)=\sqrt{W^{\prime}(x)^2+
b_{k-1}x^{k-1}+\cdots}.
\nonu
\end{eqnarray}
Therefore as in \cite{Cachazo:2002pr}, by deforming contours of all $S_i$'s 
and evaluating the residue at infinity on the first sheet, we 
obtain the following relation, 
\begin{eqnarray}
\sum_{i=1}^{n}S_i=\sum_{i=1}^{k}S_i=-\frac{1}{4}b_{k-1},
\nonu
\end{eqnarray}
where we used $\sum_{i=n+1}^kS_i=0$. 

With this relation in mind, next we consider the variation 
of $W_{\text{eff}}$ with respect to $S_i$: 
\begin{eqnarray}
\frac{\partial W_{\text{eff}}(S_i,\La)}{\partial S_i}=0, \quad 
i=1\cdots n. \label{beom}
\end{eqnarray}
Solving these equations, we obtain the expectation values, $\langle S_i
\rangle$. Of course, these vacuum expectation values depend on $\La$, $g_r$ and $z_I$. Thus
when we evaluate the variation of $W_{\text{eff}}(\langle S_i \rangle ,\La
)$ with respect to $\La$, we have to pay attention to implicit
dependence on $\La$. However the implicit dependence does not contribute
because of the equation of motion (\ref{beom}):
\begin{eqnarray}
\frac{d W_{\text{eff}}(\langle S_i \rangle, \La )}{d \La }&=&
\sum_{i=1}^n \frac{ \partial \langle S_i \rangle }{\partial 
\La}\cdot \frac{\partial W_{\text{eff}}(\langle S_i \rangle , 
\La ) }{\partial \langle S_i \rangle }+\frac{\partial W_{\text{
eff}}(\langle S_i \rangle , \La ) }{\partial  \La  } \nonu \\
&=&\frac{\partial W_{\text{eff}}(\langle S_i \rangle , \La ) }
{\partial  \La  }.
\nonu
\end{eqnarray}
On the other hand, explicit dependence on $\La$ can be easily obtained
by monodromy analysis. Here let us recall the fact that the presence of
fundamentals does not change the Riemann surface.  In fact, looking at
(\ref{Wexact-y}) we can read off the dependence from the term $2\pi i
\tau_0= \log \left( \frac{B_L \La^{2{N_c}-N_f}}{\La_0^{2{N_c}-L}}
\right)$,
\begin{eqnarray}
\frac{d W_{\text{eff}}(\langle S_i \rangle, \La )}{d \log 
\La^{2{N_c}-N_f} }=S=-\frac{b_{k-1}}{4}. \label{mar63}
\end{eqnarray}

To finish our proof, we have to pay attention to $f_{k-1}(x)$,
on-shell. Namely putting $\langle S_i \rangle$ into $f_{k-1}(x)$ what
kind of property does it have? To see it, let us consider change of
variables from $S_i$'s to $b_i$'s.  As discussed in \cite{Cachazo:2002ry} the
Jacobian of the change is non-singular if $0\le j \le k-2$,
\begin{eqnarray}
\frac{\partial S_i}{\partial b_j}=-\frac{1}{8\pi i}
\oint_{A_i}dx \frac{x^j}{\sqrt{W^{\prime}(x)^2+f(x)}}.
\nonu
\end{eqnarray}
In our present case, since only $n$ of $k$ $S_i$'s are 
dynamical variable, we use $b_i$, $i=0,\cdots n-1$ in a function 
$f_{k-1}(x)=\sum b_ix^i$ as new variables, instead of $S_i$'s. 
As discussed in \cite{Cachazo:2002pr,Ookouchi:2002be,Cachazo:2003yc}, by using Abel's 
theorem, the equation of motions for $b_i$'s is interpreted 
as an existence condition of a meromorphic function that has an 
${N_c}$-th order pole at infinity on the first sheet and an $({N_c}-N_f)$-th 
order zero at infinity on the second sheet of $\Sigma$ and a 
first order zero at $\widetilde{q}_I$. For a theory with 
$N_f\le 2{N_c}$, such a function can be constructed as follows 
\cite{Naculich:2002hr,Cachazo:2003yc}:
\begin{eqnarray}
\psi (x)=P_{N_c}(x)+\sqrt{P^2_{N_c}(x)-4\La^{2{N_c}-N_f}\widehat{B}(x)}.
\nonu
\end{eqnarray}
For this function to be single valued on the matrix model curve 
$y(x)$, the following condition must be satisfied,
\begin{eqnarray}
P_{N_c}^2(x)-
4\La^{2{N_c}-N_f}\widehat{B}(x)&=&F_{2n}(x){H^2_{{N_c}-n}}(x) \nonu
\\
W^{\prime}(x)^2+f(x)&=&F_{2n}(x)Q^2_{k-n}(x)
\nonu
\end{eqnarray}
This is exactly the same as the factorization form we already 
see in the field theory analysis. Therefore the value 
$b_{k-1}$ of on-shell matrix model curve in dual theory is 
the same one for field theory analysis. Comparing two results, 
(\ref{db_{n-1}}) and (\ref{Apr1}) with
corresponding results for the dual geometry analysis, 
(\ref{sqb}) and
(\ref{mar63}) we have shown the equivalence between these
two descriptions of
effective superpotentials.

\section{Computation of superpotential --- gauge theory side}


\label{comp_spot-gt}

In this Appendix, we demonstrate the factorization method used in
subsection \ref{QuantumMassless} to compute the low energy
superpotential, taking the $N_f=4$ case as an example.  Therefore there
are two kinds of solutions for the factorization problem
\eqref{factorization} and \eqref{p3h1}.

\newpage
\noindent
$\bullet$ {\bf The breaking pattern $\widehat{U(2)}\times U(1)$}

The first kind of solution for the factorization problem is given by
 \begin{eqnarray}
A=0 ,\qquad
B =  2a,
 \qquad 
 C= a^2-4\La^2,
\qquad 
  D= 0,\qquad 
   c = 0,
\qquad
  b= 0.
\nonu
\end{eqnarray} 
In the classical limit $\Lambda\to 0$, we can see the 
characteristic function goes as $P_3(x)\to x^2 \left(x+a\right)$, 
which means that the breaking pattern is $\widehat{U(2)}
\times U(1)$.  Note that since we are assuming $m_f=0$,
the notation ``$~~\widehat{}~~$'' should be used for the gauge group
that corresponds to the cut near the critical point at $x=0$.
Inserting these solutions into (\ref{constraint}) we obtain 
one constraint,
\begin{eqnarray}
\Delta^2=a^2+8\La^2. 
\nonu
\end{eqnarray}
We can easily represent $a$ as a Taylor expansion of $\La$: 
\begin{eqnarray}
a=-1 + 4\,T + 8\,T^2 + 32\,T^3 + 160\,T^4 + 896\,T^5 + 5376\,
T^6+\cdots,
\nonu
\end{eqnarray}
where we put $\Delta=1$ and defined $T\equiv \La^2$.
\footnote{If we take care  of a phase factor of $\La$, 
we will obtain the effective superpotentials corresponding 
to each vacuum. However in our present calculation, we want 
to check whether the effective superpotentials of two method, 
field theory and dual geometry, agree with each other. 
Therefore, we have only to pay attention to the coefficients in 
$W_{\text{low}}$, neglecting the phase factor.}
 The coefficients of $P_3(x)$ are related to the
Casimirs $U_j=\frac{1}{ j}\ev{\Tr[\Phi^j]}$ as follows.  
For $N_c=3$,
$N_f=4$, \eqref{PN2} reads
\begin{align*}
 P_3(x)
 &=
 x^3 \exp\biggl(-\sum_{j=1}^{\infty}\frac{U_j}{x^j} \biggr)
 +\La^{5}\frac{x^{4}}{x^3}\exp\biggl(\sum_{j=1}^{\infty}
\frac{U_j}{x^j} \biggr)\cr
 &=
 x^3-U_1 x^2+\left(-U_2+\frac{U_1^2}{ 2}+\La^2 \right)x\\
 &\qquad\qquad\qquad
 +\left(-U_3+U_1U_2-\frac{U_1^3}{6}-\La^2 U_1 \right)+\cdots.\ 
\nonu
\end{align*}
Comparing the coefficients, we obtain
\begin{align}
 U_1=-a,\qquad
 U_2=-b+\frac{a^2}{ 2}+\La^2 ,\qquad
 U_3=-c+ab-\frac{a^3}{ 3}-a \La^2 .
\nonu
\end{align}
Furthermore, one can compute $a_{1,2}$ from \eqref{constraint}.
Plugging all these into \eqref{W_low_U(3)}, we finally obtain
\begin{gather*}
W_{\text{low}}=W_{\text{cl}} + T - 6T^2 - \frac{40T^3}{3} - 
         56T^4 - 288T^5 - \frac{4928T^6}{3}+
        \cdots,\\
         T\equiv \Lambda^2,
 \qquad W_{\text{cl}} =-\frac{1}{6}.\nonu
\end{gather*}

\noindent
$\bullet$ {\bf The breaking pattern $\widehat{U(1)}\times U(2)$}

The other kind of solution for the factorization problem is given by
 \begin{eqnarray}
 A&=& \frac{1}{2}(-a-2\eta \La), \qquad 
B = a-2\eta \La ,
 \qquad 
 C= \frac{1}{4}(a+2\eta \La)^2, \nonu \\
&&
  D= 0,\qquad 
   c = 0,
\qquad
  b= \frac{1}{4}(a+2\eta \La)^2
\nonu
\end{eqnarray}
where $\eta \equiv \pm1$. These solutions correspond to the 
breaking pattern $\widehat{U(1)}\times U(2)$ in the classical 
limit. Inserting these solutions into (\ref{constraint}) we 
obtain one constraint,
\begin{eqnarray}
\Delta^2=\frac{1}{4}(a^2-20a\, \eta\, \La +4\La^2). 
\nonu
\end{eqnarray}
Again, let us represent $a$ as a Taylor series of $\La$: 
\begin{eqnarray}
a=-2 + 10\,T - 24\,T^2 + 144\,T^4 - 1728\,T^6+\cdots,
\nonu
\end{eqnarray}
where we put $\Delta=1$, $\eta=1$ and defined $T\equiv \La$.
Doing the same way as previous breaking pattern, 
we can compute the effective superpotential as 
\begin{gather*}
        W_{\text{low}} =W_{\text{cl}} + 2T - 13T^2 + \frac{176T^3}{3} - 
         138T^4 + 792T^6 +\cdots,\\
         T\equiv \Lambda, \qquad W_{\text{cl}}=-\frac{1}{3}.  \nonu
\end{gather*}

The other cases with $N_f=1,2,3$ and $5$ can be done analogously.

\bibliographystyle{uclathes}
\end{document}